\documentclass[twoside]{report} 
\usepackage{graphicx}
\usepackage{subfig}  
\usepackage[section]{placeins} 
\usepackage{caption}

\usepackage{amsfonts}
\usepackage{amsmath}
\usepackage{amssymb}
\usepackage{latexsym}
\usepackage{amsthm, mathrsfs}
\usepackage{lmodern} 

\usepackage{tikz}
\usetikzlibrary{decorations.pathmorphing}
\usetikzlibrary{decorations.markings} 
\usepgflibrary{arrows}

\usepackage{fancyhdr}
\usepackage[Sonny]{fncychap}

\usepackage{a4}
\usepackage{setspace} 
\usepackage{cite} 

\newcommand{\BbbR}{\mathbb{R}}
\newcommand{\BbbZ}{\mathbb{Z}}

\DeclareMathOperator{\Realpart}{Re}
\DeclareMathOperator{\Imagpart}{Im}
\newcommand{\x}{\mathsf{x}}
\newcommand{\U}{\mathsf{U}}
\newcommand{\X}{\mathsf{X}}
\DeclareMathOperator{\sgn}{sgn}
\DeclareMathOperator{\expo}{e}
\newcommand{\scri}{\mathscr{I}}
\newcommand{\scrh}{\mathscr{H}}
\newcommand{\be}{\begin{equation}}
\newcommand{\ee}{\end{equation}}
\newcommand{\ba}{\begin{eqnarray}}
\newcommand{\ea}{\end{eqnarray}}
\newcommand{\bea}{\begin{equation}\begin{aligned}}
\newcommand{\eea}{\end{aligned}\end{equation}}
\newcommand{\baa}{\begin{array}}
\newcommand{\eaa}{\end{array}}

\newcommand{\bfx}{{\bf x}}

\newcommand{\Etilde}{\tilde{E}}

\DeclareMathOperator{\Ci}{Ci}

\DeclareMathOperator{\si}{si}
\DeclareMathOperator{\ci}{ci}
\DeclareMathOperator{\arctanh}{arctanh}

\DeclareMathOperator{\arcsinh}{arcsinh}
\DeclareMathOperator{\arccosh}{arccosh}

\newcommand{\huOne}{h_1\left(\frac{u-\tau_0+\delta}{\delta}\right)}
\newcommand{\hsOne}{h_1\left(\frac{u-s-\tau_0+\delta}{\delta}\right)}
\newcommand{\dhuOne}{h_1^{\prime}\left(\frac{u-\tau_0+\delta}{\delta}\right)}

\newcommand{\ddhuOne}{h_1^{\prime\prime}\left(\frac{u-\tau_0+\delta}{\delta}\right)}

\newcommand{\huTwo}{h_2\left(\frac{-u+\tau+\delta}{\delta}\right)}
\newcommand{\hsTwo}{h_2\left(\frac{-u+s+\tau+\delta}{\delta}\right)}
\newcommand{\dhuTwo}{h_2^{\prime}\left(\frac{-u+\tau+\delta}{\delta}\right)}

\newcommand{\ddhuTwo}{h_2^{\prime\prime}\left(\frac{-u+\tau+\delta}{\delta}\right)}

\newcommand{\hvOne}{h_1\left(v\right)}
\newcommand{\hvrOne}{h_1\left(v-r\right)}

\newcommand{\hbvrOne}{h_1\left(b+1-v-r\right)}
\newcommand{\dhvOne}{h_1^{\prime}\left(v\right)}
\newcommand{\ddhvOne}{h_1^{\prime\prime}\left(v\right)}
\newcommand{\hvTwo}{h_2\left(v\right)}
\newcommand{\dhvTwo}{h_2^{\prime}\left(v\right)}
\newcommand{\hbvTwo}{h_2\left(b+1-v\right)}
\newcommand{\hbvrTwo}{h_2\left(b+1-v-r\right)}
\newcommand{\ddhvTwo}{h_2^{\prime\prime}\left(v\right)}
\newcommand{\dhbvTwo}{h_2^{\prime}\left(b+1-v\right)}
\newcommand{\ddhbvTwo}{h_2^{\prime\prime}\left(b+1-v\right)}
\newcommand{\ddxxZero}{\ddot{\x}^2\left[(v-1)\delta+\tau_0\right]}
\newcommand{\xTwoThreeZero}{\ddot{\x}\left[(v-1)\delta+\tau_0\right]\x^{(3)}\left[(v-1)\delta+\tau_0\right]}
\newcommand{\ddxx}{\ddot{\x}^2\left[(1-v)\delta+\tau\right]}
\newcommand{\xTwoThree}{\ddot{\x}\left[(1-v)\delta+\tau\right]\x^{(3)}\left[(1-v)\delta+\tau\right]}
\newcommand{\DxT}{\left(\Delta\x\right)^2}
\newcommand{\DxF}{\left[\left(\Delta\x\right)^2\right]^2}
\newcommand{\dx}{\dot{\x}}
\newcommand{\ddx}{\ddot{\x}}
\newcommand{\dddx}{\x^{(3)}}
\newcommand{\ddddx}{\x^{(4)}}
\newcommand{\dddddx}{\x^{(5)}}
\DeclareMathOperator{\phin}{\phi^{\text{in}}_{\omega\ell}}
\DeclareMathOperator{\phup}{\phi^{\text{up}}_{\omega\ell}}
\DeclareMathOperator{\Phin}{\Phi^{\text{in}}_{\omega\ell}}
\DeclareMathOperator{\Phup}{\Phi^{\text{up}}_{\omega\ell}}
\newcommand{\Rin}{R^{\text{in}}_{\omega\ell}}
\newcommand{\Rup}{R^{\text{up}}_{\omega\ell}}
\DeclareMathOperator{\rin}{\mathsf{\rho^{\text{in}}_{\omega\ell}}}
\DeclareMathOperator{\rup}{\rho^{\text{up}}_{\omega\ell}}
\DeclareMathOperator{\tPhin}{\widetilde{\Phi}^{\text{in}}_{\omega\ell}}
\DeclareMathOperator{\tPhup}{\widetilde{\Phi}^{\text{up}}_{\omega\ell}}
\newcommand{\rinf}{r_{\text{inf}}}
\DeclareMathOperator{\ninf}{ninf}
\newcommand{\rh}{r_{H}}
\newcommand{\nh}{n_{H}}
\newcommand{\Aup}{A^{\text{up}}_{\omega\ell}}
\newcommand{\Ain}{A^{\text{in}}_{\omega\ell}}
\newcommand{\Bup}{B^{\text{up}}_{\omega\ell}}
\newcommand{\Bin}{B^{\text{in}}_{\omega\ell}}

\begin{document}

\newpage
\thispagestyle{empty} 

\pagestyle{fancy} 
\renewcommand{\chaptermark}[1]{\markboth{\chaptername%
\ \thechapter:\,\ #1}{}}
\renewcommand{\sectionmark}[1]{\markright{\thesection\,\ #1}}

\singlespacing 

\vspace*{1cm}    

\begin{center}
{\huge\bf Particle detectors in curved spacetime quantum field theory \\}

\vspace{2cm}

{\Large 
Lee Hodgkinson, MPhys.}

\vspace{3cm}

{\large
Thesis submitted to the University of Nottingham\\
for the Degree of Doctor of Philosophy\\
\vspace{0.3cm}
October 2013
\vspace{0.3cm}
}
\end{center}

\pagenumbering{roman}  

\newpage
\doublespacing  
\begin{abstract}
\thispagestyle{plain}   
Unruh-DeWitt particle detector models are studied in a variety of time-dependent and time-independent settings. We work within the framework
of first-order perturbation theory and couple the detector to a massless scalar field. The necessity of switching on (off) the detector smoothly is emphasised throughout, and the transition rate is found by taking the sharp-switching limit of the regulator-free and finite response function.

The detector is analysed on a variety of spacetimes: $d$-dimensional Minkowski, the
Ba\~nados-Teitelboim-Zanelli (BTZ) black hole, the two-dimensional Minkowski half-plane, two-dimensional Minkowski with a receding mirror, and the two- and four-dimensional Schwarzschild black holes.

In $d$-dimensional Minkowski spacetime, the transition rate is found to be finite up to dimension five. In dimension six, the transition rate diverges unless the detector is on a trajectory of constant proper acceleration, and the implications of this divergence to the global embedding spacetime (GEMS) methods are studied.

In three-dimensional curved spacetime, the transition rate for the scalar field in an arbitrary Hadamard state 
is found to be finite and regulator-free. Then on the Ba\~nados-Teitelboim-Zanelli (BTZ) black hole spacetime, we analyse the detector coupled to the field in the Hartle-Hawking vacua, under both transparent and reflective boundary conditions at infinity. Results are presented for the co-rotating detector, which responds thermally, and for the radially-infalling detector.

Finally, detectors on the Schwarzschild black hole are considered. 
We begin in two dimensions, in an attempt to gain insight by exploiting the conformal triviality,
and where we apply a temporal cut-off to regulate the infrared divergence. 
In four-dimensional Schwarzschild spacetime, we proceed numerically, and the Hartle-Hawking, Boulware and Unruh vacua rates are compared. Results are presented for the case of the static detectors, which respond thermally, and also for the case of co-rotating detectors.

\end{abstract}
\newpage
\newpage

\chapter*{Acknowledgements}
\addcontentsline{toc}{chapter}
         {\protect\numberline{Acknowledgements\hspace{-96pt}}}
         
First, I would like to thank my supervisor, Dr Jorma Louko, for always being available to chat, suggesting interesting topics, reading this manuscript and not blushing when I asked silly questions. Sincerely, I doubt that it is possible to find a supervisor with more time for his students or more down to earth.

My PhD was funded by the Engineering and Physical Sciences Research Council (EPSRC) --- grant reference number EP/P505038/1 --- to whom I am grateful. I thank the Universitas 21 Network for funding in the form of a Universitas 21 Prize Scholarship; this award facilitated a month-long visit to University of British Columbia (UBC). I am also grateful for access to the University of Nottingham High Performance Computing Facility.

I thank the Department of Physics and
Astronomy at UBC for their hospitality; in particular, I thank Professor Bill Unruh for the interesting discussions and suggestions regarding this work. I also thank Professor Adrian Ottewill, University College Dublin, without whom the four-dimensional Schwarzschild work would not have been possible. Dr Chris Fewster, University of York, should also be thanked for suggesting a robust derivation of the transition rate formula in $2+1$ dimensions when the field is in an arbitrary Hadamard state.

I owe a debt to my family for their perpetual encouragement and for providing an environment from which my interest in science blossomed. Finally, I thank Louise for the love she has shown me, for her unwavering selflessness and for making these years happy ones.

\newpage

\newpage
\singlespacing   
\tableofcontents
\doublespacing   
\newpage

\newpage

\newlength\longest

\clearpage

\thispagestyle{empty}
\null\vfill

\settowidth\longest{\Large\itshape and the dead worlds, stars, systems, infinity xx}
{\centering
\parbox{\longest}{%
  \raggedright{\Large\itshape%
  Why I came here, I know not;\\
  where I shall go it is useless to inquire \\
  --- in the midst of myriads of the living\\
  and the dead worlds, stars, systems, infinity,\\
  why should I be anxious about an atom? \par\bigskip
  }   
  \raggedleft\large\MakeUppercase{Lord Byron}\par%
}
}

\vfill\vfill

\setlength{\oddsidemargin}{16mm}  
\setlength{\textwidth}{140mm}
\setlength{\textheight}{200mm}
\setlength{\topmargin}{8mm}

\clearpage

\pagenumbering{arabic}

\chapter{Introduction}

In this thesis we probe the particle content of a quantum field in a variety of time-independent and time-dependent situations. The coupling of quantum field theory and general relativity, the two pillars of modern physics, is a notoriously difficult challenge that is yet to yield a fully satisfactory solution. It is this impasse that has led many researchers to turn to the field of quantum field theory on curved spacetime. In this approach one treats the spacetime geometry classically; a quantum field theory is imposed onto this geometry presuming the back-reaction of the field to be negligible.
\par 
Despite the conceptual simplicity of this approach, the theory has enjoyed many successes, the most celebrated of which being Hawking's result~\cite{hawking} that black holes formed by stellar collapse emit thermal radiation with a temperature
\be 
T=\kappa/2\pi\,,
\ee
with $\kappa$ being the surface gravity of the black hole. For any Killing horizon, the surface gravity can be defined as the gradient of the norm of the associated Killing vector, $\chi^{\mu}$, evaluated at the horizon. That is to say, 
\be 
\kappa^2=-\frac{1}{2} \left(\nabla_{\mu} \chi_{\nu}\right)\left(\nabla^{\mu} \chi^{\nu}\right)\,.
\ee
The surface gravity in a static, asymptotically flat spacetime is the proper acceleration of a static observer near the horizon adjusted by the redshift factor between the static observer near the horizon and another static observer at infinity~\cite{Carroll}.
\par 
Further to Hawking's result, in 1976, Unruh~\cite{unruh} showed that even in flat spacetime the particle content of the field is contextual, and an observer moving with uniform proper acceleration through Minkowski spacetime would see the Minkowski vacuum state as a thermal bath of particles, characterised by a temperature 
\be 
T=a/2\pi\,,
\label{eq:intro1:UnruhT}
\ee
with $a$ being the proper acceleration of the observer.
\par 
Even after decades of research into these topics, the questions led to by Hawking's early work, such as the information loss paradox~\cite{Hawking:1976ra, Preskill:1992tc}, are still directly leading to questions at the forefront of modern research. One such topic is the recent suggestion that an observer falling through the event horizon of a black hole is met with a ``firewall''~\cite{Almheiri:2012rt} --- an intense thermal gas of particles near the black hole event horizon that would cause an observer to burn up. This is contrary to the long-established ``no-drama'' principle, which states that an observer falling through the event horizon of a black hole experiences nothing unusual.
\par 
Furthermore, much research is still ongoing investigating the tantalising possibility of experimentally verifying the Unruh effect in the laboratory. This is a great challenge as reading an Unruh temperature~\eqref{eq:intro1:UnruhT} of the order 1 Kelvin requires accelerations with magnitudes $10^{21} \mathrm{m}/\mathrm{s}^2$, and one also needs to take into account the background noise from the conventional Larmor radiation. Perhaps the most promising methods to date involve using high-intensity lasers to accelerate electrons to vast accelerations~\cite{Chen:1998kp} or, more recently, by making use of the Berry phase~\cite{MartinMartinez:2010sg}, which may permit measurement of the effect at vastly smaller-magnitude accelerations, which need be maintained over shorter time frames.
\par 
Additionally, much research has taken place into investigating analogue systems, where simulated Hawking radiation could possibly be observed. These investigations began with the so-called ``dumb holes'', which are black hole analogues using sound waves in a fluid flow~\cite{Unruh:1980cg}, but investigations have been carried out also in shallow water waves~\cite{Weinfurtner:2013cd} and finally in fiber-optic systems~\cite{Philbin:2007ji}, for which some researchers now claim to have experimental evidence that simulated Hawking radiation has been observed~\cite{Belgiorno:2010wn}, although this is currently highly disputed~\cite{Schutzhold:2010am,Belgiorno:2010hk}. 
\par
Going beyond the Hawking effect and Unruh effect to more general settings, we find many situations that give rise to particle production.  If one has a spacetime with isometries, it is possible to define the particle content of the field using solutions to the field equations that are positive frequency with respect to the associated timelike Killing vector that generates the isometry. For example, in Minkowski spacetime one has Poincare symmetry, and it is precisely because of this symmetry that all inertial observers can agree on the particle content of a given state of the field, and in particular all inertial observers can agree that the Minkowski vacuum state is a state devoid of particles. Such observers may disagree on the magnitude of the energy of a particle, but Lorentz-invariance will ensure that it is positive frequency and thus still indeed a particle.
The issue of particles becomes thornier when considering a curved spacetime. Owing to the general absence of any symmetries and timelike Killing vectors,  we have nothing to define the solutions of the field equations as being positive-frequency with respect to.
\par 
One way to cut through this ambiguity is to define particles operationally; that is to say, we couple our quantum field to some simple quantum-mechanical system, which could be anything from a hydrogen atom to a simple harmonic oscillator, and we think of it as our `detector'. In other words, we take the conceptual view that the particle content of the field is not well defined when the field is isolated but only by the field's interactions with such a detector. Hence, a particle is simply something that a particle detector detects~\cite{UnruhPubCom}. In this model, an upward transition in the quantum-mechanical system by the absorption of a quantum of energy from the field is interpreted as the detection of a particle, whilst the downward transition is interpreted as the emission of a particle. The simplest of these detector models is the Unruh-DeWitt detector~\cite{unruh,deWitt}, in which the quantum-mechanical system is coupled to the field by a monopole-moment operator. 
\par 
The transition rate associated with such a detector is of primary interest, but it can be difficult to compute in a precise manner. In all cases but for the stationary detector, one must take extreme care with the way the detector is switched on (off) if the conventional regularisation is to be maintained~\cite{satz:smooth}. Alternatively, the point-like coupling of the detector to the field can be replaced by one that is smeared~\cite{schlicht}. Throughout this thesis, we shall adopt the smooth-switching approach of Satz~\cite{satz:smooth}. Using this smooth-switching approach will enable us to investigate the response of a detector in a host of situations, some of which will be time-dependent.
\par 
We shall now give a detailed outline of this thesis. In Chapter~\ref{ch:techIntro}, we shall begin with an overview of the Unruh-DeWitt detector model, giving the mathematical definition of such a detector in terms of first-order perturbation theory and introducing the key concepts of the detector response function and the transition rate. We shall stress the need to remove the regulator of the Wightman function before trying to take the sharp-switching limit. 
\par 
In Chapter~\ref{ch:by4d}, we analyse the response of an Unruh-DeWitt detector coupled to a massless scalar field that is arbitrarily accelerated in Minkowski spacetime of dimension up to and including six. The first step is to obtain the detector response function in a regulator-free form before we take the sharp-switching limit to obtain the transition rate. We shall find that the transition probability will diverge for dimensions greater than three, but the transition rate will remain finite up to and including dimension five. In dimension six, the transition rate develops a logarithmic divergence on all trajectories except those with constant scalar proper acceleration. The chapter ends with a discussion of the implications of this divergence for the global embedding spacetime (GEMS) method's suitability for investigating the detector response in curved spacetime.
\par 
One may wonder if it makes sense to use first-order perturbation theory when the response rate it leads to diverges in dimensions greater than three. We shall see in Chapter~\ref{ch:techIntro} that the total transition probability consists of a factor depending only on the internal details of the detector as well as $c^2$, where $c$ is the small detector to field coupling-constant. In four dimensions, in the limit of the detector switch-on (off) time, $\delta$, going to zero, the response diverges as $\log{\delta}$. Thus, provided that for a theory parametrised by $\delta$, $c$ is also chosen to ensure that it is bounded in absolute value by $k/\sqrt|\log{\delta}|$ then we can choose the positive constant $k$ to be small enough that the perturbative treatment remains valid. Similarly, in five and six dimensions the response in the sharply-switched limit diverges as $\delta^{-1}$ and $\delta^{-2}$ respectively, and we are required to choose $c$ such that it is bounded in absolute value by $k\sqrt{\delta}$ and $k\delta$ respectively.
\par 
In Chapter~\ref{ch:btz}, we examine the Unruh-DeWitt detector coupled to the scalar field in three-dimensional curved spacetime, following a similar analysis to~\cite{satz-louko:curved}. We first obtain a regulator-free expression for the transition probability in an arbitrary Hadamard state, working within first-order perturbation
theory and assuming smooth switching, and we show that both the transition probability and the instantaneous transition rate remain well defined in the sharp-switching limit. We then specialise the spacetime to the Ba\~nados-Teitelboim-Zanelli black hole and to a massless, conformally-coupled scalar field in the Hartle-Hawking vacua, using both transparent and reflective boundary conditions at the spatial infinity. We then analyse the co-rotating trajectory, finding a thermal response, and also the case of a detector freely-falling into the hole on a geodesic. A host of numerical results are presented, and these are complemented by good agreement from analytic results in a variety of asymptotic regimes.
\par 
Chapters~\ref{ch:2DSchw} and~\ref{ch:4DSchw} constitute our investigations into detectors on the Schwarzschild black hole. We start in Chapter~\ref{ch:2DSchw} by considering the $(1+1)$-dimensional Schwarzschild black hole; that is to say, we drop the angular components of the four-dimensional Schwarzschild spacetime. The first issue we face with this approach is that for a massless scalar field the Wightman function has an infrared divergence, in addition to the ultraviolet divergence that is found in all dimensions. We would prefer to work with a detector coupled to a massive scalar field in $(1+1)$-Schwarzschild, which does not suffer such infrared divergences, obtain the transition rate and then only at the very end of the calculation take the $m\to 0$ limit, but the calculations involved prove prohibitive. Nevertheless, this infrared divergence can be regularised by employing a temporal-window function in our detector response and considering a detector switched on in the asymptotic past~\cite{Langlois-thesis}. Langlois showed that the use of these temporal-window functions gave the expected results for the inertial- and uniformly-accelerated detectors in $(1+1)$-Minkowski spacetime. 
\par 
First, with hope to gain confidence in employing the Langlois cut-off to the real $(1+1)$-Schwarzschild spacetime of interest, we investigate the static detector in the $(1+1)$-Minkowski half-plane. We analyse both the case of our detector coupled to massive scalar field, with $m\to 0$ taken at the end, and the case of our detector coupled to a massless scalar field from the outset, where a Langlois temporal-window function is employed to deal with the infrared divergence. We find that the transition rates computed from the two approaches agree exactly.
\par Next, using the Langlois cut-off procedure, we consider a static detector sat external to the $(1+1)$-Schwarzschild black hole and coupled to a massless scalar field in the Hartle-Hawking vacuum. Encouragingly, the transition rate found has the expected thermal character in the local Hawking temperature.
\par Bolstered by these successes, we next consider the transition rate for a static detector coupled to a massless scalar field in the Unruh vacuum on the $(1+1)$-Schwarzschild spacetime. The transition rate we find in this case has a part that is expected --- half the Boulware rate plus half the Hartle-Hawking rate --- but also contains a rather odd term of the form $T/2\omega^2$, with $T$ being the local Hawking temperature and $\omega$ being the detector's energy gap. To investigate this unexpected term, we turn to the mirrors analogy in $(1+1)$-Minkowski spacetime~\cite{byd}; the Unruh vacuum mocks up the outgoing radiation from a collapsing star, and the receding mirror in $(1+1)$-Minkowski spacetime makes a good comparison. Once again the expected transition rate plus a strange term are found, as in the Unruh case, and we deduce that such a term is an artefact caused by treating the infrared divergence with the temporal-window cut-off.
\par
In Chapter~\ref{ch:4DSchw}, we investigate the four-dimensional Schwarzschild black hole. Results for the static and circular-geodesic detectors in the exterior of the hole are presented for a massless scalar field in the Hartle-Hawking, Boulware and Unruh vacuum states. The response of a static detector coupled to a field in the Hartle-Hawking vacuum is seen to be thermal in the KMS sense, with the expected local Hawking temperature. The radial part of the mode solution to the Klein-Gordon equation is not known analytically on this spacetime, and we use Mathematica code to compute these modes numerically and perform the mode sums and integrals. The result for the static detector's transition rate in the Hartle-Hawking vacuum is compared with the transition rate of a Rindler detector in the Minkowski vacuum state in Rindler spacetime, where the Rindler detector is given the appropriate proper scalar acceleration. The result for a circular-geodesic detector's transition rate is compared with a Rindler detector given a constant drift-velocity in the direction transverse to the Rindler plane, in order to simulate the circular motion. Both comparisons give good agreement as the radius of the detector increases. Finally, we attempt to improve the comparison of the circular-geodesic detector in the Hartle-Hawking vacuum, with the Rindler detector given a transverse drift-velocity, by making the transverse drift dimension periodic. This additional periodicity is seen not to improve the comparison and in fact makes it much worse, leading to a transition rate with an oscillatory de-excitation response, somewhat reminiscent of the BTZ black hole results of Chapter~\ref{ch:btz}. Additionally, we present the necessary analytic work, which complements our Mathematica code, to enable the computation of the transition rate for a detector radially infalling on a geodesic to the hole. At the time of writing the data for the radial-infall case was still in the process of being gathered at the University of Nottingham High Performance Cluster (HPC).
\par In Chapter~\ref{ch:discussion}, we summarise the work completed in this thesis and discuss potential future research directions.
\par 
For each dimension considered, our metric signature is of the form $({-}{+}{+}\ldots{+})$, and we use units in which 
$c=\hbar=G=1$. 
Spacetime points are denoted by sans-serif letters.
Lorentz $d$-vectors are denoted with sans-serif letters ($\x$) and Euclidean three-vectors with bold letters ($\bfx$). For the
Minkowski or Euclidean product of two vectors of the respective kind we use a dot notation, $\x\cdot\x$ or $\bfx\cdot\bfx$. 
$O(x)$ denotes a quantity for which $O(x)/x$ is bounded as
$x\to0$, 
$o(x)$ denotes a quantity for which $o(x)/x \to0$ as
$x\to0$, 
$O(1)$ denotes a quantity that is bounded in the limit under
consideration, 
and 
$o(1)$ denotes a quantity that vanishes in the limit under
consideration. 
\chapter[The Unruh-DeWitt particle detector model]{The Unruh-DeWitt particle detector model}
\chaptermark{Unruh-DeWitt model}
\label{ch:techIntro}

\section{Unruh-DeWitt model}
\label{sec:UDWmod}

We wish to a consider a simple model particle detector, which we take to be point-like and consist of an idealised two-state quantum-mechanical system. This two-state system consists of an initial state
$|0_d\rangle$ having energy $0$, and the state $|E_d\rangle$ having energy $E$, where $E$ may be positive or negative. Occasionally, we shall also use $\omega$ to denote the detector's energy gap, although in Chapter~\ref{ch:4DSchw} the symbol $\omega$ is reserved to denote the radial mode's frequency. This simple quantum-mechanical system that we think of as our detector is coupled to the quantum field $\phi$ in a way we shall soon make precise. Generally, as the detector moves through spacetime it will absorb (emit) quanta of energy from (to) the field, (de-)exciting it from its initial state to alternative state. The first question we must address is ``what is the probability of such a transition occurring?''; in the case of excitation, ``what is the probability of detecting a \emph{particle}?''. We answer this question within the framework of first-order perturbation theory. 
\par 
If the path of the detector through spacetime is specified by $\x(\tau)$, where $\tau$ is the detector's proper time, and the path is assumed to be smooth, then the interaction Hamiltonian for the detector-field system takes the form
\be
H_{\text{int}}=c\chi(\tau)\mu(\tau)\phi(\x(\tau))\,,
\label{ch:techIntro:Hint} 
\ee
where here $c$ is a small coupling-constant, $\chi$ is known as the switching-function and $\mu$ is the monopole-moment operator of our `atom'. In order for transitions to occur, we must assume that the matrix form of the monopole-moment operator, when expressed in the basis of our energy eigenstates $|0_d\rangle$ and $|E_d\rangle$, is not diagonal.  We can think of the switching function $\chi$ as turning on (off) our detector; in other words, as $\chi$ goes to zero the detector and field are decoupled, so no particles in the field are detected. Many of the results in the following chapters will make use of the compact support of $\chi$, namely that only over a finite range of proper time is $\chi$  non-zero and our detector switched on. Finally, it is extremely important that we switch on (off) the interaction smoothly, as we shall soon see, and thus we insist that $\chi$ is a smooth function .
\par 
In the framework of first-order perturbation theory, we seek to answer the question: ``What is the probability of observing the detector in the state $|E_d\rangle$, at some time long after the interaction has ceased?''. The S-matrix to first order is given by
\be 
S^{(1)}=-i\int^{\infty}_{-\infty}\mathrm{d}\tau H_{\text{int}}(\tau)\,.
\label{ch:techIntro:S1}
\ee
We shall assume the field is initially in some arbitrary Hadamard state~\cite{kay-wald}. Hadamard states have many desirable properties. In a Hadamard state, the stress-energy tensor is guaranteed to be renormalisable, and the singularity structure of the Wightman function in the coincidence limit is well defined~\cite{kay-wald}. The exact characterisation of this singularity will depend on the dimension~\cite{Decanini:2005gt}, as we shall make explicit in the chapters that follow. All the quantum states that we consider in this thesis, from the $d$-dimensional Minkowski vacuum to say the Hartle-Hawking vacuum on some black hole spacetime, are Hadamard states. We shall denote this initial Hadamard state of the field as $|\Psi\rangle$, then, by using~\eqref{ch:techIntro:Hint}, the amplitude of a transition from our initial state $|0_d\rangle \otimes |\Psi\rangle$ to final state $|E_d\rangle \otimes |\Psi^{\prime}\rangle$ is
\be 
-i c \langle E_d, \Psi^{\prime}|\int^{\infty}_{-\infty}\,\mathrm{d}\tau\,\chi(\tau)\mu(\tau)\phi(\x(\tau))|\Psi,0_d\rangle\,.
\label{ch:techIntro:amplitude}
\ee
In the interaction picture, in which we work, the monopole-moment operator evolves according to the free-field Hamiltonian:
\be 
\mu(\tau)=\expo^{iH_0 \tau}\mu(0)\expo^{-iH_0 \tau}\,,
\label{ch:techIntro:muInt}
\ee
where $H_0 |E_d\rangle = E |E_d\rangle$ and $H_0 |0_d\rangle = 0$. Using these and substituting~\eqref{ch:techIntro:muInt} into~\eqref{ch:techIntro:amplitude}, we are led to the amplitude:
\be 
-i c \langle E_d|\mu(0)|0_d\rangle \int^{\infty}_{-\infty}\,\mathrm{d}\tau\,\expo^{iE\tau} \chi(\tau)\langle \Psi^{\prime}|\phi(\x(\tau))|\Psi\rangle\,.
\label{ch:techIntro:amplitude-alt}
\ee
Now we take the modulus squared of~\eqref{ch:techIntro:amplitude-alt} and, owing to the fact that we are uninterested in the final state of the field $|\Psi^{\prime}\rangle$, we sum over the complete set of states to get the total probability for the field ending in any arbitrary state. The result is
\bea 
P(E)&=c^2|\langle 0_d|\mu(0)|E_d \rangle|^2\times\\
&\int^{\infty}_{-\infty}\,\mathrm{d}\tau^{\prime}\int^{\infty}_{-\infty}\,\mathrm{d}\tau^{\prime\prime}\,\expo^{-iE\left(\tau^{\prime}-\tau^{\prime\prime}\right)}\chi\left(\tau^{\prime}\right)\chi\left(\tau^{\prime\prime}\right)
\langle \Psi|\phi\left(\x\left(\tau^{\prime}\right)\right)\phi\left(\x\left(\tau^{\prime\prime}\right)\right)|\Psi\rangle\,.
\label{ch:techIntro:transProb}
\eea
The first factor on the right-hand side of~\eqref{ch:techIntro:transProb} only depends on the internal details of the detector, such as if we had taken a simple harmonic oscillator as our detector versus say a hydrogen atom; we drop this internal factor because we consider this portion of the probability to be uninteresting. The interesting part of the probability, which encodes the trajectory of the detector through spacetime along with the quantum state that the field is in, is what remains. This factor is called the detector response function, and it is defined by
\be 
\mathcal{F}\left(E\right)=\int^{\infty}_{-\infty}\,\mathrm{d}\tau^{\prime}\int^{\infty}_{-\infty}\,\mathrm{d}\tau^{\prime\prime}\,\expo^{-iE\left(\tau^{\prime}-\tau^{\prime\prime}\right)}\chi\left(\tau^{\prime}\right)\chi\left(\tau^{\prime\prime}\right)
\langle \Psi|\phi\left(\x\left(\tau^{\prime}\right)\right)\phi\left(\x\left(\tau^{\prime\prime}\right)\right)|\Psi\rangle\,.
\label{eq:techIntro:respFunc}
\ee
The two-point correlation function that occurs in~\eqref{eq:techIntro:respFunc} is known as the Wightman function, and it is defined by
\be
W\left(\x\left(\tau^{\prime}\right),\x\left(\tau^{\prime\prime}\right)\right):=\langle \Psi|\phi\left(\x\left(\tau^{\prime}\right)\right)\phi\left(\x\left(\tau^{\prime\prime}\right)\right)|\Psi\rangle\,.
\label{ch:techIntro:W}
\ee
Technically, $W$ is really a distribution, and although it is suppressed in~\eqref{ch:techIntro:W}, we regularise the Wightman function by the usual $i\epsilon$ prescription; this regularisation consists of replacing the spacetime interval $\sigma$ by $\sigma_{\epsilon}$, where $\sigma_{\epsilon}:=\sigma+2i\epsilon\left[T(\x)-T(\x^{\prime})\right]+\epsilon^2$ and $T$ is any globally-defined, future-increasing function. The resulting Wightman function, $W_{\epsilon}$, is then integrated against smooth, compactly-supported functions of $\x$ and $\x^{\prime}$, and afterwards the limit $\epsilon\to 0$ is finally taken. We shall frequently use the notation $W(\tau^{\prime},\tau^{\prime\prime})$ for  $W(\x\left(\tau^{\prime}\right),\x\left(\tau^{\prime\prime}\right))$.
\par 
It is helpful at this point to make a change of variables in the detector response function~\eqref{eq:techIntro:respFunc}. Using $\overline{W(\tau',\tau'')} = W(\tau'',\tau')$ and changing the integration variables from $(\tau',\tau'')$ to $(u,s)$ --- where $u := \tau'$ and $s:=\tau'-\tau''$ when $\tau''<\tau'$, and $u := \tau''$ and $s:=\tau''-\tau'$ when $\tau'<\tau''$ --- a useful alternative expression for the response function is~\cite{schlicht}
\begin{equation}
\mathcal{F}\left(E\right)=\lim_{\epsilon\to 0_+} 2 \Realpart 
\int^{\infty}_{-\infty}\,\mathrm{d}u\,\chi(u)\int^{\infty}_0\,\mathrm{d}s\,\chi(u-s)\,\expo^{-iEs} \, W_{\epsilon}(u,u-s)
\,,
\label{eq:techIntro:respFunc-alt}
\end{equation}
where $W_{\epsilon}$ denotes the $i\epsilon$-regularised Wightman function.
It is out of the distributional character of the Wightman function that arises the need to integrate it against smooth-switching functions, $\chi$, in order to obtain a mathematically well-defined result for the response function. 
\section{The transition rate}
\label{sec:TR}
\par 
Another quantity of interest, which we shall frequently make use of in this thesis, is the detector's \emph{transition rate}. Heuristically, the transition rate represents the ``number of particles detected per unit proper time''. We are led to this quantity by asking the question ``what is the probability of detecting a particle at some time \emph{during} the detector field interaction?''. 
\par Some early investigations into the transition rate~\cite{byd} analysed only stationary situations, and some investigations~\cite{Sriramkumar:1994pb,Svaiter:1992xt} took the response function and effectively inserted theta-type sharp-switching functions. In all but the stationary situations, this procedure leads to issues such as Lorentz-noncovariant terms or divergent terms~\cite{Schlicht:thesis, schlicht,louko-satz:profile}. Only if the trajectory is stationary can this smooth switching be neglected; on a stationary trajectory, the Wightman function only depends on the proper-time difference, $W(\tau',\tau'')=W(\tau'-\tau'')$, and with a change of variables the response function may be written as
\be 
\mathcal{F}\left(E\right)= \int^{\infty}_{-\infty}\,\mathrm{d}\tau^{\prime}\int^{\infty}_{-\infty}\,\mathrm{d}s\,\expo^{-iEs} W_{\epsilon}(s)\,.
\label{eq:techIntro:respFunc-stat}
\ee
One can then define the transition rate by simply dropping the external integral:
\be 
\mathcal{\dot{F}}\left(E\right)= \int^{\infty}_{-\infty}\,\mathrm{d}s\,\expo^{-iEs} W_{\epsilon}(s)\,.
\label{eq:techIntro:transRate-stat}
\ee
We show in Appendix~\ref{ch:appendix:4dStatTR} that~\eqref{eq:techIntro:transRate-stat} is equivalent to the transition rate found in~\cite{satz-louko:curved} for the case of a detector
on a stationary trajectory, switched on in the asymptotic past.
\begin{figure}[t]  
  \centering
\includegraphics[width=0.89\textwidth]{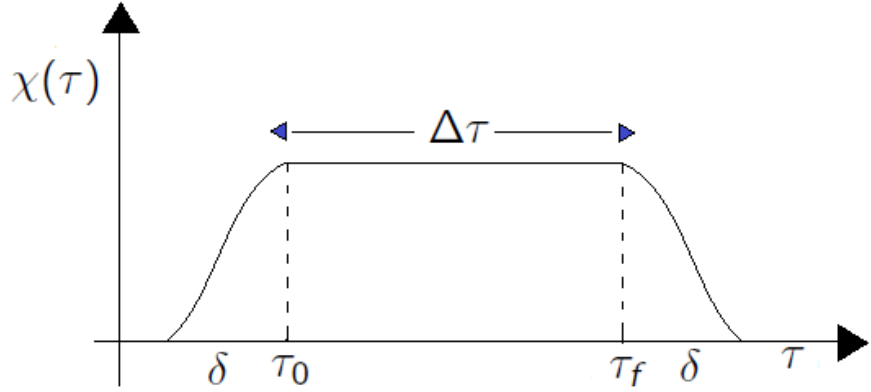}            
\caption{Satz sharp-switching limit. We first obtain a regulator-free response function and only then take the limit $\delta/\Delta\tau\to 0$.}
\label{fig:sslim}
\end{figure}
The form of the transition rate~\eqref{eq:techIntro:transRate-stat} can occasionally be useful for certain stationary situations, as we shall see in Chapter~\ref{ch:4DSchw}, but more often than not in this thesis, we shall employ a more general procedure to obtain the~\emph{instantaneous transition rate}. This procedure is that of smooth-switching, first investigated by Satz~\cite{satz:smooth}. One first considers the detector response function~\eqref{eq:techIntro:respFunc-alt}, maintaining $\chi$ to be a switching function whose only properties we invoke are that it must be smooth and of compact support. In dimensions $d>2$, the limit $\epsilon\to 0_+$ cannot be taken point-wise under the integral, and we remove the $i\epsilon$-regulator by methods to be outlined in the chapters that follow. With this regulator-free response function, one then takes the \emph{sharp-switching limit}, which is the limit of our switching functions tending to theta-like switching functions in some controlled manner, see Figure~\ref{fig:sslim}. Finally, we differentiate with respect to the proper time. 
\par The operational meaning of the instantaneous transition rate is somewhat subtle. It should be clear that $\mathcal{F}_{\tau}\left(E\right)$ represents the fraction of detectors from some ensemble (e.g. of atoms or ions in some experiment) that have undergone a transition at time $\tau$. The intricacy in interpretation comes in the fact that post-measurement the quantum-mechanical system has been altered, and $\mathcal{F}_{\tau}\left(E\right)$ no longer represents this fraction for that particular ensemble at later times. Thus, in order to measure the number of transitions per unit proper time, $\dot{\mathcal{F}}_{\tau}\left(E\right)$, one would need to use an additional, identical ensemble, at which a measurement could be made at some infinitesimal moment later. In other words, each ensemble is used to measure $\mathcal{F}_{\tau}\left(E\right)$ at a single value of $\tau$ only.
\section{The Schlicht approach}
\label{sec:Schlicht}
\par An alternative method to the Satz smooth-switching approach is that of Schlicht~\cite{Schlicht:thesis}, who was the first to notice the problems that occur when analysing the transition rate of the conventionally-regularised, point-like-coupled detector. Schlicht took the view that the response function~\eqref{eq:techIntro:respFunc} --- only with sharp-switching functions inserted, which had until then been the primary means of analysing the response of an Unruh-DeWitt detector --- had some unfavourable features. The response function does not express time-dependence nor causality; for a non-stationary detector we expect a time-dependent response and, moreover, that the reaction at a given instant should only depend on the past-trajectory of the detector. Schlicht's criticism was that in~\eqref{eq:techIntro:respFunc} there could be no room for these features because $\tau',\tau''$ are integrated out. This dissatisfaction led Schlicht to investigate what we call the instantaneous transition rate. However, when initially obtaining this quantity using the conventionally-regularised, point-like detector that is switched on (off) sharply he encountered issues. Even for the case of a uniformly linearly accelerated detector, Schlicht found a transition rate that was time-dependent and had negative values, despite the stationarity of the trajectory. Building on the work of Takagi~\cite{takagi}, Schlicht used, instead of the point-like detector, a detector that was ``smeared'' spatially. Using this ``smeared detector'' leads to the expected results for the transition rate of detectors on a variety of trajectories, such as time-independent transition rates for the six stationary classes of trajectory in Minkowski spacetime, and gives the usual thermal response for the Rindler detector.
\par 
As mentioned, in this thesis we shall use the smooth-switching method of Satz whilst retaining the point-like coupled detector with conventional regularisation. The primary reason for adopting the smooth-switching method is that it is easier to extend this approach to curved spacetimes; the approach of Schlicht requires a globally-defined Fermi-Walker co-ordinate system, which is difficult to construct on a general curved spacetime.  Satz pin-pointed that the failure that Schlicht observed when considering the point-like detector was due to the fact that the distributional nature of the Wightman function was not being fully appreciated, and it was shown in~\cite{Fewster:1999gj,junker,hormander-vol1,hormander-paper1} that one needs to integrate the Wightman distribution against smooth compactly-supported functions in order to get an unambiguous number as a result. 

\chapter[Beyond four-dimensional Minkowski spacetime]{Response function beyond four-dimensional Minkowski spacetime }
\chaptermark{Beyond four dimensions}
\label{ch:by4d}

In this chapter we investigate the response of an Unruh-DeWitt particle detector traversing a general timelike trajectory in Minkowski spacetime of dimension other than four. Our first task, regardless of the dimension, will be to take the $\epsilon\to 0$ limit of the response function,~\eqref{eq:techIntro:respFunc-alt}. The method we use is an adaptation of that introduced in \cite{satz:smooth} for $d=4$.  We shall then make precise the notion of the \emph{sharp-switching limit}, mentioned in Chapter~\ref{ch:techIntro}. Having successfully taken this limit, we shall then be free to differentiate with respect to the proper time in order to obtain the instantaneous transition rate. 
\par The case $d=2$ is exceptional. The Wightman function of a massless scalar field in two dimensions is infrared divergent, and it should be understood in some appropriate limiting sense, such as the $m\to0$ limit of a scalar field of mass~$m>0$. Given this understanding, the singularity in the correlation function $W(\tau',\tau'')$ is logarithmic in $\tau'-\tau''$, and therefore it is integrable. In this case, it follows by dominated convergence that the sharp-switching limit in \eqref{eq:techIntro:respFunc-alt} can be taken immediately by setting $\chi(u)=\Theta(u-\tau_0)\Theta(\tau-u)$, where $\Theta$ is the Heaviside function, $\tau_0$~is the moment of switch-on and $\tau$ is the moment of switch-off. The result is 
\begin{equation}
\label{eq:tottranprob}
\mathcal{F}_{\tau}\left(E\right)= 
2 \Realpart 
\int^{\tau}_{\tau_0}\,\mathrm{d}u\,
\int^{u-\tau_0}_0\,\mathrm{d}s\,\,\expo^{-iEs} \, W(u,u-s) 
\ . 
\end{equation}
The instantaneous transition rate can then be defined as the derivative of $\mathcal{F}_{\tau}\left(E\right)$ (\ref{eq:tottranprob}) with respect to $\tau$, with the result \cite{Langlois-thesis,white-dissertation} 
\begin{equation}
\label{eq:tranrate}
\dot{\mathcal{F}}_{\tau}\left(E\right)=2 \Realpart 
\int^{\Delta\tau}_0\,\mathrm{d}s\,\,\expo^{-iEs} \, W(\tau,\tau-s)
\ , 
\end{equation}
where $\Delta\tau:=\tau-\tau_0$. 
\par For $d>2$, the singularity in $W(\tau',\tau'')$ is proportional to ${(\tau'-\tau'')}^{2-d}$, and the regulator must be removed more carefully. The case $d=4$ was addressed in~\cite{satz:smooth}, and in the following sections we shall address the cases $d=6$, $d=3$ and $d=5$ in turn. 
\par The chapter ends with a discussion of the application of the results we obtain in six-dimensional Minkowski spacetime to the global embedding spacetime (GEMS) methods for investigating the detector response in curved spacetime. 
\par The results of this chapter were published in~\cite{Hodgkinson:2011pc}.
\section{Response function for $d=6$}
\label{sec:6d}

In this section, we remove the regulator from the response function formula~\eqref{eq:techIntro:respFunc-alt} in six-dimensional Minkowski spacetime. 

The regularised $d=6$ Wightman function 
reads \cite{kay-wald,Decanini:2005gt,Langlois-thesis}
\begin{equation}
\label{eq:w6d}
W_\epsilon(u,u-s)=\frac{1}{4\pi^3}
\frac{1}{\left[{(\Delta\x)}^2+2i\epsilon\Delta t+\epsilon^2\right]^2}
\ , 
\end{equation}
where $\epsilon>0$ is the regulator, 
$\Delta\x:=\x(u)-\x(u-s)$ and $\Delta t:=t(u)-t(u-s)$. 
By substituting~\eqref{eq:w6d} into~\eqref{eq:techIntro:respFunc-alt},  we obtain the response function
\begin{equation}
\begin{aligned}
\label{eq:detres6d}
\mathcal{F}(E)&=\lim_{\epsilon\to 0}
\frac{1}{2\pi^3} 
\int^{\infty}_{-\infty}\,\mathrm{d}u\,\chi (u)\int^{\infty}_{0}\,
\mathrm{d}s\,
\frac{\chi(u-s)}{R^4} \, \times 
\\[1ex]
&\times 
\left[
\cos{(Es)}\bigl[\bigl((\Delta\x)^2+\epsilon^2\bigr)^2-4\epsilon^2{(\Delta t)}^2\bigr] 
-4\epsilon\sin{(Es)}\Delta t \bigl({(\Delta\x)}^2+\epsilon^2\bigr) 
\right] \,,
\end{aligned}
\end{equation}
with 
\begin{align}
\label{eq:R-def}
R := 
\sqrt{\left[{(\Delta\x)}^2+\epsilon^2\right]^2+4\epsilon^2{(\Delta t)}^2}
\ , 
\end{align}
where in \eqref{eq:R-def} the quantity under the square root is
positive, and the positive branch of the square root is taken. 
\par Before continuing any further, we record here  
inequalities that will be used repeatedly throughout this chapter. First, because
geodesics maximise the proper time on 
timelike curves in Minkowski spacetime, it follows 
that $|(\Delta \x)^2| \geq s^2$. Second, because 
$\chi$ has compact support,  
the contributing interval of $s$ in \eqref{eq:detres6d} is bounded above, 
uniformly under the integral over~$u$. From 
the small-$s$ expansions $(\Delta\x)^2=-s^2+O\left(s^4\right)$
and $\Delta t = O\left(s\right)$
, it hence follows that
$|(\Delta\x)^2| \leq Ks^2$ 
and 
$|\Delta t| \leq sM$, where $K$ and $M$ 
are positive constants, independent of~$u$.
\par First, we need to address the integral over $s$ in~\eqref{eq:detres6d}. 
Working under the expression 
${(2\pi^3)}^{-1} \int^{\infty}_{-\infty}\,\mathrm{d}u\,\chi (u)$, 
we write this integral over $s$ as the sum 
$I^{\text{even}}_{<} + I^{\text{odd}}_{<} + I^{\text{even}}_{>} + I^{\text{odd}}_{>}$, 
where the superscript even (odd) refers to the factor 
$\cos(Es)$ (respectively $\sin(Es)$), and the subscript $<$ ($>$) 
indicates that the range of integration is 
$(0,\eta)$ (respectively $(\eta,\infty)$), where 
$\eta:=\epsilon^{1/4}$. 
We remark that this choice for 
$\eta$ differs from the choice 
$\eta=\epsilon^{1/2}$ that was 
made for $d=4$ in \cite{louko-satz:profile,satz:smooth} 
and will be made for $d=3$
in Section~\ref{sec:3d} below, 
for reasons that stem from the increasing singularity of the 
Wightman function with increasing~$d$. 

We consider the two intervals of $s$ in the next two subsections. 

\subsection{Subinterval $\eta < s < \infty$}
\label{sec:6d:larges}

We shall first consider $I^{\text{even}}_{>}$. When the regulator is set to zero, the integrand in $I^{\text{even}}_{>}$
reduces to $\chi(u-s)\cos{(Es)}/\left[(\Delta\x)^2\right]^2$. 
This replacement creates an error in $I^{\text{even}}_{>}$ that can be arranged in the form 
\begin{equation}
\begin{aligned}
\label{eq:error6d}
\int^{\infty}_{\eta}\,\mathrm{d}s\, &\chi(u-s) \, 
\frac{\cos{(Es)}}{\left[(\Delta\x)^2\right]^2} \, \times\\
& \times 
\left[\frac{\left(\left(1+\frac{\epsilon^2}{(\Delta\x)^2}\right)^2-4\epsilon^2\frac{(\Delta t)^2}{\left[(\Delta\x)^2\right]^2} \right)-\left(\left(1+\frac{\epsilon^2}{(\Delta\x)^2}\right)^2+4\epsilon^2\frac{(\Delta t)^2}{\left[(\Delta\x)^2\right]^2} \right)^2}{\left(\left(1+\frac{\epsilon^2}{(\Delta\x)^2}\right)^2+4\epsilon^2\frac{(\Delta t)^2}{\left[(\Delta\x)^2\right]^2} \right)^2}\right] .
\end{aligned}
\end{equation}
Using $|(\Delta \x)^2| \geq s^2$ and $s\ge \eta = \epsilon^{1/4}$, we have 
$|\epsilon^2/(\Delta\x)^2 |\leq \epsilon^2/s^2 \leq \epsilon^2/\sqrt{\epsilon} =O\left(\epsilon^{3/2}\right)$. Using $|\Delta t| \leq sM$, we similarly have 
$\epsilon^2 (\Delta t)^2/\left[(\Delta\x)^2\right]^2 =O\left(\epsilon^{3/2}\right)$.
Hence, the integrand in 
\eqref{eq:error6d} is bounded in absolute value by 
a constant times $\epsilon^{3/2}/\left[(\Delta\x)^2\right]^2 \le \epsilon^{3/2}/s^4$. 
It follows that the integral is of order $O\left(\epsilon^{3/2}/\eta^3\right) = O\left(\eta^3\right)$. 

Now concentrating on $I^{\text{odd}}_{>}$, we have
\begin{align}
I^{\text{odd}}_{>}&:=-4\epsilon\int^{\infty}_{\eta}\,\mathrm{d}s\,\frac{\chi(u-s)\sin{\left(Es\right)}\Delta t\left((\Delta\x)^2+\epsilon^2\right)}{\left[\left[{(\Delta\x)}^2+\epsilon^2\right]^2+4\epsilon^2{(\Delta t)}^2\right]^2}\nonumber\\
&=4\int^{\infty}_{\eta}\,\mathrm{d}s\,\frac{\chi(u-s)\sin{(Es)}}{\sqrt{-(\Delta\x)^2}}\left(\frac{\epsilon\Delta t}{\left[-(\Delta\x)^2\right]^{5/2}}\right)\left(\frac{1+\frac{\epsilon^2}{\Delta\x^2}}{\left(\left(1+\frac{\epsilon^2}{(\Delta\x)^2}\right)^2+4\epsilon^2\frac{(\Delta t)^2}{\left[(\Delta\x)^2\right]^2} \right)^2}\right)\,.
\end{align}
Using the bounds computed previously along with $\epsilon (\Delta t)/\left[-(\Delta\x)^2\right]^{5/2} \leq \epsilon M/s^4$, the fact that the switching function $\chi(u-s) \leq 1$ and that it bounds the upper limit of the $s$-integral from above, by virtue of its compact support, we can write
\begin{align}
|I^{\text{odd}}_{>}|&\leq 4 M \epsilon \int^{s_c}_{\eta}\,\mathrm{d}s\,\left|\frac{\sin{(Es)}}{s^5}\right| \left[1+O\left(\epsilon^{3/2}\right)\right]\nonumber\\
&\leq 4 M \epsilon |E| \int^{s_c}_{\eta}\,\mathrm{d}s\,\frac{1}{s^4} \left[1+O\left(\epsilon^{3/2}\right)\right]\nonumber\\
&=O\left(\eta\right)\,,
\end{align}
where $s_c>0$ and is a real constant large enough such that $\chi(u-s_c)=0$.

Collecting, we have 
\begin{equation}
\label{eq:6d:upper}
I^{\text{even}}_{>}+I^{\text{odd}}_{>}
=\int^{\infty}_{\eta}\,\mathrm{d}s\,
\frac{\chi(u-s)\cos{(Es)}}{\left[(\Delta\x)^2\right]^2}+O\left(\eta\right) 
\ . 
\end{equation}
\subsection{Subinterval $0 < s < \eta$}
\label{sec:6d:smalls}
Consider $I^{\text{odd}}_{<}$, which is defined by 
\begin{equation}
\label{eq:6dlessodd}
I^{\text{odd}}_{<}: =-4\epsilon\int^{\eta}_{0}\,\mathrm{d}s\,\chi(u-s)
\, 
\frac{\sin{(Es)}\Delta t\bigl({(\Delta\x)}^2+\epsilon^2\bigr)}
{R^4}
\ . 
\end{equation}
The delicate task is to estimate the 
denominator in~\eqref{eq:6dlessodd}. 

By Taylor's theorem, 
$(\Delta t)^2$, $(\Delta\x)^2$ and $\left[(\Delta\x)^2\right]^2$ have the asymptotic 
small-$s$ expansions 
\begin{subequations}
\label{eq:tDeltaandSq-expansions}
\begin{align}
(\Delta t)^2 & =\sum^{n_1-1}_{n=0} T_n s^n 
+ O(s^{n_1})
\ , 
\\
(\Delta\x)^2 & =\sum^{n_2-1}_{n=0} X_n s^n 
+ O(s^{n_2})
\ , 
\label{eq:Deltaxexp}
\\
\left[(\Delta\x)^2\right]^2 & =\sum^{n_3-1}_{n=0} F_n s^n 
+ O(s^{n_3})
\ .
\label{eq:Deltax4exp}
\end{align}
\end{subequations}
The expansion coefficients 
$T_n$, 
$X_n$ and 
$F_n$ are functions of~$u$, and they satisfy
$T_0 = T_1 = X_0 = X_1 = X_3=0$, $X_2 = -1$, $T_2=\dot{t}^2$ and $T_3=-\dot{t}\ddot{t}$, where the dots indicate derivatives with respect to $u$.
The consequences for~$F_n$ are easily computed, in particular $F_0=F_1=F_2=F_3=F_5=0$ and $F_4=1$. 
The positive integers $n_1$, $n_2$ and $n_3$ may be chosen arbitrarily. 
(Note, because the trajectory is assumed smooth but not necessarily analytic, 
the error terms in \eqref{eq:tDeltaandSq-expansions} 
are not guaranteed to vanish for fixed $s$ as $n_i\to\infty$.)  
With this notation established, we rearrange the denominator:
\begin{align}
R^2&:= 
\left[{(\Delta\x)}^2+\epsilon^2\right]^2+4\epsilon^2{(\Delta t)}^2\nonumber\\
&=\epsilon^4+2\epsilon^2{(\Delta\x)}^2+\left[(\Delta\x)^2\right]^2+4\epsilon^2(\Delta t)^2\nonumber\\
&=\epsilon^4+2\epsilon^2\left[-s^2+\sum^{n_1-1}_{n=4} X_n s^n +2\sum^{n_1-1}_{n=2} T_n s^n \right]+s^4+\sum^{n_3-1}_{n=6} F_n s^n\nonumber\\
&\qquad\qquad\qquad\qquad\qquad\qquad+ O\left(\epsilon^2 s^{n_1}\right)+ O\left(s^{n_3}\right)\nonumber\\
&=\epsilon^4+2\epsilon^2\left[-s^2+\sum^{n_1-1}_{n=4} X_n s^n+2\dot{t}^2 s^2-2\dot{t}\ddot{t} s^3 +2\sum^{n_1-1}_{n=4} T_n s^n \right]+s^4+\sum^{n_3-1}_{n=6} F_n s^n\nonumber\\
&\qquad\qquad\qquad\qquad\qquad\qquad+ O\left(\epsilon^2 s^{n_1}\right)+ O\left(s^{n_3}\right)\nonumber\\
&=\epsilon^4+2\epsilon^2\left[(2\dot{t}^2-1)s^2-2\dot{t}\ddot{t} s^3+\sum^{n_1-1}_{n=4} (X_n+2T_n) s^n \right]+s^4+\sum^{n_3-1}_{n=6} F_n s^n\nonumber\\
&\qquad\qquad\qquad\qquad\qquad\qquad+ O\left(\epsilon^2 s^{n_1}\right)+ O\left(s^{n_3}\right)\nonumber\,.
\end{align}
Now we make the change of variables $s=\epsilon r$ in integral~\eqref{eq:6dlessodd} (meaning the range of integration over $r$ becomes $0< r < \epsilon^{-3/4}$), and we see that $R^2$ can be expressed as
\begin{align}
R^2&=\epsilon^4+2\epsilon^2\left[(2\dot{t}^2-1)\epsilon^2 r^2-2\dot{t}\ddot{t} \epsilon^3 r^3+\sum^{n_1-1}_{n=4} (X_n+2T_n) \epsilon^n r^n \right]+\epsilon^4 r^4+\sum^{n_3-1}_{n=6} F_n \epsilon^n r^n\nonumber\\
&\qquad\qquad\qquad\qquad\qquad\qquad+ O\left(\epsilon^{n_1+2} r^{n_1}\right)+ O\left(\epsilon^{n_3} r^{n_3}\right)\nonumber\\
&=\epsilon^4\left[1+2(2\dot{t}^2-1)r^2+r^4-4\dot{t}\ddot{t}\epsilon r^3+\sum^{n_1-1}_{n=4} (2X_n+4T_n) \epsilon^{n-2} r^n
+\sum^{n_3-1}_{n=6} F_n \epsilon^{n-4} r^n\right]\nonumber\\
&\qquad\qquad\qquad\qquad\qquad\qquad+ O\left(\epsilon^{n_1+2} r^{n_1}\right)+ O\left(\epsilon^{n_3} r^{n_3}\right)\nonumber\,.
\end{align}
Next we define 
\begin{align}
P :=  1+2(2\dot{t}^2-1)r^2+r^4  
\ .
\end{align}
Owing to the fact that $\dot{t} \geq 1$, $P$ is positive for $r \ge0$.  
Finally, we make the rearrangement 
\begin{align}
\label{eq:den6d-raw}
R^2 
& = 
\epsilon^{4}P\Bigg[1-\frac{4\dot{t}\ddot{t}\epsilon r^3}{P}
+\sum^{n_4-1}_{n=0}\left(2X_{(n+4)}+4T_{(n+4)}\right)\frac{\epsilon^{n+2}r^{n+4}}{P}
+\sum^{n_5-1}_{n=0}F_{(n+6)}\frac{\epsilon^{n+2}r^{n+6}}{P}
\nonumber 
\\[1ex]
& \hspace{10ex}
+ \frac{O\bigl(\epsilon^{n_4+2}r^{n_4+4}\bigr)}{P}
+ \frac{O\bigl(\epsilon^{n_5+2}r^{n_5+6}\bigr)}{P}
\Biggr]
\ , 
\end{align}
where the positive integers $n_4$ and $n_5$ may be chosen arbitrarily. 

We wish to regard the external factor $\epsilon^4 P$ 
in \eqref{eq:den6d-raw} as the dominant part and the 
terms in the square brackets as a leading 1 plus 
sub-leading corrections. 
To this end, we rewrite \eqref{eq:den6d-raw} as 
\begin{align}
&R^2 
= 
\epsilon^{4}P\Bigg[1-\frac{4\dot{t}\ddot{t}\epsilon r^3}{P}z
+\sum^{n_4-1}_{n=0}\left(2X_{(n+4)}+4T_{(n+4)}\right)\frac{\epsilon^{n+2}r^{n+4}}{P}z^{2+(n/4)}
\nonumber
\\[1ex]
& 
+\sum^{n_5-1}_{n=0}F_{(n+6)}\frac{\epsilon^{n+2}r^{n+6}}{P}z^{(n+2)/4}
+ \frac{O\bigl(\epsilon^{n_4+2}r^{n_4+4}\bigr)}{P}z^{2+(n_4/4)}
+ \frac{O\bigl(\epsilon^{n_5+2}r^{n_5+6}\bigr)}{P}z^{(n_5+2)/4}
\Bigg] \ , 
\label{eq:den6d}
\end{align}
where the book-keeping parameter $z$, with numerical value~$1$, 
indicates what order in $\epsilon$ the term 
in question is \emph{uniformly\/} over the full range of~$r$, 
$0\le r \le \epsilon^{-3/4}$.  
Remembering that $P\to r^4$ as $r\to\infty$, the term 
$-4\dot{t}\ddot{t}\epsilon r^3/P$
is assigned the factor $z$ because 
$r^3/P$ is bounded by a constant. 
The $z$-factors in the other terms follow because 
$r^{n+4}/P$ and $r^{n+6}/P$ are respectively 
bounded by 
a constant times 
$\epsilon^{-3n/4}$ and a constant times $\epsilon^{-3(n+2)/4}$. 

We can now insert~\eqref{eq:den6d} in the denominator of~\eqref{eq:6dlessodd}; similarly, in the numerator we use the following expansions:
\begin{align}
\label{eq:Dt6d}
\Delta t & :=t(u)-t(u-\epsilon r)\nonumber\\
&=\epsilon r \left[\dot{t}-\frac{1}{2}\ddot{t}\epsilon r z^{1/4}+\cdots+\frac{1}{9!}t^{(9)}\epsilon^8 r^8 z^2+O\left(z^{9/4}\right)\right]\,,
\end{align}
\begin{align}
\label{eq:chi6d}
\chi(u-\epsilon r)& =\chi(u)-\epsilon r\dot{\chi}(u)z^{1/4}+\cdots+\frac{1}{8!}\chi^{(8)}(u)\epsilon^8 r^8 z^2+O\left(z^{9/4}\right)\,,
\,
\end{align}
and
\begin{align}
\label{eq:sin6d}
\sin{(\epsilon E r)}&=\epsilon E r-\frac{1}{3!}\epsilon^3E^3r^3+\cdots+\frac{1}{9!}\epsilon^9 E^9 r^9 +O\left(\epsilon^{11 } E^{11} r^{11}\right)\nonumber\\
&=\epsilon E r \left[1-\frac{1}{3!}\epsilon^2E^2r^2\sqrt{z}+\cdots+\frac{1}{9!}\epsilon^8 E^8 r^8 z^2+O\left(z^{5/2}\right)\right] .
\end{align}
It also proves convenient to bring in a factor of $P^{-1}$ from the denominator expansion~\eqref{eq:den6d} and couple it with the
$\left((\Delta\x)^2+\epsilon^2\right)$ factor in~\eqref{eq:6dlessodd}:
\begin{align}
\label{eq:Dx2pe2}
\frac{\left((\Delta\x)^2+\epsilon^2\right)}{P}&=\epsilon^2\left[\frac{1-r^2}{P}+X_4\frac{\epsilon^2 r^4}{P}z^2+O\left(z^{9/4}\right)\right]\,.
\end{align}
In each of the expansions made so far, we have expanded up to and inclusive of order~$z^2$ for reasons that will become apparent momentarily.
Remembering to change the integration variable to $r$, we substitute~\eqref{eq:den6d},~\eqref{eq:Dt6d},~\eqref{eq:chi6d},~\eqref{eq:sin6d} and~\eqref{eq:Dx2pe2} into~\eqref{eq:6dlessodd} (remembering we have already used one of the $P^{-1}$ factors from the denominator) to obtain
\begin{equation}
\begin{aligned}
\label{eq:6d:odd:premult}
& 
I^{\text{odd}}_{<} = -\frac{1}{\epsilon^2}\int^{\eta^{-3}}_{0}\,\mathrm{d}r\,\frac{4Er^2}{P} \left[\chi(u)-\epsilon r\dot{\chi}(u)z^{1/4}+\cdots+\frac{1}{8!}\chi^{(8)}(u)\epsilon^8 r^8 z^2+O\left(z^{9/4}\right)\right]\times \\
&\ \times \left[1-\frac{1}{3!}\epsilon^2E^2r^2\sqrt{z}+\cdots+\frac{1}{9!}\epsilon^8 E^8 r^8 z^2+O\left(z^{5/2}\right)\right]\times \\
&\ \times \left[\dot{t}-\frac{1}{2}\ddot{t}\epsilon r z^{1/4}+\cdots+\frac{1}{9!}t^{(9)}\epsilon^8 r^8 z^2+O\left(z^{9/4}\right)\right]\left[\frac{1-r^2}{P}+X_4\frac{\epsilon^2 r^4}{P}z^2+O\left(z^{9/4}\right) \right] \times\\
&\ \times \Bigg[1-\frac{4\dot{t}\ddot{t}\epsilon r^3}{P}z+\left(2X_4+4T_4\right)\frac{\epsilon^{2}r^{4}}{P}z^{2}+\sum^{6}_{n=0}F_{(n+6)}\frac{\epsilon^{n+2}r^{n+6}}{P}z^{(n+2)/4}+O\left(z^{9/4}\right)\Bigg]^{-2} \,.
\end{aligned}
\end{equation}
In each square bracket factor of the integrand we have kept terms to order $z^2$ because 
of the factor $\epsilon^{-2}$ outside the integral and because 
$\int^{\eta^{-3}}_{0} (r^2/P)\, \mathrm{d}r$ remains bounded as $\eta\to0$. 
We can now Taylor expand the integrand 
in \eqref{eq:6d:odd:premult} in~$z^{1/4}$. 
Keeping terms to order~$z^2$,
the dropped terms are of order $z^{9/4}$ and their contribution to 
$I^{\text{odd}}_{<}$ is~$O(\eta)$. In practice this is done with the computer algebra package Maple (TM).
After this expansion, $z$ can be replaced by its numerical value~$1$, and we obtain
for $I^{\text{odd}}_{<}$ a lengthy expression that 
consists of elementary integrals of rational functions,
plus the error term~$O(\eta)$.
\par Consider next $I^{\text{even}}_{<}$, 
which can be rearranged as 
\begin{equation}
\begin{aligned}
\label{eq:6dlesseven}
I^{\text{even}}_{<}
&=
\int^{\eta}_{0}\,\mathrm{d}s\,
\chi(u-s) \, \frac{\cos{(Es)}}{R^2}
\\[1ex]
&
\hspace*{3ex}
-8\epsilon^2\int^{\eta}_{0}\,\mathrm{d}s\,
\chi(u-s) \, \frac{\cos{(Es)} {(\Delta t)}^2}{R^4}
\ . 
\end{aligned}
\end{equation}
Proceeding as above, we find 
\begin{align}
&I^{\text{even}}_{<}
=\frac{1}{\epsilon^3}\int^{\eta^{-3}}_{0}\,\frac{\mathrm{d}r}{P}\,\left[\chi(u)-\epsilon r\dot{\chi}(u)z^{1/4}+\cdots+\frac{1}{12!}\chi^{(12)}(u)\epsilon^{12} r^{12} z^3+O\left(z^{13/4}\right)\right]\times 
\notag 
\\
&\hspace{17ex}
\times \left[1-\frac{1}{2!}\epsilon^2E^2r^2\sqrt{z}+\cdots+\frac{1}{12!}\epsilon^{12} E^{12} r^{12} z^3+O\left(z^{7/2}\right)\right]\times 
\notag 
\\
&\hspace{17ex}
\times \Bigg[1-\frac{4\dot{t}\ddot{t}\epsilon r^3}{P}z+\sum^{4}_{n=0}\left(2X_{(n+4)}+4T_{(n+4)}\right)\frac{\epsilon^{(n+2)}r^{(n+4)}}{P}z^{2+(n/4)}
\notag 
\\
&\hspace{23ex}
+\sum^{10}_{n=0}F_{(n+6)}\frac{\epsilon^{n+2}r^{n+6}}{P}z^{(n+2)/4}+O\left(z^{13/4}\right)\Bigg]^{-1} 
\notag 
\\
&\hspace{8ex}
-\frac{8}{\epsilon^3}\int^{\eta^{-3}}_{0}\,\mathrm{d}r\,\left[\chi(u)-\epsilon r\dot{\chi}(u)z^{1/4}+\cdots-\frac{1}{15!}\chi^{(15)}(u)\epsilon^{15} r^{15} z^{15/4}+O\left(z^{4}\right)\right]\times 
\notag 
\\
&\hspace{20ex}
\times \left[1-\frac{1}{2!}\epsilon^2E^2r^2\sqrt{z}+\cdots-\frac{1}{14!}\epsilon^{14} E^{14} r^{14} z^{7/2}+O\left(z^{4}\right)\right]\times 
\notag 
\\
&\hspace{20ex}
\times \left[\frac{T_2r^2}{P^2}+\frac{T_3\epsilon r^3}{P^2}z+\frac{T_4\epsilon^2 r^4}{P^2}z^2+\frac{T_5\epsilon^3 r^5}{P^2}z^3+O\left(z^{4}\right)\right]\times 
\notag 
\\
&\hspace{20ex}
\times \Bigg[1-\frac{4\dot{t}\ddot{t}\epsilon r^3}{P}z+\sum^{7}_{n=0}\left(2X_{(n+4)}+4T_{(n+4)}\right)\frac{\epsilon^{(n+2)}r^{(n+4)}}{P}z^{2+(n/4)}
\notag 
\\
&\hspace{25.5ex}
+\sum^{13}_{n=0}F_{(n+6)}\frac{\epsilon^{n+2}r^{n+6}}{P}z^{(n+2)/4}+O\left(z^{4}\right)\Bigg]^{-2}
\ . 
\label{eq:6d:even:premult}
\end{align}
We Taylor expand the integrands in \eqref{eq:6d:even:premult}
in~$z^{1/4}$, 
keeping in the first (respectively second) integrand terms to order 
$z^3$~($z^{15/4}$), 
at the expense of an error of order $O(\eta)$ in~$I^{\text{even}}_{<}$. 
Replacing $z$ 
by its numerical value $1$, we then obtain 
for $I^{\text{even}}_{<}$ a lengthy expression that 
consists of elementary integrals of rational functions 
plus the error term~$O(\eta)$.

\subsection{Combining the subintervals}

Evaluating the numerous elementary integrals 
obtained from \eqref{eq:6d:odd:premult} and 
\eqref{eq:6d:even:premult}
and combining the results with~\eqref{eq:6d:upper}, 
we find from \eqref{eq:detres6d} that the response function takes the form 
\begin{equation}
\begin{aligned}
\label{eq:6d:result}
\mathcal{F}(E)&=\lim_{\eta\to 0}\frac{1}{2\pi^3} \int^{\infty}_{-\infty}\,\mathrm{d}u\,\chi (u)\Bigg[ -\frac{\chi(u)}{3\eta^3}-\frac{E\pi}{12}\left[\chi(u)(E^2+\ddot{\x}^2)-3\ddot{\chi}(u)\right]\\
&+\frac{1}{6\eta}\left[\chi(u)(3E^2+\ddot{\x}^2)-3\ddot{\chi}(u)\right] +\int^{\infty}_{\eta}\,\mathrm{d}s\,\frac{\chi(u-s)\cos{(Es)}}{\left[(\Delta\x)^2\right]^2}\Bigg]
\ , 
\end{aligned}
\end{equation}
where $\ddot{\x}^2$ is evaluated at~$u$. 
The uniformity of the $O(\eta)$ error terms in 
$u$ has been used to control the errors, and 
all terms involving the Lorentz-noncovariant quantities 
$T_n$ have cancelled on integration 
over $u$ (cf.~Section 3 of \cite{satz:smooth} 
for a similar cancellation in four dimensions). 
Taking the inverse powers of $\eta$ 
under the $s$-integral, we have 
\begin{equation}
\label{eq:6d:result2}
\begin{aligned}
&\mathcal{F}(E)=-\frac{E}{24\pi^2}\int^{\infty}_{-\infty}\,\mathrm{d}u\left[\chi^2(u)(E^2+\ddot{\x}^2)+3\dot{\chi}^2(u)\right]\\
&+\lim_{\eta\to 0}\frac{1}{2\pi^3}\int^{\infty}_{-\infty}\,\mathrm{d}u
\, 
\chi(u)\int^{\infty}_{\eta}\,\mathrm{d}s\Bigg(\frac{\chi(u-s)\cos{(Es)}}{\left[(\Delta \x)^2\right]^2}
-\frac{\chi(u)}{s^4}+\frac{\chi(u) (3E^2+\ddot{\x}^2)}{6s^2}-\frac{\ddot{\chi}(u)}{2s^2}\Bigg). 
\end{aligned}
\end{equation}

To take the limit $\eta\to0$ in~\eqref{eq:6d:result2}, 
we add and 
subtract under the $s$-integral terms that disentangle 
the small-$s$ divergences of 
$\cos(Es)/\left[(\Delta \x)^2\right]^2$ from the small-$s$ behaviour of $\chi(u-s)$ following \cite{satz:smooth}.
First note that
\be 
\frac{\cos{(Es)}}{\left[(\Delta \x)^2\right]^2}=\frac{1}{s^4}-\frac{3E^2+\ddot{\x}^2}{6s^2}+\frac{\ddot{\x}\cdot\x^{(3)}}{6s}+O\left(s^0\right)\,,
\ee
which characterises the small-$s$ divergence $\cos(Es)/\left[(\Delta \x)^2\right]^2$, as can be obtained using~\eqref{eq:Deltax4exp}. After subtracting these terms from the  $\cos(Es)/\left[(\Delta \x)^2\right]^2$ part of the integrand in~\eqref{eq:6d:result2}, we obtain
\bea
\label{eq:6d:predetresp}
&\mathcal{F}(E)  =
-\frac{E}{24\pi^2}\int^{\infty}_{-\infty}\,\mathrm{d}u
\left[
\chi^2(u)
\left(E^2+\ddot{\x}^2\right)
+3\dot{\chi}^2(u)
\right]
\\
&\ +\lim_{\eta\to 0}\frac{1}{2\pi^3}\int^{\infty}_{-\infty}\,\mathrm{d}u\,\chi(u)\int^{\infty}_{\eta}\,\mathrm{d}s\,\Bigg[\chi(u-s)\left(\frac{\cos{(Es)}}{\left[(\Delta \x)^2\right]^2}
-\frac{1}{s^4}+\frac{3E^2+\ddot{\x}^2}{6s^2}-\frac{\ddot{\x}\cdot\x^{(3)}}{6s}\right)\\
&+\frac{\chi(u-s)}{s^4}-\frac{\chi(u-s)(3E^2+\ddot{\x}^2)}{6s^2}+\frac{\chi(u-s)\ddot{\x}\cdot\x^{(3)}}{6s}-\frac{\chi(u)}{s^4}
+\frac{\chi(u) (3E^2+\ddot{\x}^2)}{6s^2}-\frac{\ddot{\chi}(u)}{2s^2}  \Bigg]
\ .
\eea
After a regrouping, this can be written as
\bea
\label{eq:6d:detresp}
\mathcal{F}(E) & =
-\frac{E}{24\pi^2}\int^{\infty}_{-\infty}\,\mathrm{d}u
\left[
\chi^2(u)
\left(E^2+\ddot{\x}^2\right)
+3\dot{\chi}^2(u)
\right]
\\
&\ -\frac{E^2}{4\pi^3}\int^{\infty}_{0}\,\frac{\mathrm{d}s}{s^2}\int^{\infty}_{-\infty}\,\mathrm{d}u\, \chi(u)\left[\chi(u-s)-\chi(u)\right]\\
&\ +\frac{1}{2\pi^3}\int^{\infty}_{0}\,\frac{\mathrm{d}s}{s^4}\int^{\infty}_{-\infty}\,\mathrm{d}u\, \chi(u)
\bigl[
\chi(u-s)-\chi(u)
-\tfrac12 s^2\ddot{\chi}(u)
\bigr]
\\
&\ -\frac{1}{12\pi^3}\int^{\infty}_{0}\,\frac{\mathrm{d}s}{s^2}
\int^{\infty}_{-\infty}\,\mathrm{d}u\,\chi(u)
\Bigl\{
\left[\chi(u-s)-\chi(u)\right] \ddot{\x}^2
-s \chi(u-s) \, \ddot{\x}\cdot\x^{(3)}
\Bigr\}
\\
&\ +\frac{1}{2\pi^3}\int^{\infty}_{-\infty}\,\mathrm{d}u\,\chi(u)\int^{\infty}_{0}\,\mathrm{d}s\,\chi(u-s)\left(\frac{\cos{(Es)}}{\left[(\Delta \x)^2\right]^2}
-\frac{1}{s^4}+\frac{3E^2+\ddot{\x}^2}{6s^2}-\frac{\ddot{\x}\cdot\x^{(3)}}{6s}\right)
\ , 
\eea
where 
$\ddot{\x}^2$ and $\ddot{\x}\cdot\x^{(3)}$ are evaluated at~$u$. 
The interchanges of the integrals before taking 
the limit $\eta\to0$ are 
justified by absolute convergence of the double integrals, 
and taking the limit $\eta\to0$ under the outer integral is 
justified by dominated convergence: 
in each integral over $s$ in~\eqref{eq:6d:detresp},  
the integrand is regular as $s\to0$, and we show an example of how
this is determined in Appendix~\ref{ch:appendix:6dgroups}

Equation \eqref{eq:6d:detresp} is our final, 
regulator-free, expression for the 
response function. In Section~\ref{sec:sharp}
we shall consider its behaviour when the switching approaches the step-function. 
\section{Response function for $d=3$}
\label{sec:3d}
In this section we remove the regulator from the response function 
formula~\eqref{eq:techIntro:respFunc-alt} for $d=3$.
The qualitatively new feature is that the techniques of 
Section \ref{sec:6d} 
need to be adapted to the fractional power in the Wightman function. 
\subsection{Regularisation}
\label{sec:3d:reg}

The regularised $d=3$ Wightman function 
reads \cite{kay-wald,Decanini:2005gt,Langlois-thesis}
\begin{equation}
\label{eq:w3d:1}
W_\epsilon (u,u-s)=\frac{1}{4\pi}\frac{1}{\sqrt{{(\Delta\x)}^2+2i\epsilon\Delta t+\epsilon^2}}
\ , 
\end{equation}
where the branch of the square root is chosen such that 
the $\epsilon\to0$ limit of the square root is positive when ${(\Delta\x)}^2>0$. 
Separating the real and imaginary parts gives 
\begin{equation}
\label{eq:w3d:3}
W_\epsilon (u,u-s) = 
\frac{\sqrt{R+{(\Delta\x)}^2+\epsilon^2}-i\sqrt{R-{(\Delta\x)}^2-\epsilon^2}}{4\sqrt{2}\,\pi \, R}\,,
\end{equation}
where $R$ is given by~\eqref{eq:R-def}. The quantities under the square roots in~\eqref{eq:w3d:3} are positive,
and the square roots are taken positive. 
From \eqref{eq:techIntro:respFunc-alt}, we now obtain 
\begin{equation}
\begin{aligned}
\label{eq:3d:detresp}
\mathcal{F}(E)
&=
\lim_{\epsilon\to 0}\frac{1}{2\sqrt{2}\,\pi} \int^{\infty}_{-\infty}\,\mathrm{d}u\,\chi (u)\int^{\infty}_{0}\,\mathrm{d}s\,
\frac{\chi(u-s)}{R}\times 
\\[1ex]
& \hspace{4ex}
\times \left[\cos{(Es)}\sqrt{R+{(\Delta\x)}^2+\epsilon^2}-\sin{(Es)}\sqrt{R-{(\Delta\x)}^2-\epsilon^2}\right]
\ . 
\end{aligned}
\end{equation}

We proceed as in Section~\ref{sec:6d}. 
Working under the expression
${\left(2\sqrt{2}\,\pi\right)}^{-1}
\int^{\infty}_{-\infty}\,\mathrm{d}u\,\chi (u)$, we write the integral
over $s$ as the sum $I^{\text{even}}_{<} + I^{\text{odd}}_{<} +
I^{\text{even}}_{>} + I^{\text{odd}}_{>}$, where the notation follows
Section \ref{sec:6d} with the exception that we now choose
$\eta:=\epsilon^{1/2}$. We consider the two intervals of $s$ in the
next two subsections. 

\subsection{Subinterval $\eta < s < \infty$}
\label{sec:3d:larges}

Consider $I^{\text{odd}}_{>}$. 
When $\epsilon$ is set to zero, the integrand in 
$I^{\text{odd}}_{>}$ reduces to 
$- \chi(u-s)\sin{(Es)}\sqrt{-2/{(\Delta\x)}^2}$, 
where the quantity under the square root 
is positive and the square root is taken positive. 
This replacement creates an error in $I^{\text{odd}}_{>}$ that can be arranged in the form 
\begin{equation} 
\begin{aligned}
\label{eq:3d:error:2}
&\int^{\infty}_{\eta}\,\mathrm{d}s\,\frac{\chi(u-s)\epsilon^2}{{(\Delta\x)}^2\sqrt{-{(\Delta\x)}^2}}\times
\\[1ex]
&\hspace{6ex}
\times \left[\frac{\left(2+\frac{\epsilon^2}{(\Delta\x)^2}+4\frac{(\Delta t)^2}{(\Delta\x)^2}-2\frac{S}{(\Delta\x)^2}-\epsilon^2\frac{S^2}{\left[(\Delta\x)^2\right]^3}\right)}{\frac{R}{(\Delta\x)^2}
\left(\sqrt{1+\frac{\epsilon^2}{(\Delta\x)^2}-\frac{R}{(\Delta\x)^2}}-\frac{\sqrt{2}R}{(\Delta\x)^2}\right)
\left(\frac{R}{(\Delta\x)^2}-1-\frac{\epsilon^2S}{\left[(\Delta\x)^2\right]^2}\right)} \right] 
\ , 
\end{aligned}
\end{equation} 
where $S := 3{(\Delta\x)}^2+2\epsilon^2+8{(\Delta t)}^2$.  

Using bounding arguments similar to those in Section~\ref{sec:6d}, 
we find that $R/(\Delta\x)^2=- 1 + O\left(\eta^2\right)$, 
$\epsilon^2/(\Delta\x)^2=O\left(\eta^2\right)$
and 
$(\Delta t)^2/(\Delta\x)^2=O\left(1\right)$, 
and as a consequence 
$S/(\Delta\x)^2=O\left(1\right)$. 
The integrand in \eqref{eq:3d:error:2}
is hence bounded in absolute value by a constant times 
\be
\epsilon^{2}/\left[-{(\Delta\x)}^2\right]^{3/2} \le \epsilon^{2}/s^3\,,
\ee
from which it follows that the integral is of order 
$O\left(\epsilon^{2}/\eta^2\right) = O\left(\eta^2\right)$. 
\par Similarly, we can write $I^{\text{even}}_{>}$ in the form
\bea 
&-2\epsilon\int^{\infty}_{\eta}\,\mathrm{d}s\,\frac{\chi(u-s)\cos{(Es)}\Delta t}{\DxT\sqrt{-\DxT}}\frac{1}{\left(\left(1+\frac{\epsilon^2}{\DxT}\right)^2+4\epsilon^2\frac{\Delta t^2}{\DxF}\right)^{1/2}}\times\\
&\frac{1}{\sqrt{1+\epsilon^2/\DxT+\sqrt{\left(1+\epsilon^2/\DxT\right)^2+4\epsilon^2\Delta t^2/\DxF}}}\,.
\label{eq:3d:IGreaterEven}
\eea
Once again, using the bounding arguments, we find
\bea
\epsilon^2/(\Delta\x)^2&=O\left(\eta^2\right)\,,\\
\epsilon^2\Delta t^2/\DxF &\leq \epsilon^2 M^2/s^2 =O\left(\eta^2\right)\,.
\eea
Hence, the integrand of~\eqref{eq:3d:IGreaterEven} is bounded by a constant times
\be 
\epsilon \Delta t/\left[-\DxT\right]^{3/2}\leq \epsilon/s^2\,,
\ee 
from which it follows the integral is of order $O\left(\eta\right)$.

Collecting, we have 
\begin{equation}
\label{eq:3d:allupper}
I^{\text{even}}_{>}+I^{\text{odd}}_{>}=-\int^{\infty}_{\eta}\,\mathrm{d}s\,
\chi(u-s)\sin{(Es)}\sqrt{\frac{-2}{{(\Delta\x)}^2}}
\ \ +\ O\left(\eta\right)
\ . 
\end{equation}

\subsection{Subinterval $0 < s < \eta$}
\label{sec:3d:smalls}

Consider $I^{\text{odd}}_{<}$, which is given by 
\begin{equation}
\label{eq:3d:lessodd:1}
I^{\text{odd}}_{<}=-\int^{\eta}_{0}\,\mathrm{d}s\,\chi(u-s)
\, \frac{\sin{(Es)}}{R}\sqrt{R-{(\Delta\x)}^2-\epsilon^2}
\ . 
\end{equation}
Writing $s = \epsilon r$
and introducing the book-keeping parameter $z$, as in 
Section~\ref{sec:6d}, 
the counterpart of \eqref{eq:den6d} reads 
\begin{equation}
\begin{aligned}
\label{eq:den3d}
&R^2 
= 
\epsilon^{4}P\Bigg[1-\frac{4\dot{t}\ddot{t}\epsilon r^3}{P}z
+\sum^{n_4-1}_{n=0}\left(2X_{(n+4)}+4T_{(n+4)}\right)\frac{\epsilon^{n+2}r^{n+4}}{P}z^{2+(n/2)}
\\[1ex]
& 
+\sum^{n_5-1}_{n=0}F_{(n+6)}\frac{\epsilon^{n+2}r^{n+6}}{P}z^{1+(n/2)}
+ \frac{O\bigl(\epsilon^{n_4+2}r^{n_4+4}\bigr)}{P}z^{2+(n_4/2)}
+ \frac{O\bigl(\epsilon^{n_5+2}r^{n_5+6}\bigr)}{P}z^{1+(n_5/2)}
\Bigg] \ , 
\end{aligned}
\end{equation}
where the powers of $z$ differ from those in 
\eqref{eq:den6d} because the range of $r$ is now 
$0\le r \le \epsilon^{-1/2}$. 
It follows that in the denominator of 
\eqref{eq:3d:lessodd:1}
we have $R = \epsilon^2 \sqrt{P} \left[1 + O(z)\right]$, and in the numerator we have the factor $\chi(u-s) \sin(Es) = 
\epsilon E r \left[\chi(u) +  O\left(\sqrt{z}\right) \right]$. 

To estimate the square root in the 
numerator in~\eqref{eq:3d:lessodd:1},
we note first that all the terms with a positive power of $z$ in 
\eqref{eq:den3d} are at small $r$ asymptotic to a power of $r$ that is greater than~2. It follows that the same powers of $z$ are retained if these terms are multiplied by any positive function of $r$ that is bounded at small $r$ by a constant times $r^{-3}$ and at large $r$ by a constant. 

Now, we rearrange the quantity under the square root 
in \eqref{eq:3d:lessodd:1} as 
\begin{align}
R-{(\Delta\x)}^2-\epsilon^2 
& = 
\epsilon^2 Q \Biggl[1 
-\sum^{n_6-1}_{n=0} X_{(n+4)} \frac{\epsilon^{n+2} r^{n+4}}{Q}
+ \frac{\sqrt{P}}{Q}
\left(\frac{R}{\epsilon^2 \sqrt{P}} \, -1\right)
\nonumber
\\[1ex]
& \hspace{8ex}
+ \frac{O\bigl(\epsilon^{n_6+2} r^{n_6+4}\bigr)}{Q}
\Biggr]
\ , 
\label{eq:3d:tricky}
\end{align}
where 
\begin{align}
Q :=  \sqrt{P} + r^2 - 1
\end{align}
and the positive integer $n_6$ may be chosen arbitrarily. 
Note that $Q$ is positive for $r>0$, its small-$r$ behaviour is $Q = 2{\dot{t}}^2 r^2 + O\left(r^4\right)$ 
, where the coefficient of $r^2$ is positive, and its behaviour at large $r$ 
is 
$Q/r^2 = 2+ O\left(r^{-2}\right)$. 
We wish to regard the external factor $\epsilon^2 Q$ 
in \eqref{eq:3d:tricky} as the dominant part and the 
terms in the square brackets as a leading 1 plus 
sub-leading corrections. In the terms proportional to  
$X_{(n+4)}$, this is accomplished by 
inserting the book-keeping factors  
$z^{1 + (n/2)}$. From the asymptotic behaviour of $\sqrt{P}/Q$ 
at small and large $r$, we see that in the 
term involving $\sqrt{P}/Q$ 
this is accomplished by taking $R^2$ to be given by~\eqref{eq:den3d}, with the $z$-factors therein. 
A~Taylor expansion in $z^{1/2}$ then shows that 
\be
R-{(\Delta\x)}^2-\epsilon^2 = \epsilon^2 Q\left[1 + O(z)\right]\,.
\ee
%
\par Collecting, we find 
\begin{align}
I^{\text{odd}}_{<}
& =
-\epsilon E\int^{1/\eta}_0\,\mathrm{d}r\,\frac{r \sqrt{Q}}{\sqrt{P}} 
\left[\chi(u)+O\left(\sqrt{z}\,\right)\right] 
\notag
\\
\label{eq:3d:loweroddfinal}
& =O\left(\eta\right)
\ , 
\end{align}
where the final form follows because the integrand 
asymptotes to a constant at large~$r$. 

Consider then $I^{\text{even}}_{<}$, given by 
\begin{equation}
\label{eq:3d:lesseven}
I^{\text{even}}_{<}=
\int^{\eta}_{0}\,\mathrm{d}s\,\chi(u-s)
\, \frac{\cos{(Es)}}{R}\sqrt{R+{(\Delta\x)}^2+\epsilon^2}
\ . 
\end{equation}
We now rearrange the quantity under the square root in 
\eqref{eq:3d:lesseven} as 
\begin{align}
R+{(\Delta\x)}^2+\epsilon^2 
& = 
\epsilon^2 N \Biggl[1 
+\sum^{n_7-1}_{n=0} X_{(n+4)} \frac{\epsilon^{n+2} r^{n+4}}{N}
+ \frac{\sqrt{P}}{N}
\left(\frac{R}{\epsilon^2 \sqrt{P}} \, -1\right)
\nonumber
\\[1ex]
& 
\hspace{8ex} 
+ \frac{O\bigl(\epsilon^{n_7+2} r^{n_7+4}\bigr)}{N}
\Biggr]
\ , 
\label{eq:3d:tricky2}
\end{align}
where 
\begin{align}
N :=  \sqrt{P} + 1 - r^2
\end{align}
and the positive 
integer $n_7$ may be chosen arbitrarily. 
Note that $N$ is positive, 
its small-$r$ behaviour is $N = 2 + O\left(r^2\right)$ 
and its large-$r$ 
behaviour is 
$N = 2{\dot{t}}^2 + O\left(r^{-2}\right)$. 
We wish to regard the external factor $\epsilon^2N$ 
in \eqref{eq:3d:tricky2} as the dominant part. 
In the square brackets, the terms proportional 
to $X_{(n+4)}$ can be given the
book-keeping factors  
$z^{n/2}$, 
while in the term involving $\sqrt{P}/N$, the large-$r$ behaviour of 
$\sqrt{P}/N$ implies that the powers of $z$ inherited from 
\eqref{eq:den3d} must be appropriately decreased. 
Using $F_6 = - 2 X_4$, we find 
\begin{align}
\label{eq:3d:tricky3}
R+{(\Delta\x)}^2+\epsilon^2 
= 
\epsilon^2 N \left[1 
+X_4 \frac{\epsilon^{2} r^{4}}{N}
\left(1 - \frac{r^2}{\sqrt{P}}\right)
+ O\left(\sqrt{z}\right)
\right]
\ . 
\end{align}
Although the term proportional to $X_4$ in the square brackets in 
\eqref{eq:3d:tricky3} has arisen as a combination of two individual terms that came with $z$-factors $z^0$, 
a cancellation between these individual terms at large $r$ implies that the term as a whole can now be reassigned the factor~$z$. We hence have 
$\sqrt{R+{(\Delta\x)}^2+\epsilon^2} 
= 
\epsilon \sqrt{N} \left[1 
+ O\left(\sqrt{z}\right)
\right]$. 
Using this and \eqref{eq:den3d} in $\eqref{eq:3d:lesseven}$, 
we obtain 
\begin{align}
I^{\text{even}}_{<}
& = \int^{1/\eta}_0\,\mathrm{d}r\, \frac{\sqrt{N}}{\sqrt{P}}
\left[\chi(u) +O\left(\sqrt{z}\,\right) \right]
\notag
\\[1ex]
\label{eq:3d:lowerevenfinal}
&=
\frac{\pi\chi(u)}{\sqrt{2}} +O\left(\eta\right)
\ , 
\end{align}
where the final 
form comes by 
 extending the 
 upper limit to $\infty$, at the expense of an 
 error of order $O\left(\eta\right)$, and 
evaluating the elementary integral using~\eqref{eq:App3dMinkInt:I-fin} from Appendix~\ref{ch:appendix:3dint}.

\subsection{Combining the subintervals}

Combining (\ref{eq:3d:allupper}), (\ref{eq:3d:loweroddfinal}) and 
(\ref{eq:3d:lowerevenfinal}), 
we obtain the response function in the final form 
\begin{align}
\label{eq:3d:detrespfinal}
\mathcal{F}\left(E\right)
=
\frac{1}{4}\int^{\infty}_{-\infty}\,\mathrm{d}u\,\chi^2(u)
\ - \ 
\frac{1}{2\pi}\int^{\infty}_{-\infty}\,\mathrm{d}u\,\chi(u)\int^{\infty}_{0}\,\mathrm{d}s \, 
\frac{\chi(u-s) \sin(Es)}{\sqrt{-{(\Delta\x)}^2}}
\ .  
\end{align}
The limit $\eta\to0$ has been taken by just setting the lower limit of the $s$-integral to zero, as the small-$s$ behaviour of the numerator cancels the singularity in the denominator. 
\section{Response function for $d=5$}
\label{sec:5d}

In this section, we remove the regulator from the response function 
formula~\eqref{eq:techIntro:respFunc} for $d=5$, extending 
the technique of Section~\ref{sec:3d}. 

\subsection{Regularisation}
\label{sec:5d:reg}

The regularised $d=5$ Wightman function 
reads \cite{kay-wald,Decanini:2005gt,Langlois-thesis}
\begin{equation}
\label{eq:w5d:1}
W_\epsilon (u,u-s)=\frac{1}{8\pi^2}\frac{1}{{\left[{(\Delta\x)}^2+2i\epsilon\Delta t+\epsilon^2\right]}^{3/2}}
\ . 
\end{equation}
Separating the real and imaginary parts and substituting~\eqref{eq:w5d:1} into 
\eqref{eq:techIntro:respFunc-alt}, the result is 
\begin{equation}
\begin{aligned}
\label{eq:5d:detres}
&\mathcal{F}\left(E\right)=\lim_{\epsilon\to 0}
\frac{1}{4\sqrt{2}\,\pi^2}\int^{\infty}_{-\infty}\,\mathrm{d}u\,\chi(u)\,\int^{\infty}_{0}\,\mathrm{d}s\,\frac{\chi(u-s)}{R^3} \, \times\\
&\times 
\Bigg[\cos(Es) 
\left( \bigl({(\Delta\x)}^2+\epsilon^2\bigr)\sqrt{R+{(\Delta\x)}^2+\epsilon^2}-2\epsilon(\Delta t)\sqrt{R-{(\Delta\x)}^2-\epsilon^2} \, \right) 
\\
&\hspace{4ex}
-\sin(Es) 
\left( \bigl({(\Delta\x)}^2+\epsilon^2\bigr)\sqrt{R-{(\Delta\x)}^2-\epsilon^2}+2\epsilon(\Delta t)\sqrt{R+{(\Delta\x)}^2+\epsilon^2} \, \right)\Bigg]
\ . 
\end{aligned}
\end{equation}
Working under the expression 
${(4\sqrt{2}\,\pi^2)}^{-1} \int^{\infty}_{-\infty}\,\mathrm{d}u\,\chi(u)$, we write the integral 
over $s$ as the sum $I^{\text{even}}_{<} + I^{\text{odd}}_{<} +
I^{\text{even}}_{>} + I^{\text{odd}}_{>}$, 
choosing $\eta:=\epsilon^{1/4}$ as in Section~\ref{sec:6d}. 

\subsection{Subinterval $\eta < s < \infty$}
\label{sec:5d:larges}
Moreover, we split the $I^{\text{even}}_{>}$ part into two parts, which are defined as
\bea
I^{\text{even}}_{>\,1}&:=\int^{\infty}_{\eta}\,\mathrm{d}s\,\frac{\chi(u-s)}{R^3}\cos(Es)\left( \bigl({(\Delta\x)}^2+\epsilon^2\bigr)\sqrt{R+{(\Delta\x)}^2+\epsilon^2}\right)\,,\\
I^{\text{even}}_{>\,2}&:=-\int^{\infty}_{\eta}\,\mathrm{d}s\,\frac{\chi(u-s)}{R^3}\cos(Es)\left( 2\epsilon(\Delta t)\sqrt{R-{(\Delta\x)}^2-\epsilon^2} \, \right)\,.
\eea
$I^{\text{even}}_{>\,1}$ can then be expressed as
\bea
I^{\text{even}}_{>\,1}&=-2\int^{\infty}_{\eta}\,\mathrm{d}s\,\chi(u-s)\cos{(Es)}\frac{\epsilon \Delta t}{[(\Delta\x)^2]^2\sqrt{-(\Delta\x)^2}}\left(1+\frac{\epsilon^2}{(\Delta\x)^2}\right)\times\\
&\left[\frac{1}{\left(\left(1+\frac{\epsilon^2}{\DxT}\right)^2+4\epsilon^2\frac{\Delta t^2}{\DxF}\right)^{3/2}}\frac{1}{\sqrt{1+\frac{\epsilon^2}{\DxT}+\sqrt{\left(1+\frac{\epsilon^2}{\DxT}\right)^2+\frac{4\epsilon^2\Delta t^2}{\DxF}}}}\right]\,,
\eea
and $I^{\text{even}}_{>\,2}$ can be expressed as
\bea
I^{\text{even}}_{>\,2}&=-2\epsilon\int^{\infty}_{\eta}\,\mathrm{d}s\,\chi(u-s)\cos{(Es)}\frac{\Delta t \sqrt{-\DxT}}{[-\DxT]^3}\times\\
&\left[\frac{\sqrt{1+\epsilon^2/\DxT+\sqrt{\left(1+\epsilon^2/\DxT\right)^2+4\epsilon^2\Delta t^2/\DxF}}}{\left(\left(1+\frac{\epsilon^2}{\DxT}\right)^2+4\epsilon^2\frac{\Delta t^2}{\DxF}\right)^{3/2}}\right]\,.
\eea
Using the bounding arguments from Section~\ref{sec:6d}, we see that $I^{\text{even}}_{>}=O(\eta)$.
\par When we set the regulator to zero in $I^{\text{odd}}_{>}$, the integrand reduces to the form $\chi(u-s)\sin{(Es)}\sqrt{-2/[\DxT]^3}$, and this replacement creates an error which can be arranged in the form
\bea
&-\int^{s_c}_{\eta}\,\mathrm{d}s\,\frac{\chi(u-s)\sin{(Es)}}{R^3\DxF}\Bigg[\DxF\left(\DxT+\epsilon^2\right)\sqrt{R-\DxT-\epsilon^2}\\
&\qquad\qquad\qquad\qquad\qquad+2\epsilon\DxF\Delta t\sqrt{R+\DxT+\epsilon^2}-R^3\sqrt{-2\DxT}\Bigg]\,,
\eea
where $s_c>0$ and is a real constant that is large enough such that $\chi(u-s_c)=0$. With some algebra, this can be shown to be equal to
\bea 
&4\epsilon^2\int^{s_c}_{\eta}\,\mathrm{d}s\,\frac{\chi(u-s)\sin{(Es)}\Delta t^2}{[\DxT]^3\sqrt{-\DxT}}\times\\
&\left[\frac{1}{\sqrt{1+\frac{\epsilon^2}{\DxT}+\sqrt{\left(1+\frac{\epsilon^2}{\DxT}\right)^2+\frac{4\epsilon^2\Delta t^2}{\DxF}}}}\frac{1}{\left(\left(1+\frac{\epsilon^2}{\DxT}\right)^2+4\epsilon^2\frac{\Delta t^2}{\DxF}\right)^{3/2}}\right]\\
&+\epsilon^2\int^{s_c}_{\eta}\,\mathrm{d}s\,\frac{\chi(u-s)\sin{(Es)}\sqrt{-\DxT}}{[\DxT]^3}\times\\
&\qquad\qquad\qquad\qquad\left[\frac{\sqrt{1+\frac{\epsilon^2}{\DxT}+\sqrt{\left(1+\frac{\epsilon^2}{\DxT}\right)^2+\frac{4\epsilon^2\Delta t^2}{\DxF}}}}{\left(\left(1+\frac{\epsilon^2}{\DxT}\right)^2+4\epsilon^2\frac{\Delta t^2}{\DxF}\right)^{3/2}}\right]\\
&-\epsilon^2\int^{s_c}_{\eta}\,\mathrm{d}s\,\frac{\chi(u-s)\sin{(Es)}}{\DxF\sqrt{-\DxT}}\left(2+\frac{\epsilon^2}{\DxT}+\frac{4\Delta t^2}{\DxT}-\frac{2V}{[\DxT]^5}-\frac{\epsilon^2V^2}{[\DxT]^{11}}\right)\times\\
&\Bigg[\frac{1}{\left(\frac{R}{\DxT}\right)^3\left(\sqrt{1+\epsilon^2/\DxT+\sqrt{\left(1+\epsilon^2/\DxT\right)^2+4\epsilon^2\Delta t^2/\DxF}}+\sqrt{2}\left(\frac{R}{\DxT}\right)^3\right)}\\
&\qquad\qquad\qquad\qquad\qquad\qquad\qquad\qquad\qquad\qquad\qquad\qquad\times\frac{1}{\left(\frac{R}{\DxT}-1-\frac{\epsilon^2V}{[\DxT]^6}\right)}\Bigg]\,,
\label{eq:5d:OddErrorAlt}
\eea
with
\bea
V:&=\left(11[\DxT]^5+24[\DxT]^4\Delta t^2\right)\\
&+\left(30[\DxT]^4+96\DxF\Delta t^4+96[\DxT]^3\Delta t^2\right)\epsilon^2\\
&+\left(144\DxF\Delta t^2+40[\DxT]^3+128\Delta t^6+192\DxT\Delta t^4\right)\epsilon^4\\
&+\left(96\DxT\Delta t^2+96\Delta t^4+30\DxF\right)\epsilon^6+\left(12\DxT+24\Delta t^2\right)\epsilon^8+2\epsilon^{10}\,,
\eea
and estimates similar to those used above show that~\eqref{eq:5d:OddErrorAlt} is bounded by a term of order $O(\eta^3)$.
Collecting both the even and odd pieces, we have
\begin{align}
\label{eq:5d:upper}
I^{\text{odd}}_{>}+I^{\text{even}}_{>}=\int^{\infty}_{\eta}\,\mathrm{d}s\,\chi(u-s)\sin{(Es)}\sqrt{\frac{-2}{\left[{(\Delta\x)}^2\right]^3}}
\ \ +\ O\left(\eta\right)
\ .
\end{align}
\subsection{Subinterval $0 < s < \eta$}
\label{sec:5d:smalls}

Consider $I^{\text{even}}_{<}$ 
and $I^{\text{odd}}_{<}$, given by 
\begin{subequations}
\label{eq:5dlessboth}
\begin{align}
I^{\text{even}}_{<}&=\int^{\eta}_{0}\,\mathrm{d}s\,\chi(u-s)\frac{\cos{(Es)}}{R^3} \, \times 
\notag 
\\[1ex]
& \hspace{2ex}
\times 
\left[ 
\bigl({(\Delta\x)}^2+\epsilon^2\bigr)\sqrt{R+{(\Delta\x)}^2+\epsilon^2}
-2\epsilon
(\Delta t)\sqrt{R-{(\Delta\x)}^2-\epsilon^2} \, 
\right]
\ , 
\label{eq:5dlesseven}
\\
I^{\text{odd}}_{<}&= 
- \int^{\eta}_{0}\,\mathrm{d}s\,\chi(u-s)\frac{\sin{(Es)}}{R^3} \, \times 
\notag 
\\[1ex]
& \hspace{4ex}
\times 
\left[ 
\bigl({(\Delta\x)}^2+\epsilon^2\bigr)\sqrt{R-{(\Delta\x)}^2-\epsilon^2}
+2\epsilon
(\Delta t)\sqrt{R+{(\Delta\x)}^2+\epsilon^2} \, 
\right]
\ .
\label{eq:5dlessodd}
\end{align}
\end{subequations}
In the $R^3$ in the denominators, we use~\eqref{eq:den6d}. 
In the square root 
$\sqrt{R-{(\Delta\x)}^2-\epsilon^2}$ in the numerators, 
we use~\eqref{eq:3d:tricky}, 
inserting the factors $z^{(n+2)/4}$ 
in the terms proportional to $X_{(n+4)}$ and using \eqref{eq:den6d} in the last term of~\eqref{eq:3d:tricky} for~$R$.
By the asymptotic behaviour of $\sqrt{P}/Q$ and the observations made in Section~\ref{sec:3d}, this makes $z$ into an appropriate parameter for organising the square brackets in \eqref{eq:3d:tricky} 
into a Taylor expansion in $z^{1/4}$ with the leading term~$1$. 
\par The full expansion of $R-\DxT-\epsilon^2$  up to and including terms of order $z^2$ is
\bea
&R-\DxT-\epsilon^2=\epsilon^2 Q\Bigg[1+\frac{\ddot{\x}^2\epsilon^2 r^4}{12Q}\left(1+\frac{r^2}{\sqrt{P}}\right)\sqrt{z}+\frac{\dx\cdot\ddddx\epsilon^3 r^5}{36 Q}\left(1+\frac{r^2}{\sqrt{P}}\right)z^{3/4}\\
&+\left[\left(\frac{\ddx\cdot\ddddx}{40}+\frac{[\dddx]^2}{45}\right)\left(1+\frac{r^2}{\sqrt{P}}\right)\frac{\epsilon^4 r^6}{Q}+\frac{2\epsilon r^3 T_3}{\sqrt{P} Q}\right]z\\
&-\left(\frac{\ddx\cdot\dddddx}{180}+\frac{\dddx\cdot \ddddx}{72 }\right)\left(1+\frac{r^2}{\sqrt{P}}\right)\frac{\epsilon^5 r^7}{Q}z^{5/4}\\
&+\left(\frac{[\ddddx]^2}{448}+\frac{\ddx\cdot\x^{(6)}}{1008}+\frac{\dddx\cdot\dddddx }{315}\right)\left(1+\frac{r^2}{\sqrt{P}}\right)\frac{\epsilon^6 r^8}{Q}z^{3/2}\\
&-\left[\left(\frac{\dddx\cdot\x^{(6)}}{1728 }+\frac{\ddx\cdot\x^{(7)}}{6720}+\frac{\ddddx\cdot\x^{(5)}}{960 }\right)\left(1+\frac{r^2}{\sqrt{P}}\right)\frac{\epsilon^7 r^9}{Q}\right]z^{7/4}\\
&+\Bigg[\left(2T_4-\frac{\ddx^2}{12}\right)\frac{\epsilon^2 r^4}{Q\sqrt{P}}-\frac{2\epsilon^2 r^6 T_3^2}{Q P^{3/2}}\\
&+\left(\frac{\dddx\cdot\x^{(7)}}{11340}+\frac{\ddx\cdot\x^{(8)}}{51840 }+\frac{[\x^{(5)}]^2}{8100}+\frac{\ddddx\cdot\x^{(6)}}{5184}\right)\left(1+\frac{r^2}{\sqrt{P}}\right)\frac{\epsilon^8 r^{10}}{Q}\Bigg]z^2+O\left(z^{9/4}\right)\Bigg]\,.
\label{eq:5d:RminusDX2E2}
\eea
\par In the square root 
$\sqrt{R+{(\Delta\x)}^2+\epsilon^2}$ in the numerator of~\eqref{eq:5dlessboth}, 
we wish to use~\eqref{eq:3d:tricky2}. 
Attempting to regard the $1$ in the square brackets as the dominant term can at first sight seem problematic because the terms 
proportional to $X_{(n+4)}$ acquire the $z$-factors 
$z^{-1 + (n/4)}$, where the exponent is non-positive for $n\le 4$, and 
when the last term is Taylor expanded in $z^{1/4}$ 
using~\eqref{eq:den6d}, the asymptotic behaviour of the factor 
$\sqrt{P}/N$ implies that the exponents of $z$ must be appropriately decreased and some of these decreased exponents are non-positive. 
However, the non-positive powers of $z$ coming from the last term and from the terms proportional to $X_{(n+4)}$ 
can be grouped into combinations that can be reassigned 
positive powers of~$z$, similarly to what happened for $d=3$ in~\eqref{eq:3d:tricky3}. 
After these reassignments, we obtain for 
$\sqrt{R+{(\Delta\x)}^2+\epsilon^2}$ a Taylor expansion in $z^{1/4}$
that starts as 
$\epsilon\sqrt{N}\left[1+O\left(z^{1/4}\right)\right]$; owing to the size of this result, we do not reproduce the full expression up to order $z^2$ here.
We use Maple (TM) to Taylor expand the square roots of~\eqref{eq:5d:RminusDX2E2} and the large, $z$-ordered expression obtained for $\sqrt{R+{(\Delta\x)}^2+\epsilon^2}$, using $z$ as the expansion parameter and keeping powers of $z$ up to and including $z^2$. 
\par We next split $I^{\text{even}}_{<}$ in two, as follows
\bea
I^{\text{even}}_{<\,1} &:=\int^{\eta}_0\,\mathrm{d}s\,\frac{\chi(u-s)\cos{(Es)}}{R^3}\left(\DxT+\epsilon^2\right)\sqrt{R+\DxT+\epsilon^2}\,,\\
I^{\text{even}}_{<\,2} &:=-2\epsilon\int^{\eta}_0\,\mathrm{d}s\,\frac{\chi(u-s)\cos{(Es)}}{R^3}\Delta t\sqrt{R-\DxT-\epsilon^2}\,.
\eea
After a change of variables $s=\epsilon r$ and using the expansions previously discussed for the terms in the denominator and numerator, we obtain
\bea
&I^{\text{even}}_{<\,1}=\frac{1}{\epsilon^2}\int^{\epsilon^{-3/4}}_0\,\mathrm{d}r\,\sqrt{\frac{N}{P}}\times\\
& \Bigg[1-\frac{4\dot{t}\ddot{t}\epsilon r^3}{P}z+\left(2X_4+4T_4\right)\frac{\epsilon^{2}r^{4}}{P}z^{2}+\sum^{6}_{n=0}F_{(n+6)}\frac{\epsilon^{n+2}r^{n+6}}{P}z^{(n+2)/4}+O\left(z^{9/4}\right)\Bigg]^{-3/2}\times\\
&\left[\chi(u)-\epsilon r\dot{\chi}(u)z^{1/4}+\cdots+\frac{1}{8!}\chi^{(8)}(u)\epsilon^8 r^8 z^2+O\left(z^{9/4}\right)\right]\times \\
&\left[1-\frac{1}{2!}\epsilon^2E^2r^2\sqrt{z}+\cdots+\frac{1}{8!}\epsilon^8 E^8 r^8 z^2+O\left(z^{5/2}\right)\right]\times \\
&\left[\frac{1-r^2}{P}+\frac{X_4 \epsilon^2 r^4}{P}z^2+O\left(z^{9/4}\right)\right]\times\\
&\left[1+\frac{\epsilon r^3 T_3}{N\sqrt{P}}z^{1/4}+\cdots+\Big(\cdots\Big) z^2+O\left(z^{9/4}\right)\right]\,.
\label{eq:5d:lesseven1Exp}
\eea
If we analyse the integral
\be 
\int^{\epsilon^{-3/4}}_0\,\mathrm{d}r\,\frac{\sqrt{N}}{\sqrt{P}}\,,
\ee 
we find that this is of the form
\be 
\int^{\epsilon^{-3/4}}_0\,\mathrm{d}r\,\frac{\sqrt{N}}{\sqrt{P}}=B_{+}+\mathcal{O}\left(\epsilon^{3/4}\right)\,,
\ee
with $B_+>0$ and constant, which is the justification for expanding the remaining integrand of~\eqref{eq:5d:lesseven1Exp} up to and including order $z^2$.
Similarly, we analyse $I_{<\,2}^{\text{even}}$ finding
\bea
&I^{\text{even}}_{<\,2}=-\frac{2}{\epsilon^2}\int^{\epsilon^{-3/4}}_0\,\mathrm{d}r\,\sqrt{\frac{Q}{P}}\times\\
& \Bigg[1-\frac{4\dot{t}\ddot{t}\epsilon r^3}{P}z+\left(2X_4+4T_4\right)\frac{\epsilon^{2}r^{4}}{P}z^{2}+\sum^{6}_{n=0}F_{(n+6)}\frac{\epsilon^{n+2}r^{n+6}}{P}z^{(n+2)/4}+O\left(z^{9/4}\right)\Bigg]^{-3/2}\times\\
&\left[\chi(u)-\epsilon r\dot{\chi}(u)z^{1/4}+\cdots+\frac{1}{8!}\chi^{(8)}(u)\epsilon^8 r^8 z^2+O\left(z^{9/4}\right)\right]\times \\
&\left[1-\frac{1}{2!}\epsilon^2E^2r^2\sqrt{z}+\cdots+\frac{1}{8!}\epsilon^8 E^8 r^8 z^2+O\left(z^{5/2}\right)\right]\times \\
&\left[\frac{\dot{t}r}{P}-\frac{\ddot{t} \epsilon r^2}{2P}z+\frac{t^{(3)}\epsilon^2r^3}{6P}z^2+O\left(z^{9/4}\right)\right]\times\\
&\left[1+\frac{\ddot{\x}^2\epsilon^2 r^4}{24 Q}\left(1+\frac{r^2}{\sqrt{P}}\right)\sqrt{z}+\cdots+\Big(\cdots\Big)z^2+O\left(z^{9/4}\right)\right]\,.
\label{eq:5d:lesseven2Exp}
\eea
If we consider the integral
\be
\int^{\epsilon^{-3/4}}_0\,\mathrm{d}r\,\sqrt{\frac{Q}{P}}\,,
\ee
we find that is of the form
\be 
\int^{\epsilon^{-3/4}}_0\,\mathrm{d}r\,\sqrt{\frac{Q}{P}}=C_{+}+O\left(\log{\left(\epsilon\right)}\right)\,,
\ee
with $C_{+}>0$ and constant, which justifies expanding the remaining integrand of~\eqref{eq:5d:lesseven2Exp} up to and including terms of order $z^2$.

We use Maple (TM) to perform the algebraic manipulations necessary to multiply the factors in the integrands of~\eqref{eq:5d:lesseven1Exp} and~\eqref{eq:5d:lesseven2Exp}, dropping powers of $z$ that are too high to contribute in the $\epsilon\to0$ limit, and then finally setting $z=1$. The evaluation of $I_{<}^{\text{odd}}$ proceeds almost identically and we obtain for 
$I^{\text{even}}_{<}$ 
and $I^{\text{odd}}_{<}$ formulas that consist of 
sums of finitely many elementary integrals 
plus an error term that vanishes in the $\epsilon\to0$ limit. 
The elementary integrals are 
of the form 
$\int^{\eta^{-3}}_{0}\,\mathrm{d}r\,r^n N^{\pm 1/2} P^{-m}$, 
$\int^{\eta^{-3}}_{0}\,\mathrm{d}r\,r^n N^{-3/2} P^{-m}$, 
$\int^{\eta^{-3}}_{0}\,\mathrm{d}r\,r^n Q^{\pm 1/2} P^{-m}$ 
and 
$\int^{\eta^{-3}}_{0}\,\mathrm{d}r\,r^n Q^{-3/2} P^{-m}$, 
where $n$ is a positive integer and $m$ is a positive 
integer or half-integer.

\subsection{Combining the subintervals}

Evaluating the numerous elementary integrals 
that came from \eqref{eq:5dlessboth}, 
combining the results with 
\eqref{eq:5d:upper} 
and proceeding as in Section~\ref{sec:6d}, we 
find from 
\eqref{eq:5d:detres}
that the response function is given by  
\begin{align}
\mathcal{F}\left(E\right)
&=\frac{1}{64\pi}\int^{\infty}_{-\infty}\,\mathrm{d}u\,\left[\chi^2\left(4E^2+\ddot{\x}^2\right)+4\dot{\chi}^2\right]
\notag
\\
&\hspace{2ex}
+\lim_{\eta\to 0}\frac{1}{4\pi^2}\int^{\infty}_{-\infty}\,\mathrm{d}u \, \chi(u)\,\int^{\infty}_{\eta}\,\mathrm{d}s\,
\left(
\frac{\chi(u-s) \sin{\left(Es\right)}}{\sqrt{\bigl[-{(\Delta\x)}^2\bigr]^3}}-\frac{E\chi(u)}{s^2}
\right)
\,.
\end{align}
To take the limit 
$\eta\to0$, 
we add and subtract under the $s$-integral terms that 
disentangle the small-$s$ divergence of 
$\sin{\left(Es\right)} \bigl[-{(\Delta\x)}^2\bigr]^{-3/2}$ 
from the small-$s$ behaviour of $\chi(u-s)$. 
Proceeding as in Section~\ref{sec:6d}, we find 
\begin{equation}
\begin{aligned}
\label{eq:5d:detresp:ssform}
\mathcal{F}\left(E\right)&=\frac{1}{64\pi}\int^{\infty}_{-\infty}\,\mathrm{d}u\,\left[\chi^2(u)\left(4E^2+\ddot{\x}^2\right)+4\dot{\chi}^2(u)\right]
\\[1ex]
&\ +\frac{E}{4\pi^2}\int^{\infty}_{0}\,\frac{\mathrm{d}s}{s^2}\int^{\infty}_{-\infty}\,\mathrm{d}u\, \chi(u)\left[\chi(u-s)-\chi(u)\right]\\
&\ +\frac{1}{4\pi^2}\int^{\infty}_{-\infty}\,\mathrm{d}u\, \,\chi(u) 
\int^{\infty}_{0}\,\mathrm{d}s\,\chi(u-s)
\left(\frac{\sin{(Es)}}{\sqrt{\left[-{(\Delta\x)}^2\right]^3}}-\frac{E}{s^2}\right)
\ . 
\end{aligned}
\end{equation}
\section{Sharp-switching limit}
\label{sec:sharp}

In this section we consider the limit in which the switching function approaches a step-function of unit height and fixed duration. Concretely, we take \cite{satz:smooth,satz-louko:curved}
\begin{equation}
\label{eq:ss:switchingfunc}
\chi(u)=h_1\left(\frac{u-\tau_0+\delta}{\delta}\right)
\times 
h_2\left(\frac{-u+\tau+\delta}{\delta}\right)
\ , 
\end{equation}
where the parameters 
$\tau$, $\tau_0$ and $\delta$ satisfy 
$\tau>\tau_0$ and $\delta>0$, 
and 
$h_1$ and $h_2$ are smooth non-negative functions satisfying 
$h_i(x)=0$ for $x\le0$ and $h_i(x)=1$ for $x\ge1$. 
In words, the detector
is switched on over an 
interval of duration $\delta$ just before 
proper time~$\tau_0$, 
it stays on until proper time~$\tau$, 
and it is switched off over an interval of duration 
$\delta$ just after proper time~$\tau$. 
The manner of the switch-on and switch-off 
is specified respectively by the functions $h_1$ and~$h_2$. 
The limit of sharp switching is then $\delta\to0$, 
with $\tau_0$ and $\tau$ fixed. 

We denote the response function by~$\mathcal{F}_\tau$, 
where the subscript serves as an explicit reminder of the dependence on the switch-off moment~$\tau$. We are interested both in $\mathcal{F}_\tau$ and in its derivative with respect to~$\tau$, which we denote by $\mathcal{\dot{F}}_{\tau}$. As mentioned in Chapter~\ref{ch:techIntro}, 
$\mathcal{\dot{F}}_{\tau}$ can be regarded as the detector's instantaneous transition rate per unit proper time, observationally meaningful in terms of a series of measurements in identical ensembles of detectors~\cite{satz-louko:curved}. 

The case of two-dimensional Minkowski spacetime, $d=2$, 
was discussed previously. 
We shall address the cases from 
$d=3$ to $d=6$ in the following subsections. 

\subsection{$d=3$}

For $d=3$, $\mathcal{F}_\tau$ is given by~\eqref{eq:3d:detrespfinal}. 
The limit $\delta\to0$ is well defined and 
can be taken directly in~\eqref{eq:3d:detrespfinal}. To take the limit of the first term of~\eqref{eq:3d:detrespfinal}, we substitute in the switching function~\eqref{eq:ss:switchingfunc} and momentarily drop the $1/4$ pre-factor to obtain
\be 
\int^{\infty}_{-\infty}\,\mathrm{d}u \,h_1^2\left(\frac{u-\tau_0+\delta}{\delta}\right)h_2^2\left(\frac{-u+\tau+\delta}{\delta}\right)\,,
\ee
before changing variables as $v\to b+1-v$, where $b:=1+\Delta\tau/\delta$, to obtain
\be 
\delta \int^{\infty}_{-\infty}\,\mathrm{d} v\,h_1^2(v)h_2^2(b+1-v)\,.
\ee
Only the range $(0,b+1)$ can make a contribution, and to evaluate this integral we split the $v$-integral into three sub-intervals $(0,1),(1,b),(b,b+1)$, which we call $I_{1,2,3}$. For $I_1$, we have
\be 
I_1=\delta \int^1_0\,\mathrm{d} v\,h_1^2(v)\,.
\ee
For $I_2$, we get
\be 
I_2=\delta \int^{b}_1\,\mathrm{d} v\,=\Delta \tau\,,
\ee
and, finally, for $I_3$:
\bea 
I_3&=\delta\int^{b+1}_b\,\mathrm{d} v\,h_2^2(b+1-v)\\
&=\delta\int^1_0\,\mathrm{d} v\,h_2^2(v)\,,
\eea
where to obtain the second equality for $I_3$ we have changed variables as $v\to b+1-v$.
Combining $I_{1,2,3}$ and restoring the $1/4$ pre-factor, we find that in the sharp-switching limit the first term of~\eqref{eq:3d:detrespfinal} is
\be 
\frac{\Delta\tau}{4}+\frac{\delta}{4}\int^1_0\,\mathrm{d} v\,\Big(h_1^2(v)+h_2^2(v)\Big)\,.
\ee
Combining this with the second term of~\eqref{eq:3d:detrespfinal} in the sharp-switching limit, the result is
\begin{align}
\label{eq:ss:3d:detrespSS}
\mathcal{F}_\tau (E)
=\frac{\Delta\tau}{4}
-\frac{1}{2\pi}\int^{\tau}_{\tau_0}\,\mathrm{d}u\,\int^{u-\tau_0}_{0}\,\mathrm{d}s\,\frac{\sin{(Es)}}{\sqrt{-{(\Delta\x)}^2}}
\ , 
\end{align}
Differentiation with respect to $\tau$ gives 
\begin{align}
\label{eq:ss:3d:transrate}
\dot{\mathcal{F}}_{\tau}\left(E\right)
=
\frac{1}{4}-\frac{1}{2\pi}\int^{\Delta\tau}_{0}\,\mathrm{d}s\,\frac{ \sin{\left(Es\right)}}{\sqrt{-\left(\Delta\x\right)^2}}
\ . 
\end{align}
\subsection{$d=4$}

The case $d=4$ was addressed in~\cite{satz:smooth}. 
The expression for the response function with 
a general switching function reads 
\begin{align}
\mathcal{F}(E)
&= -\frac{E}{4\pi}\int_{-\infty}^{\infty}\mathrm{d}u\,\chi^2(u)\,
+ \frac{1}{2\pi^2}\int_0^{\infty}\frac{\mathrm{d}s}{s^2}\int_{-\infty}^{\infty}\mathrm{d}u\,\chi(u)\left[ \chi(u)-\chi(u-s)\right]
\notag
\\[1ex]
& 
\hspace{3ex}
+\frac{1}{2\pi^2}\int_{-\infty}^{\infty}\mathrm{d}u\,\chi(u)\int_0^{\infty}\mathrm{d}s\,\chi(u-s)\left( \frac{\cos(Es)}{{(\Delta \x)}^2}+\frac{1}{s^2}\right) 
\ . 
\label{eq:4dFraw}
\end{align}
The first and third terms in \eqref{eq:4dFraw} have well-defined limits as $\delta\to0$.  
The second term in \eqref{eq:4dFraw} takes at small $\delta$ the form ${(2\pi^2)}^{-1} \ln(\Delta\tau/\delta) + C + O(\delta/\Delta\tau)$, where $C$ is a constant determined by the functions $h_1$ and~$h_2$, and this term hence diverges logarithmically as $\delta\to0$. However, the $\tau$-derivative of this term remains finite as $\delta\to0$, and 
the transition rate has the well-defined limit 
\begin{align}
\label{eq:ss:4d:transrate}
\dot{\mathcal{F}}_{\tau}\left(E\right)
=
-\frac{E}{4\pi}+\frac{1}{2\pi^2}
\int_0^{\Delta\tau}\textrm{d}s
\left( 
\frac{\cos (Es)}{{(\Delta \x)}^2} 
+ 
\frac{1}{s^2} 
\right) 
\ \ +\frac{1}{2\pi^2 \Delta \tau}
\ . 
\end{align}

\subsection{$d=5$}

For $d=5$, 
$\mathcal{F}_\tau$ is given by~\eqref{eq:5d:detresp:ssform}. 
The last term in \eqref{eq:5d:detresp:ssform} has a well-defined 
limit as $\delta\to0$. The first term can be analysed by substituting in the switching function~\eqref{eq:ss:switchingfunc}, for now ignoring the $1/64\pi$ pre-factor, to obtain
\bea
&\int^{\infty}_{-\infty}\,\mathrm{d} u\,\Bigg[h_1^2\left(\frac{u-\tau_0+\delta}{\delta}\right)h_2^2\left(\frac{-u+\tau+\delta}{\delta}\right)\left(\ddx^2(u)+4E^2\right)\\
&+\frac{4}{\delta^2}\Bigg(h^{'2}_1\left(\frac{u-\tau_0+\delta}{\delta}\right)h^2_2\left(\frac{-u+\tau+\delta}{\delta}\right)+h^{2}_1\left(\frac{u-\tau_0+\delta}{\delta}\right)h^{'2}_2\left(\frac{-u+\tau+\delta}{\delta}\right)\\
&\qquad\qquad-2\huOne\dhuOne\huTwo\dhuTwo\Bigg)\Bigg]\,,
\eea
which can be expressed as
\bea 
&\int^{\infty}_{-\infty}\,\mathrm{d} v\,\Bigg[\delta h_1^2(v)h_2^2(b+1-v)\Big(4E^2+\ddxxZero\Big)\\
&+\frac{4}{\delta}\Big(h_1^{'2}(v)h_2^2(b+1-v)+h_1^2(v)h_2^{'2}(b+1-v)\\
&\qquad\qquad\qquad\qquad\qquad\qquad-2\hvOne\dhvOne\hbvTwo\dhbvTwo\Big)\Bigg]
\eea
through the change of variables $v=\left(u-\tau_0+\delta\right)/\delta$ and the definition $b:=1+\Delta\tau/\delta$. Recalling the definition of $h_i(x)$, only the range $(0,b+1)$ can contribute to the $v$-integral, and to evaluate this expression we split the $v$-integral into the intervals $(0,1),(1,b),(b,b+1)$, which we label as $I_{1,2,3}$ respectively. For $I_1$, we have
\be
I_1=\int^1_0\,\mathrm{d} v\left[\delta h_1^2(v)\Big(4E^2+\ddxxZero\Big)+\frac{4}{\delta}h_1^{'2}(v)\right]\,,
\label{eq:5d:ss:I1}
\ee
which we note is a constant independent of the switch-off time, $\tau$. For $I_2$, we obtain
\be  
I_2=\delta \int^{b}_1\,\mathrm{d} v\,\Big(4E^2+\ddxxZero\Big)
\label{eq:5d:ss:I2}
\ee
and thus
\be  
\frac{dI_2}{d\tau}=4E^2+\ddx^2(\tau)\,.
\label{eq:5d:ss:dI2}
\ee
Finally, for $I_3$ we find
\bea 
I_3&=\int^{b+1}_b\,\mathrm{d} v\, \left[\delta h_2^2(b+1-v)\Big(4E^2+\ddxxZero\Big)+\frac{4}{\delta}h_2^{'2}(b+1-v)\right]\\
&=\int^1_0\,\mathrm{d} v\,\left[\delta h_2^2(v)\Big(4E^2+\ddxx\Big)+\frac{4}{\delta}h_2^{'2}(v)\right]\,,
\label{eq:5d:ss:I3}
\eea
where to obtain the second equality we have changed variables as $v\to b+1-v$. This implies that
\bea 
\frac{dI_3}{d\tau}=O\left(\delta\right)\,.
\label{eq:5d:ss:dI3}
\eea
Restoring the pre-factor $1/64\pi$ and combining the derivatives,~\eqref{eq:5d:ss:dI2} and~\eqref{eq:5d:ss:dI3}, we see that in the sharp-switching limit, $\delta\to 0$, the first term of~\eqref{eq:5d:detresp:ssform} takes the form
\be 
\frac{1}{64\pi}\left[4E^2+\ddx^2(\tau)\right]\,.
\ee 
In summary, the part of the 
first term that contains~$\chi^2$ has a well-defined limit as $\delta \to 0$. 
The part of the first term that contains
${\dot\chi}^2$ equals~$C'/\delta$, 
where $C'$ is a positive constant defined by
\be 
C':=\frac{1}{16\pi}\int^1_0\,\mathrm{d} v\,\left(\left[\dhvOne\right]^2+\left[\dhvTwo\right]^2\right)\,.
\ee
This part diverges as $\delta\to0$ but is 
independent of $\tau$ and does, therefore, 
not contribute to~$\dot{\mathcal{F}}_{\tau}$. 
Finally, the second term in \eqref{eq:5d:detresp:ssform}
is similar to the second term in the $d=4$ formula 
\eqref{eq:4dFraw}, being logarithmically divergent as $\delta\to0$ 
but having a $\tau$-derivative that has a well-defined limit as 
$\delta\to0$. 

Collecting, we find that the transition rate has a 
well-defined $\delta\to0$ limit, given by  
\begin{align}
\label{eq:ss:5d:transrate}
\dot{\mathcal{F}}_{\tau}\left(E\right)
=
\frac{4E^2 + \ddot{\x}^2(\tau)}{64\pi}
+ 
\frac{1}{4\pi^2}\int^{\Delta\tau}_{0}\,\mathrm{d}s
\left(\frac{\sin{(Es)}}{\sqrt{\left[-{(\Delta\x)}^2\right]^3}}-\frac{E}{s^2}\right)
\ -\frac{E}{4\pi^2\Delta\tau}
\ . 
\end{align}

\subsection{$d=6$}

For $d=6$, 
$\mathcal{F}_\tau$ is given by~\eqref{eq:6d:detresp}.
The last term in~\eqref{eq:6d:detresp} remains finite as 
$\delta\to0$. The first and second terms are similar to those encountered in $d=5$, with contributions that diverge in the $\delta\to0$ limit 
proportionally to $1/\delta$ and $\ln\delta$, but with $\tau$-derivatives that remain finite in this limit.  
\par We shall analyse the sharp-switching limit of the third and fourth terms of~\eqref{eq:6d:detresp} in Appendix~\ref{ch:appendix:6dSSL}. 
The third and fourth terms can be handled by breaking the integrations into subintervals as in~\cite{satz:smooth}. 
The third term diverges proportionally to $\delta^{-2}$ as $\delta\to 0$, but its $\tau$-derivative has a well-defined limit as $\delta\to0$, which is $-1/6\pi^3\Delta\tau^3$.
The fourth term resembles the second term in that the divergence at $\delta\to 0$ is logarithmic in~$\delta$, but the presence of 
$\ddot{\x}^2$ and $\ddot{\x}\cdot\x^{(3)}$ in the integrand has the consequence that the coefficient of the divergent logarithm depends on the trajectory and does not vanish on differentiation with respect to~$\tau$. 
We find that the transition rate is given by 
\begin{align}
\mathcal{\dot{F}}_{\tau}(E)
&=
\frac{\ddot{\x}(\tau)\cdot\x^{(3)}(\tau)}{12\pi^3}\left(\ln{\left(\frac{\Delta\tau}{\delta}\right)+C^{'}_{+}}\right)
-\frac{E\bigl(E^2+\ddot{\x}^2(\tau)\bigr)}{24\pi^2}
\notag
\\[1ex]
&\hspace{2ex}
+\frac{1}{2\pi^3}\int^{\Delta\tau}_{0}\,\mathrm{d}s\,\left(
\frac{\cos{(Es)}}{\left[(\Delta \x)^2\right]^2}
-\frac{1}{s^4}+\frac{3E^2+\ddot{\x}^2(\tau)}{6s^2}
-\frac{\ddot{\x}(\tau)\cdot\x^{(3)}(\tau)}{6s}\right)
\notag
\\[1ex]
&\hspace{2ex}
+\frac{3E^2+\ddot{\x}^2(\tau)}{12\pi^3\Delta\tau}
-\frac{1}{6\pi^3\Delta\tau^3}
+O\left(\delta\ln{\left(\frac{\Delta\tau}{\delta}\right)}\right)
\ , 
\label{eq:ss:6d:transrate}
\end{align}
where the constant $C^{'}_{+}$ is determined 
by the switch-off function $h_2$ by 
\begin{align}
C^{'}_{+}&=-2\int^1_0\,\mathrm{d}r\,\frac{1}{r^2}
\left(\int^1_0\,\mathrm{d}v\,h_2(1-v)\left[h_2(1-v+r)-h_2(1-v)\right]-\tfrac{1}{2}r\right)
\notag 
\\
&\hspace{8ex}-2\int^1_0\,\mathrm{d}v\,h_2(v)\left[1-h_2(v)\right]
\ . 
\label{eq:Cprimeplus}
\end{align}
\par The qualitatively new feature is that the transition rate 
\eqref{eq:ss:6d:transrate} does not have a well-defined limit 
for generic trajectories as $\delta\to0$, 
because the coefficient of 
$\ddot{\x}(\tau)\cdot\x^{(3)}(\tau)$ diverges in this limit; 
further, even if $\delta$ is kept finite, 
the coefficient of this term depends on the details of the switch-off profile through the constant $C^{'}_{+}$~\eqref{eq:Cprimeplus}. 
The limit exists only for trajectories whose 
scalar proper acceleration, $\sqrt{{{\ddot\x}}^2}$, is constant over the trajectory, 
in which case the coefficient of the divergent term 
in \eqref{eq:ss:6d:transrate} vanishes. 
Note that this special class includes all trajectories that are 
uniformly accelerated, in the sense of 
following an orbit of a timelike Killing vector. 
\section[Spacetime dimension versus sharp switching]{Spacetime dimension versus sharp switching%
              \sectionmark{Spacetime dimension}}
\sectionmark{Spacetime dimension}
\label{sec:consistencychecks}
We have found that the sharp-switching limit of the detector 
response function becomes increasingly singular as the 
spacetime dimension $d$ increases from $2$ to~$6$. 
In this section we discuss further aspects of this singularity. 

First, we have seen that the sharp-switching limit of the response function diverges for~$d\ge4$. For $d=4$ and $d=5$ the divergent term is independent of the total detection time, and the limit of the instantaneous transition rate is still finite. For $d=6$, however, the instantaneous transition rate diverges for generic trajectories. We summarise this behaviour in Table~\ref{table:dimcomp}. 

\begin{table}[t]
\begin{center}
  \begin{tabular}{ | l | l | l  |}
    \hline
    $d$ & $\mathcal{F}_\tau$ & $\mathcal{\dot{F}}_{\tau} \vphantom{{\text{\Large A}}^A}$
    \\ \hline
    2 & finite & finite 
    \\ \hline
    3 & finite & finite 
    \\ \hline
    4 & $\ln\delta$ & finite 
    \\ \hline
    5 & $1/\delta$ & finite 
    \\ \hline
    6 & $1/\delta^2\vphantom{\text{\large$A^A$}}$ & $\ddot{\x}\cdot\x^{(3)} \ln\delta$ 
    \\ \hline
    \end{tabular}
\end{center}
\caption{The divergent pieces of the 
total transition probability $\mathcal{F}_\tau$ 
and the instantaneous transition rate $\mathcal{\dot{F}}_{\tau}$ 
for spacetime dimensions $d=2,\ldots,6$ 
in the sharp-switching limit.\label{table:dimcomp}}
\end{table}

Second, we re-emphasise that when the Wightman distribution $W$ in 
\eqref{eq:techIntro:respFunc} or \eqref{eq:techIntro:respFunc-alt}
is represented as the $\epsilon\to0$ limit of the regularised Wightman function~$W_\epsilon$, 
the $\epsilon\to0$ limit needs to be taken \emph{before\/} considering the sharp-switching limit: 
this is the only way one is guaranteed to be implementing the technical definition of the Wightman function correctly. 
With the regulator that we have used in this paper 
[equations
\eqref{eq:w6d}, 
\eqref{eq:w3d:1}
and~\eqref{eq:w5d:1}], it is known that 
attempting to reverse the limits na\"\i{}vely for $d=4$ 
would yield an incorrect, and even Lorentz-noncovariant, result for the transition rate for all non-inertial trajectories~\cite{schlicht,Schlicht:thesis,louko-satz:profile}. 
We have verified that attempting to reverse the limits na\"\i{}vely 
would be incorrect also for $d=3$, $d=5$ and $d=6$. 
For $d=3$, substituting the 
regularised Wightman function 
\eqref{eq:w3d:1} 
in 
\eqref{eq:tranrate} 
and evaluating the limit by the method of Section \ref{sec:3d} does 
give the correct result~\eqref{eq:ss:3d:transrate}, but attempting to take 
the limit $\epsilon\to0$ in \eqref{eq:tranrate} 
na\"\i{}vely under the integral would miss 
the first of the two terms in~\eqref{eq:ss:3d:transrate}. 
For $d=5$, substituting the regularised Wightman function 
\eqref{eq:w5d:1} in the na\"\i{}ve transition rate formula 
\eqref{eq:tranrate}
and evaluating the limit $\epsilon\to0$ 
by the methods of Section \ref{sec:5d} yields for the transition 
rate an expression that consists of 
\eqref{eq:ss:5d:transrate} plus the Lorentz-noncovariant terms 
\begin{align}
\frac{\ddot{t}(2+\dot{t})E}{8\pi^2{(1+\dot{t})}^2}
-\frac{\ddot{t}}{8\pi^2{(1+\dot{t})}^2 \epsilon}
\ , 
\end{align}
of which the second diverges as $\epsilon\to0$. 
For $d=6$, starting with the regularised Wightman function 
\eqref{eq:w6d}
yields for the transition rate a formula that is similar
to~\eqref{eq:ss:6d:transrate}, 
with the logarithmically divergent term replaced by a 
term that is logarithmically divergent in~$\epsilon$, 
plus a number of Lorentz-noncovariant terms. 

Third, because the sharp-switching divergence of $\mathcal{\dot{F}}_{\tau}$ for $d=6$ is perhaps surprising, we have verified that a similar divergence occurs also in the point-like detector model where the switching is sharp at the outset but the detector is initially spatially smeared, having the Lorentz-function spatial profile with an overall size parameter~$\epsilon$, 
and the point-like detector is recovered in the limit $\epsilon\to0$~\cite{schlicht,Langlois}. 
(The model can be alternatively regarded as that of a sharply-switched point-like detector whose Wightman function is regularised in terms of the frequency measured in the detector's instantaneous rest frame, rather than in terms of the frequency measured in an externally-specified Lorentz frame~\cite{Langlois}.) Adapting the methods of Section \ref{sec:6d} and proceeding as in~\cite{louko-satz:profile}, we find that the expression for $\mathcal{\dot{F}}_{\tau}$ is obtained from 
\eqref{eq:ss:6d:transrate} by the replacement 
$\ln(\tau/\delta) + C^{'}_{+} \to \ln(\tau/\epsilon) -\frac43 - \ln2$, 
so that the point-like detector limit 
$\epsilon\to0$ is again divergent unless the trajectory has constant scalar acceleration. 

Fourth, for a trajectory of uniform linear acceleration~$a$, 
switched on in the infinite past, 
the transition rate formulas 
\eqref{eq:ss:3d:transrate}, 
\eqref{eq:ss:4d:transrate}, 
\eqref{eq:ss:5d:transrate}
and 
\eqref{eq:ss:6d:transrate} 
yield 
\begin{align}
\label{eq:takagi:alld}
\begin{aligned}
&\dot{\mathcal{F}}_{d=3}\left(E\right)
=
\frac{1}{2}\frac{1}{e^{2\pi E/a}+1}
\ , \ \ \ 
&&
\dot{\mathcal{F}}_{d=5}\left(E\right)
=
\frac{1}{32\pi}\frac{\left(4 E^2+a^2 \right)}{e^{2\pi E/a}+1}\ , 
\\[1ex]
&\dot{\mathcal{F}}_{d=4}\left(E\right)
=
\frac{1}{2\pi}\frac{E}{e^{2\pi E/a}-1}
\ , \ \ \ 
&&\dot{\mathcal{F}}_{d=6}\left(E\right)
=\frac{1}{12\pi^2}\frac{E\left(E^2+a^2\right)}{e^{2\pi E/a}-1} 
\ . 
\end{aligned}
\end{align}
This was verified for $d=4$ in~\cite{louko-satz:profile}, 
and we have used the same contour deformation 
method for the other values of~$d$. 
The results \eqref{eq:takagi:alld} agree 
with those found in~\cite{takagi}, equation~(4.1.27), 
where they were obtained from 
a definition of transition rate that relies at the outset 
on the stationarity of the trajectory. 

Finally, we would like to speculate on how the response function and transition rate 
patterns that we have found for $d\le6$ might continue to $d>6$, and specifically to $d=7$. 

Recall that the formula \eqref{eq:techIntro:respFunc-alt} gives the response function in terms of the distributional Wightman function~$W$.
If $W$ is to be replaced by the un-regularised Wightman function under the integrals, then the negative powers of $s$ in 
$\Realpart \left[e^{-iEs} \, W(u,u-s) \right]$ 
must be subtracted. 
The last term in our formulas 
\eqref{eq:6d:detresp}, 
\eqref{eq:3d:detrespfinal}, 
\eqref{eq:5d:detresp:ssform}, 
and 
\eqref{eq:4dFraw}
is precisely of this form. 
The corresponding term can be constructed for any~$d$, 
and for $d=7$ it reads 
\begin{align}
\label{eq:7d-int-last}
\frac{3}{8\pi^3}
\int^{\infty}_{-\infty}\,\mathrm{d}u\,\chi(u)
\int^{\infty}_{0}\,\mathrm{d}s\,\chi(u-s)
\left(
\frac{\sin{(Es)}}{\sqrt{-\left[{(\Delta\x)}^2\right]^5}}-\frac{E}{s^4}+\frac{E\left(4E^2+5\ddot{\x}^2\right)}{24s^2}
-\frac{5E\,\ddot{\x}\cdot\x^{(3)}}{24s}
\right)
\ . 
\end{align}

Next, observe that our formulas 
\eqref{eq:6d:detresp}, 
\eqref{eq:5d:detresp:ssform}, 
and 
\eqref{eq:4dFraw}
contain terms in which the subtracted negative powers of $s$ are combined with similar powers of $s$ multiplied by quadratic combinations of $\chi$ and its derivatives evaluated at $u$ rather than at~$u-s$. All the negative powers of $s$ that appear in \eqref{eq:7d-int-last} have already appeared in this fashion in~\eqref{eq:6d:detresp}, and comparison of the coefficients shows that the corresponding terms for $d=7$ read 
\begin{align}
&-\frac{E^3}{16\pi^3}\int^{\infty}_{0}\,\frac{\mathrm{d}s}{s^2}\int^{\infty}_{-\infty}\,\mathrm{d}u\, \chi(u)\left[\chi(u-s)-\chi(u)\right]
\notag
\\[1ex]
&+\frac{3E}{8\pi^3}\int^{\infty}_{0}\,\frac{\mathrm{d}s}{s^4}\int^{\infty}_{-\infty}\,\mathrm{d}u\, \chi(u)
\bigl[
\chi(u-s)-\chi(u)
-\tfrac12 s^2\ddot{\chi}(u)
\bigr]
\notag
\\[1ex]
&-\frac{5}{64\pi^3}\int^{\infty}_{0}\,\frac{\mathrm{d}s}{s^2}
\int^{\infty}_{-\infty}\,\mathrm{d}u\,\chi(u)
\Bigl\{
\left[\chi(u-s)-\chi(u)\right] \ddot{\x}^2
-s \chi(u-s) \, \ddot{\x}\cdot\x^{(3)}
\Bigr\}
\ . 
\label{eq:7d-int-middle}
\end{align} 

The remaining term in 
\eqref{eq:6d:detresp}, 
\eqref{eq:3d:detrespfinal}, 
\eqref{eq:5d:detresp:ssform}, 
and 
\eqref{eq:4dFraw}
is a single integral involving derivatives of~$\x$. 
We are not aware of pattern arguments 
that might fix this term fully for general $d$, 
but we note that if this term for $d=7$ contains the piece 
\begin{align}
\frac{1}{2048\pi^2}\int^{\infty}_{-\infty}\,\mathrm{d}u
\, 
\chi^2(u) 
\left(4E^2+\ddot{\x}^2\right) 
\left(4E^2+9\ddot{\x}^2\right)
\ , 
\label{eq:7d-int-first-part}
\end{align}
then the transition rate computed from 
\eqref{eq:7d-int-last}, 
\eqref{eq:7d-int-middle}
and  
\eqref{eq:7d-int-first-part}
for a uniformly linearly accelerated trajectory agrees with that 
found in~\cite{takagi}. We further note that the power of $E$ in the single integral term in \eqref{eq:6d:detresp}, 
\eqref{eq:3d:detrespfinal}, 
\eqref{eq:5d:detresp:ssform}, 
and 
\eqref{eq:4dFraw}
fits the empirical formula 
\begin{equation}
\frac{\Gamma(d/2-1)}{(d-3)!} \frac{(-E)^{(d-3)}}{4\pi^{(d/2-1)}}
\ , 
\end{equation} 
and so does the highest power of $E$ in~\eqref{eq:7d-int-first-part}. 

We anticipate that the $d=7$ response function contains terms in addition to 
\eqref{eq:7d-int-last}, 
\eqref{eq:7d-int-middle}
and  
\eqref{eq:7d-int-first-part}; 
in particular, the pattern from $d\le6$ suggests that there should be a term proportional to $\delta^{-3}$ as $\delta\to0$, perhaps involving $\int_{-\infty}^{\infty}\mathrm{d}u \, \ddot\chi^2(u)$. 
However, if the only terms contributing to the 
transition rate are \eqref{eq:7d-int-last}, 
\eqref{eq:7d-int-middle}
and 
\eqref{eq:7d-int-first-part}, 
then a comparison with the $d=6$ case shows that the 
transition rate takes the form 
\begin{align}
\dot{\mathcal{F}}_{\tau}\left(E\right)
&=
\frac{5E \, \ddot{\x}(\tau) \cdot \x^{(3)}(\tau)}{64\pi^3}\left(\ln{\left(\frac{\Delta\tau}{\delta}\right) +C^{'}_{+}}\right)
+\frac{\bigl(4E^2+9\ddot{\x}^2(\tau)\bigr)
\bigl(4E^2+\ddot{\x}^2(\tau)\bigr)}{2048\pi^2}
\notag
\\[1ex]
&\hspace{2ex}
+\frac{3}{8\pi^3}\int^{\Delta\tau}_{0}\,\mathrm{d}s 
\left(
\frac{\sin{(Es)}}{\sqrt{-\left[{(\Delta\x)}^2\right]^5}}-\frac{E}{s^4}+\frac{E\bigl(4E^2+5\ddot{\x}^2(\tau)\bigr)}{24s^2}
-\frac{5E\,\ddot{\x}(\tau)\cdot\x^{(3)}(\tau)}{24s}
\right)
\notag
\\[1ex]
&\hspace{2ex}
+\frac{E\bigl(4E^2+5\ddot{\x}^2(\tau)\bigr)}{64\pi^3\Delta\tau}
-\frac{E}{8\pi^3\Delta\tau^3}
+O\left(\delta\ln{\left(\frac{\Delta\tau}{\delta}\right)}\right)
\ , 
\label{eq:7d:guess}
\end{align}
where $C^{'}_{+}$ is again 
given by~\eqref{eq:Cprimeplus}. 
While we must leave \eqref{eq:7d:guess} 
to the status of a conjecture, we note that it shares the 
logarithmic divergence of the 
$d=6$ transition rate \eqref{eq:ss:6d:transrate} and the 
divergent term is again proportional to $\ddot{\x} \cdot \x^{(3)}$. 
\section[Application: Schwarzschild embedded in $d=6$ Minkowski spacetime]{Application: Schwarzschild embedded in $d=6$ Minkowski spacetime%
              \sectionmark{Application: GEMS}}
\sectionmark{Application: GEMS}
\label{sec:GEMS}

The GEMS method
\cite{Deser:1997ri,Deser:1998bb,Deser:1998xb,Santos:2004ws} aims to model detector
response in four-dimensional spacetime by an embedding into a
higher-dimensional flat spacetime with an appropriately-chosen quantum
state, typically the Minkowski vacuum. The method has yielded
reasonable results for stationary trajectories in spacetimes of high
symmetry. A~review with references is given
in~\cite{Langlois,Langlois-thesis}.

We wish to discuss the prospects of GEMS modelling in non-stationary
situations in view of our results.

Recall that the $d=4$ Minkowski vacuum response function formula
\eqref{eq:4dFraw} and instantaneous transition rate formula
\eqref{eq:ss:4d:transrate} generalise to an arbitrary Hadamard state
on an arbitrary four-dimensional spacetime as \cite{satz-louko:curved}
\begin{align}
{\mathcal{F}}(E) 
&=
-\frac{E}{4\pi}\int_{-\infty}^{\infty}\mathrm{d}u\,{[\chi(u)]}^2 
\ + \ 
\frac{1}{2\pi^2}\int_0^{\infty}
\frac{\mathrm{d}s}{s^2}\int_{-\infty}^{\infty}\mathrm{d}u\,\chi(u)
\bigl[ \chi(u)-\chi(u-s)\bigr] 
\nonumber 
\\[1ex]
&\hspace{3ex}
+ 
2
\int_{-\infty}^{\infty}\mathrm{d}u\,\chi(u)
\int_0^{\infty}\mathrm{d}s\,\chi(u-s) 
\,\mathrm{Re} \left(
\mathrm{e}^{-iE s}\, W_0(u,u-s)+\frac{1}{4\pi^2s^2}
\right)
\ , 
\label{eq:4dcurvedprobability}
\\[1ex]
\label{eq:4dcurvedrate}
\dot{\mathcal{F}}_{\tau}\left(E\right)
&=
-\frac{E}{4\pi}
+2\int_0^{\Delta\tau}\mathrm{d}s
\, \mathrm{Re}
\left( \mathrm{e}^{-iE s}W_0(\tau,\tau-s)+\frac{1}{4\pi^2s^2}\right)
\ \ +\frac{1}{2\pi^2 \Delta \tau} 
\ , 
\end{align}
where $W_0$ is the point-wise 
$i\epsilon\to0$ limit of the Wightman function. 
The divergence structure at $\delta\to0$ is exactly 
as in Minkowski vacuum: 
the response function 
\eqref{eq:4dcurvedprobability} diverges logarithmically but the transition 
rate has the finite limit given by~\eqref{eq:4dcurvedrate}. 

As a concrete example, consider a detector in the extended
Schwarzschild spacetime, globally embedded in $d=6$ Minkowski space as
in \cite{fronsdal} (for further discussion see~\cite{ferraris}).  For
static trajectories in exterior Schwarzschild, GEMS modelling with
$d=6$ Minkowski vacuum predicts a thermal response in the local
Hawking temperature~\cite{Deser:1998xb}. One might hence anticipate
this modelling to extend to more general detector trajectories in the
Hartle-Hawking-Israel vacuum \cite{Hartle:1976tp,Israel:1976ur}.

Now, while the genuine $d=4$ sharp-switching transition rate
\eqref{eq:4dcurvedrate} is finite for arbitrary trajectories in the
Hartle-Hawking-Israel vacuum, the $d=6$ Minkowski vacuum transition
rate \eqref{eq:ss:6d:transrate} diverges in the sharp-switching limit
unless the $d=6$ scalar proper acceleration is constant. There are
trajectories of constant $d=6$ scalar proper acceleration through
every point in the extended Schwarzschild spacetime, and these
trajectories include all the stationary trajectories, that is, the
exterior-region circular trajectories that have constant (in general
non-inertial) angular velocity. However, we have verified by a direct
calculation that the only timelike Schwarzschild geodesics of constant
$d=6$ scalar acceleration are the exterior circular geodesics.  This
suggests that the GEMS method may not provide a viable model for
detectors on generic geodesics in Schwarzschild.

\chapter[The response of a detector on the BTZ black hole]{The response of a detector on the BTZ black hole}
\chaptermark{BTZ}
\label{ch:btz}
In this chapter, we examine the Unruh-DeWitt detector coupled to a scalar field in three-dimensional curved spacetime, following a similar analysis to~\cite{satz-louko:curved}.
\par The chapter begins with the derivation of a regulator-free expression for the transition probability when the scalar field is in an arbitrary three-dimensional Hadamard state. We then take the sharp-switching limit and obtain the instantaneous transition rate.
\par We continue by specialising the spacetime to the Ba\~nados-Teitelboim-Zanelli (BTZ) black hole and coupling our detector conformally to a massless scalar field in the Hartle-Hawking vacua, using both transparent and reflective boundary conditions at the infinity, without yet having specified the detector's trajectory. 
\par Next, we analyse the co-rotating trajectory finding a thermal response, and we also investigate the case of a detector freely-falling into the hole on a geodesic. For both cases analysed, a host of numerical results are presented, and these are complemented by good agreement from analytic results in a variety of asymptotic regimes.
\par The work this chapter represents was published in~\cite{Hodgkinson:2012mr}.
\section[Transition probability and transition rate in three spacetime dimensions]{Transition probability and transition rate in three spacetime dimensions%
              \sectionmark{Transition probability and rate}}
\sectionmark{Transition probability and rate}
\label{sec:resp3d}
Our first task is to obtain the detector response function for a detector coupled to a field in an arbitrary three-dimensional Hadamard state. We first rewrite the 
response function \eqref{eq:techIntro:respFunc-alt} in a form in which the regulator $\epsilon$ does not appear. 
We then take the sharp-switching limit and show that both the transition probability and the transition rate remain well defined in this limit. 
We follow closely the procedure developed in~\cite{satz-louko:curved,louko-satz:profile,satz:smooth,Hodgkinson:2011pc} and presented in Chapter~\ref{ch:by4d}.
\subsection{Hadamard form of $W_{\epsilon}$}
\label{sec:resp3d:HadForm}
In a three-dimensional spacetime, the Wightman distribution $W(\x,\x')$ of a real scalar field in a Hadamard state can be represented by a family of functions with the short distance
form~\cite{Decanini:2005gt}
\begin{equation}
\label{eq:3dHadamard}
W_{\epsilon}(\x,\x')=\frac{1}{4\pi}\left[\frac{U(\x,\x')}{\sqrt{\tilde\sigma_{\epsilon}(\x,\x')}}
+\frac{H(\x,\x')}{\sqrt{2}}\right]\,, 
\end{equation}
where $\epsilon$ is a positive parameter, 
$\tilde\sigma(\x,\x')$ is the squared geodesic distance between $\x$ and~$\x'$, 
$\tilde\sigma_{\epsilon}(\x,\x'):=\tilde\sigma(\x,\x')+2i\epsilon\left[T(\x)-T(\x')\right]+\epsilon^2$ 
and $T$ is any globally-defined future-increasing $C^{\infty}$ function. 
The branch of the square root is such that the $\epsilon\to 0_+$ 
limit of the square root is positive when 
$\tilde\sigma(\x,\x')>0$~\cite{kay-wald,Decanini:2005gt} and the branch cut is taken along the negative real axis.
Here $U(\x,\x')$ and $H(\x,\x')$ are symmetric biscalars that 
possess expansions of the form
\begin{subequations}
\begin{equation}
\label{eq:Udef}
U(\x,\x')=\sum^{\infty}_{n=0} U_n(\x,\x')\sigma^n(\x,\x')\,,
\end{equation}
\begin{equation}
\label{eq:Hdef}
H(\x,\x')=\sum^{\infty}_{n=0} H_n(\x,\x')\sigma^n(\x,\x')\,,
\end{equation}
\label{eq:3d:UH}
\end{subequations}
where the coefficients $U_n(\x,\x')$ satisfy the recursion relations
\begin{eqnarray}
& & (n+1)(2n+1) U_{n+1} + (2n+1) U_{n+1 ; \mu} \sigma ^{; \mu}
-(2n+1) U_{n+1} \Delta ^{-1/2}{\Delta^{1/2}}_{;\mu} \sigma^{; \mu}
\nonumber \\
&  & + \left( \Box_x -m^2 -\xi R \right) U_n =0\,, 
\quad n=0,1,2,\ldots,
\label{eq:Urec}
\end{eqnarray}
with the boundary condition
\begin{equation}
\label{eq:U0}
U_0= \Delta^{1/2}\,,
\end{equation}
and the coefficients $H_n(\x,\x')$ satisfy the recursion relations
\begin{eqnarray}
& & (n+1)(2n+3) H_{n+1} + 2(n+1) H_{n+1 ; \mu} \sigma ^{; \mu}- 
2(n+1) H_{n+1} \Delta ^{-1/2}{\Delta ^{1/2}}_{;\mu}\sigma^{; \mu}\nonumber \\
&  &  + \left( \Box_x-m^2-\xi R \right) H_n =0\,, 
\quad n=0,1,2,\ldots\,,
\label{eq:Hrec}
\end{eqnarray}
where $\sigma = \tfrac12 \tilde\sigma$, 
$\Delta(\x,\x')$ is the Van Vleck determinant, 
$m$ is the mass and $\xi$ is the curvature coupling parameter~\cite{Decanini:2005gt}.
We note that the series in~\eqref{eq:3d:UH} are defined in a convex normal
neighbourhood but they need not be defined globally, and even in a
convex normal neighbourhood, the series are asymptotic series that do not
necessarily converge, not even in the coincidence limit~\cite{kay-wald}.

The $i\epsilon$ prescription in \eqref{eq:3dHadamard} 
defines the singular part of $W(\x,\x')$: the action of the 
Wightman distribution is obtained by integrating $W_{\epsilon}(\x,\x')$ 
against test functions and taking the limit 
$\epsilon\to 0_+$ as in~\eqref{eq:techIntro:respFunc-alt}. 
This limit can be shown to be independent of the choice of 
global time function $T$~\cite{Fewster:1999gj,junker,hormander-vol1,hormander-paper1}. 

\subsection{Transition probability without $i\epsilon$-regulator}

To evaluate the $\epsilon\to 0_+$ limit in~\eqref{eq:techIntro:respFunc-alt}, the main issue is at $s=0$,
where the Hadamard expansion~\eqref{eq:3dHadamard} shows that the integrand develops a
non-integrable singularity as $\epsilon\to 0_+$. We shall work under the
assumption that any other singularities that the integrand develops as
$\epsilon\to 0_+$ are integrable. This will be the case in our applications
in Sections \ref{sec:corot} and~\ref{sec:inertial}. We note in passing that similar integrable
singularities can occur in any spacetime dimension, and the
four-dimensional results in~\cite{satz-louko:curved} should hence be understood to
involve a similar assumption.


We write the detector response function~\eqref{eq:techIntro:respFunc} as
\be 
\mathcal{F}\left(E\right)=\frac{1}{4}\int^{\infty}_{-\infty}\,\mathrm{d}\tau\,\chi^2\left(\tau\right)+\int^{\infty}_{-\infty}\,\mathrm{d}\tau\,\int^{\infty}_{-\infty}\,\mathrm{d}\tau^{\prime}\, \widetilde{W}\left(\tau,\tau^{\prime}\right)\chi(\tau)\expo^{-iE(\tau-\tau')}\chi\left(\tau^{\prime}\right)\,,
\label{eq:TRoldvars}
\ee
with
\be 
\widetilde{W}(\tau,\tau^{\prime}):=W(\tau,\tau^{\prime})-\frac{1}{4}\delta(\tau-\tau^{\prime})\,.
\ee
Note that $\widetilde{W}(\tau,\tau^{\prime})$ satisfies $\widetilde{W}(\tau,\tau^{\prime})=\overline{\widetilde{W}}(\tau^{\prime},\tau)$ because $W(\tau,\tau^{\prime})$ has this property.
\par 
Now the Hadamard form of the Wightman function implies that
\be 
W(\tau,\tau^{\prime})=\frac{1}{4}\delta(\tau-\tau^{\prime})-\frac{i}{4\pi}\mathcal{P}\left(\frac{1}{\tau-\tau^{\prime}}\right)+\text{integrable}\,,
\ee
where $\mathcal{P}$ represents the Cauchy principal value.
Thus, 
\be 
\widetilde{W}(\tau,\tau^{\prime})=-\frac{i}{4\pi}\mathcal{P}\left(\frac{1}{\tau-\tau^{\prime}}\right)+\text{integrable}\,.
\ee
We are now in a position to write~\eqref{eq:TRoldvars} as
\be 
\mathcal{F}\left(E\right)=\frac{1}{4}\int^{\infty}_{-\infty}\,\mathrm{d}\tau\,\chi^2\left(\tau\right)+\lim_{\epsilon\to 0}
\int_{|\tau-\tau^{\prime}|\geq \epsilon} \mathrm{d}\tau\,\mathrm{d}\tau^{\prime}\, W_0\left(\tau,\tau^{\prime}\right)\chi(\tau)\expo^{-iE(\tau-\tau')}\chi\left(\tau^{\prime}\right)\,,
\label{eq:TRoldvars-alt}
\ee
where $W_0$ is the point-wise $\epsilon \to 0$ limit of
$W_{\epsilon}$. 
\par
Using the property $W_0\left(\tau,\tau^{\prime}\right)=\overline{W}_0\left(\tau^{\prime},\tau\right)$,
we can write~\eqref{eq:TRoldvars-alt} as
\bea
\mathcal{F}\left(E\right)&=\frac{1}{4}\int^{\infty}_{-\infty}\,\mathrm{d}\tau\,\chi^2\left(\tau\right)+\lim_{\epsilon\to 0}2\int_{-\infty}^{\infty}\mathrm{d}u\,\chi(u)\,\int_{\epsilon}^{\infty}\mathrm{d}s\,\chi(u-s) 
\Realpart\left[\mathrm{e}^{-iEs} \, W_0(u,u-s)\right]\\
&=\frac{1}{4}\int^{\infty}_{-\infty}\,\mathrm{d}u\,\chi^2(u)
+2\int_{-\infty}^{\infty}\mathrm{d}u\,\chi(u)\,\int_{0}^{\infty}\mathrm{d}s\,\chi(u-s) 
\Realpart\left[\mathrm{e}^{-iEs} \, W_0(u,u-s)\right]\,.
\label{eq:resp3d:sharp:prob}
\eea
\par
Note that the integrals in~\eqref{eq:resp3d:sharp:prob} 
are regular, at $s=0$ by the Hadamard short-distance behaviour of~$W_0$, 
and at $s>0$ by our assumptions about the singularity structure of $W_0$ at 
timelike-separated points. 

\subsection{Sharp-switching limit and the transition rate}
\label{sec:resp3d:sharp}

Up to now we have assumed the switching function $\chi$ to be smooth. 
When $\chi$ approaches the characteristic function of the interval
$[\tau_0,\tau_0+\tau]$, as in Section~\ref{sec:3d} 
the integrands in \eqref{eq:resp3d:sharp:prob} remain regular, 
and taking the sharp-switching limit under the
integral can be justified by dominated convergence. 
The transition probability takes the form
\begin{align}
\label{eq:resp3d:sharp:SSprob}
\mathcal{F}_\tau (E)
= 
\frac{\Delta\tau}{4}
+2\int^{\tau}_{\tau_0}\,\mathrm{d}u\,\int^{u-\tau_0}_{0}\,\mathrm{d}s
\Realpart\left[\mathrm{e}^{-iEs} \, W_0(u,u-s)\right], 
\end{align}
where $\Delta \tau := \tau - \tau_0$
and the subscript $\tau$ is included as a 
reminder of the dependence on the switch-off moment. 
Differentiation with respect to $\tau$ shows that 
the transition rate is given by 
\begin{align}
\label{eq:resp3d:sharp:rate}
\dot{\mathcal{F}}_{\tau}\left(E\right)
=
\frac{1}{4}+2\int^{\Delta\tau}_{0}\,\mathrm{d}s
\Realpart\left[\mathrm{e}^{-iEs} \, W_0(\tau,\tau-s)\right]
\,. 
\end{align}

Note that both 
\eqref{eq:resp3d:sharp:SSprob}
and 
\eqref{eq:resp3d:sharp:rate}
are well defined under our assumptions, and in the 
special case of a massless scalar field in the Minkowski 
vacuum they reduce to what was found in Chapter~\ref{ch:by4d}. 
Spacetime curvature has hence not introduced new 
singularities in the sharp-switching limit. 

Note also that the pre-integral term $\frac14$ in \eqref{eq:resp3d:sharp:rate}
would have been missed if the limit $\epsilon\to0_+$ had been taken 
na\"\i{}vely under the integral in~\eqref{eq:techIntro:respFunc-alt}. 
Yet this term is essential: it was observed in Chapter~\ref{ch:by4d}
that without this term one would not recover the standard thermal response 
for a uniformly linearly accelerated detector in 
Minkowski vacuum~\cite{takagi,Ooguri:1985nv}, 
and we shall see in Section \ref{sec:corot} that without this term we would not 
recover thermality for a co-rotating detector in the BTZ spacetime. 

\section{Detector in the BTZ spacetime}
\label{sec:BTZoverview}

We now turn to a detector in the BTZ
black hole spacetime \cite{BTZ,HBTZ}, specialising to a massless
conformally-coupled scalar field in the Hartle-Hawking vacuum 
with transparent or reflective boundary conditions. In this section we briefly recall
relevant properties of the spacetime and the Wightman
function. 
More detail can be found in the review
in~\cite{Carlip:1995}.

Recall first that three-dimensional 
anti-de~Sitter 
spacetime $\text{AdS}_3$ may be defined as the submanifold 
\begin{equation}
\label{eq:AdShyperb}
-\ell^2 = 
-T_1^2-T_2^2 + X_1^2+X_2^2  
\end{equation}
in $\mathbb{R}^{2,2}$ with co-ordinates $(T_1,T_2,X_1,X_2)$ and metric
\begin{equation}
\label{eq:R22metric}
dS^2= -dT_1^2-dT_2^2 + dX_1^2+dX_2^2\,, 
\end{equation}
where $\ell$ is a positive parameter of dimension length. 
The BTZ black hole is obtained as a quotient of an open region in 
$\text{AdS}_3$ under a discrete isometry group~$\simeq\BbbZ$.
Specialising to a nonextremal black hole, 
a set of co-ordinates that are adapted to the relevant isometries
and cover the exterior region of the black hole are the 
BTZ co-ordinates $(t,r,\phi)$, defined in 
$\text{AdS}_3$ by 
\begin{eqnarray}
\label{eq:BoyerLind}
X_1=\ell\sqrt{\alpha}
  \sinh\left(\frac{r_+}{\ell}\phi-\frac{r_-}{\ell^2}t\right) &,&\quad\!\!
X_2=\ell\sqrt{\alpha-1}
  \cosh\left(\frac{r_+}{\ell^2}t-\frac{r_-}{\ell}\phi\right) , \nonumber\\[1ex]
T_1=\ell\sqrt{\alpha}
   \cosh\left(\frac{r_+}{\ell}\phi-\frac{r_-}{\ell^2}t\right) &,&\quad\!
T_2=\ell\sqrt{\alpha-1}
  \sinh\left(\frac{r_+}{\ell^2}t-\frac{r_-}{\ell}\phi\right)\,, 
\end{eqnarray}
where
\begin{equation}
\label{eq:alpha}
\alpha(r)=\left(\frac{r^2-r_-^2}{r_+^2-r_-^2}\right)  
\end{equation}
and the parameters $r_\pm$ satisfy $|r_-| < r_+$. 
The co-ordinate ranges covering the black hole exterior are 
$r_+ < r < \infty$, $-\infty < t < \infty$ and $-\infty < \phi < \infty$, 
and the $\BbbZ$ quotient is realised as the identification
$(t,r,\phi) \sim (t,r,\phi+2\pi)$. The outer 
horizon is at $r\to r_+$, 
and the asymptotically $\text{AdS}_3$ infinity is at $r\to\infty$. 
The metric takes the form
\begin{equation}
\label{eq:BTZmetric}
ds^2 = -( N^\perp)^2dt^2 + f^{-2}dr^2
  + r^2\left( d\phi + N^\phi dt\right)^2
\end{equation}
with
\begin{equation}
\label{eq:lapseshift}
N^\perp = f
= \left( -M + \frac{r^2}{\ell^2} + \frac{J^2}{4r^2} \right)^{1/2}\,,
\quad N^\phi = - \frac{J}{2r^2} \,,
\end{equation}
where the mass $M$ and the angular momentum $J$ are given by
\begin{equation}
\label{eq:MandJ}
M=(r_+^2+r_-^2)/\ell^2, \quad J=2r_+ r_-/\ell\,,
\end{equation}
and they satisfy $|J| < M \ell$.

We are interested in quantum states in which the 
Wightman function on the black hole spacetime can be expressed as an image sum of the 
corresponding $\text{AdS}_3$ Wightman function. If $G_A(\x,\x')$ denotes the 
$\text{AdS}_3$ Wightman function, the BTZ Wightman function reads \cite{Carlip:1995}
\begin{equation}
\label{eq:BTZWightmanSum}
G_{\text{BTZ}}(\x,\x')=\sum_n\,G_A(\x,\Lambda^n \x')\,,
\end{equation}
where $\Lambda \x'$ denotes the action on $\x'$ of the group element 
$(t,r,\phi)\mapsto (t,r,\phi+2\pi)$,  
and the notation suppresses the distinction between points 
on $\text{AdS}_3$ and points on the quotient spacetime. 
The scalar field is assumed untwisted so that no 
additional phase factors appear in~\eqref{eq:BTZWightmanSum}.

We consider a massless, conformally-coupled field, and the family of 
$\text{AdS}_3$ Wightman functions \cite{Carlip:1995}
\begin{equation}
\label{eq:AdSWightman}
G^{(\zeta)}_{A}(\x,\x')=\frac{1}{4\pi} \! 
\left(\frac{1}{\sqrt{{\Delta\X}^2(\x,\x')}}
-\frac{\zeta}{\sqrt{{\Delta\X}^2(\x,\x')+4\ell^2}}\right)\,,
\end{equation}
where the parameter $\zeta\in\{0,1,-1\}$ 
specifies whether the boundary condition at infinity is respectively 
transparent, Dirichlet or Neumann. The transparent boundary condition corresponds to a particular recirculation of momentum and angular momentum at spatial infinity~\cite{Avis:1977yn}.
Here $\Delta\X^2(\x,\x')$ 
is the squared geodesic distance between $\x$ and $\x'$ 
in the flat embedding spacetime $\mathbb{R}^{2,2}$, 
given by 
\begin{equation}
\label{eq:R22Interval}
{\Delta\X}^2 (\x,\x') :=
-{(T_1-T_1^{'})}^2 -{(T_2-T_2^{'})}^2 
+ {(X_1-X_1^{'})}^2 +{(X_2-X_2^{'})}^2 \ ,
\end{equation}
and we have momentarily suppressed 
the $i\epsilon$ prescription in~\eqref{eq:AdSWightman}. 

With
\eqref{eq:BTZWightmanSum}
and~\eqref{eq:AdSWightman}, 
the transition rate
\eqref{eq:resp3d:sharp:rate}
takes the form 
\begin{align}
\dot{\mathcal{F}}_{\tau}(E)
=
\frac{1}{4}+\frac{1}{2\pi\sqrt{2}}\sum_{n=-\infty}^{\infty}\int^{\Delta\tau/\ell}_{0}\,
\mathrm{d}\tilde{s} \Realpart
\left[\mathrm{e}^{-iE\ell \tilde{s}}
\left(\frac{1}{\sqrt{\Delta\tilde{\X}^2_n}}-\frac{\zeta}{\sqrt{\Delta\tilde{\X}^2_n+2}}\right)
\right] , 
\label{eq:BTZ:rate}
\end{align}
where we have 
introduced the dimensionless integration variable 
$\tilde{s}:=s/\ell$ 
and written 
\begin{align}
\Delta\tilde{\X}_n^2 & :={\Delta\X}^2\bigl(\x(\tau),\Lambda^n\x(\tau-\ell\tilde{s})\bigr)
/\bigl(2\ell^2\bigr)
\nonumber 
\\[1ex]
& \hspace{.8ex} = -1+\sqrt{\alpha(r)\alpha(r')}
\cosh\!\left[(r_+/\ell)\left(\phi-\phi'-2\pi n \right)-(r_-/\ell^2)\left(t-t'\right)\right]
\nonumber
\\[1ex]
&\hspace{3.5ex}
-\sqrt{\bigl(\alpha(r)-1\bigr)\bigl(\alpha(r')-1\bigr)}
\cosh\!\left[(r_+/\ell^2)\left(t-t'\right)-(r_-/\ell)\left(\phi-\phi'-2\pi n \right)\right]\,,
\label{eq:DXN2}
\end{align}
where the unprimed co-ordinates are evaluated at $\x(\tau)$ 
and the primed co-ordinates 
at $\x(\tau-\ell\tilde{s})$. 

What remains is to specify the branches of the square roots in~\eqref{eq:BTZ:rate}. 
As $s$ extends to a global time function in the relevant part of $\text{AdS}_3$, 
the prescription \eqref{eq:3dHadamard} 
implies that the square roots in \eqref{eq:BTZ:rate} are positive 
when the arguments are positive, 
and the square roots are analytically continued to 
negative values of the arguments 
by giving $s$ a small, negative imaginary part. 


\section{Co-rotating detector in BTZ}
\label{sec:corot}

In this section we investigate the transition rate of a 
detector that is in the exterior region of the BTZ black hole and 
co-rotating with the horizon. 
Because the detector is stationary, we take the switch-on to be in the asymptotic past. 
When the black hole is spinless, the detector is static.

\subsection{Transition rate and the KMS property}
\label{sec:corot:rate}

The angular velocity of the horizon is 
given by \cite{BTZ,HBTZ,Carlip:1995}
\begin{equation}
\label{eq:corot:Omega}
\Omega_H
= r_-/(r_+ \ell)\,, 
\end{equation}
and it has an operational meaning as the value that 
$\mathrm{d}\phi/\mathrm{d}t$ takes on any 
timelike worldline that crosses the horizon. 
The worldline of a detector that is in the exterior region and 
rigidly co-rotating with the horizon reads 
\begin{equation}
\label{eq:corot:traj}
r= \text{constant}\ , \ \ 
t=
\frac{\ell r_+ \tau}{\sqrt{r^2-r_+^2}\sqrt{r_+^2-r_-^2}} \ , 
\ \ 
\phi= \frac{r_- \tau}{\sqrt{r^2-r_+^2}\sqrt{r_+^2-r_-^2}} \ , 
\end{equation}
where
the value of $r$ specifies the radial location and $\tau$ is the proper time. 
We have set the additive constants in $t$ and $\phi$ 
to zero without loss of generality. 

Substituting \eqref{eq:corot:traj} into \eqref{eq:DXN2} and taking the 
switch-on to be in the asymptotic past, 
the transition rate \eqref{eq:BTZ:rate} takes the form
\begin{align}
\dot{\mathcal{F}}
(E)
& = 
\frac{1}{4}+\frac{1}{4\pi\sqrt{\alpha(r)-1}}
\sum_{n=-\infty}^{\infty}
\int^{\infty}_{0}\,\mathrm{d}\tilde{s} 
\Realpart \left[\mathrm{e}^{-iE\ell \tilde{s}}
\left(\frac{1}{\sqrt{K_n-\sinh^2\bigl(\Xi \tilde{s}+n\pi r_-/\ell \bigr)}} \right. \right. 
\nonumber\\[1ex]
& \hspace{20ex}
\left. \left. 
-\frac{\zeta}{\sqrt{Q_n-\sinh^2 \bigl(\Xi \tilde{s}+n\pi r_-/\ell\bigr)}}\right)\right]\,,
\label{eq:corot:rate}
\end{align}
where 
\begin{subequations}
\label{eq:corot:pars}
\begin{align}
\label{eq:corot:Kn}
K_n & := {\bigl(1-\alpha^{-1}\bigr)}^{-1}\sinh^2 \bigl( n \pi r_+/\ell \bigr) \,, 
\\
\label{eq:corot:Qn}
Q_n & := K_n + {(\alpha-1)}^{-1} \,, 
\\
\label{eq:corot:Xi}
\Xi & := {\left(2\sqrt{\alpha-1} \, \right)}^{-1}\,, 
\end{align} 
\end{subequations}
$\alpha$ is given by~\eqref{eq:alpha}, 
and we have dropped the subscript $\tau$ from $\dot{\mathcal{F}}$ 
because the situation is stationary and 
the transition rate is independent of~$\tau$. 
The square roots in \eqref{eq:corot:rate} 
are positive 
for positive values of the argument, 
and they are analytically continued to 
negative values of the argument 
by giving $\tilde{s}$ a small, negative imaginary part. 
Note that the integrand in \eqref{eq:corot:rate} has 
singularities at $\tilde{s}>0$, 
at places where the quantity under a square root changes sign, 
but all of these singularities are integrable. 

We show in Appendix \ref{app:A} that \eqref{eq:corot:rate} can be written as 
\begin{align}
\dot{\mathcal{F}}(E)
&= \frac{\mathrm{e}^{-\beta E\ell /2}}{2\pi}\sum^{\infty}_{n=-\infty}
\,\cos\bigl(n\beta E r_-\bigr)
\times
\nonumber 
\\
&
\hspace{6ex}
\times 
\int^{\infty}_0\,\mathrm{d}y\,\cos\bigl(y\beta E\ell /\pi\bigr)\Bigg(\frac{1}{\sqrt{K_n+\cosh^2 \! y}}
-\frac{\zeta}{\sqrt{Q_n+\cosh^2 \! y}}\Bigg) \,, 
\label{eq:corot:rate.evald.alt}
\end{align}
or alternatively as 
\begin{align}
&
\dot{\mathcal{F}}(E)
= \frac{1}{2(\mathrm{e}^{\beta E\ell }+1)}
-\frac{\zeta \mathrm{e}^{-\beta E\ell /2}}{2\pi}\int^{\infty}_0\,
\mathrm{d}y\,\frac{\cos\bigl(y\beta E\ell /\pi\bigr)}{\sqrt{Q_0+\cosh^2 \! y}}
\nonumber 
\\[1ex]
&+\frac{\mathrm{e}^{-\beta E\ell /2}}{\pi}\sum^{\infty}_{n=1}\,\cos\bigl(n\beta E  r_-\bigr)
\int^{\infty}_0\,\mathrm{d}y\,\cos\bigl(y\beta E\ell /\pi\bigr)
\Bigg(\frac{1}{\sqrt{K_n+\cosh^2 \! y}}
-\frac{\zeta}{\sqrt{Q_n+\cosh^2 \! y}}\Bigg)\,, 
\label{eq:corot:rate.evald}
\end{align}
where 
\begin{align}
\beta:=2\pi \sqrt{\alpha-1}
\ . 
\label{eq:beta-def}
\end{align}
It is evident from \eqref{eq:corot:rate.evald.alt} or \eqref{eq:corot:rate.evald} 
that $\dot{\mathcal{F}}$ depends on $E$ only via the dimensionless 
combination $\ell\beta E$. 
It is further evident that 
$\dot{\mathcal{F}}$ has the 
KMS property \cite{Kubo:1957mj,Martin:1959jp} 
\begin{equation}
\label{eq:corot:KMS}
\dot{\mathcal{F}}(E) = \mathrm{e}^{-\ell\beta E}\dot{\mathcal{F}}(-E)\, . 
\end{equation}
The transition rate is hence thermal in the temperature~${(\ell\beta)}^{-1}$. 

It can be verified that 
${(\ell\beta)}^{-1} = (-g_{00})^{-1/2}T_0$, where $T_0=\kappa_0/(2\pi)$, 
$\kappa_0$ is the surface gravity of the black hole with respect to the 
horizon-generating Killing vector $\partial_t + \Omega_H \partial_\phi$, 
and $g_{00}$ is the time-time component of the metric in 
co-ordinates adapted to the co-rotating observers. 
This means that the temperature ${(\ell\beta)}^{-1}$ of the detector response 
is the local Hawking temperature, 
obtained by renormalising the conventional  
Hawking temperature $T_0$ by the 
Tolman redshift factor at the detector's location. 
This is the temperature one would have expected 
by general properties of the Hartle-Hawking state 
\cite{Hartle:1976tp,Israel:1976ur,Lifschytz:1993eb}, 
including the periodicity of an 
appropriately-defined imaginary time co-ordinate~\cite{Gibbons:1976ue}, 
and also by global embedding Minkowski spacetimes (GEMS) considerations
\cite{Deser:1997ri,Deser:1998bb,Deser:1998xb,Russo:2008gb}. 

Note that the expressions 
\eqref{eq:corot:rate.evald.alt}
and 
\eqref{eq:corot:rate.evald}
contain both terms of~\eqref{eq:resp3d:sharp:rate}, 
as shown in Appendix~\ref{app:A}. 
The pre-integral term $\tfrac14$ in \eqref{eq:resp3d:sharp:rate} 
is hence essential for recovering thermality: in \eqref{eq:corot:rate.evald} 
it can be regarded as having been grouped in the term $\tfrac12{(\mathrm{e}^{\beta E\ell }+1)}^{-1}$, 
which gives the transition rate in pure $\text{AdS}_3$ with the transparent boundary condition. 
The superficial Fermi-Dirac appearance of this pure $\text{AdS}_3$ term is a general feature of 
linearly-coupled scalar fields in 
odd spacetime dimensions~\cite{takagi,Lifschytz:1993eb,Ooguri:1985nv,Sriramkumar:2002nt}.

\subsection{Asymptotic regimes}
\label{sec:corot:asymptotics}

We consider the behaviour of the transition rate 
\eqref{eq:corot:rate.evald} in three asymptotic regimes. 

First, suppose $r_+\to\infty$ so that $r_-/r_+$ and $r/r_+$ are fixed. 
Physically, this is the limit of a large black hole with fixed~$J/M$, 
and the detector is assumed not to be close to the 
black hole horizon. Note that 
$\alpha$ and $\beta$ remain fixed in this limit. 
It follows from 
\eqref{eq:alpha} and \eqref{eq:corot:pars}
that in \eqref{eq:corot:rate.evald} 
this is the limit in which $K_n$ and $Q_n$ with $n\ge1$ are large. 
Assuming that $E$ is fixed and non-zero, and using formula \eqref{eq:Jall3}
in Appendix~\ref{ch:appendixBTZ}, we find 
\begin{align}
& 
\dot{\mathcal{F}}
(E)
= \frac{1}{2(\mathrm{e}^{\beta E\ell }+1)}
-\frac{\zeta \mathrm{e}^{-\beta E\ell /2}}{2\pi}\int^{\infty}_0\,
\mathrm{d}y\,\frac{\cos\bigl(y\beta E\ell /\pi\bigr)}{\sqrt{Q_0+\cosh^2 \! y}}
\nonumber 
\\[1ex]
&\hspace{1ex}
+\frac{\mathrm{e}^{-\beta E\ell /2}
\cos\bigl(\beta E r_- \bigr)}{\sqrt{\pi}\beta E\ell }
\times
\nonumber\\[1ex]
&\hspace{3ex}
\times 
\Bigg\{\Imagpart 
\Bigg[
\left(\frac{{(4K_1})^{i\beta E\ell /(2\pi)}}{\sqrt{K_1}}-
\frac{\zeta{(4Q_1})^{i\beta E\ell /(2\pi)}}{\sqrt{Q_1}}
\right)
\Gamma\left(1+\frac{i\beta E\ell }{2\pi}\right)
\Gamma\left(\frac12 - \frac{i\beta E\ell }{2\pi}\right)
\Bigg]
\nonumber\\[1ex]
&\hspace{8ex}
+O\left(\mathrm{e}^{-2\pi r_+/\ell}\right)\Bigg\}
\ , 
\label{eq:corot:larger+}
\end{align}
where the displayed next-to-leading term 
comes from the $n=1$ term in 
\eqref{eq:corot:rate.evald}
and
is of order~$\mathrm{e}^{-\pi r_+/\ell}$. 
The corresponding formula for $E=0$ can be obtained from formula \eqref{eq:Jall30} in 
Appendix \ref{ch:appendixBTZ} and has a next-to-leading term 
of order $r_+\mathrm{e}^{-\pi r_+/\ell}$. 

Next, suppose that $r_+\to0$ so that $r_-/r_+$ and $r/r_+$ are again fixed. 
This is the limit of a small black hole. 
Note that 
$\alpha$ and $\beta$ are again fixed. 
The dominant behaviour comes now from the sum over $n$ 
and can be estimated by the Riemann sum technique of Appendix~\ref{app:C}. 
We find 
\begin{align}
\dot{\mathcal{F}}(E)
&=\frac{\ell \mathrm{e}^{-\beta E\ell /2}}{\pi^2r_+}
\int^{\infty}_0 \,\mathrm{d}v
\int^{\infty}_0\,\mathrm{d}y\,
\cos\!\left(\frac{v \beta E\ell  r_-}{\pi r_+}\right)
\cos\!\left(\frac{y\beta E\ell }{\pi}\right)
\times
\nonumber\\[1ex]
&\hspace{2ex} \times 
\left[ 
{\left(\frac{\alpha\sinh^2 \! v}{(\alpha-1)}+\cosh^2{y}\right)}^{\!-1/2}
-\zeta {\left(\frac{1 + \alpha \sinh^2 \! v}{(\alpha-1)}+\cosh^2 \! y \right)}^{\!-1/2}
\right]
+\frac{o(1)}{r_+}. 
\label{eq:corot:smallr+}
\end{align}
The leading term is proportional to $1/r_+$ and it hence diverges in the limit of a small black hole. 

Finally, suppose that $E \to \pm \infty$ with the other quantities fixed. 
The analysis of Appendix \ref{app:D} shows that each integral term in \eqref{eq:corot:rate.evald}
is oscillatory in~$E$, with an envelope that falls off as 
$1/\sqrt{-E}$ 
at $E \to -\infty$ but exponentially at $E \to +\infty$. 
Applying this estimate to the lowest few values of $n$ in \eqref{eq:corot:rate.evald} 
should be a good estimate to the whole sum when $r_+/\ell$ is large. 
We have not attempted to estimate the whole sum at $E \to \pm \infty$ when $r_+/\ell$ is small.

\subsection{Numerical results}
\label{sec:corot:results}

We now turn to numerical evaluation of the transition rate~\eqref{eq:corot:rate.evald}. 
We are particularly interested in the interpolation between 
the asymptotic regimes identified in Subsection~\ref{sec:corot:asymptotics}. 

$\dot{\mathcal{F}}$ \eqref{eq:corot:rate.evald} 
depends on five independent variables. 
Two of these are the mass and the angular momentum of the black hole, 
encoded in the dimensionless parameters $r_+/\ell$ and $r_-/\ell$. 
The third is the location of the detector, entering 
$\dot{\mathcal{F}}$ only in the dimensionless 
combination $\alpha$~\eqref{eq:alpha}. 
The fourth is the detector's energy gap~$E$, 
entering $\dot{\mathcal{F}}$ only in the dimensionless combination $\beta E\ell$ 
where $\beta$ was given in~\eqref{eq:beta-def}. 
The last one is the discrete parameter $\zeta \in \{0,1,-1\}$ 
which specifies the boundary condition at infinity. 

We plot $\dot{\mathcal{F}}$ as a function of $\ell\beta E$, grouping the plots in 
triplets where $\zeta$ runs over its three values and the other three parameters are fixed. 
We proceed from large $r_+/\ell$ towards small~$r_+/\ell$. 

\begin{figure}[p]  
  \centering
  \subfloat[$\zeta=0$]{\label{fig:Num_bc0_rp10_rm0_a4_and_a100_k3}\includegraphics[width=0.5\textwidth]{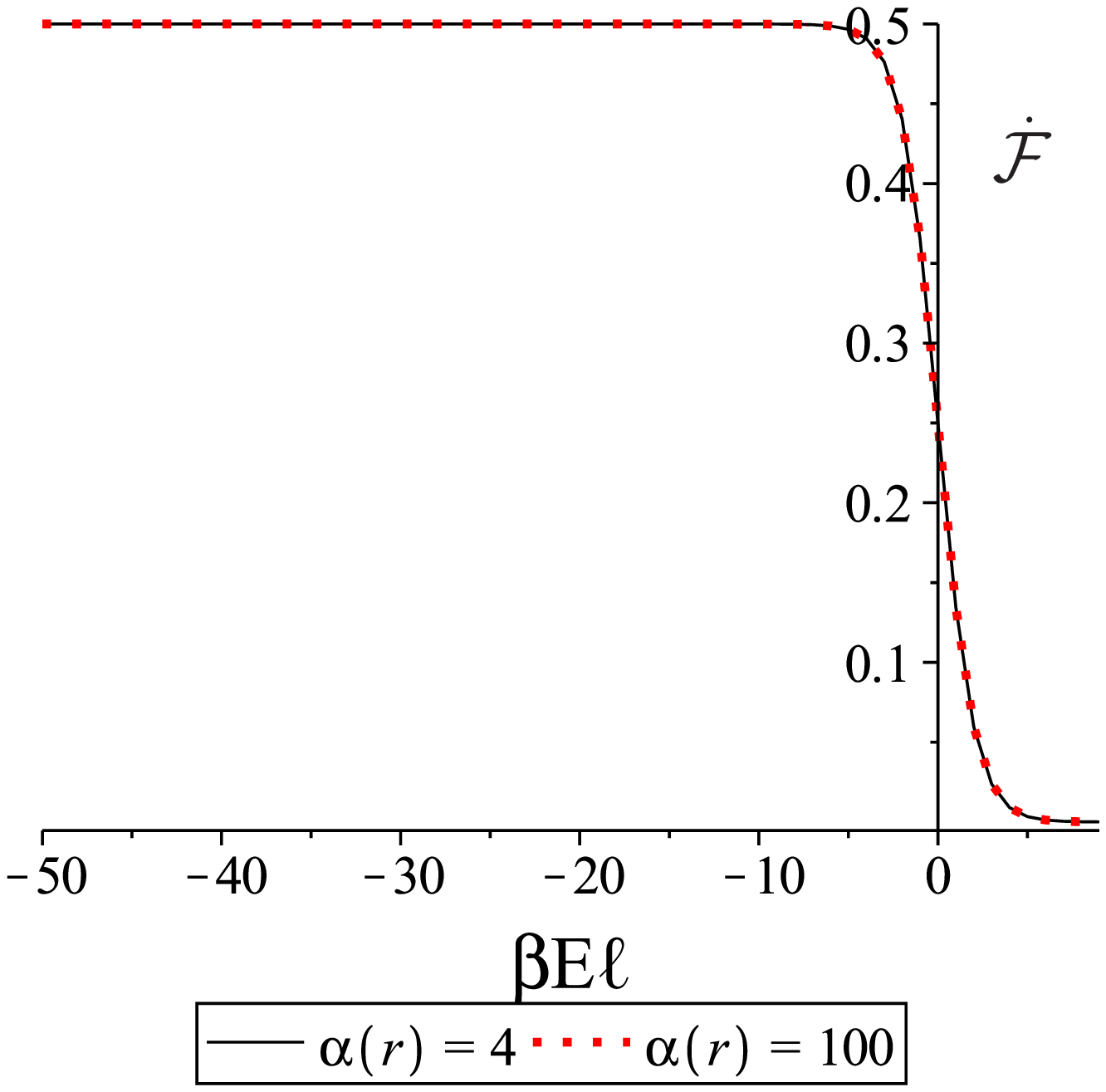}}            
  \subfloat[$\zeta=1$]{\label{fig:Num_bc1_rp10_rm0_a4_and_a100_k3}\includegraphics[width=0.5\textwidth]{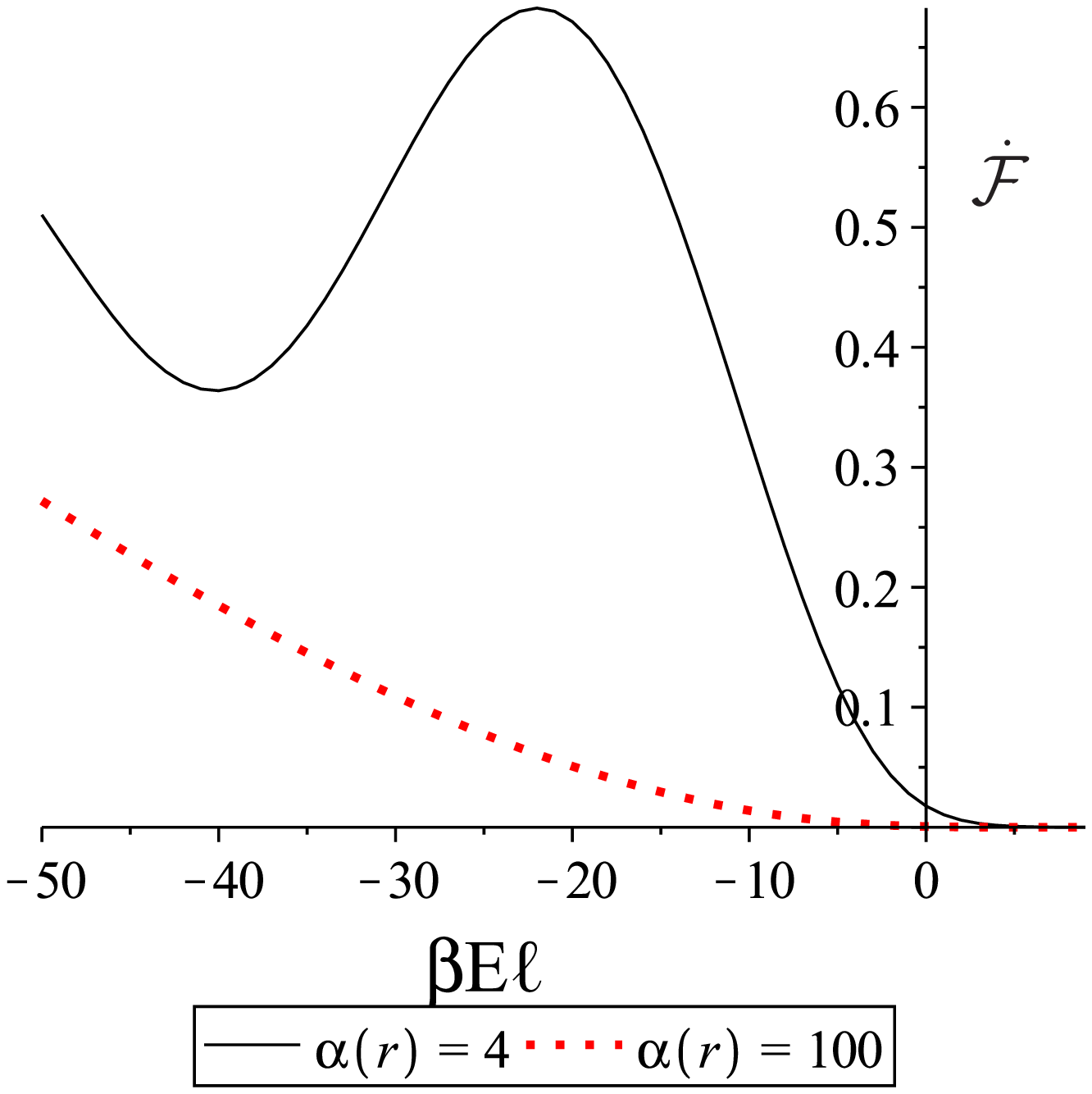}}  \\[2em]
  \subfloat[$\zeta=-1$]{\label{fig:Num_bc-1_rp10_rm0_a4_and_a100_k3}\includegraphics[width=0.5\textwidth]{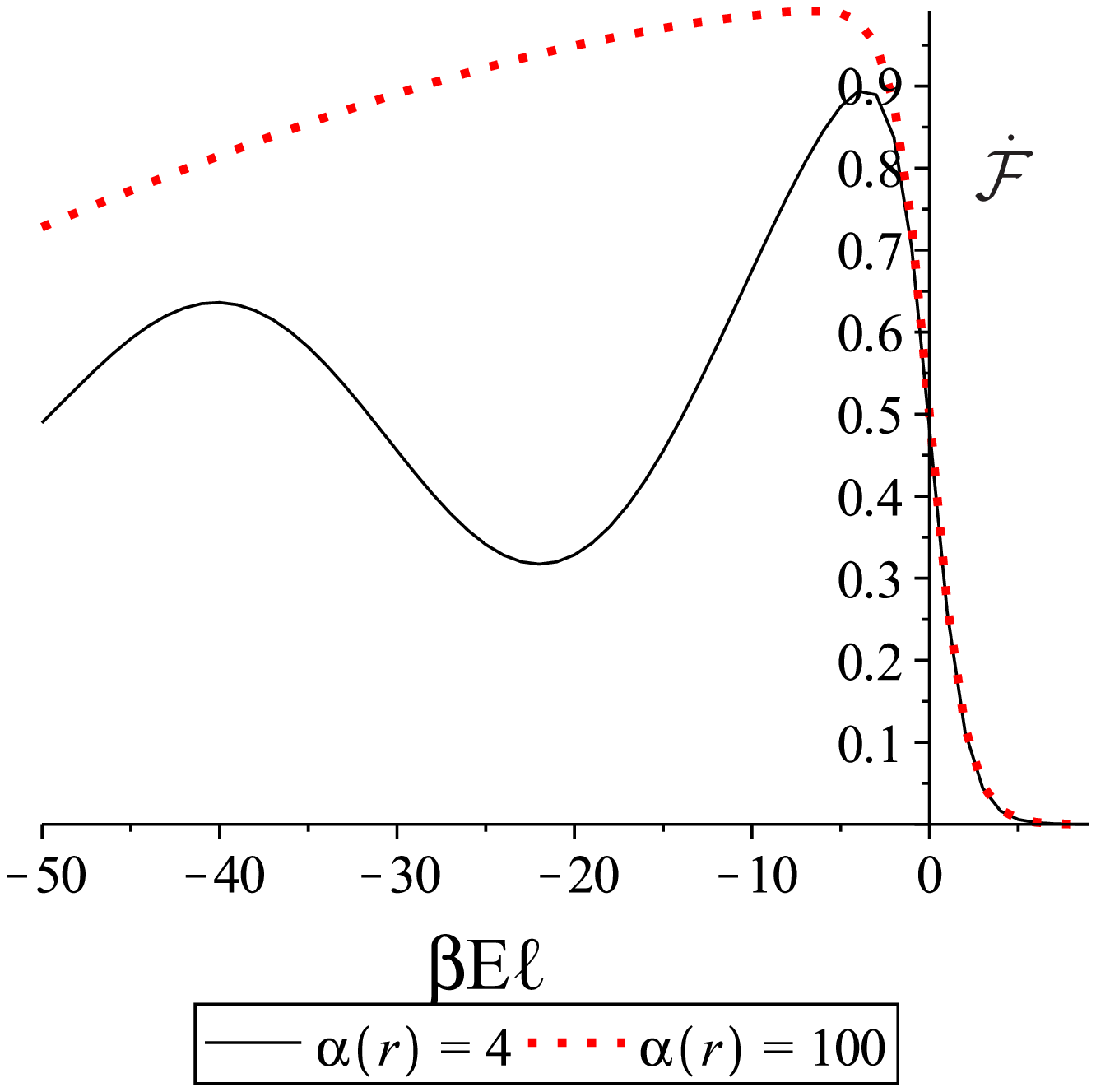}}  
\caption{$\dot{\mathcal{F}}$ as a function of $\beta E\ell$ for 
$r_+/\ell=10$ and $r_-/\ell=0$, with $\alpha=4$ (solid) and $\alpha=100$ (dotted). 
Numerical evaluation from \eqref{eq:corot:rate.evald} with $n\le3$.}
\label{fig:transrate_rp10}
\end{figure}

In the regime $r_+/\ell \gtrsim 3$, numerics confirms that 
the $n\ge1$ terms in \eqref{eq:corot:rate.evald} are small. 
$\dot{\mathcal{F}}$ therefore depends on 
$r_+/\ell$ and $r_-/\ell$ significantly only through~$\beta$, 
that is, through the local temperature. 
The detector's location enters 
$\dot{\mathcal{F}}$ in part via 
$\beta$ \eqref{eq:beta-def}, but also via $Q_0$ 
in \eqref{eq:corot:rate.evald}, 
and the latter affects only the boundary 
conditions $\zeta=1$ and $\zeta=-1$, in opposite directions. 
Plots for $r_+/\ell =10$ are shown in Figure~\ref{fig:transrate_rp10}. 

\begin{figure}[p!]
  \centering
  \subfloat[$\zeta=0$]{\label{fig:Num_bc0_rp1_rm0_rm0pt99_a2_k3}\includegraphics[width=0.5\textwidth]{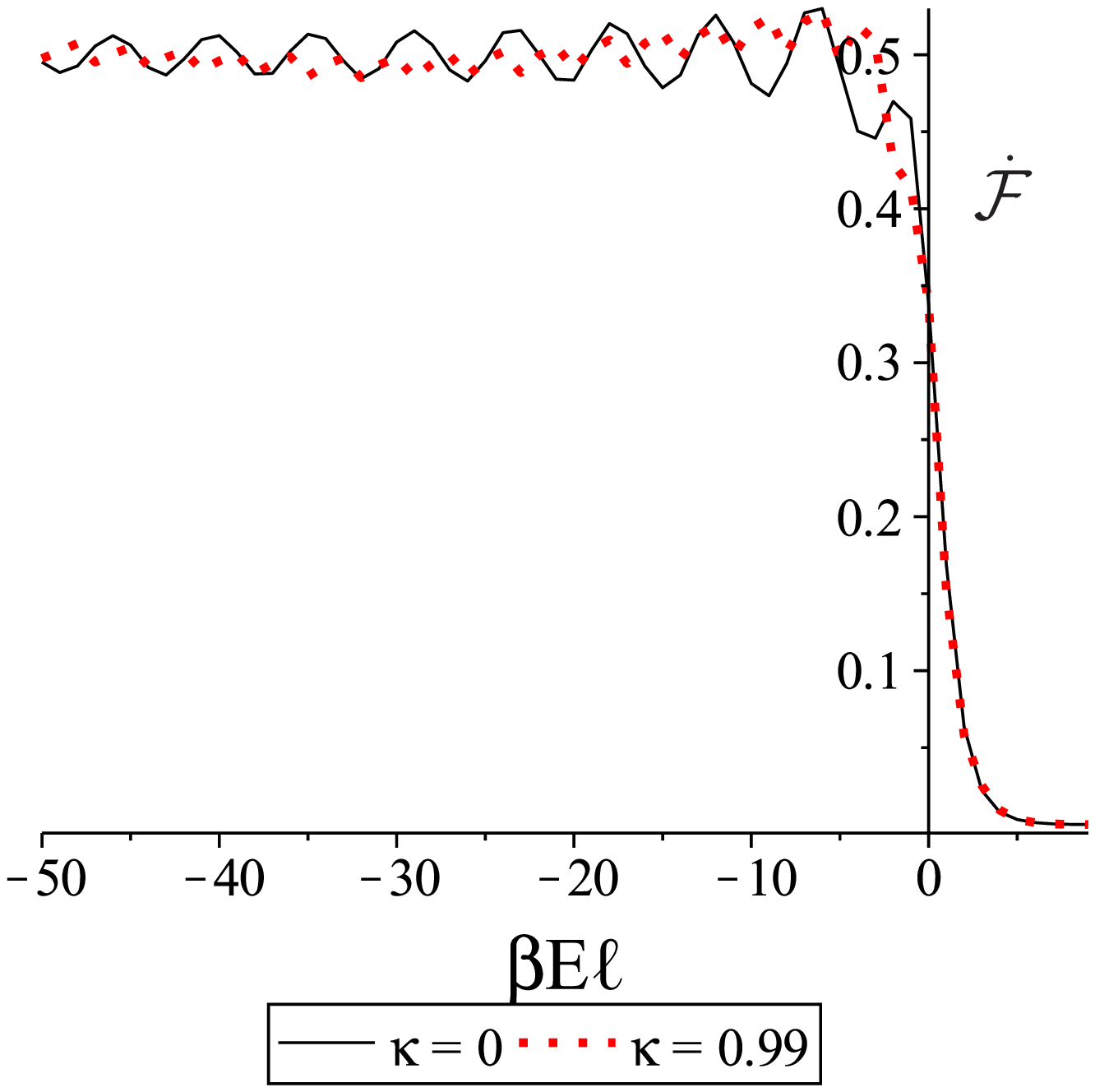}}            
  \subfloat[$\zeta=1$]{\label{fig:Num_bc1_rp1_rm0_rm0pt99_a2_k3}\includegraphics[width=0.5\textwidth]{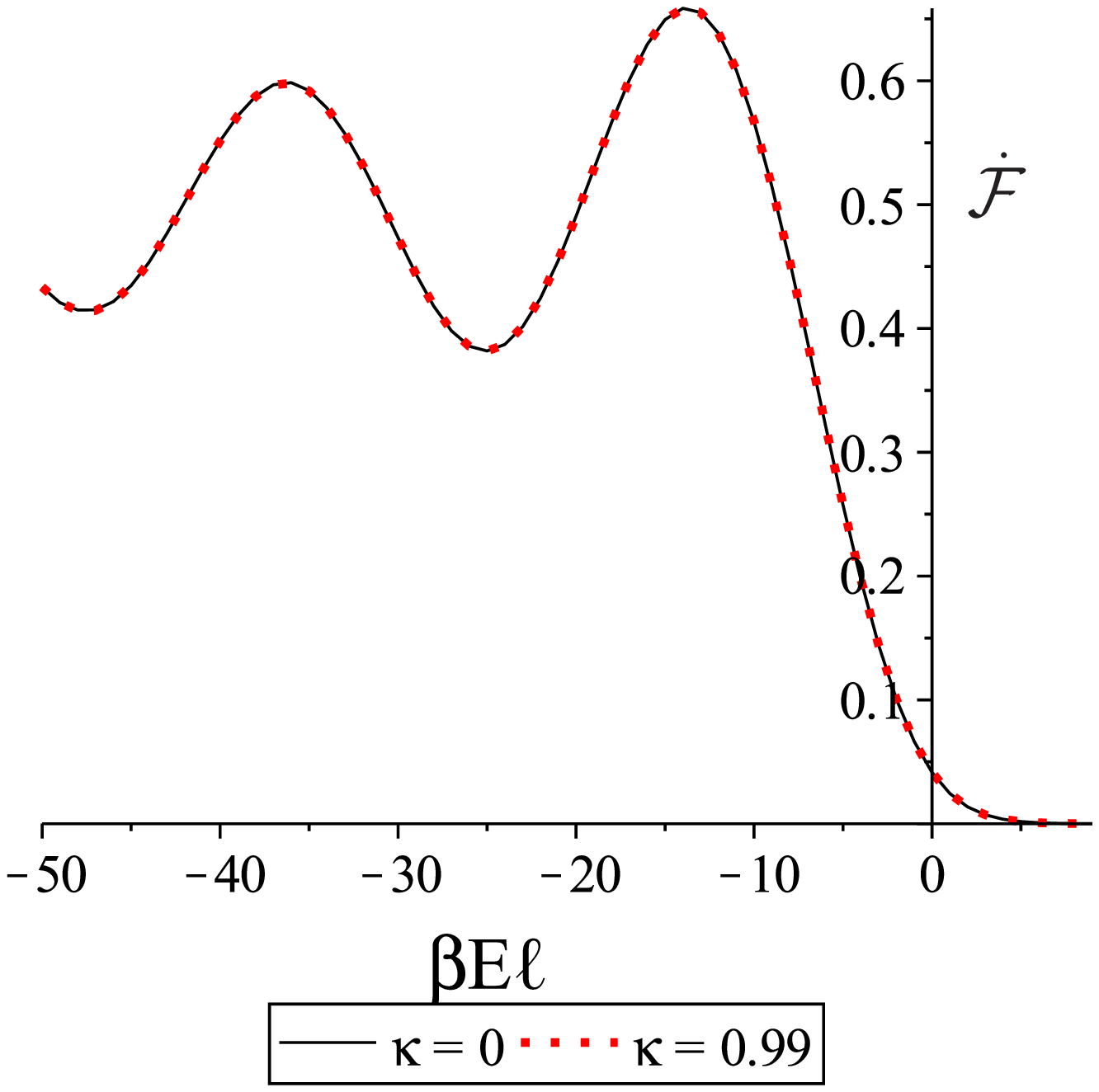}}  \\[2em]
  \subfloat[$\zeta=-1$]{\label{fig:Num_bc-1_rp1_rm0_rm0pt99_a2_k3}\includegraphics[width=0.5\textwidth]{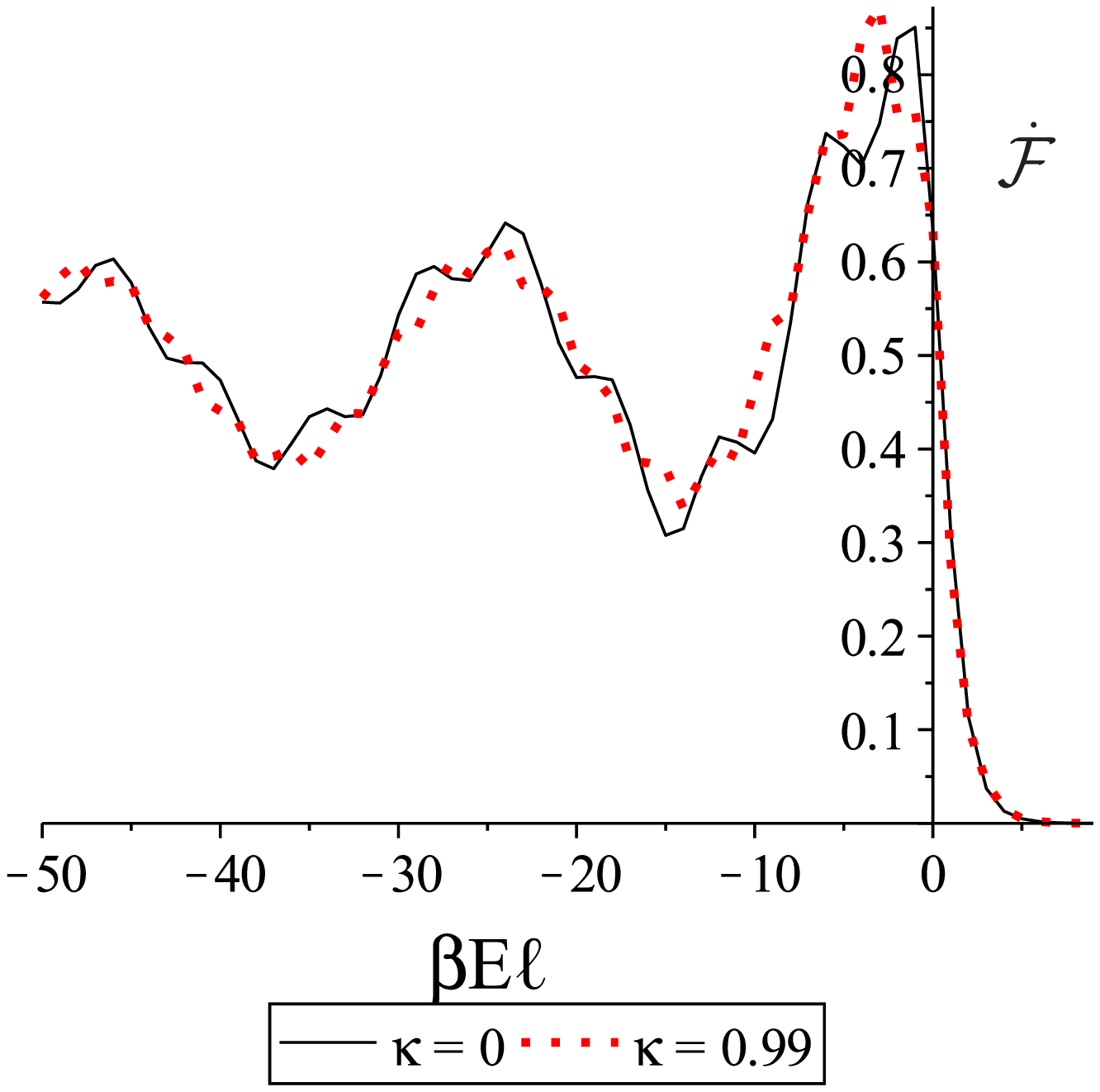}}  
\caption{$\dot{\mathcal{F}}$ as a function of $\beta E\ell$ for 
$r_+/\ell=1$ and $\alpha=2$, with $\kappa =0$ (solid) and $\kappa =0.99$ (dotted) where $\kappa := r_-/r_+$. 
Numerical evaluation from \eqref{eq:corot:rate.evald} with $n\le3$.}
\label{fig:transrate_rp1_a2_rm0_rm0pt99}
\end{figure}

\begin{figure}[p!]
    \centering
  \subfloat[$\zeta=0$]{\label{fig:Num_bc0_rp1_rm0_rm0pt99_a100_k3}\includegraphics[width=0.5\textwidth]{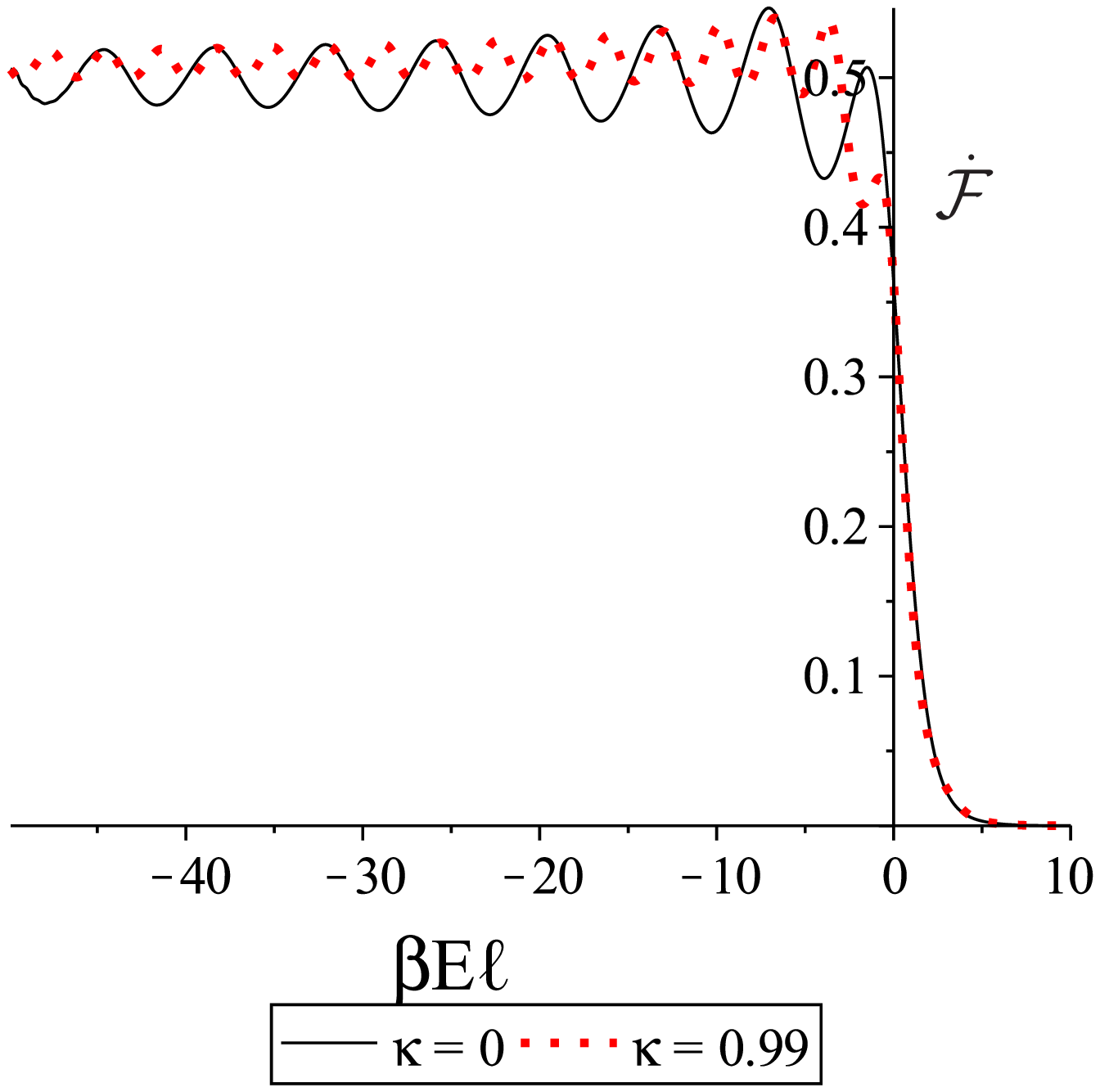}}            
  \subfloat[$\zeta=1$]{\label{fig:Num_bc1_rp1_rm0_rm0pt99_a100_k3}\includegraphics[width=0.5\textwidth]{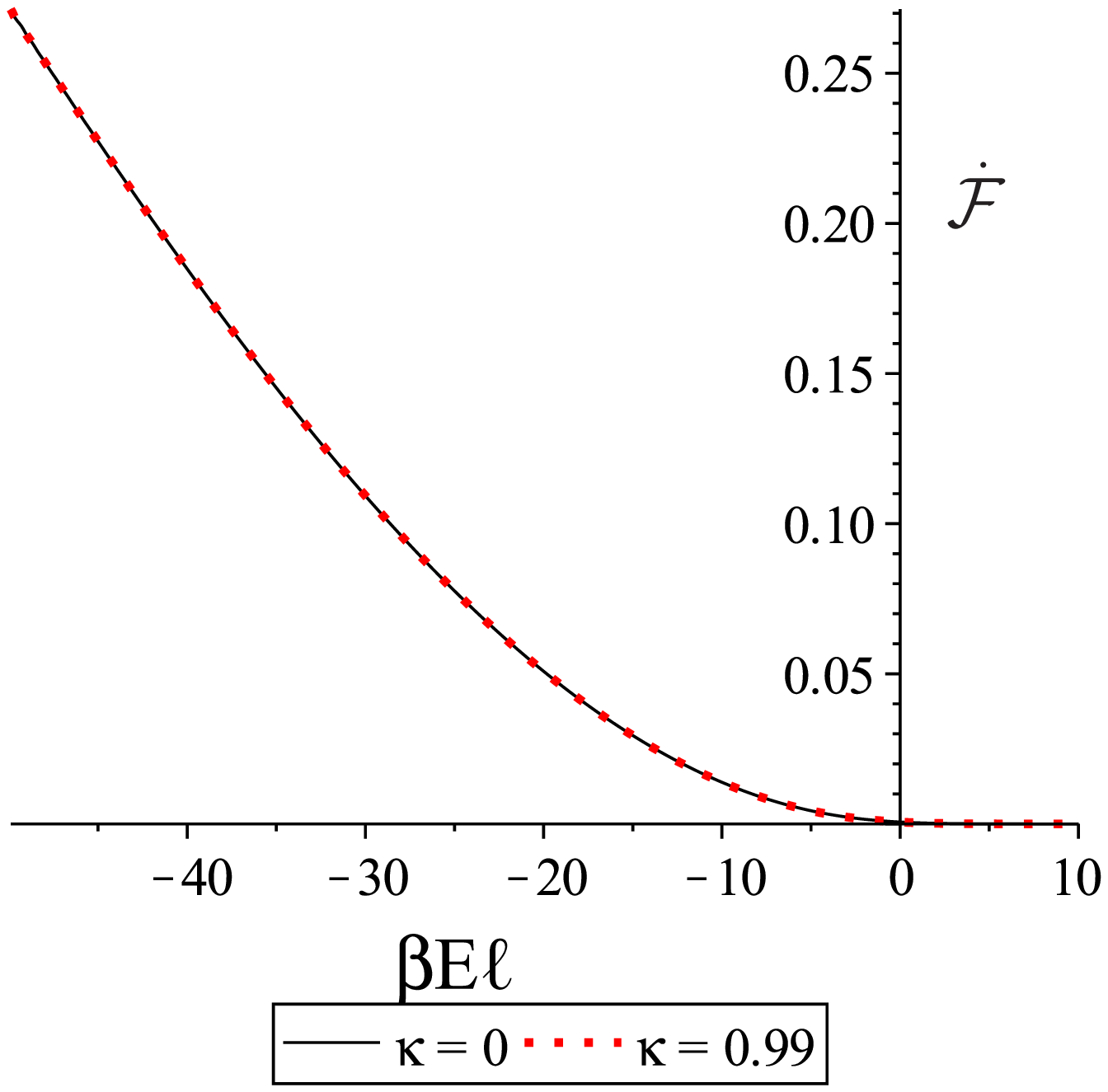}} \\[2em] 
  \subfloat[$\zeta=-1$]{\label{fig:Num_bc-1_rp1_rm0_rm0pt99_a100_k3}\includegraphics[width=0.5\textwidth]{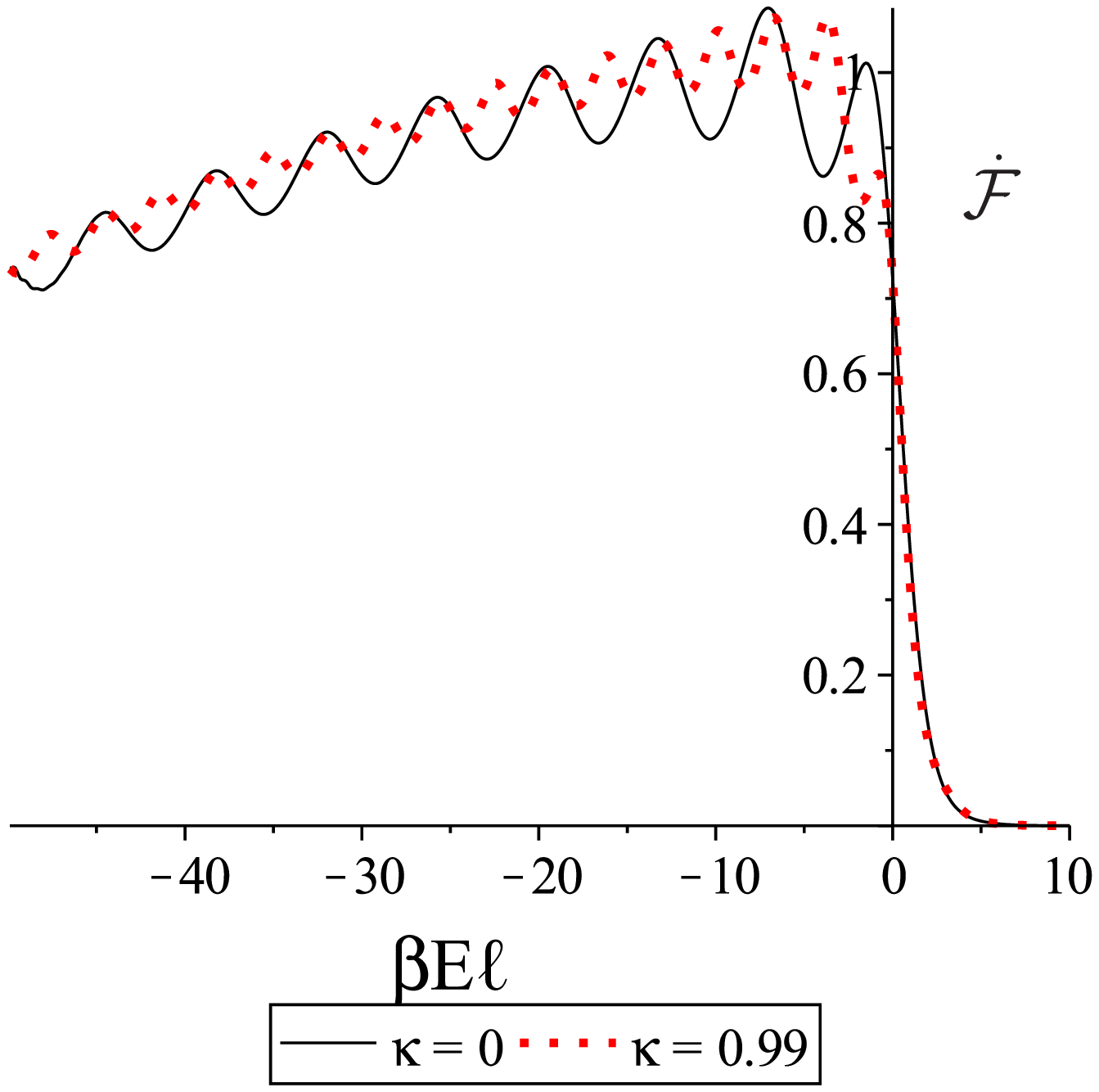}}  
\caption{As in Figure \ref{fig:transrate_rp1_a2_rm0_rm0pt99} but for $\alpha=100$.}
\label{fig:transrate_rp1_a100_rm0_rm0pt99}
\end{figure}

As $r_+/\ell$ decreases, the $n=1$ term in 
\eqref{eq:corot:rate.evald} starts to become appreciable near $r_+/\ell \approx 1$. 
The dependence on $r_-/\ell$ is then no longer exclusively through~$\beta$, and the  
effect is largest for $\zeta=0$ and $\zeta=-1$ but smaller for $\zeta=1$, 
owing to a partial cancellation between the two terms under 
the integral in \eqref{eq:corot:rate.evald} for $\zeta=1$. 
Plots for $r_+/\ell =1$ are shown in Figures
\ref{fig:transrate_rp1_a2_rm0_rm0pt99}
and~\ref{fig:transrate_rp1_a100_rm0_rm0pt99}. 

\begin{figure}[p]
    \centering
  \subfloat[$\zeta=0$]{\label{fig:Num_vs_largeAsy_bc0_rp0pt3_rm0pt299_a2_k3}\includegraphics[width=0.5\textwidth]{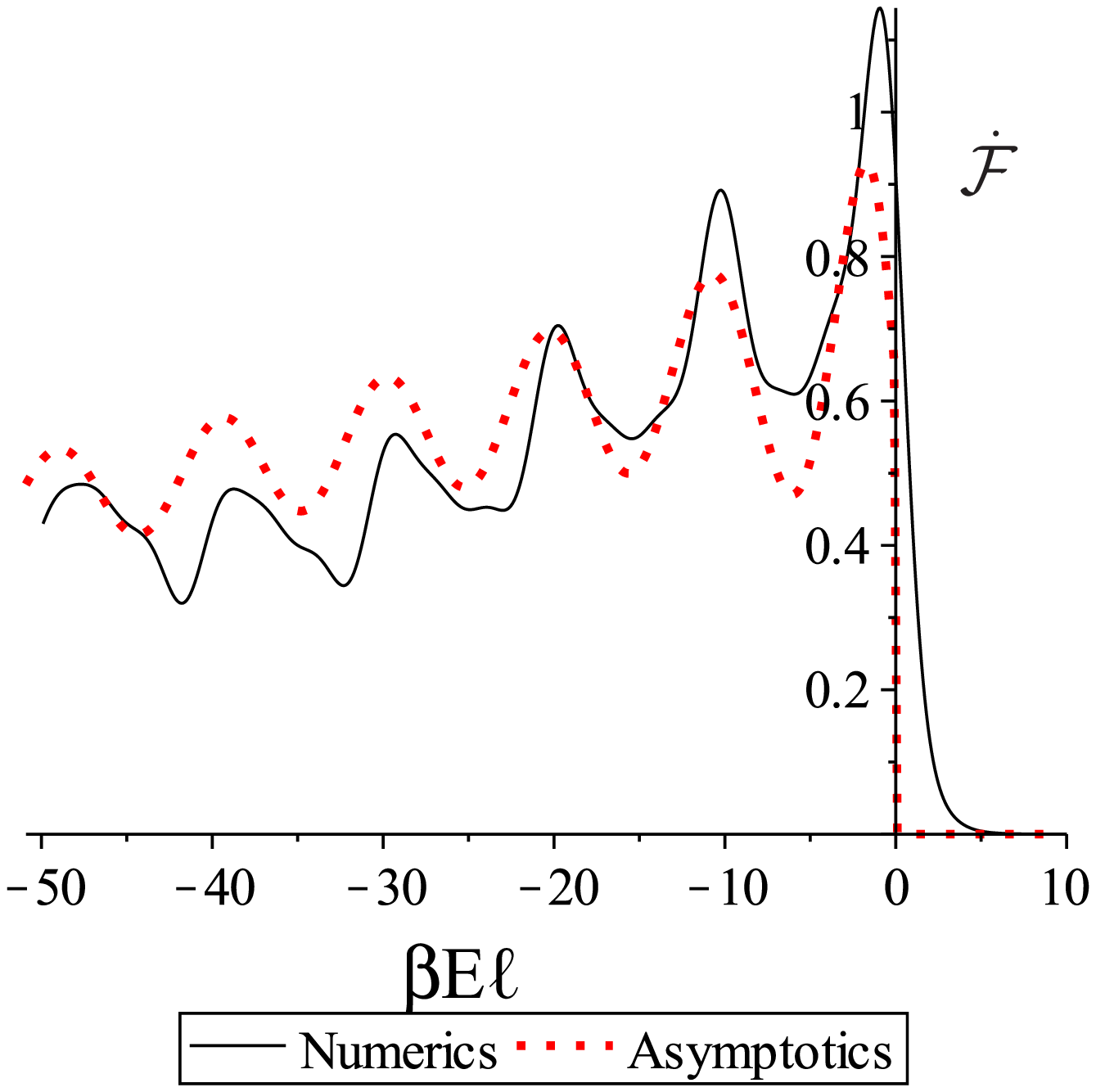}}            
  \subfloat[$\zeta=1$]{\label{fig:Num_vs_largeAsy_bc1_rp0pt3_rm0pt299_a2_k3}\includegraphics[width=0.5\textwidth]{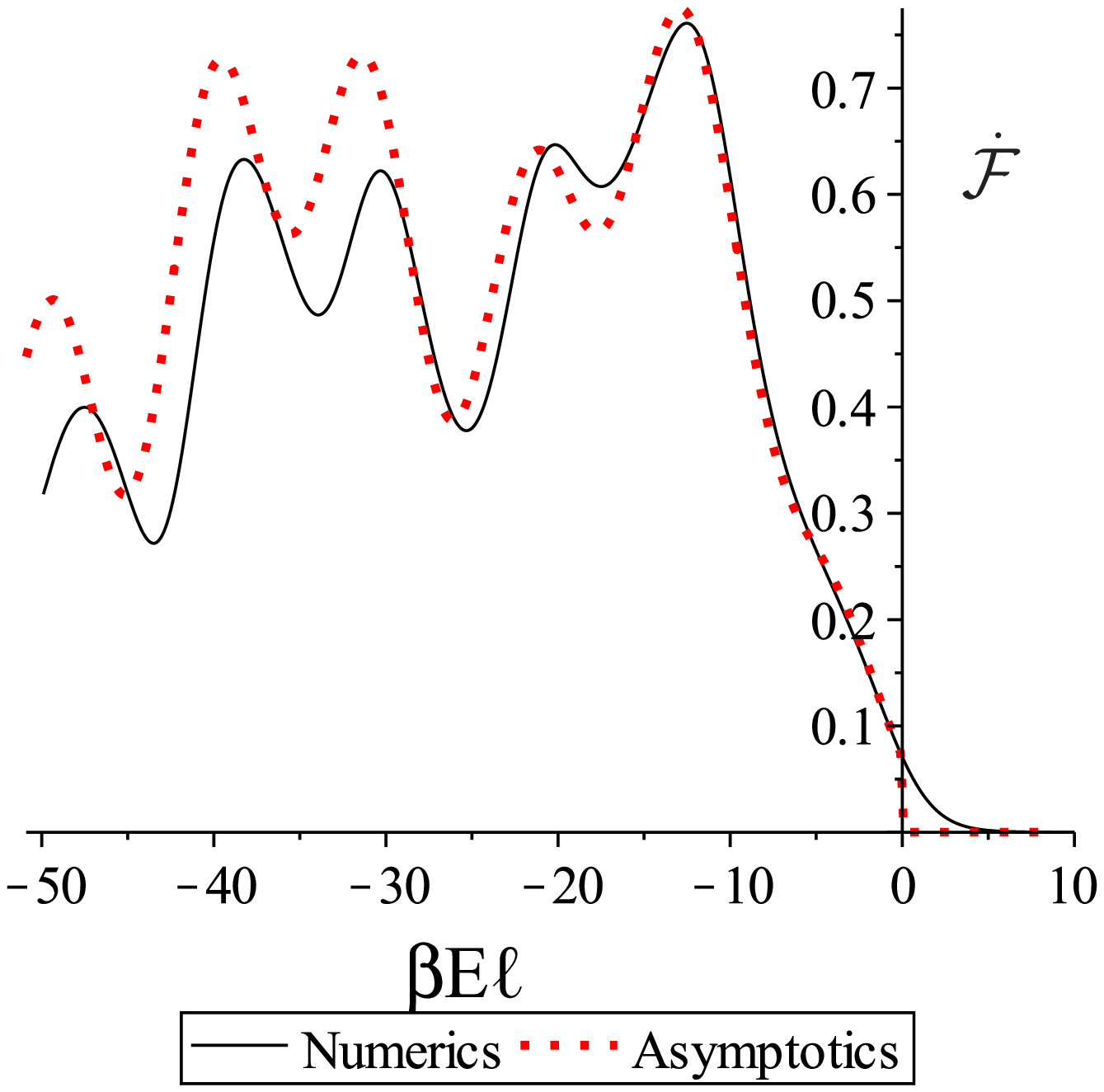}}  \\[2em]
  \subfloat[$\zeta=-1$]{\label{figNum_vs_largeAsy_bc-1_rp0pt3_rm0pt299_a2_k3}\includegraphics[width=0.5\textwidth]{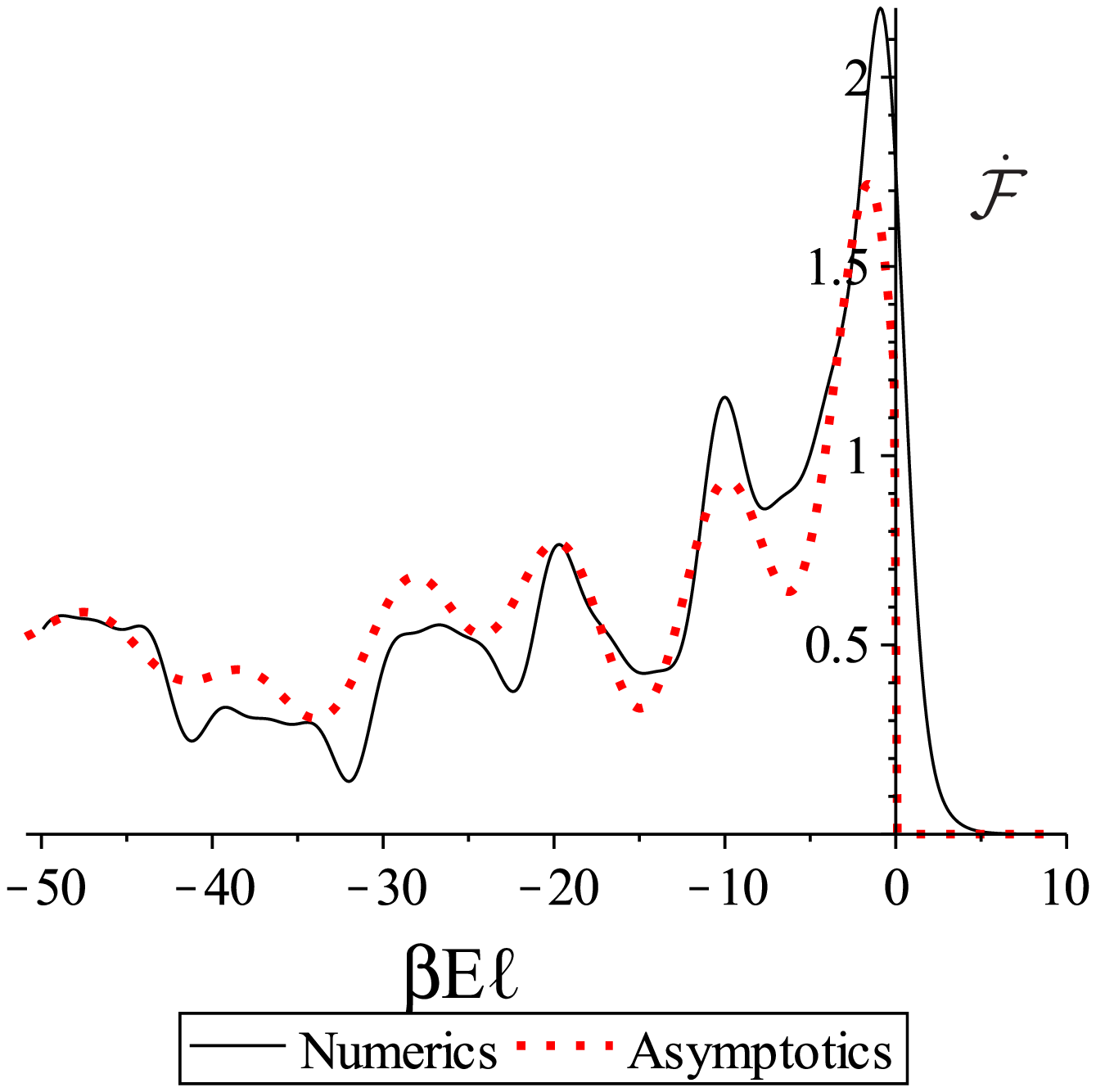}}  
\caption{$\dot{\mathcal{F}}$ as a function of $\beta E\ell$ for 
$r_+/\ell=0.3$ and $r_-/\ell=0.299$, with $\alpha=2$. 
Solid curve shows numerical evaluation from \eqref{eq:corot:rate.evald} with $n\le3$. 
Dotted curve shows the asymptotic, large-$r_+/\ell$ approximation~\eqref{eq:corot:larger+}.}
\label{fig:Num_vslargeAsy_rp0pt3_a2_rm0pt299}
\end{figure}

\begin{figure}[p]
    \centering
  \subfloat[$\zeta=0$]{\label{fig:bc0.eps}\includegraphics[width=0.5\textwidth]{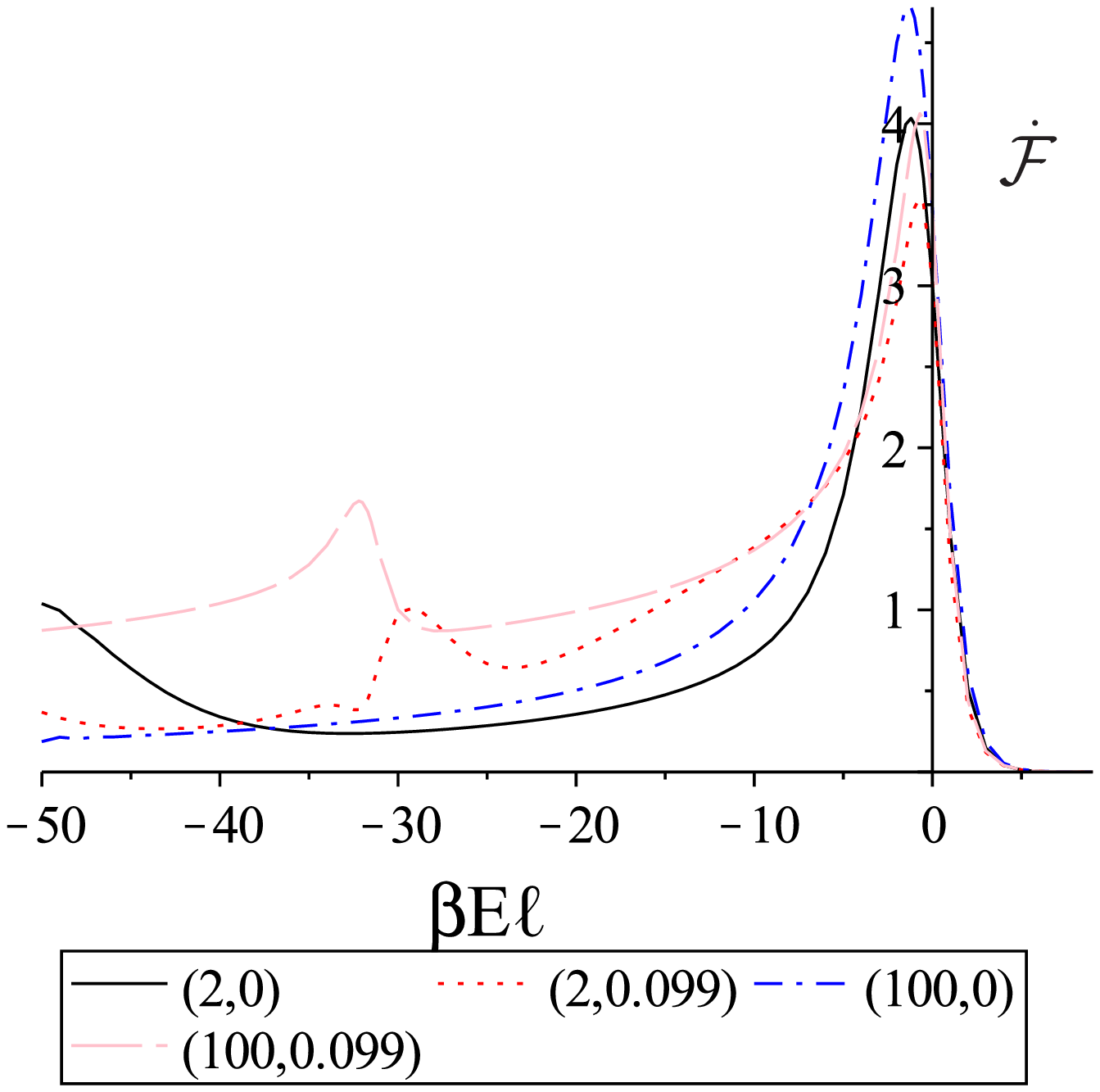}}            
  \subfloat[$\zeta=1$]{\label{fig:Num_bc1_rp0pt01_rm0_rm0pt99_a2_a100_k3}\includegraphics[width=0.5\textwidth]{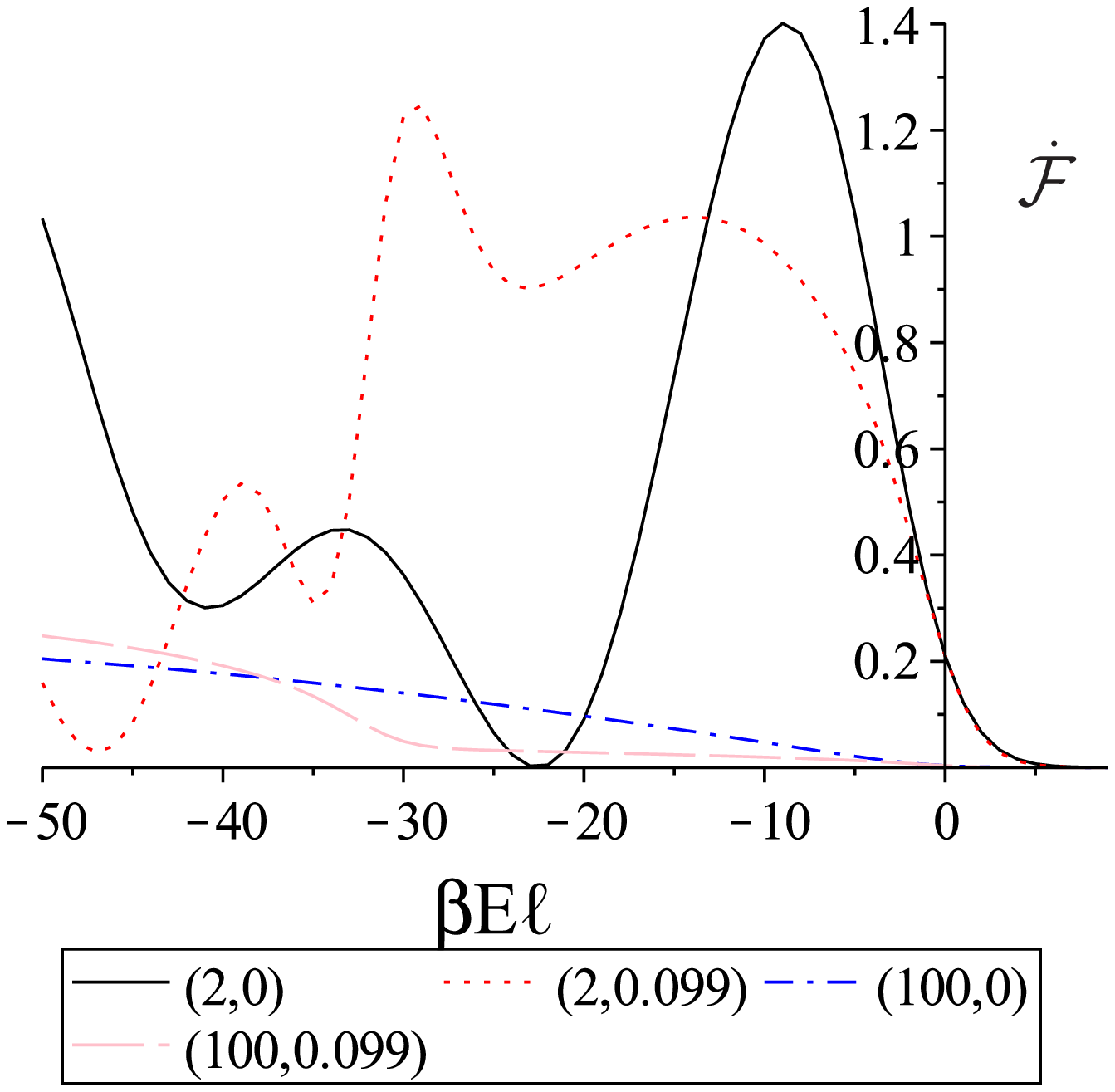}}  \\[2em]
  \subfloat[$\zeta=-1$]{\label{fig:Num_bc-1_rp0pt01_rm0_rm0pt99_a2_a100_k3}\includegraphics[width=0.5\textwidth]{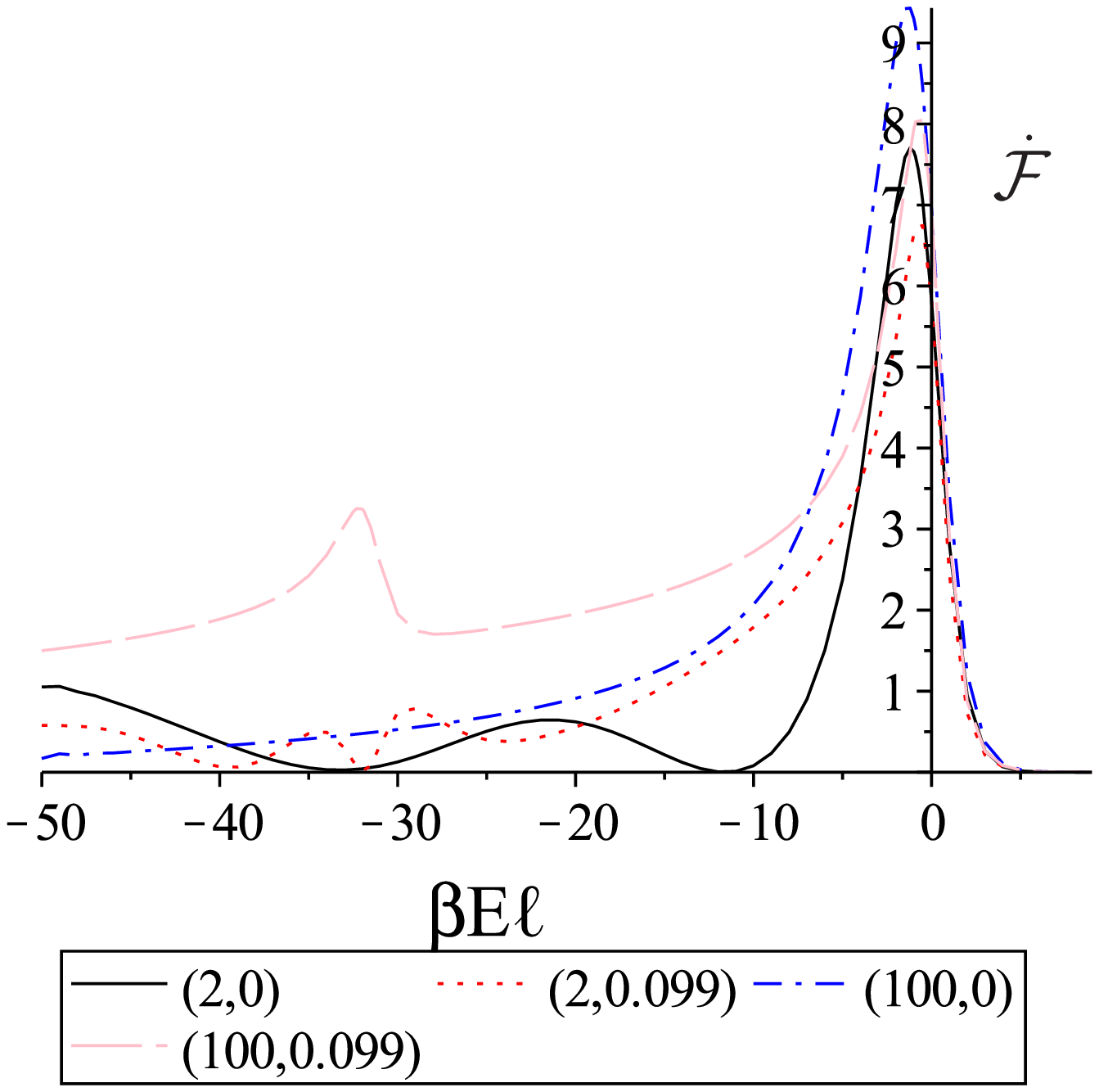}}  
\caption{$\dot{\mathcal{F}}$ as a function of $\beta E\ell$ for 
$r_+/\ell=0.1$, with selected values of the pair $(\alpha,r_-/\ell)$ as shown in the legend. 
Numerical evaluation from \eqref{eq:corot:rate.evald} with $n\le35$.}
\label{fig:transrate_rp0pt1_a2_a100_rm0_rm0pt99}
\end{figure}

As $r_+/\ell$ decreases below~$1$, 
the next-to-leading asymptotic formula \eqref{eq:corot:larger+} 
starts to become inaccurate near $r_+/\ell = 0.3$, 
as shown in Figure~\ref{fig:Num_vslargeAsy_rp0pt3_a2_rm0pt299}, 
although the partial cancellation between the two terms under the integral 
in \eqref{eq:corot:rate.evald} and the similar partial cancellation in 
\eqref{eq:corot:larger+} moderates the effect for $\zeta=1$. 
At $r_+/\ell = 0.1$, shown in Figure \ref{fig:transrate_rp0pt1_a2_a100_rm0_rm0pt99}, 
$\dot{\mathcal{F}}$ is sensitive to changes in both $r_-/\ell$ and~$\alpha$. 
When $\alpha\gg1$, the $\zeta=-1$ curves in Figure 
\ref{fig:transrate_rp0pt1_a2_a100_rm0_rm0pt99} have 
approximately the same profile as the $\zeta=0$ 
curves but at twice the magnitude: from 
\eqref{eq:corot:Kn} and \eqref{eq:corot:Qn} 
we see that this indicates the regime 
where the $n\ge1$ terms in~\eqref{eq:corot:rate.evald} give the 
dominant contribution to~$\dot{\mathcal{F}}$. 

\begin{figure}[p]\centering

  \subfloat[$\zeta=0$]{\label{fig:Num_vs_smallAsy_bc0_rp0pt01_rm0_a4_k300}\includegraphics[width=0.5\textwidth]{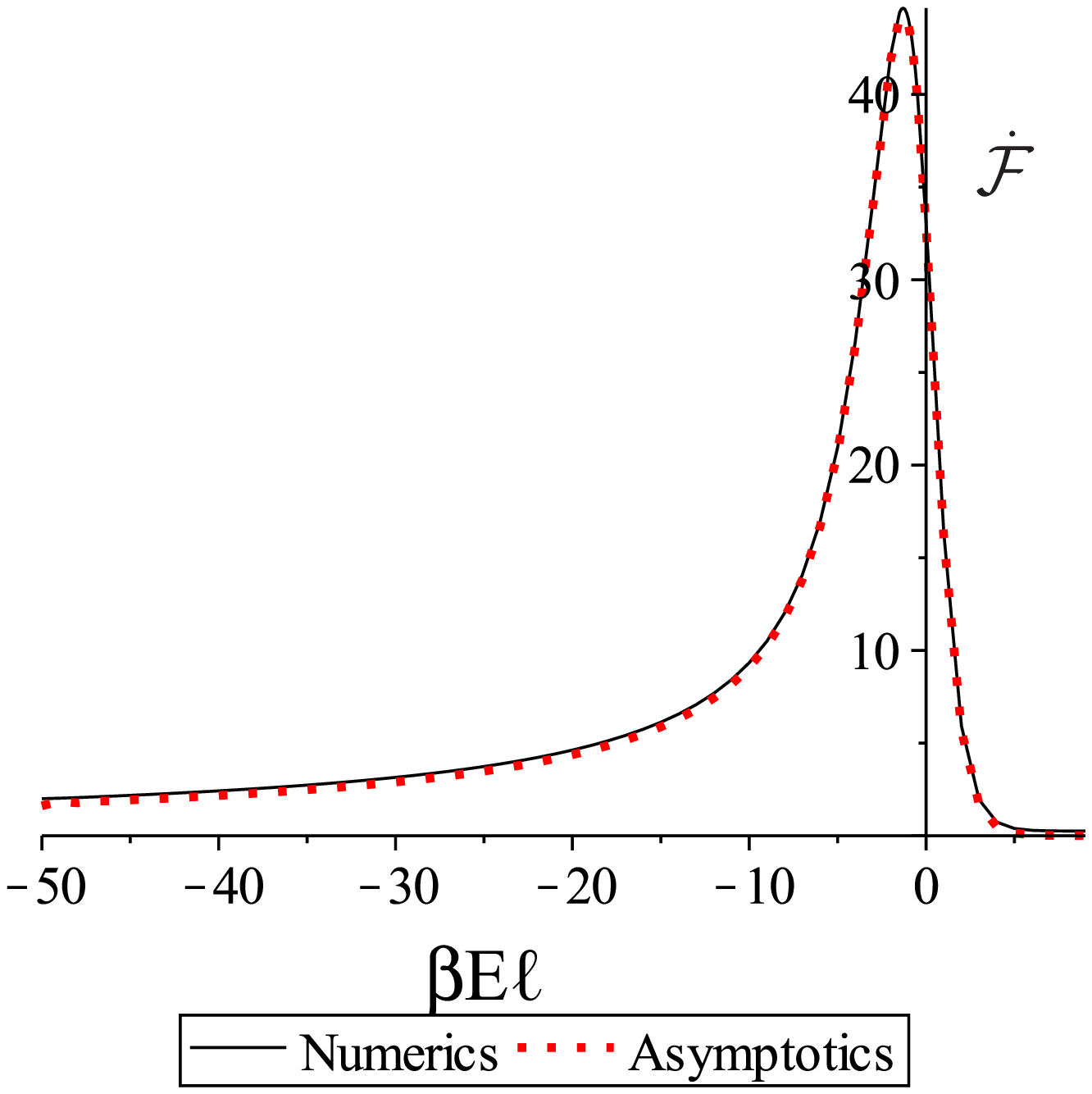}}          
  \subfloat[$\zeta=1$]{\label{fig:Num_vs_smallAsy_bc1_rp0pt01_rm0_a4_k300}\includegraphics[width=0.5\textwidth]{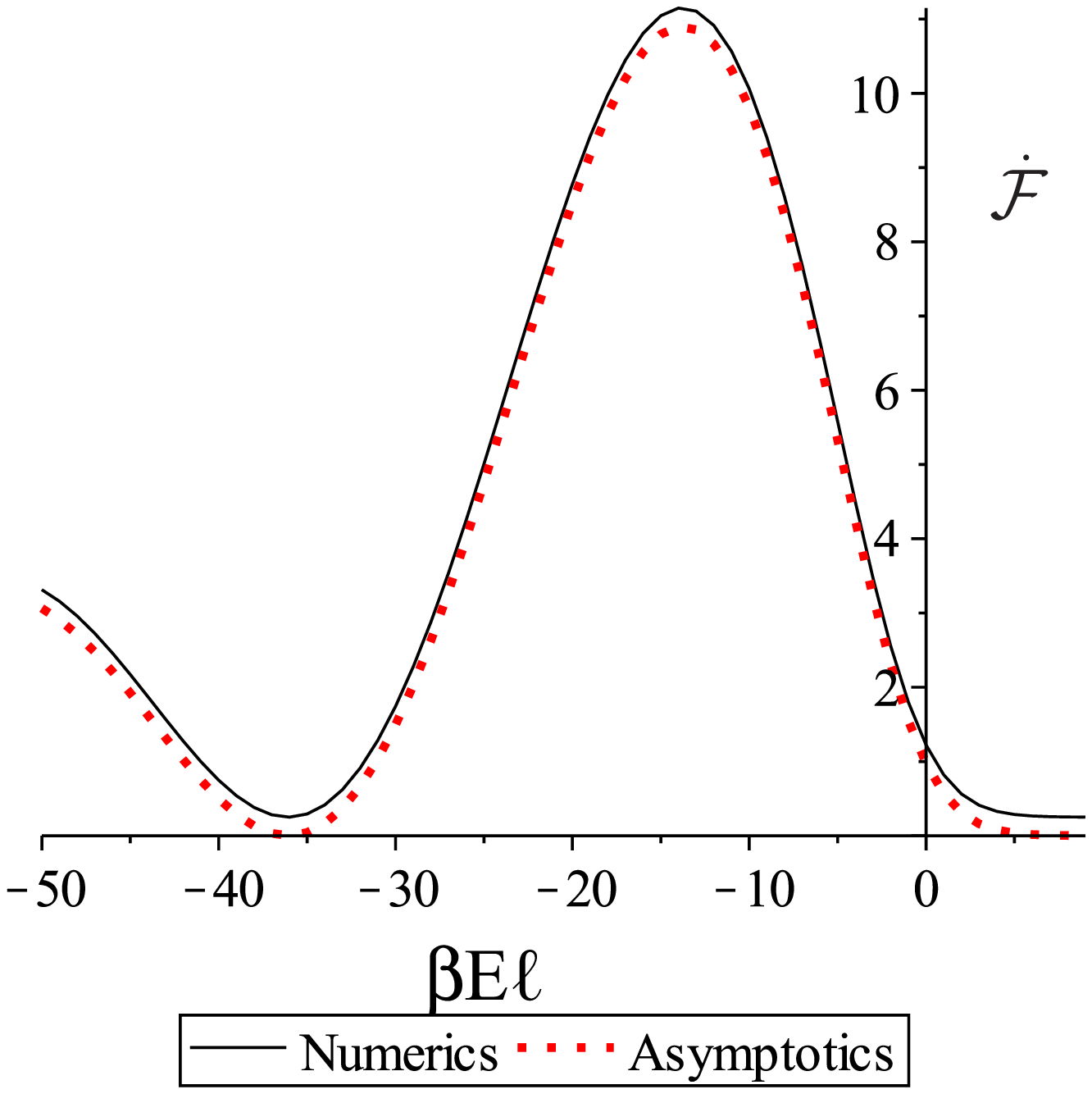}}  \\[2em]
  \subfloat[$\zeta=-1$]{\label{fig:Num_vs_smallAsy_bc-1_rp0pt01_rm0_a4_k300}\includegraphics[width=0.5\textwidth]{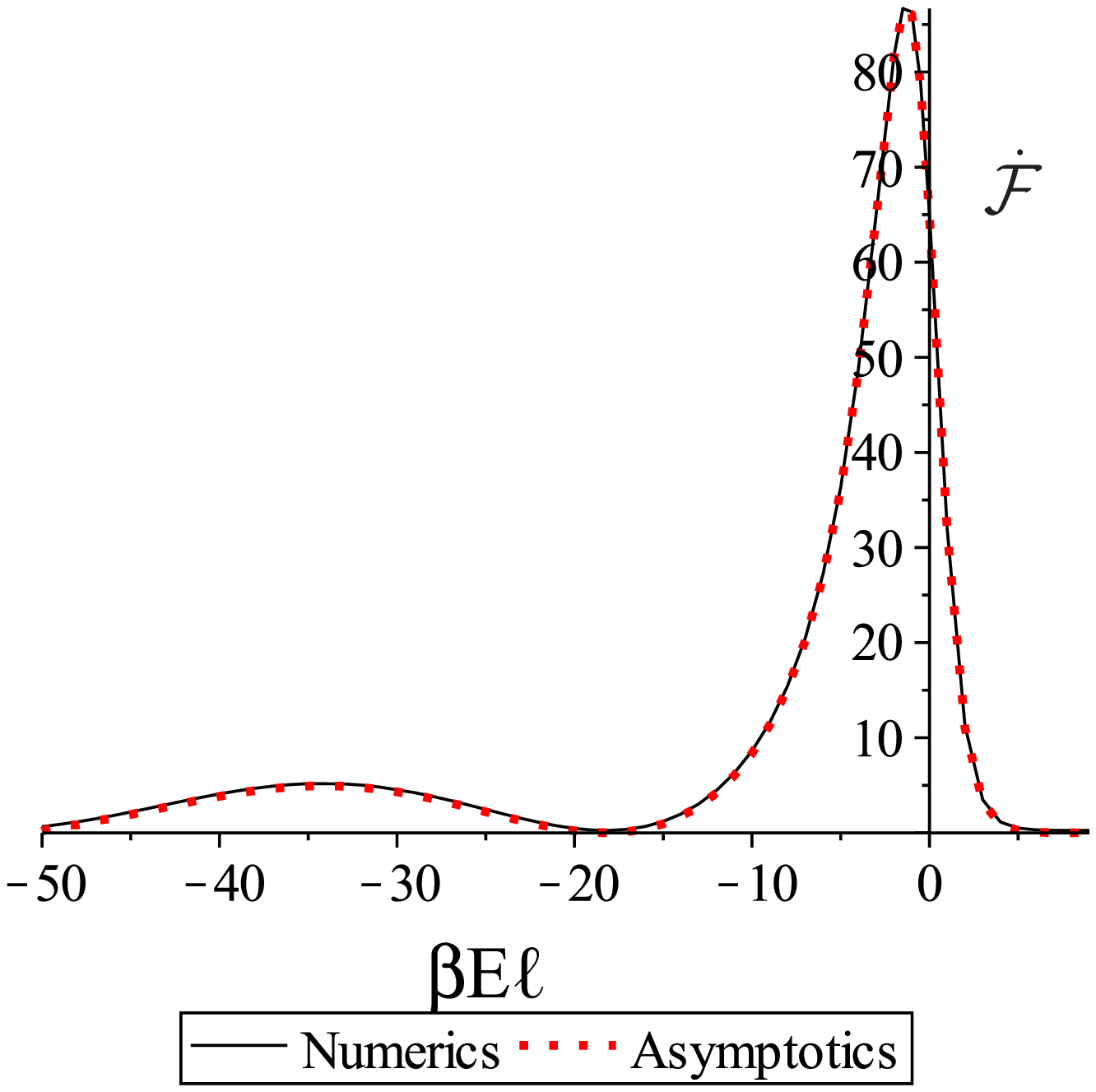}}  
\caption{$\dot{\mathcal{F}}$ as a function of $\beta E\ell$ for 
$r_+/\ell=0.01$ and $r_-=0$, with $\alpha=4$. 
Solid curve shows numerical evaluation from \eqref{eq:corot:rate.evald} with $n\le300$. 
Dotted curve shows the asymptotic small-$r_+/\ell$ approximation~\eqref{eq:corot:smallr+}. 
Qualitatively similar graphs ensue for $r_-/r_+=0.99$.}
\label{fig:Num_vs_smallAsy_rp0pt01_rm0_a4_k300}
\end{figure}

As $r_+/\ell$ decreases further, we enter the validity regime of the asymptotic 
formula \eqref{eq:corot:smallr+}, as shown in 
Figure~\ref{fig:Num_vs_smallAsy_rp0pt01_rm0_a4_k300} for $r_+/\ell=0.01$. 
Note that again 
the $\zeta=-1$ curve has approximately the same profile as the $\zeta=0$ 
curve but at twice the magnitude, indicating that the dominant contribution 
comes from the $n\ge1$ terms in~\eqref{eq:corot:rate.evald}.

\section{Radially-infalling detector in spinless BTZ}
\label{sec:inertial} 

In this section we consider a detector on a radially-infalling
geodesic in a spinless BTZ spacetime. 

\subsection{Transition rate}
\label{sec:inertial:transrate}

Recall from Section
\ref{sec:BTZoverview} that for a spinless hole $r_-=0$ and $r_+ =
M\ell >0$, and the horizon is at $r= r_+$. To begin with, we assume that at least
part of the trajectory is in the exterior region, $r>r_+$. Working in
the exterior BTZ co-ordinates~\eqref{eq:BoyerLind}, the
radial timelike geodesics take the form 
\begin{align}
&r=\ell\sqrt{M} q \cos\tilde\tau\,, 
\nonumber\\
&t=\bigl(\ell/\sqrt{M} \, \bigr)
\arctanh\left(\frac{\tan\tilde\tau}{\sqrt{q^2-1}} \right)\,, 
\nonumber\\
&\phi=\phi_0\, , 
\label{eq:inertial:traj}
\end{align}
where $q>1$, $\phi_0$ denotes the
constant value of~$\phi$, and $\tilde\tau$ is an affine parameter such that the proper 
time equals~$\tilde\tau\ell$. 
The additive constants in $\tilde\tau$ and $t$ 
have been chosen so that $r$ reaches its maximum value 
$\ell\sqrt{M}q$ at $\tilde\tau=0$ with $t=0$. 

Substituting \eqref{eq:inertial:traj} in \eqref{eq:BTZ:rate}
and~\eqref{eq:DXN2}, we find that the transition rate is given by 
\begin{align}
&\dot{\mathcal{F}}_{\tau}(E) 
=
1/4
\nonumber\\[1ex]
&\hspace{2ex}
+\frac{1}{2\pi\sqrt{2}}
\sum_{n=-\infty}^{\infty}
\int^{\Delta\tilde\tau}_{0}\,\mathrm{d}\tilde{s}
\Realpart\Biggl[\frac{\mathrm{e}^{-i\Etilde
\tilde{s}}}{\sqrt{-1+K_n\cos\tilde{\tau} \cos( \tilde{\tau}-\tilde{s})
+\sin \tilde{\tau} \sin (\tilde{\tau}-\tilde{s})}}
\nonumber\\[1ex]
&\hspace{19ex}
-\zeta\frac{\mathrm{e}^{-i \Etilde
\tilde{s}}}{\sqrt{1+K_n\cos\tilde{\tau} \cos(\tilde{\tau} -\tilde{s})
+\sin \tilde{\tau} \sin (\tilde{\tau}-\tilde{s})}}
\Biggr]\,, 
\label{eq:inertial:rate}
\end{align}
where 
\begin{equation}
\label{eq:inertial:Kn}
K_n:= 1+2q^2\sinh^2\bigl(n\pi\sqrt{M} \, \bigr)\, . 
\end{equation}
The detector is switched off at proper time $\tau$ and switched on at proper time 
$\tau_0 = \tau - \Delta\tau$, and we have written 
$\tilde\tau := \tau/\ell$, $\Delta\tilde\tau := \Delta\tau/\ell$
and $\Etilde := E \ell$. 
The square roots in \eqref{eq:inertial:rate} are positive when the
arguments are positive, and they are analytically continued to
negative values of the arguments by giving $\tilde{s}$ a small
negative imaginary part. 

Although the above derivation of \eqref{eq:inertial:rate} proceeded using the exterior BTZ
co-ordinates, the result \eqref{eq:inertial:rate} holds by analytic
continuation even if the geodesic enters the black or white hole regions. 
The ranges of the parameters are 
$-\pi/2 < \tilde\tau - \Delta\tilde\tau < \tilde\tau < \pi/2$, 
so that the 
detector is switched on after emerging from the white hole singularity 
and switched off before hitting the black hole singularity.

\subsection{The $n=0$ term and KMS}
\label{sec:inertial:zeroth}

We write \eqref{eq:inertial:rate} as 
\begin{align}
\dot{\mathcal{F}}_{\tau} = 
\dot{\mathcal{F}}_{\tau}^{n=0} + \dot{\mathcal{F}}_{\tau}^{n\ne0}
\ , 
\end{align}
where $\dot{\mathcal{F}}_{\tau}^{n=0}$ consists of the $n=0$ term and 
$\dot{\mathcal{F}}_{\tau}^{n\ne0}$ consists of the sum $\sum_{n\ne0}$. 
We consider first $\dot{\mathcal{F}}_{\tau}^{n=0}$. 

$\dot{\mathcal{F}}_{\tau}^{n=0}$ gives the transition rate of 
a detector on a geodesic in pure~$\text{AdS}_3$. 
$\dot{\mathcal{F}}_{\tau}^{n=0}$
does not depend on $M$ or~$q$, 
and it depends on the switch-on and switch-off moments 
only through~$\Delta\tilde\tau$, the total detection time. 
Using 
\eqref{eq:inertial:rate} and~\eqref{eq:inertial:Kn}, we find 
\begin{align}
\label{eq:inertial:n0}
\dot{\mathcal{F}}_{\tau}^{n=0}(E) 
=
\frac14
- \frac{1}{4\pi}
\int^{\Delta\tilde{\tau}}_{0}\,\mathrm{d}\tilde{s}\,
\Biggl[\frac{\sin\bigl(\Etilde\tilde{s}\bigr)}{\sin(\tilde{s}/2)}
+\zeta\frac{\cos\bigl(\Etilde\tilde{s}\bigr)}{\cos(\tilde{s}/2)}\Biggr]\, , 
\end{align}
where $\Etilde := E\ell$. 
As 
$0<\Delta\tilde{\tau}< \pi$, 
\eqref{eq:inertial:n0} is well defined. 

Numerical examination shows that $\dot{\mathcal{F}}_{\tau}^{n=0}$ 
does not satisfy the KMS condition. 
This is compatible with the embedding space
discussion of~\cite{Deser:1997ri,Deser:1998bb,Deser:1998xb,Russo:2008gb}, 
according to which a stationary detector in $\text{AdS}_3$ should respond 
thermally only when its scalar proper 
acceleration exceeds~$1/\ell$. 

The asymptotic behaviour of $\dot{\mathcal{F}}_{\tau}^{n=0}$ 
at large positive and negative energies for fixed $\Delta\tilde{\tau}$
can be found 
by the method of Appendix~\ref{app:zeroterm-largenegenergy}. We find 
\begin{align}
\dot{\mathcal{F}}_{\tau}^{n=0}(E)
&= 
\frac{\Theta\bigl(-\Etilde\bigr)}{2}
+\frac{1}{4\pi \Etilde}
\left(\frac{\cos\bigl(\Etilde\Delta\tilde{\tau}\bigr)}{\sin(\Delta\tilde{\tau}/2)}
-\zeta\frac{\sin\bigl(\Etilde\Delta\tilde{\tau}\bigr)}{\cos(\Delta\tilde{\tau}/2)}\right)
+ O\bigl(1/\Etilde^2\bigr)\,, 
\label{eq:inertial:negE}
\end{align}
where $\Theta$ is the Heaviside step-function.

\subsection{The $n\neq 0$ terms and large-$M$ asymptotics}
\label{sec:inertial:nonzero}

We now turn to $\dot{\mathcal{F}}_{\tau}^{n\neq 0}$, which contains the 
dependence of $\dot{\mathcal{F}}_{\tau}$ on $M$ and~$q$. 

We consider $\dot{\mathcal{F}}_{\tau}^{n\neq 0}$ in the limit of large~$M$. 
We introduce a positive constant $c \in (0,\pi/2)$, and we assume that 
the switch-on and switch-off moments are separated from the 
initial and final singularities at least by proper time~$c\ell$. 
In terms of $\tilde\tau$ and $\Delta\tilde\tau$, this means that we assume 
\begin{align}
-\pi/2 + c < \tilde\tau < \pi/2 - c \ , \ 
0< \Delta\tilde\tau < \tilde\tau +\pi/2 - c \ . 
\label{eq:awaysinguniform}
\end{align}

Owing to $K_n=K_{-n}$, we can replace the sum $\sum_{n\ne0}$ in \eqref{eq:inertial:rate} 
by $2\sum_{n=1}^{\infty}$. Given~\eqref{eq:awaysinguniform}, 
the expression 
$\cos\tilde{\tau} \cos( \tilde{\tau}-\tilde{s})$ 
is bounded below by a positive constant. 
Using~\eqref{eq:inertial:Kn}, this implies that the quantities under 
the $n\ne0$ square roots in \eqref{eq:inertial:rate} are
dominated at large $M$ by the term that involves~$K_n$, 
and we may write 
\begin{align}
\label{eq:inertial:LargeM.prep}
\dot{\mathcal{F}}_{\tau}^{n\neq 0} (E)
&=
\frac{1}{\pi\sqrt{2\cos\tilde{\tau}}}
\sum_{n=1}^\infty \frac{1}{\sqrt{K_n}}
\int^{\Delta\tilde{\tau}}_{0}\,\frac{\cos\bigl(\Etilde \tilde{s}\bigr) \, 
\mathrm{d}\tilde{s}}{\sqrt{\cos(\tilde{\tau}-\tilde{s})}}
\left(
\frac{1}{\sqrt{1+ f_-/K_n}}
- 
\frac{\zeta}{\sqrt{1+ f_+/K_n}}
\right)\,,
\end{align}
where 
\begin{align}
f_{\pm} := \frac{\sin \tilde{\tau} \sin (\tilde{\tau}-\tilde{s}) \pm1}{\cos\tilde{\tau} 
\cos(\tilde{\tau} -\tilde{s})} \ . 
\end{align}
The large-$M$ expansion of $\dot{\mathcal{F}}_{\tau}^{n\neq 0}$ is then obtained by 
a binomial expansion of the square roots in \eqref{eq:inertial:LargeM.prep} 
at $K_n\to\infty$ and using~\eqref{eq:inertial:Kn}. 
The expansion is uniform in $\tilde\tau$ and $\Delta\tilde\tau$ 
within the range~\eqref{eq:awaysinguniform}, 
and by \eqref{eq:inertial:Kn} it is also uniform in~$q$. 
The first few terms are 
\begin{align}
\dot{\mathcal{F}}_{\tau}^{n\neq 0}(E)
& =
\frac{1}{\pi \sqrt{2\cos\tilde{\tau}}}
\int^{\Delta\tilde{\tau}}_{0}\,\frac{\cos\bigl(\Etilde \tilde{s}\bigr) \, 
\mathrm{d}\tilde{s}}{\sqrt{\cos(\tilde{\tau}-\tilde{s})}}
\Bigg[(1-\zeta) \left(\frac{1}{\sqrt{K_1}} + \frac{1}{\sqrt{K_2}}\right)
+ 
\frac{\zeta f_+ - f_-}{2 K_1^{3/2}}
\Bigg]
\nonumber\\[1ex]
&\hspace{3ex}
+ O \bigl(\mathrm{e}^{-5\pi\sqrt{M}}\bigr)
\ . 
\label{eq:inertial:LargeM.33}
\end{align}
For $\zeta\ne1$, the dominant contribution comes from the term proportional to $(1-\zeta)$ 
and is of order $\mathrm{e}^{-\pi\sqrt{M}}$.

\subsection{Numerical results}
\label{sec:inertial:results}

\begin{figure}[!b]
\centering
\includegraphics[scale=1.5]{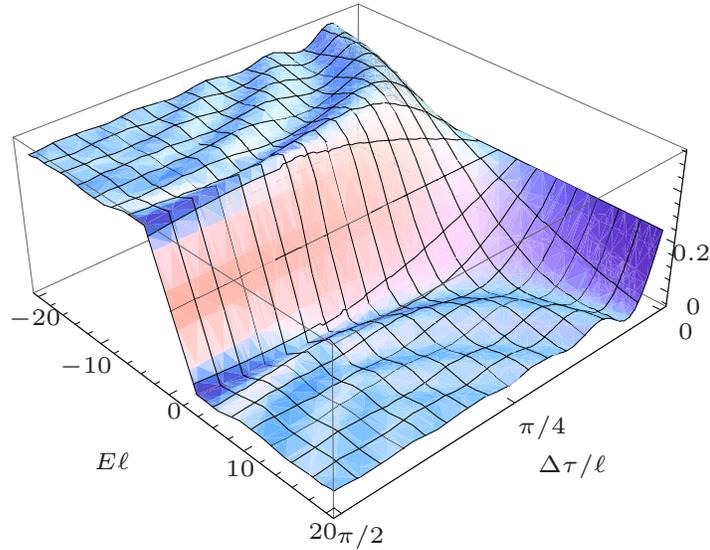}
\caption{$\dot{\mathcal{F}}_{\tau}^{n=0}$ 
\eqref{eq:inertial:n0} as a function of $E\ell$ and 
$\Delta\tau/\ell$ for $\zeta=0$.\label{fig:zeroth3DBC0}}  
\end{figure}

\begin{figure}[p]
\centering
\subfloat[$E\ell=-100$]{\label{fig:zeroth_E-100}\includegraphics[width=0.5\textwidth]{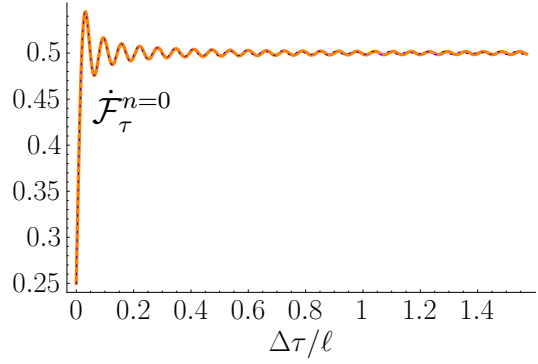}}
\\[3em]
\subfloat[$E\ell=-5$]{\label{fig:zeroth_E-5}\includegraphics[width=0.5\textwidth]{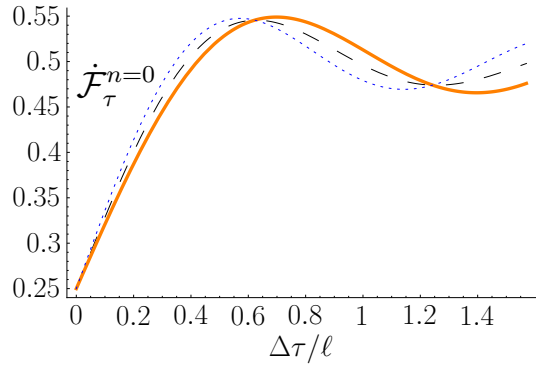}} 
\\[3em]
\subfloat[$E\ell=20$]{\label{fig:zeroth_E20}\includegraphics[width=0.5\textwidth]{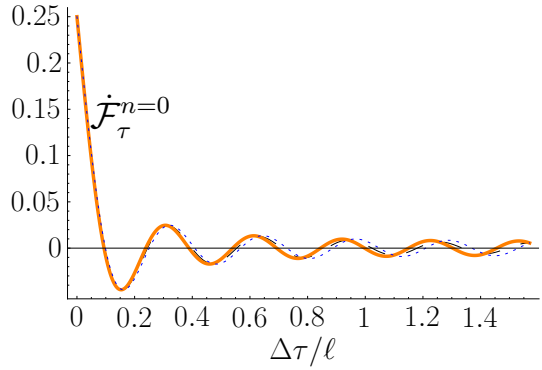}}
\caption{$\dot{\mathcal{F}}_{\tau}^{n=0}$ 
\eqref{eq:inertial:n0} as a function 
$\Delta\tau/\ell$ for selected values of $E\ell$, with
$\zeta=0$ (dashed line), $\zeta=1$ (thick line) 
and $\zeta=-1$ (dotted line).\label{fig:zeroth}}
\end{figure}

At large~$M$, 
the dominant contribution to $\dot{\mathcal{F}}_{\tau}$ 
comes from
$\dot{\mathcal{F}}_{\tau}^{n=0}$~\eqref{eq:inertial:n0}, 
which depends only on $E\ell$ and~$\Delta\tau/\ell$. 
Plots are shown in Figures \ref{fig:zeroth3DBC0} and~\ref{fig:zeroth}.
When $|E\ell|$ is large, the oscillatory dependence on 
$\Delta\tau/\ell$ shown in the plots 
is in agreement with the asymptotic formula~\eqref{eq:inertial:negE}. 

\begin{figure}[p]
\centering
\subfloat[$\zeta=0$]{\label{fig:zerothVSfdot200_E-5_M0pt0001_q100:BC0}\includegraphics[width=0.5\textwidth]{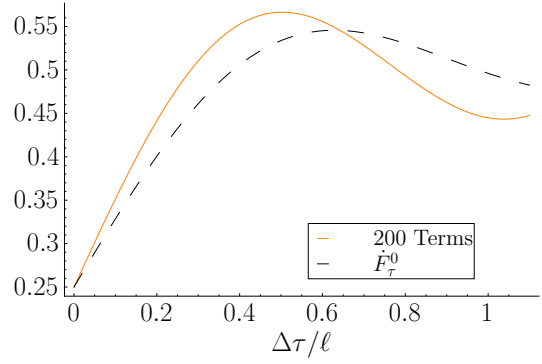}}
\\[3em]
\subfloat[$\zeta=1$]{\label{fig:zerothVSfdot200_E-5_M0pt0001_q100:BC1}\includegraphics[width=0.5\textwidth]{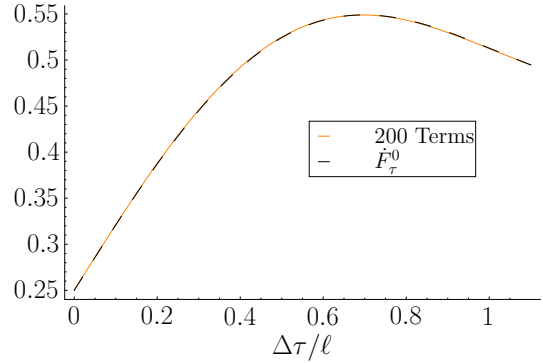}}
 \\[3em]
\subfloat[$\zeta=-1$]{\label{fig:zerothVSfdot200_E-5_M0pt0001_q100:BC-1}\includegraphics[width=0.5\textwidth]{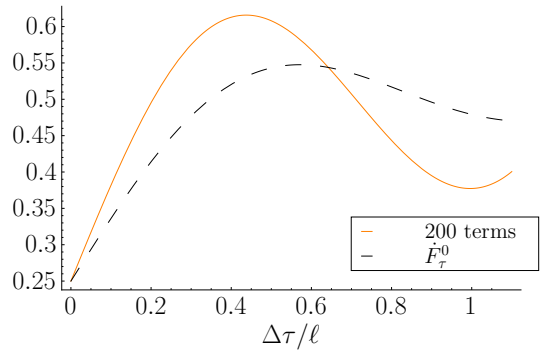}}
\caption{$\dot{\mathcal{F}}_{\tau}$ \eqref{eq:inertial:rate}
with $M=10^{-4}$, 
$q=100$, 
$\tau_0=0$ and $E\ell=-5$. 
Solid curve shows numerical evaluation from \eqref{eq:inertial:rate} 
with $200$ terms 
and dashed curve shows the individual $n=0$ term 
$\dot{\mathcal{F}}_{\tau}^{n=0}$~\eqref{eq:inertial:n0}.
The horizon-crossing occurs outside the plotted range, at 
$\Delta\tau/\ell = \arccos(0.01) \approx 1.56$.\label{fig:zerothVSfdot200_E-5_M0pt0001_q100}}
\end{figure}

\begin{figure}[p]
\centering
\subfloat[$\zeta=0$]{\label{fig:zerothVSfdot200_E20_M0pt0001_q100:BC0}\includegraphics[width=0.5\textwidth]{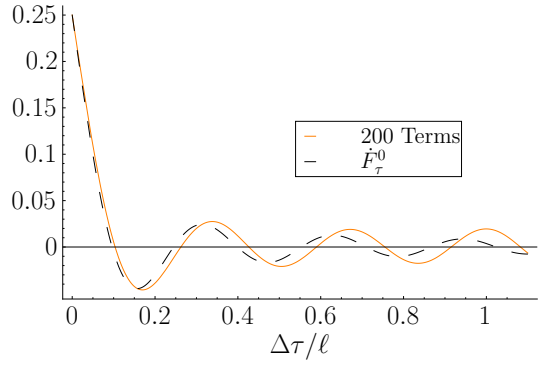}}
 \\[3em]
\subfloat[$\zeta=1$]{\label{fig:zerothVSfdot200_E20_M0pt0001_q100:BC1}\includegraphics[width=0.5\textwidth]{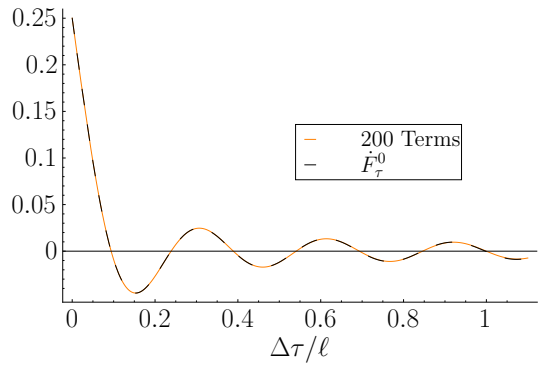}}
 \\[3em]
\subfloat[$\zeta=-1$]{\label{fig:zerothVSfdot200_E20_M0pt0001_q100:BC-1}\includegraphics[width=0.5\textwidth]{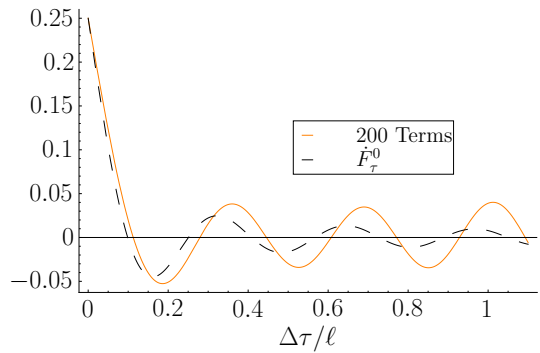}}
\caption{As in Figure \ref{fig:zerothVSfdot200_E-5_M0pt0001_q100} 
but with $E\ell=20$.\label{fig:zerothVSfdot200_E20_M0pt0001_q100}}
\end{figure}

When $M$ decreases, the contribution to 
$\dot{\mathcal{F}}_{\tau}$ from 
$\dot{\mathcal{F}}_{\tau}^{n\ne0}$ becomes significant. For $M=0.1$, 
the terms shown in \eqref{eq:inertial:LargeM.33} are still a good fit to the numerics 
provided both the switch-on and the switch-off are in the exterior region. 
For smaller~$M$, the number of terms that need to be included in 
$\dot{\mathcal{F}}_{\tau}^{n\ne0}$ increases rapidly. 
A~set of plots is shown in Figures 
\ref{fig:zerothVSfdot200_E-5_M0pt0001_q100}
and 
\ref{fig:zerothVSfdot200_E20_M0pt0001_q100}
for $M=10^{-4}$ with $q=100$, taking the detector to be switched on at the moment
where $r$ reaches its maximum and following the detector over a significant fraction 
of its fall towards the horizon. 
$\dot{\mathcal{F}}_{\tau}^{n\ne0}$ turns out to be still insignificant at large, negative~$E\ell$, 
but it starts to become significant at $E\ell\gtrsim-5$, and its effect 
then depends strongly on the boundary condition parameter $\zeta$, being the smallest for $\zeta=1$. 

For fixed~$M$, following the detector close to the future singularity numerically would pose two complications. First, an increasingly large number of terms would need to be included in $\dot{\mathcal{F}}_{\tau}^{n\ne0}$. Second, the evaluation of the individual terms to sufficient accuracy would need to handle numerically integration over an integrable singularity in~$\tilde{s}$. This singularity arises because the quantity under the first square root in \eqref{eq:inertial:rate} can change sign within the integration interval. We have not pursued this numerical problem.

\chapter[Two-dimensional Schwarzschild spacetime]{Two-dimensional Schwarzschild spacetime}
\chaptermark{$2d$ Schwarzschild}
\label{ch:2DSchw}

This chapter marks the beginning of our investigation into detectors in Schwarzschild spacetime. We start in this chapter by considering two-dimensional Schwarzschild spacetime, i.e. by dropping the angular co-ordinates, and we couple the detector to a massless, conformally-coupled scalar field. The reason for doing so is that studying detectors on the full four-dimensional Schwarzschild spacetime is only possible numerically; before we undertake this task in Chapter~\ref{ch:4DSchw}, we hope to gain analytical insight from the two-dimensional case, where the conformal triviality can be exploited to obtain an explicit solution for the scalar field modes and, hence, the Wightman function.
\par Despite the solvable form of the mode equation, other issues present themselves in two dimensions: the Wightman function is ill-defined for a massless scalar field owing to infrared divergences. This means that the limit of infinite total detection time is problematic for a detector coupled to a massless scalar field. Ideally, when investigating a detector on a two-dimensional spacetime, we would like to work initially with a massive scalar field and only at the very end take the massless limit. In the two-dimensional Schwarzschild black hole spacetime, however, working with the massive scalar field is prohibitively complicated. Hence, we shall attempt to regulate the massless field with an exponential cut-off procedure introduced by Langlois~\cite{Langlois-thesis}. 
\par In order to install confidence in the massless theory with Langlois regularisation, we first compare its results to the $m\to 0$ limit of the massive theory for a simpler situation. For a static detector in the Minkowski half-space, we demonstrate that the transition rate obtained by working with a massless scalar field and using the Langlois method agrees with the transition rate obtained by working with a massive scalar field --- from the outset --- and taking the massless limit at the end.
\par Using the Langlois method, we then address a static detector coupled to a massless scalar field in two-dimensional Schwarzschild spacetime, first in the Hartle-Hawking vacuum and then in the Unruh vacuum. In the Hartle-Hawking case, the expected Planckian transition rate and local Hawking temperature are recovered. In the Unruh case an additional term to those expected is obtained.
\par To investigate the unexpected term found in the Unruh vacuum further, we next look to an inertial detector coupled to a massless scalar field in the Minkowski spacetime with receding mirror, and we take the quantum field to be in the `in' vacuum state~\cite{byd}. The `in' vacuum in the receding-mirror spacetime is a close analogue of the Unruh vacuum, which is designed to mimic the geometric effects of stellar collapse~\cite{byd}; indeed, we find that in the late-time limit the receding-mirror transition rate, also, has an unexpected term of the same form as that found in the Unruh vacuum. In addition to this term and the other expected terms, we also obtain a cosine term in the late-time limit, which was unanticipated.  In the early-time limit, we find the rate to agree exactly with that of the static detector in the Minkowski half-space, as one would expect.

\section{Static detector in the Minkowski half-space}
\label{sec:staticMirror}
In this section, we investigate detectors in two-dimensional Minkowski spacetime with an infinite, static boundary at the space origin, known as the Minkowski half-space. We initially compute the transition rate of a static detector coupled to a massive scalar field, and we take the massless limit of the resulting transition rate at the end of the calculation. Finally, we use the Langlois cut-off method to compute the transition rate of a static detector coupled to a massless scalar field from the outset.
\subsection{Wightman function for a massive scalar field}
The first quantity we need to compute before calculating the transition rate of a detector in the Minkowski half-space is the Wightman function.
\par We can use the method of images to calculate the Wightman function of a massive scalar field. Assuming Dirichlet boundary conditions, this leads to
\begin{equation}
\langle 0|\tilde{\phi}(\tau)\tilde{\phi}(\tau')|0\rangle =\langle 0|\phi(\tau)\phi(\tau')|0\rangle-\langle 0|\phi(\tau)\phi(\Lambda \tau')|0\rangle\, ,
\label{eq: Schw:2d:methIm}
\end{equation}
where $\Lambda: (t,x) \mapsto (t,-x)$, and $\phi$ satisfies the massive Klein-Gordon equation in Minkowski spacetime: 
\begin{equation}
\left(-\Box+m^2 \right)\phi=0\,,
\label{eq: Schw:2d: massKG}
\end{equation}
where $m>0$, and $\tilde{\phi}$ satisfies the massive Klein-Gordon equation on the Minkowski half-space.
Thus, if we compute the Wightman function for a massive scalar field on the whole of the two-dimensional Minkowski spacetime, we get the Wightman function for a massive scalar field on the half-space, $x>0$, instantly from~\eqref{eq: Schw:2d:methIm}. 
\par The Wightman function for the field in the full Minkowski spacetime is given by
\begin{equation}
W(\x(\tau),\x(\tau')):=\frac{1}{4\pi}\int^{\infty}_{-\infty}\,\frac{\mathrm{d}k}{\sqrt{k^2+m^2}}\exp{\left[-i\sqrt{k^2+m^2}(\Delta t-i\epsilon)+ik\Delta x\right]}\,.
\end{equation} 
One can evaluate this integral analytically by first changing variables as $k=m\sinh{r}$, to obtain
\begin{equation}
W(\x,\x')=\frac{1}{4\pi}\int^{\infty}_{-\infty}\,\mathrm{d}r\,\exp{\Bigg[-im\left[(\cosh{r})(\Delta t-i\epsilon)-(\sinh{r})\Delta x\right]\Bigg]}\,,
\end{equation} 
and this can be written as
\begin{equation}
W(\x,\x') = \frac{1}{4\pi}\int^{\infty}_{-\infty}\,\mathrm{d}r
\begin{cases} 
 \exp{\left[-im\, \text{sgn}{(\Delta t)}\sqrt{(\Delta t-i\epsilon)^2-(\Delta x)^2}\cosh{(r)}\right]}\,, &  |\Delta t|>|\Delta x|\,, \\ 
\exp{\left[+im\, \text{sgn}{(\Delta x)}\sqrt{(\Delta x)^2-(\Delta t-i\epsilon)^2}\sinh{(r)}\right]}\,, &  |\Delta x|>|\Delta t|\,.
\end{cases}
\label{eq:Schw:2d:massKGWboth}
\end{equation}
Next, we use (10.32.7) of~\cite{dlmf} to evaluate the spacelike case of~\eqref{eq:Schw:2d:massKGWboth}, from which we obtain
\begin{equation}
W(\x,\x') = 
\frac{1}{2\pi}K_0[m\sqrt{(\Delta x)^2-(\Delta t-i\epsilon)^2}]\,,\qquad   |\Delta x|>|\Delta t|\,.
\label{eq: Schw:2d:massKGWspacelike}
\end{equation}
Analytically continuing~\eqref{eq: Schw:2d:massKGWspacelike} to the timelike case, we find 
\begin{equation}
W(\x,\x') = 
\frac{1}{2\pi}K_0[i m\, \text{sgn}(\Delta t)\sqrt{\Delta t^2-\Delta x^2}]\,,\qquad  |\Delta x|<|\Delta t|\,,
\label{eq: Schw:2d:massKGWtimelike}
\end{equation}
where the branch taken is defined by
\begin{equation}
W(\x,\x') = 
\begin{cases} 
\frac{-i}{4}H^{(2)}_0[ m \sqrt{\Delta t^2-\Delta x^2}], & \qquad |\Delta x|<|\Delta t|,\quad \Delta t>0\,,\\
\frac{i}{4}H^{(1)}_0[m \sqrt{\Delta t^2-\Delta x^2}], &\qquad  |\Delta x|<|\Delta t|,\quad \Delta t<0\,.
\label{eq: Schw:2d:massKGWtimelikeBranches}
\end{cases}
\end{equation}
\par In summary, the Wightman function for a massive scalar field on the \emph{full} Minkowski spacetime can be written as
\begin{equation}
W(\x,\x') = 
\begin{cases} 
\frac{1}{2\pi}K_0[m\sqrt{\Delta x^2-\Delta t^2}]\,, &\qquad  |\Delta x|>|\Delta t|\,,\\
\frac{-i}{4}H^{(2)}_0[ m \sqrt{\Delta t^2-\Delta x^2}]\,, & \qquad |\Delta x|<|\Delta t|,\quad \Delta t>0\,,\\
\frac{i}{4}H^{(1)}_0[ m \sqrt{\Delta t^2-\Delta x^2}]\,, &\qquad  |\Delta x|<|\Delta t|\,,\quad \Delta t<0\,.
\label{eq: Schw:2d:massKGW}
\end{cases}
\end{equation}
\par We can now employ the method of images, substituting~\eqref{eq: Schw:2d:massKGW} into~\eqref{eq: Schw:2d:methIm}, and the result for the Wightman function on the Minkowski half-space reads
\be 
\tilde{W}(\x,\x') = 
\frac{1}{2\pi}K_0\left(m\sqrt{\left(\Delta x\right)^2-(\Delta t-i\epsilon)^2}\right)-\frac{1}{2\pi}K_0\left(m\sqrt{P^2-\left(\Delta t-i\epsilon\right)^2}\right)\,,
\label{eq: Schw:2d:massKGWfull}
\ee
where $P:=x+x'$, and where the branches are as in~\eqref{eq: Schw:2d:massKGWspacelike},~\eqref{eq: Schw:2d:massKGWtimelike} and~\eqref{eq: Schw:2d:massKGWtimelikeBranches}: each square root is positive when the quantity under the square root is positive, and the analytic continuation is specified by the $i\epsilon$-regulator.

\subsection[Coupling to a massive scalar field]{Transition rate for a static detector coupled to a massive scalar field}
\label{sec:2d:massive}
Now that we have the Wightman function for a massive scalar field in the Minkowski half-space, equation~\eqref{eq: Schw:2d:massKGWfull}, we are in a position to compute the transition rate for a detector on a trajectory of our choosing. We look at the static detector, sat eternally at a fixed distance, $x=x_0>0$, from the boundary at $x=0$.
\par We shall first deal with the contribution to the transition rate from the Minkowski piece in~\eqref{eq: Schw:2d:massKGWfull}, and then the contribution from the boundary, or image term, in~\eqref{eq: Schw:2d:massKGWfull} will be considered. We denote the contributions respectively by $\dot{\mathcal{F}}^{\text{Mink}}$ and $\Delta\dot{\mathcal{F}}$. 
\par Taking the switch-on time to the asymptotic past (the situation is time-independent), the non-image part of the transition rate is
\begin{equation}
\begin{aligned}
\dot{\mathcal{F}}^{\text{Mink}}\left(\omega\right)&=\frac{1}{2} \Imagpart \int^{\infty}_0\,\mathrm{d}s\, \expo^{-i\omega s} H_0^{(2)}(m s)\\
&=\frac{1}{2m} \Imagpart \int^{\infty}_0\,\mathrm{d}y\, \expo^{-i\omega y/m} H_0^{(2)}(y)\,.
\label{eq: Schw:2d:staticMassiveMink}
\end{aligned}
\end{equation}
To evaluate~\eqref{eq: Schw:2d:staticMassiveMink}, we use the standard integral (6.611.7) of~\cite{gradshteyn}:
\be 
\int^{\infty}_0\,\mathrm{d}x\,\expo^{-\alpha x} H_0^{(2)}(x)=\frac{1}{\sqrt{\alpha^2+1}}\left[1+\frac{2i}{\pi}\log{\left(\alpha+\sqrt{1+\alpha^2}\right)}\right]\,,\qquad \Realpart\alpha>0\,.
\label{eq:Schw:2d:6.611.7}
\ee
We need to continue~\eqref{eq:Schw:2d:6.611.7} to $\alpha$ on the imaginary axis, and the result depends on whether $|\alpha|>1$ or $|\alpha|<1$.
\par If $|\alpha|<1$, then the external $1/\sqrt{\alpha^2+1}$ in~\eqref{eq:Schw:2d:6.611.7} continues to something real, whilst the logarithm becomes pure imaginary. Thus, the right-hand side of~\eqref{eq:Schw:2d:6.611.7} is entirely real in the $|\alpha|<1$ case, and upon taking the imaginary part, as required by~\eqref{eq: Schw:2d:staticMassiveMink}, it vanishes.  If, on the other hand, $|\alpha|>1$, then when $\alpha\to i\beta$, with $\beta \in \BbbR$ and $|\beta|>1$, the right-hand side of~\eqref{eq:Schw:2d:6.611.7} is analytically continued to
\be 
\frac{1}{i\sgn{(\beta)}\sqrt{\beta^2-1}}\left[1-\sgn{(\beta)}+\text{imaginary}\right]=\frac{i}{\sqrt{\beta^2-1}}\left(1-\sgn{\left(\beta\right)}\right)+\text{real}\,.
\label{eq:Schw:2d:6.611.7-alphagt1}
\ee
\par If we apply these insights to~\eqref{eq: Schw:2d:staticMassiveMink}, we find
\be
\dot{\mathcal{F}}^{\text{Mink}}\left(\omega\right)=\frac{\Theta (-m-\omega)}{\sqrt{\omega^2-m^2}}\,.
\label{eq: Schw:2d:staticMassiveMink-fin}
\ee
%
%
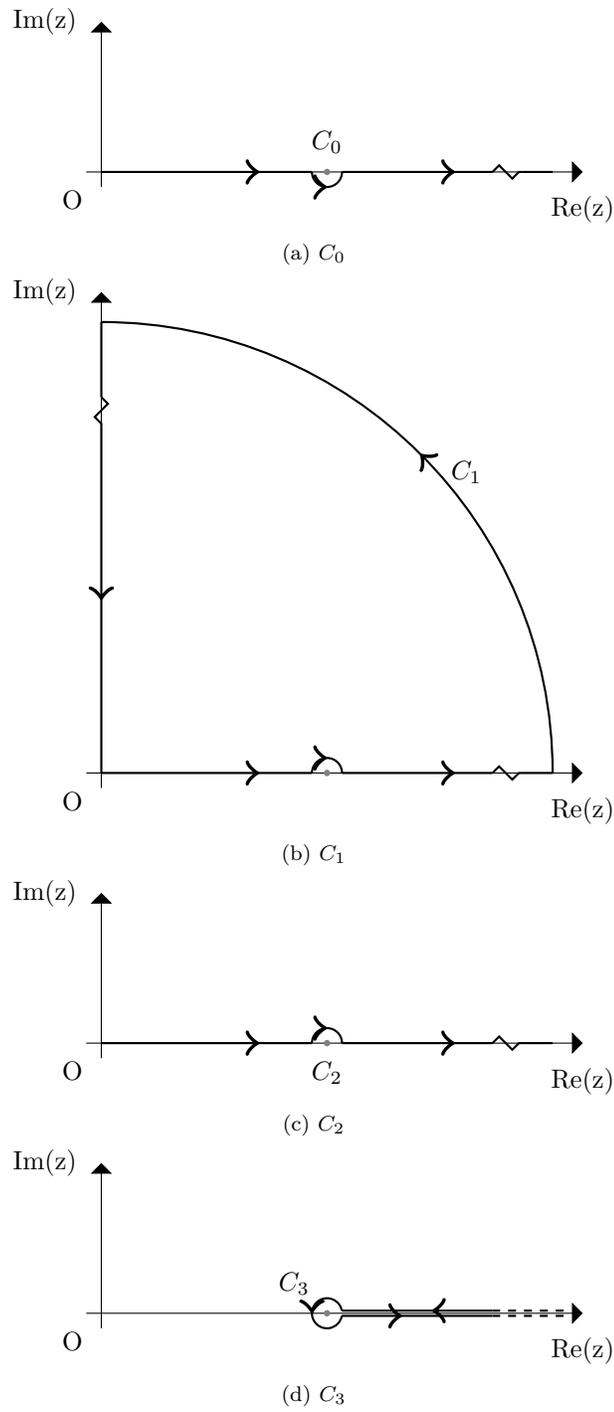
\begin{figure}[p]
\tikzset{-->-/.style={decoration={
  markings,
  mark=at position #1 with {\arrow[scale=2]{>}}},postaction={decorate}}}
\begin{center}
\subfloat[$C_0$]{\label{fig:2DSchw:MassiveStatic:C0}
\begin{tikzpicture}[scale=2]
\node (x)    at (1.5,0)  []   {};
\node (C0)  at  (1.5,0)   [label={[label distance=0.004cm]90:$C_0$}]  {};

\node (orig)   at (0,0)  [label={[label distance=0.001cm]-135:O}]    {};  
\node (B)  at (3.0,0)   {};

\path  
  (x) +(180:1mm)  coordinate   (xl)
       +(0:1mm) coordinate  (xr)
  (xr) +(0:1)  coordinate  (Zigl)
  (Zigl) +(0:2mm) coordinate  (Zigr)
      ;           
\draw[-triangle 90] (-0.1,0) -- (3.2,0);
\draw[-triangle 90] (0,-0.1) -- (0,1);

\node[below=0.2cm] at (3.2,0) {Re(z)};
\node[left=0.2cm] at (0,1.) {Im(z)};

\draw [-->-=.75,thick] (orig.center)--(xl);
\draw[-->-=.5,thick] (xl) arc (-180:0: 1mm);
\draw [-->-=.75,thick] (xr)--(Zigl);
\draw[decorate,decoration=zigzag, thick] (Zigl)--(Zigr) ; 
\draw [thick] (Zigr)--(B.center);

\filldraw [gray] (x) circle (0.5pt)
                  ;
\end{tikzpicture}
}
\hspace{.1ex}
\subfloat[$C_1$]{\label{fig:2DSchw:MassiveStatic:C1}
\begin{tikzpicture}[scale=2]
\node (x)    at (1.5,0)  []   {};
\node (C1)  at  (2.2,2)   [label={[label distance=0.002cm]0:$C_1$}]  {};

\node (orig)   at (0,0)  [label={[label distance=0.001cm]-135:O}]    {};  
\node (B)  at (3.0,0)   {};
\node (C)  at  (0,3.0)   {};

\path  
  (x) +(180:1mm)  coordinate   (xl)
       +(0:1mm) coordinate  (xr)
  (xr) +(0:1)  coordinate  (Zigl)
  (Zigl) +(0:2mm) coordinate  (Zigr)
  (C) +(-90:0.5) coordinate (Zigt)
  (Zigt) +(-90:2mm) coordinate (Zigb)
         ;      
     
\draw[-triangle 90] (-0.1,0) -- (3.2,0);
\draw[-triangle 90] (0,-0.1) -- (0,3.2);

\node[below=0.2cm] at (3.2,0) {Re(z)};
\node[left=0.2cm] at (0,3.2) {Im(z)};

\draw [-->-=.75,thick] (orig.center)--(xl);
\draw[-->-=.5,thick] (xl) arc (180:0: 1mm);
\draw [-->-=.75,thick] (xr)--(Zigl);
\draw[decorate,decoration=zigzag, thick] (Zigl)--(Zigr) ; 
\draw [thick] (Zigr)--(B.center);
\draw[-->-=.5,thick] (B.center) arc (0:90: 30mm);
\draw [thick] (C.center)--(Zigt);
\draw[decorate,decoration=zigzag, thick] (Zigt)--(Zigb) ; 
\draw [-->-=.5,thick] (Zigb)--(orig.center);

\filldraw [gray] (x) circle (0.5pt)
                  ;
\end{tikzpicture}
}
\hspace{.1ex}
\subfloat[$C_2$]{\label{fig:2DSchw:MassiveStatic:C2}
\begin{tikzpicture}[scale=2]
\node (x)    at (1.5,0)  []   {};
\node (C2)  at  (1.5,0)   [label={[label distance=0.002cm]-90:$C_2$}]  {};

\node (orig)   at (0,0)  [label={[label distance=0.001cm]-135:O}]    {};  
\node (B)  at (3.0,0)   {};

\path  
  (x) +(180:1mm)  coordinate   (xl)
       +(0:1mm) coordinate  (xr)
  (xr) +(0:1)  coordinate  (Zigl)
  (Zigl) +(0:2mm) coordinate  (Zigr)
      ;           
\draw[-triangle 90] (-0.1,0) -- (3.2,0);
\draw[-triangle 90] (0,-0.1) -- (0,1.);

\node[below=0.2cm] at (3.2,0) {Re(z)};
\node[left=0.2cm] at (0,1.) {Im(z)};

\draw [-->-=.75,thick] (orig.center)--(xl);
\draw[-->-=.5,thick] (xl) arc (180:0: 1mm);
\draw [-->-=.75,thick] (xr)--(Zigl);
\draw[decorate,decoration=zigzag, thick] (Zigl)--(Zigr) ; 
\draw [thick] (Zigr)--(B.center);

\filldraw [gray] (x) circle (0.5pt)
                  ;
\end{tikzpicture}
}
\hspace{.1ex}
\subfloat[$C_3$]{\label{fig:2DSchw:MassiveStatic:C3}
\begin{tikzpicture}[scale=2]

\node (zI)    at (1.5,0)   {};
\node (zIII)  at (2.1,0)   {};

\node (orig)   at (0,0)  [label={[label distance=0.001cm]-135:O}]    {};  

\node (C3)  at  (1.5,0)   [label={[label distance=0.004cm]130:$C_3$}]  {};

\path  
  (zI) +(10:1mm)  coordinate  (zItr)
       +(170:1mm) coordinate  (zItl)
       +(-10:1mm)  coordinate  (zIbr)
       +(-170:1mm) coordinate  (zIbl)
  (zIII) +(10:1mm)  coordinate  (zIIItr)
       +(170:1mm) coordinate  (zIIItl)
       +(-10:1mm)  coordinate  (zIIIbr)
       +(-170:1mm) coordinate  (zIIIbl)
         ;
       
\path (zIIItr) +(0:.40)  coordinate  (zigTopLeft); 
\path (zIIIbr) +(0:.40)  coordinate  (zigBottLeft);
\path (zigTopLeft) +(0:0.5)  coordinate  (zigTopRight);
\path (zigBottLeft) +(0:0.5)  coordinate  (zigBottRight);

\draw[-triangle 90] (-0.1,0) -- (3.2,0);
\draw[-triangle 90] (0,-0.1) -- (0,1.);

\node[below=0.2cm] at (3.2,0) {Re(z)};
\node[left=0.2cm] at (0,1.) {Im(z)};

\filldraw [gray] (zI) circle (0.5pt)
                 ;
\draw[-->-=.5,thick] (zItr) arc (10:350: 1mm);

\draw[-->-=.4,thick] (zigTopLeft)--(zItr) ;
\draw[-->-=.4,thick] (zIbr)--(zigBottLeft);

\draw[dashed,thick] (zigTopLeft)--(zigTopRight);
\draw[dashed,thick] (zigBottLeft)--(zigBottRight);

\end{tikzpicture}
}
\end{center}
\caption{Contour deformations aiding in the evaluation of~\eqref{eq: Schw:2d:image}.}  
\end{figure}
%
\par Next, we calculate the contribution of the image term of~\eqref{eq: Schw:2d:massKGWfull} to the transition rate:
\begin{equation}
\begin{aligned}
\Delta\dot{\mathcal{F}}\left(\omega\right)&=-\frac{1}{\pi}\Realpart\int^{\infty}_0\,\mathrm{d}s\,\expo^{-i\omega s} K_0\left(m\sqrt{-s^2+2i\epsilon s+\epsilon^2+P^2}\right)\\
&=-\frac{1}{m\pi}\Realpart\int^{\infty}_0\,\mathrm{d}y\,\expo^{-i\omega y/m} K_0\left(\sqrt{-(y-i\epsilon)^2+(mP)^2}\right)\\
&=-\frac{1}{m\pi}\Realpart\int_{C_0}\,\mathrm{d}z\,\expo^{-i\omega z/m} K_0\left(\sqrt{-z^2+(mP)^2}\right)\,,
\label{eq: Schw:2d:image}
\end{aligned}
\end{equation}
where the contour $C_0$ is shown in Figure~\ref{fig:2DSchw:MassiveStatic:C0}.
\par A contour argument, using (10.40.2) in~\cite{dlmf}, shows that $\Delta\dot{\mathcal{F}}$ vanishes for $\omega>-m$.
For $\omega<-m$, we first note that 
\be 
0=\Realpart\int_{C_1}\,\mathrm{d}z\,\expo^{-i\omega z/m} K_0\left(\sqrt{-z^2+(mP)^2}\right)\,,
\label{eq: Schw:2d:imageC1}
\ee
where the contour $C_1$ is shown in Figure~\ref{fig:2DSchw:MassiveStatic:C1}. It can be shown, again using (10.40.2) in~\cite{dlmf}, that the contribution from the arc and imaginary axis on the contour $C_1$ are vanishing after the taking of the real part, and so we can also write
\be 
0=\Realpart\int_{C_2}\,\mathrm{d}z\,\expo^{-i\omega z/m} K_0\left(\sqrt{-z^2+(mP)^2}\right)\,,
\label{eq: Schw:2d:imageC2}
\ee
where the contour $C_2$ is shown in Figure~\ref{fig:2DSchw:MassiveStatic:C2}. The original contour that we actually wish to evaluate is $C_0$, but because the integral is vanishing over $C_2$ we are free to subtract the contribution over this path from $\Delta\dot{\mathcal{F}}$. 
\par Doing so gives
\bea
\Delta\dot{\mathcal{F}}\left(\omega\right)
&=-\frac{1}{m\pi}\Realpart\Bigg[\int_{C_0}\,\mathrm{d}z\,\expo^{-i\omega z/m} K_0\left(\sqrt{-z^2+(mP)^2}\right)\\
&\qquad\qquad\qquad\qquad-\int_{C_2}\,\mathrm{d}z\,\expo^{-i\omega z/m} K_0\left(\sqrt{-z^2+(mP)^2}\right)\Bigg]\\
&=-\frac{1}{m\pi}\Realpart\int_{C_3}\,\mathrm{d}z\,\expo^{-i\omega z/m} K_0\left(\sqrt{-z^2+(mP)^2}\right)\,,
\label{eq: Schw:2d:imageC3}
\eea
where the path $C_3$ is shown in Figure~\ref{fig:2DSchw:MassiveStatic:C3}. In order to obtain the final equality in equation~\eqref{eq: Schw:2d:imageC3}, we use the fact that for $z<mP$ the square roots in the argument of the Bessel functions in the integrand have positive argument; thus, the square roots are themselves positive, and the contributions from the $z<mP$ parts of $C_0$ and $C_2$ just cancel.

\par 
For $z>mP$, on the lower lip of $C_3$, the Bessel function in the integrand is analytically continued to
\bea 
 K_0\left(\sqrt{-z^2+(mP)^2}\right)&\to  K_0\left(i\sqrt{z^2-(mP)^2}\right)\\
 &=-i\frac{\pi}{2}H_0^{(2)}\left(\sqrt{z^2-m^2P^2}\right)\,,
 \label{eq: Schw:2d:lowerlip}
\eea
whereas on the upper lip, it is continued as
\bea 
 K_0\left(\sqrt{-z^2+(mP)^2}\right)&\to  K_0\left(-i\sqrt{z^2-(mP)^2}\right)\\
 &=i\frac{\pi}{2}H_0^{(1)}\left(\sqrt{z^2-m^2P^2}\right)\,.
 \label{eq: Schw:2d:upperlip}
\eea
Using~\eqref{eq: Schw:2d:lowerlip} and~\eqref{eq: Schw:2d:upperlip} in~\eqref{eq: Schw:2d:imageC3}, we are led to
\be 
\Delta\dot{\mathcal{F}}\left(\omega\right)=-\frac{1}{2m}\Imagpart\int^{\infty}_{mP}\,\mathrm{d}y\,\expo^{-i\omega y/m} \left(H^{(1)}_0\left(\sqrt{y^2-(mP)^2}\right)
+H^{(2)}_0\left(\sqrt{y^2-(mP)^2}\right)\right)\,,
\ee
which after a change of variables, $y=mPx$, yields
\be
\Delta\dot{\mathcal{F}}\left(\omega\right)=-P\Imagpart\int^{\infty}_1\,\mathrm{d}x\,\expo^{-iP\omega x} J_0\left(mP\sqrt{x^2-1}\right)\,.
\label{eq: Schw:2d:staticMassiveBound}
\ee
Now, we use the standard integral (6.645.2) from~\cite{gradshteyn} along with (10.25.3) of~\cite{dlmf}:
\begin{equation}
\int^{\infty}_1 \mathrm{d}x\,\expo^{-\alpha x} J_0(\beta\sqrt{x^2-1})=\frac{1}{\sqrt{\alpha^2+\beta^2}}\expo^{-\sqrt{\alpha^2+\beta^2}},\qquad \alpha>0\,, \, \beta>0\,.
\label{eq: Schw:2d:GRstand1}
\end{equation}

We are in the regime $\omega<-m$, so in order to use~\eqref{eq: Schw:2d:GRstand1} to evaluate~\eqref{eq: Schw:2d:staticMassiveBound}, we continue $\alpha\to-i\gamma$, where $\gamma=P|\omega|>0$. In this regime $|\gamma|>mP$, and the result we obtain is
\begin{equation}
\Delta\dot{\mathcal{F}}\left(\omega\right)=
P\Imagpart \frac{1}{i\sqrt{\gamma^2-m^2P^2}}\expo^{i\sqrt{\gamma^2-m^2P^2}}\,,
\label{eq: Schw:2d:staticMassiveBoundEval}
\end{equation}
which can be rewritten as
\begin{equation}
\Delta\dot{\mathcal{F}}\left(\omega\right)=-\frac{\Theta(-m-\omega)}{\sqrt{\omega^2-m^2}} \cos{\left(P\sqrt{\omega^2-m^2}\right)}\,.
\label{eq: Schw:2d:staticMassiveImage}
\end{equation}
\par Combining~\eqref{eq: Schw:2d:staticMassiveMink-fin} and~\eqref{eq: Schw:2d:staticMassiveImage}, we finally arrive at the transition rate for a static detector coupled to a massive scalar field in the two-dimensional Minkowski half-space:
\begin{equation}
\dot{\mathcal{F}}\left(\omega\right)=\frac{\Theta (-m-\omega)}{\sqrt{\omega^2-m^2}}\left[1- \cos{\left(P\sqrt{\omega^2-m^2}\right)}\right]\,.
\label{eq: Schw:2d:staticMassiveTR}
\end{equation}
In the $m\to 0$ limit,~\eqref{eq: Schw:2d:staticMassiveTR} reduces to the well-defined expression 
\be
\dot{\mathcal{F}}\left(\omega\right)\to\frac{\Theta (-\omega)}{\omega}\left[1- \cos{\left(P\omega\right)}\right]\,.
\label{eq: Schw:2d:staticMassiveTm0lim}
\ee
\subsection[Coupling to a massless scalar field]{Transition rate for a static detector coupled to a massless scalar field}
In this section, using the Langlois cut-off method to control the infrared divergences, we shall compute the transition rate of a static detector coupled to a massless scalar field in the two-dimensional Minkowski half-plane.
\par As mentioned in Section~\ref{sec:2d:massive}, in two dimensions the Wightman function is ill-defined for a massless scalar field in Minkowski spacetime. In Section~\ref{sec:2d:massive}, we took $\Delta t \to \infty$ before, finally, taking $m\to 0$; attempting to take these limits in the opposite order would lead to problems~\cite{Langlois-thesis}. Nevertheless, Langlois~\cite{Langlois-thesis} showed that if the sharp switching of the detector in the infinite past is replaced by an exponential cut-off, namely the regulator
\be 
\expo^{-s/\Delta\tau}\,
\ee
is employed, then the procedure of taking the $m\to 0$ limit point-wise, before taking $\Delta\tau \to \infty$ and then, finally, removing the cut-off at the end of the calculation, leads to the expected results for the transition rate of an inertial detector and a uniformly accelerated detector in two-dimensional Minkowski spacetime. 
\par What we shall establish is that this procedure also works for the transition rate of a static detector in the two-dimensional Minkowski half-plane.
\subsubsection{The non-image term in the Wightman function massless limit}
First, we take the $m\to 0$ limit of the timelike and $\Delta t>0$ piece of~\eqref{eq: Schw:2d:massKGW}.  We have
\begin{equation}
\begin{aligned}
W(\x(\tau),\x(\tau-s)) &=-\frac{i}{4} H_0^{(2)}(m s)\\
&\to-\frac{i}{4}-R-\frac{1}{2\pi}\log{(s)}+O\left(m^2 s^2 \log{\left(ms\right)}\right)\,,\qquad\qquad m\to 0\,,
\end{aligned}
\label{eq:Schw:2d:MasslessLim}
\end{equation}
where $\gamma$ is the Euler Gamma function, and where $R:=\left(\gamma+\log{\left(m/2\right)}\right)/2\pi$ is a formally-infinite, real constant.  
\par We shall implement the Langlois cut-off procedure, which replaces the sharp switching of the detector in the infinite past by an exponential cut-off. 
\par Consider some finite, real constant $\rho$ in the Wightman function. Next, using the Langlois cut-off procedure, we shall show that $\rho$ gives a vanishing contribution to the transition rate:
\bea 
&\Realpart\int^{\infty}_0\,\mathrm{d}s\,\expo^{-i\omega s-s/\Delta\tau}\rho\\
&=\rho\Realpart\left[-\frac{1}{i\omega+1/\Delta\tau}\expo^{-i\omega s-s/\Delta\tau}\right]_0^{\infty}\\
&=\rho\Realpart\left[\frac{1}{i\omega+1/\Delta\tau}\right]\\
&\to\rho \Realpart\left[-i/\omega\right]=0\,,\qquad \Delta\tau\to\infty\,.
\eea
Thus, the prescription is just to drop $R$ in~\eqref{eq:Schw:2d:MasslessLim}.
\par Therefore, the transition rate we need to evaluate is 
\begin{equation}
\begin{aligned}
\mathcal{\dot{F}}^{\text{Mink}}(\omega)&= 2 \Realpart \int^{\infty}_{0}\, \mathrm{d}s \,\expo^{-i\omega{s}-s/\Delta\tau} \left(-i/4-\log{(s)}/2\pi\right)\\
&=\frac{\Imagpart}{2}\left[\frac{1}{i\omega+1/\Delta\tau}\right] -\frac{\Realpart}{\pi} \int^{\infty}_{0}\, \mathrm{d}s \,\expo^{-i\omega{s}-s/\Delta\tau} \log{(s)}\,.
\label{eq:Schw:2d: minkMasslessTR1}
\end{aligned}
\end{equation}
To evaluate the second term of the second line in~\eqref{eq:Schw:2d: minkMasslessTR1}, we use (4.331.1) of~\cite{gradshteyn}. Finally, we take the cut-off to infinity, $\Delta\tau \to \infty$. The result is 
\begin{equation}
\begin{aligned}
\mathcal{\dot{F}}^{\text{Mink}}(\omega)&= \frac{1}{2|\omega|}-\frac{1}{2\omega}\\
&=-\frac{\Theta(-\omega)}{\omega}\,.
\label{eq:Schw:2d: minkMasslessResult}
\end{aligned}
\end{equation}
This is essentially the calculation done in~\cite{Langlois-thesis}.
\subsubsection{The image term in the Wightman function massless limit}
First, we need to take the $m\to 0$ limit of the image term of the massive scalar Wightman function~\eqref{eq: Schw:2d:massKGWfull}, and doing so yields 
\begin{equation}
\begin{aligned}
\Delta W(\x(\tau),\x(\tau-s)) &=-\frac{1}{2\pi} K_0( m \sqrt{P^2-(s-i\epsilon)^2})\\
&\to R+\frac{1}{4\pi}\log{(P^2-(s-i\epsilon)^2)}\,,\qquad\qquad\,m\to0\,,
\label{eq:Schw:2d:imageAsym}
\end{aligned}
\end{equation}
where $R$ is the formally-infinite constant defined in~\eqref{eq:Schw:2d:MasslessLim}.
\par Substituting~\eqref{eq:Schw:2d:imageAsym} into the transition rate and dropping the contribution from $R$, we obtain 
\bea
\Delta\mathcal{\dot{F}}(\omega)&= \frac{\Realpart}{2\pi}\int^{\infty}_0\,\mathrm{d}s\,\expo^{-i\omega{s}-s/\Delta\tau}\log{\left[P^2-(s-i\epsilon)^2\right]}\,.
\label{eq:Schw:2d: minkMasslessImage}
\eea
The integral~\eqref{eq:Schw:2d: minkMasslessImage} can be evaluated by first integrating by parts to obtain
\bea
\Delta\mathcal{\dot{F}}(\omega)&=\frac{\Realpart}{2\pi}\Bigg[\frac{1}{i\omega+1/\Delta\tau}\log{\left(P^2+\epsilon^2\right)}\\
&-\frac{2}{i\omega+1/\Delta\tau}\int^{\infty}_0\,\mathrm{d}s\,\expo^{-s(i\omega+1/\Delta\tau)}\frac{(s-i\epsilon)}{\left(P+i\epsilon-s\right)\left(P+s-i\epsilon\right)}\Bigg]\,.
\label{eq:Schw:2d: minkMasslessImage-parts}
\eea
After taking the $\Delta\tau\to\infty$ limit, the first term in~\eqref{eq:Schw:2d: minkMasslessImage-parts} is pure imaginary and vanishes upon taking the real part. The second term can be evaluated by performing a contour deformation around the pole on the positive real axis. 
%
%
\begin{figure}
\tikzset{-->-/.style={decoration={
  markings,
  mark=at position #1 with {\arrow[scale=2]{>}}},postaction={decorate}}}
\begin{center}
\subfloat[$\omega>0$]{\label{fig:2DSchw:MasslessStatic:omegaPos}
\begin{tikzpicture}[scale=1.45]
\node (x)    at (1.5,0)  []   {};
\node (C1)  at  (2.2,-2)   [label={[label distance=0.002cm]0:$C_1$}]  {};

\node (orig)   at (0,0)  [label={[label distance=0.001cm]-135:O}]    {};  
\node (B)  at (3.0,0)   {};
\node (C)  at  (0,-3.0)   {};

\path  
  (x) +(180:1mm)  coordinate   (xl)
       +(0:1mm) coordinate  (xr)
  (xr) +(0:1)  coordinate  (Zigl)
  (Zigl) +(0:3mm) coordinate  (Zigr)
  (C) +(90:0.5) coordinate (Zigt)
  (Zigt) +(90:3mm) coordinate (Zigb)
         ;      
     
\draw[-triangle 90] (-0.1,0) -- (3.2,0);
\draw[-triangle 90] (0,-3.5) -- (0,0.5);

\node[above=0.2cm] at (3.,0) {Re(z)};
\node[left=0.2cm] at (0,0.5) {Im(z)};

\draw [-->-=.75,thick] (orig.center)--(xl);
\draw[-->-=.5,thick] (xl) arc (-180:0: 1mm);
\draw [-->-=.75,thick] (xr)--(Zigl);
\draw[decorate,decoration=zigzag, thick] (Zigl)--(Zigr) ; 
\draw [thick] (Zigr)--(B.center);
\draw[-->-=.5,thick] (B.center) arc (0:-90: 30mm);
\draw [thick] (C.center)--(Zigt);
\draw[decorate,decoration=zigzag, thick] (Zigt)--(Zigb) ; 
\draw [-->-=.5,thick] (Zigb)--(orig.center);

\filldraw [gray] (x) circle (0.5pt)
                  ;
\end{tikzpicture}
}
\subfloat[$\omega<0$]{\label{fig:2DSchw:MasslessStatic:omegaNeg}
\begin{tikzpicture}[scale=1.45]
\node (x)    at (1.5,0)  []   {};
\node (C2)  at  (2.2,2)   [label={[label distance=0.002cm]0:$C_2$}]  {};

\node (orig)   at (0,0)  [label={[label distance=0.001cm]-135:O}]    {};  
\node (B)  at (3.0,0)   {};
\node (C)  at  (0,3.0)   {};

\path  
  (x) +(180:1mm)  coordinate   (xl)
       +(0:1mm) coordinate  (xr)
  (xr) +(0:1)  coordinate  (Zigl)
  (Zigl) +(0:3mm) coordinate  (Zigr)
  (C) +(-90:0.5) coordinate (Zigt)
  (Zigt) +(-90:3mm) coordinate (Zigb)
         ;      
     
\draw[-triangle 90] (-0.1,0) -- (3.2,0);
\draw[-triangle 90] (0,-0.1) -- (0,3.5);

\node[below=0.2cm] at (3.1,0) {Re(z)};
\node[left=0.2cm] at (0,3.5) {Im(z)};

\draw [-->-=.75,thick] (orig.center)--(xl);
\draw[-->-=.5,thick] (xl) arc (-180:0: 1mm);
\draw [-->-=.75,thick] (xr)--(Zigl);
\draw[decorate,decoration=zigzag, thick] (Zigl)--(Zigr) ; 
\draw [thick] (Zigr)--(B.center);
\draw[-->-=.5,thick] (B.center) arc (0:90: 30mm);
\draw [thick] (C.center)--(Zigt);
\draw[decorate,decoration=zigzag, thick] (Zigt)--(Zigb) ; 
\draw [-->-=.5,thick] (Zigb)--(orig.center);

\filldraw [gray] (x) circle (0.5pt)
                  ;
\end{tikzpicture}
}
\end{center}
\caption{Contour deformations aiding in the evaluation of the second term of~\eqref{eq:Schw:2d: minkMasslessImage-parts}.}  
\end{figure}
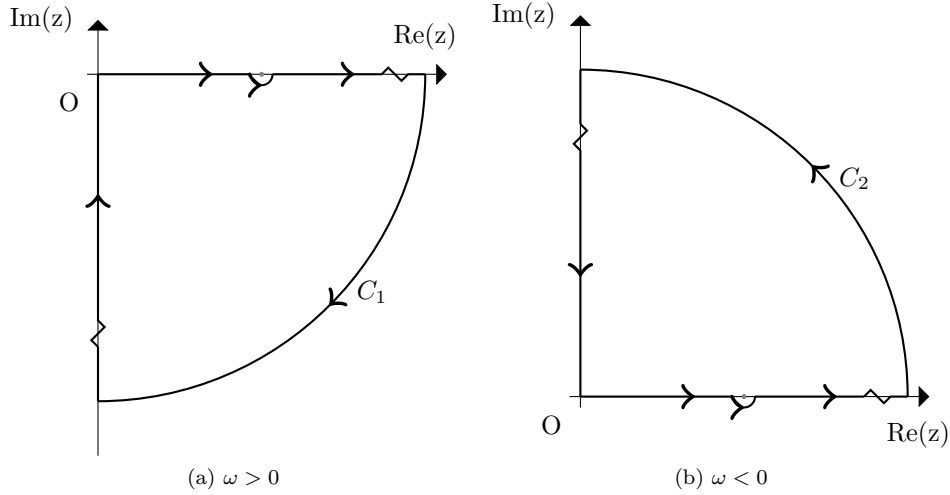
%
%
\par When $\omega>0$ the contour is deformed as shown in Figure~\ref{fig:2DSchw:MasslessStatic:omegaPos}; the contribution from the arc is vanishing, and the contribution from the part of the path along the imaginary axis also vanishes in the $\Delta\tau\to\infty$ limit after taking the real part. Because no poles are enclosed, we can conclude that the second term in~\eqref{eq:Schw:2d: minkMasslessImage-parts} is zero when $\omega>0$.
\par When $\omega<0$, we must close the contour in the upper half-space, as shown in Figure~\ref{fig:2DSchw:MasslessStatic:omegaNeg}, in order for the contribution from the arc to vanish. The contribution from the part of the path along the imaginary axis is once again vanishing, but this time we pick up a residue from the simple pole enclosed. The result is
\begin{equation}
\Delta\mathcal{\dot{F}}(\omega)= \frac{\cos{(P\omega)}}{\omega}\Theta(-\omega)\,,
\label{eq:Schw:2d: minkMasslessImage2}
\end{equation}
and combining this with~\eqref{eq:Schw:2d: minkMasslessResult}, we find finally 
\begin{equation}
\mathcal{\dot{F}}(\omega)= -\frac{\Theta(-\omega)}{\omega}\left[1-\cos{(P\omega)}\right]\,.
\label{eq:Schw:2d: masslessTRfinal}
\end{equation}
\par Comparing \eqref{eq:Schw:2d: masslessTRfinal}  with \eqref{eq: Schw:2d:staticMassiveTm0lim}, we see that we have agreement. Thus, our hypothesis that one can work with the strictly ill-defined massless Wightman function in two dimensions provided one uses a suitable cut-off procedure is seen to hold, at least in this simple spacetime. 

\section{Two-dimensional Schwarzschild spacetime}
Bolstered by the success of the Langlois cut-off procedure in the two-dimensional Minkowski half-space, in this section, we shall use the cut-off procedure to look at a static detector coupled to a massless scalar field in two-dimensional Schwarzschild spacetime. 
\par The metric for two-dimensional Schwarzschild is given by
\begin{equation}
\mathrm{d}s^2=-\left(1-\frac{2M}{r}\right)\mathrm{d}t^2+\left(1-\frac{2M}{r}\right)^{-1}\mathrm{d}r^2\,,\qquad M>0\,.
\label{eq:Schw:2d:metric}
\end{equation}
If we introduce the Kruskal co-ordinates as
\begin{equation}
\begin{aligned}
\bar{u}&=-4M\expo^{-u/4M}\,,\\
\bar{v}&=4M\expo^{v/4M}\,,
\label{eq: Schw:2d:Kruskals}
\end{aligned}
\end{equation}
with $u=t-r^{*}$, $v=t+r^{*}$ and $r^{*}=r+2M\log{|r/2M-1|}$, then the metric takes the form
\be 
\mathrm{d}s^2=-(2M/r)\expo^{-r/2M}\mathrm{d}\bar{u}\mathrm{d}\bar{v}\,.
\ee
\par In two dimensions, the singularity in the Wightman function is integrable, and the transition rate for a detector coupled to scalar field in an arbitrary Hadamard state can easily be obtained by taking the regulator to zero point-wise under the integral, the result is
\begin{equation}
\mathcal{\dot{F}}_{\tau}\left(\omega\right)=2\Realpart\int^{\Delta \tau}_0\,\mathrm{d}s\,\expo^{-i\omega s} D(\tau,\tau-s)\,,
\label{eq:Schw:2d:trHadamard}
\end{equation}
where here $D$ denotes the Wightman function in the massless limit.
\subsection[Hartle-Hawking vacuum]{The Hartle-Hawking vacuum}
\label{sec:2d:Schw:HH}  
The Wightman function for a massless scalar field in the two-dimensional Schwarzschild spacetime and in the Hartle-Hawking vacuum state is~\cite{byd}
\begin{equation}
D_K(\x(\tau),\x(\tau'))=-\frac{1}{4\pi}\log{\left[(\Delta\bar{u}-i\epsilon)(\Delta\bar{v}-i\epsilon)\right]}\,,
\label{eq:Schw:2d:D_K}
\end{equation}
where $\Delta\bar{u}:=\bar{u}(\tau)-\bar{u}(\tau')$ and $\Delta\bar{v}:=\bar{v}(\tau)-\bar{v}(\tau')$.
\par To simulate what we would obtain if we took the $m\to 0$ limit of the massive scalar field Wightman function in two-dimensional Schwarzschild spacetime, we add a finite imaginary constant to the Wightman function~\eqref{eq:Schw:2d:D_K}. We make the assumption that this finite imaginary constant can be chosen to be the same as one obtains when taking the $m\to0$ limit of the Wightman function of a massive scalar field in two-dimensional Minkowski spacetime,~\eqref{eq: Schw:2d:massKGW}. Including this constant, the Wightman function reads
\begin{equation}
D_K(\x(\tau),\x(\tau'))=-\frac{1}{4\pi}\log{\left[(\Delta\bar{u}-i\epsilon)(\Delta\bar{v}-i\epsilon)\right]}-i/4\,.
\label{eq:Schw:2d:D_K-fin}
\end{equation}
\par Implementing the Langlois type of cut-off to regulate the infrared divergence and using~\eqref{eq:Schw:2d:D_K-fin}, the transition rate is
\begin{equation}
\mathcal{\dot{F}}\left(\omega\right)=-\frac{\Realpart}{2\pi}\int^{\infty}_0\,\mathrm{d}s\,\expo^{-i\omega s-s/\Delta\tau} \left(\log{\left[(\Delta\bar{u}-i\epsilon)(\Delta\bar{v}-i\epsilon)\right]}+i\pi\right)\,,
\label{eq:Schw:2d:trHadamardAlt0}
\end{equation}
where we have used~\eqref{eq:Schw:2d:trHadamard} but taken the total detection time to infinity, which is valid for static situations.
\par For a static detector, the radial co-ordinate is fixed, at say $r=R>2M$, and we observe from the metric~\eqref{eq:Schw:2d:metric} that
\begin{equation}
\mathrm{d}\tau=\sqrt{1-2M/R}\,\mathrm{d}t \, .
\label{eq:Schw:2d:propTimeToSchwTime}
\end{equation}
If we substitute~\eqref{eq: Schw:2d:Kruskals} and~\eqref{eq:Schw:2d:propTimeToSchwTime} into~\eqref{eq:Schw:2d:trHadamardAlt0}, we arrive at
\begin{equation}
\mathcal{\dot{F}}\left(\omega\right)=-\frac{\Realpart}{2\pi}\int^{\infty}_0\,\mathrm{d}s\,\expo^{-i\omega s-s/\Delta\tau} \left(\log{\left[64M^2 \expo^{R^{*}/2M}\sinh^2{\left(\frac{s}{8M\sqrt{1-\frac{2M}{R}}}\right)}\right]}+i\pi\right)\,,
\label{eq:Schw:2d:trHadamardAlt1}
\end{equation}
where $R^{*}:=R+2M\log{\left(2M/R-1\right)}$.  
\par The contribution from the $i\pi$ term in~\eqref{eq:Schw:2d:trHadamardAlt1} is easily evaluated, and again using the fact that real constants in the integrand vanish against the cut-off, this allows us to rewrite~\eqref{eq:Schw:2d:trHadamardAlt1} as 
\begin{equation}
\mathcal{\dot{F}}\left(\omega\right)=-\frac{1}{2 \omega}-\frac{\Realpart}{\pi}\int^{\infty}_0\,\mathrm{d}s\,\expo^{-i\omega s-s/\Delta\tau} \log{\left[\sinh{\left(\frac{s}{C}\right)}\right]}\,,
\label{eq:Schw:2d:trHadamardAlt2}
\end{equation}
where $C:=8M\sqrt{1-2M/R}$. 
\par Next, we change variables to $y=s/C$ in~\eqref{eq:Schw:2d:trHadamardAlt2}, which leads to
\begin{equation}
\begin{aligned}
\mathcal{\dot{F}}\left(\omega\right)&=-\frac{1}{2 \omega}-\frac{C}{\pi}\Realpart\int^{\infty}_0\,\mathrm{d}y\,\expo^{-C y(i\omega+1/\Delta\tau)} \log{\left[\sinh{\left(y\right)}\right]}\\
&=-\frac{1}{2 \omega}-\frac{b}{\omega\pi}\Realpart\int^{\infty}_0\,\mathrm{d}y\,\expo^{-y(a+ib)} \log{\left[\sinh{\left(y\right)}\right]}\,,
\end{aligned}
\label{eq:Schw:2d:trHadamardAlt3}
\end{equation}
where $a:=C/\Delta\tau$ and $b:=\omega C$.
\par This integral is now in a form where the results of Appendix~\ref{ch:appendixB} are applicable, and using~\eqref{eq:App2dSchw:ReIRes} we find
\begin{equation}
\begin{aligned}
\mathcal{\dot{F}}\left(\omega\right)&=-\frac{1}{2 \omega}+\frac{1}{2\omega}\left(\frac{1+\expo^{-b\pi}}{1-\expo^{-b\pi}}\right)\\
&=\frac{1}{\omega}\frac{1}{\expo^{\omega/T}-1}\,,
\label{eq:Schw:2d:TRstaticFin}
\end{aligned}
\end{equation}
where $T:=1/\left(8\pi M\sqrt{1-2M/R}\right)$.
\par The transition rate~\eqref{eq:Schw:2d:TRstaticFin} matches that given in~\cite{byd}, and it has the expected thermal character: the Hawking temperature at infinity $T_0=\kappa/2\pi$, with $\kappa:=1/4M$ being the surface gravity, has been shifted by the Tolman factor, $T=(g_{00})^{-1/2} T_0$. 
\subsection[Unruh Vacuum]{The Unruh Vacuum}
\label{sec:2d:Schw:Unruh}
The Wightman function for a massless scalar field in two-dimensional Schwarzschild spacetime in the Unruh vacuum is given by~\cite{byd}
\begin{equation}
D_U(\x(\tau),\x(\tau'))=-\frac{1}{4\pi}\log{\left[(\Delta\bar{u}-i\epsilon)(\Delta v-i\epsilon)\right]}-\frac{i}{4}\,,
\label{eq:Schw:2d:D_U}
\end{equation}
where we are assuming $\Delta\bar{u}>0$ and $\Delta v>0$, and the reason for the constant imaginary contribution, $-i/4$, is as in Section~\ref{sec:2d:Schw:HH}.
\par Substituting~\eqref{eq:Schw:2d:D_U} into~\eqref{eq:Schw:2d:trHadamard} and implementing the Langlois type of cut-off gives
\begin{equation}
\mathcal{\dot{F}}\left(\omega\right)=-\frac{\Realpart}{2\pi}\int^{\infty}_0\,\mathrm{d}s\,\expo^{-i\omega s-s/\Delta\tau} \left(\log{\left[(\Delta\bar{u}-i\epsilon)(\Delta v-i\epsilon)\right]}+i\pi\right)\,.
\label{eq:Schw:2d:trHadamardUnruhAlt0}
\end{equation}
\par Specialising to the static detector at radius $r=R>2M$ and substituting~\eqref{eq: Schw:2d:Kruskals} and \eqref{eq:Schw:2d:propTimeToSchwTime} into~\eqref{eq:Schw:2d:trHadamardUnruhAlt0}, we arrive at
\begin{equation}
\begin{aligned}
&\mathcal{\dot{F}}\left(\omega\right)=-\frac{1}{2\omega}-\frac{\Realpart}{2\pi}\int^{\infty}_0\,\mathrm{d}s\,\expo^{-i\omega s-s/\Delta\tau} \log{\left[\frac{64M^2}{C}\expo^{R^{*}/4M}\expo^{-2\tau/C}s\expo^{s/C}\sinh{\left(\frac{s}{C}\right)}\right]}\,,
\end{aligned}
\label{eq:Schw:2d:trHadamardUnruhAlt1}
\end{equation}
where $C:=8M\sqrt{1-2M/R}$.
\par Notice that the Wightman function in the integrand of~\eqref{eq:Schw:2d:trHadamardUnruhAlt1} does not exhibit the time-independence that one would expect for the static detector; we believe this is a result of the infrared pathology in two dimensions.
\par Recall that any real parts of the Wightman function will vanish against the Langlois cut-off, and this allows us to write~\eqref{eq:Schw:2d:trHadamardUnruhAlt1} as
\begin{equation}
\mathcal{\dot{F}}\left(\omega\right)=-\frac{1}{2\omega}-
\frac{\Realpart}{2\pi}\int^{\infty}_0\,\mathrm{d}s\,\expo^{-i\omega s-s/\Delta\tau} \left[\log{(s)}+\frac{s}{C}+\log{\left(\sinh{\left(\frac{s}{C}\right)}\right)}\right]\,.
\label{eq:Schw:2d:trHadamardUnruhAlt2}
\end{equation}
\par The final logarithmic term under the integral in~\eqref{eq:Schw:2d:trHadamardUnruhAlt2} is the same form that we evaluated in the previous section for the Hartle-Hawking vacuum. Thus, using~\eqref{eq:App2dSchw:ReIRes}, we can immediately write
\begin{equation}
\mathcal{\dot{F}}\left(\omega\right)=-\frac{1}{2\omega}+\frac{1}{4\omega}\left(\frac{1+\expo^{-C\pi\omega}}{1-\expo^{-C\pi\omega}}\right)-
\frac{\Realpart}{2\pi}\int^{\infty}_0\,\mathrm{d}s\,\expo^{-i\omega s-s/\Delta\tau} \left[\log{(s)}+\frac{s}{C}\right]\,.
\label{eq:Schw:2d:trHadamardUnruhAlt3}
\end{equation}
\par To evaluate the remaining  $\log{(s)}$ term in the integrand of~\eqref{eq:Schw:2d:trHadamardUnruhAlt3}, we use the standard integral from~\cite{gradshteyn}:
\begin{equation}
\int^{\infty}_0\,\mathrm{d}z\,\expo^{-\mu z}\log{(z)}=-\frac{1}{\mu}\left(\gamma+\log{(\mu)}\right)\,,\qquad\qquad \Realpart{\mu}>0\,,
\label{eq:Schw:2d:stdLogInt}
\end{equation}
where $\gamma$ is the Euler Gamma function.
In the case of~\eqref{eq:Schw:2d:trHadamardUnruhAlt3}, we have $\mu=1/\Delta\tau+i\omega$, and after taking $\Delta\tau\to \infty$, we get a contribution from this term of
\begin{equation}
\begin{aligned}
&-\frac{1}{2\pi}\Realpart\left[\frac{i}{\omega}\left(\gamma+\log{\left(i\text{sgn}(\omega)|\omega|\right)}\right)\right]\\
&=-\frac{1}{2\pi}\Realpart\left[\frac{i}{\omega}\left(\gamma+\log{|\omega|}+\text{sgn}(\omega)i\frac{\pi}{2}\right)\right]\\
&=\frac{1}{4|\omega|}\,.
\end{aligned}
\end{equation}
\par The final term from the $s/C$ in the integrand of~\eqref{eq:Schw:2d:trHadamardUnruhAlt3} is elementary; it leads to the
contribution
\begin{equation}
\frac{1}{2\pi C \omega^2}\,.
\end{equation}
\par Combining these contributions, we find that the transition rate is
\begin{equation}
\begin{aligned}
\mathcal{\dot{F}}\left(\omega\right)&=\frac{1}{2\pi C \omega^2}+\frac{1}{4|\omega|}-\frac{1}{2\omega}+\frac{1}{4\omega}\left(\frac{1+\expo^{-C\pi\omega}}{1-\expo^{-C\pi\omega}}\right)\\
&=\frac{T}{2\omega^2}+\frac{1}{2\omega}\frac{1}{\expo^{\omega/T}-1}-\frac{1}{2\omega}\Theta(-\omega)\,,
\label{eq:Schw:2d:TransRateUnruh}
\end{aligned}
\end{equation}
where $T:=1/\left(8\pi M\sqrt{1-2M/R}\right)$.
The second and third terms in~\eqref{eq:Schw:2d:TransRateUnruh} are exactly what we would expect for the Unruh vacuum; namely, the right-movers contribute half the Hartle-Hawking result, and the left-movers contribute half the Boulware result. The first term is, however, unexpected. We shall discuss this term in Section~\ref{sec:2d:summary}.
\section[The receding mirror in two-dimensional Minkowski spacetime]{The receding mirror in two-dimensional Minkowski spacetime%
              \sectionmark{The receding mirror in $2d$ Minkowski spacetime}}
\sectionmark{The receding mirror in $2d$ Minkowski spacetime}
\label{sec:2d:recMirr}
The Unruh vacuum mocks up the state of a star after collapse, and similarly, the receding mirror in the `in-vacuum' at late times is analogous to a collapsing star~\cite{byd}, so it is natural to draw comparison between these two cases. 
\par Motivated by the rather unexpected additive constant in the transition rate~\eqref{eq:Schw:2d:TransRateUnruh}, $T/2\omega^2$, we shall carry out the calculation of the transition rate of an inertial detector in two-dimensional Minkowski spacetime with a receding mirror, to see if a similar term occurs there in the late-time limit. 
\begin{figure}[t]
  \centering
  \includegraphics[scale=.75]{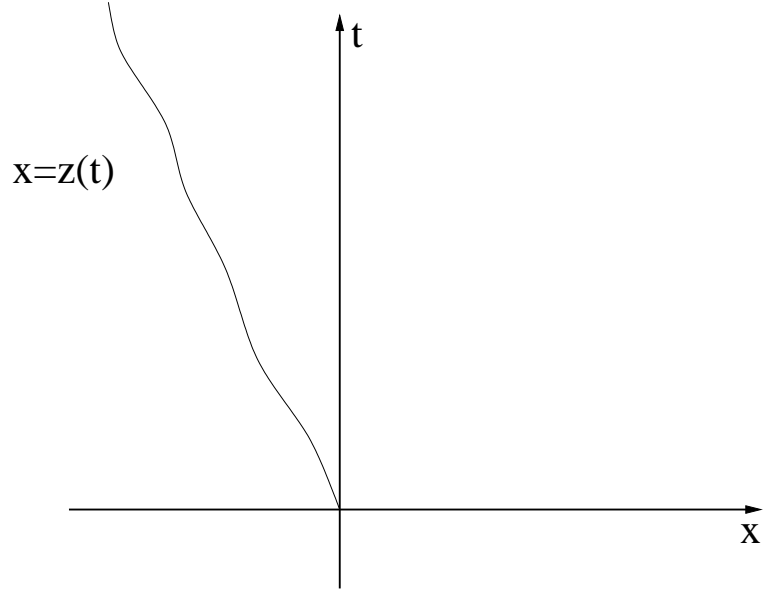}
  \caption{Moving mirror. The boundary conditions ensure that the scalar field vanishes on the boundary $x=z(t)$.}  
  \label{fig:Schw:2d:movMirr:mirrPath}
 \end{figure}
\par We denote the mirror's timelike trajectory through two-dimensional Minkowski spacetime by
\be
x=z(t) \,,
\ee
where $(t,x)$ are the Minkowski co-ordinates.
The massless scalar field satisfies the Klein-Gordon equation
\begin{equation}
\Box\phi=\frac{\partial^2\phi}{\partial u\partial v}=0\,,
\label{eq:Schw:2d:movMirr:KG}
\end{equation}
with the boundary condition at the mirror
\begin{equation}
\phi(t,z(t))=0\,.
\label{eq:Schw:2d:movMirr:BC}
\end{equation}
Equations~\eqref{eq:Schw:2d:movMirr:KG} and~\eqref{eq:Schw:2d:movMirr:BC} have the set of mode solutions
\begin{equation}
u_k^{\text{in}}(u,v)=\frac{i}{\sqrt{4\pi\omega}}\left(\expo^{-i\omega v}-\expo^{-i\omega(2\tau_u-u)}\right)\,,
\label{eq:Schw:2d:movMirr:inModes}
\end{equation}
where $\omega=|k|$, and where $\tau_u$ is determined implicitly by the trajectory through
\begin{equation}
\tau_u-z(\tau_u)=u\,.
\end{equation}
\begin{figure}[t]
  \centering
  \includegraphics[scale=.5]{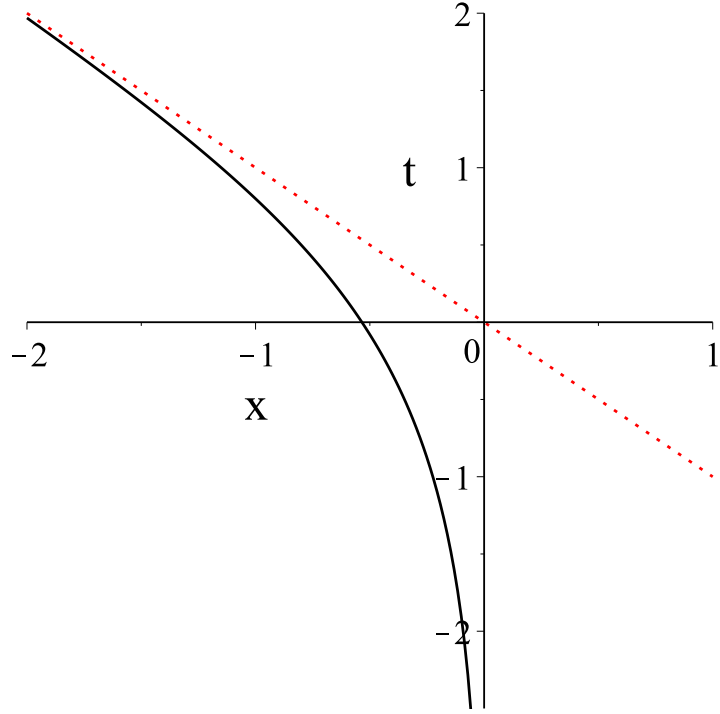}
  \caption{Trajectory we choose for the mirror. The black solid curve is equation~\eqref{eq:Schw:2d:movMirr:traj} for $\kappa=0.9$, and the dashed line represents the $u$-axis, $t=-x$.}  
  \label{fig:Schw:2d:movMirr:mirrLogPath}
 \end{figure}
\par The modes~\eqref{eq:Schw:2d:movMirr:inModes} apply to the right of the mirror. We label them as in-modes because the incoming waves correspond to simple exponential waves from $\scri^{-}$, but the right moving waves are complicated owing to the Doppler shift suffered during the reflection from the moving mirror. 
\par The vacuum corresponding to these modes is a close analogue to that of the Unruh vacuum, described in the previous section. The Wightman function corresponding to this vacuum is~\cite{byd}
\begin{equation}
D(u,v;u',v')=-\frac{1}{4\pi}\log{\left[\frac{(p(u)-p(u')-i\epsilon)(\Delta v-i\epsilon)}{(v-p(u')-i\epsilon)(p(u)-v'-i\epsilon)}\right]}\,,
\label{eq:Schw:2d:movMirr:inW}
\end{equation}
where $p(u):=2\tau_u-u$, and we have, for the moment, suppressed the additive imaginary constant.
\par We consider the mirror trajectory 
\begin{equation}
v=-\frac{1}{\kappa}\log{\left(1+\expo^{-\kappa u}\right)}\,,
\label{eq:Schw:2d:movMirr:traj}
\end{equation}
where $\kappa>0$. This trajectory asymptotes to the $u-$axis as $u\to\infty$ and to $u=v$ for $u\to -\infty$.
This leads to $p(u)=-1/\kappa \log{\left[1+\expo^{-\kappa u}\right]}$, and we substitute this into~\eqref{eq:Schw:2d:movMirr:inW}.
\par 
We want to consider a detector on a general inertial trajectory that neither collides with the mirror nor asymptotes to the mirror in the distant past. Such a trajectory has the form
\begin{equation}
x=x_0-\nu t\,,
\label{eq:Schw:2d:movMirr:detTraj}
\end{equation}
where the constant (leftward) velocity $\nu$ satisfies $0\leq \nu <1$, and where the constant $x_0$ is positive for $\nu=0$ and satisfies
\begin{equation}
x_0>\frac{1}{\kappa}\left[(1-\nu)\log{\sqrt{1-\nu}}-(1+\nu)\log{\sqrt{1+\nu}}+\nu\log{(2\nu)}\right]\,
\label{eq:Schw:2d:movMirr:x0bound}
\end{equation}
for $0<\nu<1$. We show the derivation of the no-collision bound~\eqref{eq:Schw:2d:movMirr:x0bound} in Appendix~\ref{ch:Appendix:x0bound}.
\subsection{The inertial detector}
We shall adopt the Langlois cut-off procedure to analyse the response rate of a detector with trajectory~\eqref{eq:Schw:2d:movMirr:detTraj} subject to~\eqref{eq:Schw:2d:movMirr:x0bound} in the Minkowski spacetime with moving mirror on the trajectory~\eqref{eq:Schw:2d:movMirr:traj} and with the quantum state in the in-vacuum.
\par We start by substituting the equations of motion of the mirror into the Wightman function~\eqref{eq:Schw:2d:movMirr:inW}. We factorise the logarithm in this Wightman function, working on each of the four pieces obtained in turn in the two-dimensional transition rate~\eqref{eq:Schw:2d:trHadamard}. Because the situation is stationary, we take the total detection time to infinity first.
\subsubsection{Part-I}
Consider the contribution to the transition rate from the factor
\begin{equation}
\mathcal{\dot{F}}^{1}(\omega)=-\frac{\Realpart}{2\pi}\int^{\infty}_0\,\mathrm{d}s\,\expo^{-i\omega s-s/\Delta\tau}\log{\left[p(u(\tau))-p(u(\tau-s))\right]}\,.
\end{equation}
Using~\eqref{eq:Schw:2d:movMirr:traj} and~\eqref{eq:Schw:2d:movMirr:detTraj}, this can be written as
\begin{equation}
\begin{aligned}
\mathcal{\dot{F}}^{1}(\omega)&=-\frac{\Realpart}{2\pi}\int^{\infty}_0\,\mathrm{d}s\,\expo^{-i\omega s-s/\Delta\tau}\log{\left[  \frac{1}{\kappa}\log{\left(\frac{ 1+P \expo^{\kappa\Gamma s}}{ 1+P }\right)} \right]}\\
&=-\frac{\Realpart}{2\pi}\int^{\infty}_0\,\mathrm{d}s\,\expo^{-i\omega s-s/\Delta\tau}\left(\log{\left(\frac{1}{\kappa}\right)}+\log{\left[ \log{\left(\frac{ 1+P \expo^{\kappa\Gamma s}}{ 1+P }\right)} \right]}\right)\\
&=-\frac{\Realpart}{2\pi}\int^{\infty}_0\,\mathrm{d}s\,\expo^{-i\omega s-s/\Delta\tau}\log{\left[ \log{\left(\frac{ 1+P \expo^{\kappa\Gamma s}}{ 1+P }\right)} \right]}\,,
\label{eq:Schw:2d:movMirr:P1_alt0}
\end{aligned}
\end{equation}
where
\begin{equation}
\begin{aligned}
P&:=\expo^{-\kappa(\Gamma \tau-x_0)}\,,\\
\Gamma&:=\sqrt{\frac{1+\nu}{1-\nu}}\,,
\end{aligned}
\end{equation}
and where $P$ is manifestly non-negative. The final equality in~\eqref{eq:Schw:2d:movMirr:P1_alt0} is obtained by recalling that $\kappa>0$ and that real constants in the integrand of the transition rate are vanishing when the cut-off is taken to infinity at the end of the calculation. One can also note that the argument of the inner logarithm on the final line of~\eqref{eq:Schw:2d:movMirr:P1_alt0} is unity or greater. 
\par We are most interested in the asymptotic form of the transition rate at early and late times, $\tau\to\pm\infty$. First, consider $\tau\to-\infty$. In this limit $P\to\infty$, and we can use the large-$x$ expansion $\log{(1+x)}= \log{(x)}+x^{-1}+O(x^{-2})$ to write
\bea
\mathcal{\dot{F}}^{1}(\omega)&=-\frac{\Realpart}{2\pi}\int^{\infty}_0\,\mathrm{d}s\,\expo^{-i\omega s-s/\Delta\tau}\log{\left[ \kappa\Gamma s-\expo^{\kappa\left(\Gamma\tau-x_0\right)}\left(1-\expo^{-\kappa\Gamma s}\right)+O\left(\expo^{2\kappa(\Gamma\tau-x_0)}\right)\right]}\,.
\label{eq:Schw:2d:movMirr:P1_alt1}
\eea
Changing variables to $y=\kappa\Gamma s$, we have
\begin{equation}
\begin{aligned}
\mathcal{\dot{F}}^{1}(\omega)&=-\frac{\Realpart}{2\pi\kappa\Gamma}\int^{\infty}_0\,\mathrm{d}y\,\expo^{-y(\rho+i\sigma)}\log{\left(y\right)}+O\left(\expo^{\kappa\left(\Gamma\tau-x_0\right)}\right)\,,
\label{eq:Schw:2d:movMirr:P1_alt2}
\end{aligned}
\end{equation}
where $\rho:=1/(\kappa\Gamma\Delta\tau)$ and $\sigma:=\omega/\kappa\Gamma$.
The dominant contribution can be evaluated using a standard integral in the same manner as~\eqref{eq:Schw:2d:trHadamardUnruhAlt3}. After taking the cut-off to infinity, the result is
\begin{equation}
\mathcal{\dot{F}}^{1}(\omega)\to \frac{1}{4|\omega|}\,, \qquad\qquad\, \tau\to-\infty\,.
\label{eq:Schw:2d:movMirr:P1_result}
\end{equation}
\par Next, we shall consider the $\tau \to+\infty$ limit of the part-I factor,~\eqref{eq:Schw:2d:movMirr:P1_alt0}. Consider, first, evaluating the integrand at fixed $s$, we have
\begin{equation}
\begin{aligned}
\mathcal{\dot{F}}^{1}(\omega)&=-\frac{\Realpart}{2\pi}\int^{\infty}_0\,\mathrm{d}s\,\expo^{-i\omega s-s/\Delta\tau}\log{\left[\log{\left(1+Q\left(\expo^{\alpha s}-1\right)\right)}\right]}\,,
\label{eq:Schw:2d:movMirr:P1_tauinf_alt1}
\end{aligned}
\end{equation}
where
\begin{equation}
\begin{aligned}
Q:&=\frac{P}{1+P}\,,\\
\alpha:&=\kappa\Gamma\,.
\end{aligned}
\end{equation}
In this limit, $Q\to 0$ and we can make use of the small-$x$ expansion $\log{(1+x)}=x-x^2/2+O(x^3)$. Doing so, we obtain
\begin{equation}
\begin{aligned}
&\mathcal{\dot{F}}^{1}(\omega)=\\
&-\frac{\Realpart}{2\pi}\int^{\infty}_0\,\mathrm{d}s\,\expo^{-i\omega s-s/\Delta\tau}\left(\log{(Q)}+\log{\left(\expo^{\alpha s}-1\right)}+\log{\left[1-\frac{1}{2}Q\left(\expo^{\alpha s}-1\right)+O\left(Q^2\right)\right]}\right)\,.
\label{eq:Schw:2d:movMirr:P1_tauinf_alt2}
\end{aligned}
\end{equation}
Using the fact that the real constant will vanish when integrated against the Langlois cut-off, we rewrite~\eqref{eq:Schw:2d:movMirr:P1_tauinf_alt2} as
\begin{equation}
\begin{aligned}
\mathcal{\dot{F}}^{1}(\omega)&=-\frac{\Realpart}{2\pi}\int^{\infty}_0\,\mathrm{d}s\,\expo^{-i\omega s-s/\Delta\tau}\log{\left(\expo^{\alpha s}-1\right)}+O\left(Q\right)\\
&=-\frac{\Realpart}{2\pi}\int^{\infty}_0\,\mathrm{d}s\,\expo^{-i\omega s-s/\Delta\tau}\log{\left(2 \expo^{\alpha s/2} \sinh{(\alpha s/2)}\right)}+O\left(Q\right)\,.
\label{eq:Schw:2d:movMirr:P1_tauinf_alt3}
\end{aligned}
\end{equation}
Changing variables to $y=\alpha s$, we have
\begin{equation}
\begin{aligned}
\mathcal{\dot{F}}^{1}(\omega)&=-\frac{\Realpart}{2\pi\kappa\Gamma}\int^{\infty}_0\,\mathrm{d}y\,\expo^{-y(\rho+i\sigma)}\log{\left[\expo^{y/2}\sinh{\left(\frac{y}{2}\right)}\right]}+O\left(Q\right)\\
&=-\frac{\Realpart}{2\pi\kappa\Gamma}\int^{\infty}_0\,\mathrm{d}y\,\expo^{-y(\rho+i\sigma)}\left(\frac{y}{2}+\log{\left[\sinh{\left(\frac{y}{2}\right)}\right]}\right)+O\left(Q\right)\,.
\label{eq:Schw:2d:movMirr:P1_tauinf_alt4}
\end{aligned}
\end{equation}
We have already evaluated both pieces of the integral~\eqref{eq:Schw:2d:movMirr:P1_tauinf_alt4} in Sections~\ref{sec:2d:Schw:HH} and~\ref{sec:2d:Schw:Unruh}, and we can immediately write the result
\begin{equation}
\mathcal{\dot{F}}^{1}(\omega)\to\frac{\kappa \Gamma}{4\pi\omega^2}+\frac{1}{4\omega}\coth{\left(\frac{\pi\omega}{\kappa\Gamma}\right)}\,,\qquad\qquad \tau\to\infty\,.
\label{eq:Schw:2d:movMirr:P1_tauinf_result}
\end{equation}
\par The preceding approach is not strictly valid owing to the fact it only holds for fixed $s$, and if we allow $s\to\infty$, the series approximations we used break down. To show that the result is nevertheless true, we refine this line of reasoning further by using the monotone convergence theorem.
\par We can write
\begin{equation}
\begin{aligned}
&\mathcal{\dot{F}}^{1}(\omega)=\frac{\kappa \Gamma}{4\pi\omega^2}+\frac{1}{4\omega}\coth{\left(\frac{\pi\omega}{\kappa\Gamma}\right)}\\
&-\frac{\Realpart}{2\pi}\int^{\infty}_0\,\mathrm{d}s\expo^{-i\omega s-s/\Delta\tau}\left( \log{\left[ \log{\left(1+Q\left(\expo^{\alpha s}-1\right)\right)}\right]} -\log{\left(2Q\right)}-\frac{\alpha s}{2}-\log{\left(\sinh{\left(\frac{\alpha s}{2}\right)}\right)}\right)\,,
\label{eq:Schw:2d:movMirr:P1_monConv1}
\end{aligned}
\end{equation}
where under the integral we have subtracted from the original logarithmic piece of the transition rate integrand,~\eqref{eq:Schw:2d:movMirr:P1_tauinf_alt1}, its limit to order $O\left(Q\right)$ as $\tau\to\infty$ for \emph{fixed} $s$, as computed in~\eqref{eq:Schw:2d:movMirr:P1_tauinf_alt2}. Our aim is to show that this integral does not contribute, and to this end we define
\begin{equation}
\begin{aligned}
f(Q):&= \log{\left[ \log{\left(1+Q\left(\expo^{\alpha s}-1\right)\right)}\right]} -\log{\left(2Q\right)}-\frac{\alpha s}{2}-\log{\left(\sinh{\left(\frac{\alpha s}{2}\right)}\right)}\,,\\
f'(Q)&= \frac{1}{Q\log{\left(1+Q\left(\expo^{\alpha s}-1\right)\right)}}\left[-\log{\left(1+Q\left(\expo^{\alpha s}-1\right)\right)}+\frac{Q\left(\expo^{\alpha s}-1\right)}{1+Q\left(\expo^{\alpha s}-1\right)}\right]\,.
\end{aligned}
\end{equation}
\par Owing to the fact that $s>0$ and $Q>0$, the factor external to the square brackets in $f'(Q)$ is uniformly positive, and we show next that the term inside the square brackets is uniformly negative for all values of $s$. 
\par To this end, we define $h:=Q\left(\expo^{\alpha s}-1\right)$ and
\begin{equation}
\begin{aligned}
g(h):=-\log{(1+h)}+\frac{h}{1+h}\,,\\
g'(h)=-\frac{h}{(1+h)^2}\leq 0\,.
\end{aligned}
\end{equation}
The final inequality following because $h\geq 0$. Thus, the fact that $g(0)=0$ implies that $g(h)\leq 0$.

 Thus, $f'(Q)\leq 0$ uniformly across $s$, and the monotone convergence theorem can be employed; it tells us that the integral in~\eqref{eq:Schw:2d:movMirr:P1_monConv1} really is vanishing and allows us to state that
\begin{equation}
\mathcal{\dot{F}}^{1}(\omega)=\frac{\kappa \Gamma}{4\pi\omega^2}+\frac{1}{4\omega}\coth{\left(\frac{\pi\omega}{\kappa\Gamma}\right)}\,\, , \, \, \tau\to\infty\,.
\label{eq:Schw:2d:movMirr:P1_finRes}
\end{equation}
\subsubsection{Part-II}
Next, we consider the $\Delta v$ factor in the Wightman function~\eqref{eq:Schw:2d:movMirr:inW}, which gives a contribution to the transition rate of
\begin{equation}
\begin{aligned}
\mathcal{\dot{F}}^{2}(\omega)&=-\frac{\Realpart}{2\pi}\int^{\infty}_0\,\mathrm{d}s\,\expo^{-i\omega s-s/\Delta\tau}\log{\left(\frac{s}{\Gamma}\right)}\\
&=-\frac{\Realpart}{2\pi}\int^{\infty}_0\,\mathrm{d}s\,\expo^{-i\omega s-s/\Delta\tau}\log{\left(s\right)}\,.
\end{aligned}
\end{equation}
We use the standard integral~\eqref{eq:Schw:2d:stdLogInt} and obtain the result
\begin{equation}
\mathcal{\dot{F}}^{2}(\omega)=\frac{1}{4|\omega|}\,.
\label{eq:Schw:2d:movMirr:P2_res}
\end{equation}
\subsubsection{Part-III}
Now, consider the $v-p(u')$ factor in the Wightman function~\eqref{eq:Schw:2d:movMirr:inW}, which gives a contribution to the transition rate of
\begin{equation}
\begin{aligned}
\mathcal{\dot{F}}^{3}(\omega)&=\frac{\Realpart}{2\pi}\int^{\infty}_0\,\mathrm{d}s\,\expo^{-i\omega s-s/\Delta\tau}\log{\left[V+\frac{1}{\kappa}\log{\left(1+P\expo^{\alpha s}\right)}\right]}\,,
\label{eq:Schw:2d:movMirr:P3_init}
\end{aligned}
\end{equation}
where
\begin{equation}
\begin{aligned}
V&:=x_0+\frac{\tau}{\Gamma}\,,\\
P&:=\expo^{-\kappa(\Gamma\tau-x_0)}\,,\\
\alpha&:=\kappa\Gamma \,.
\end{aligned}
\end{equation}
\par We seek the $\tau\to-\infty$ limit of~\eqref{eq:Schw:2d:movMirr:P3_init}, and in this limit $P\to\infty$. This means we can use the large-$x$ expansion $\log{(1+x)}= \log{(x)}+x^{-1}+O(x^{-2})$ to write
\begin{equation}
\begin{aligned}
\mathcal{\dot{F}}^{3}(\omega)&=\frac{\Realpart}{2\pi}\int^{\infty}_0\,\mathrm{d}s\,\expo^{-i\omega s-s/\Delta\tau}\log{\left[2X+\Gamma s+\frac{1}{\kappa P}\expo^{-\alpha s}+O\left(P^{-2}\expo^{-2\alpha s}\right)\right]}\,,
\label{eq:Schw:2d:movMirr:P3_alt0}
\end{aligned}
\end{equation}
with $X:=x_0-\nu\tau/\sqrt{1-\nu^2}$. Noting that $X\to\infty$ as $\tau\to-\infty$, then~\eqref{eq:Schw:2d:movMirr:P3_alt0} can be written as 
\begin{equation}
\begin{aligned}
\mathcal{\dot{F}}^{3}(\omega)&=\frac{\Realpart}{2\pi}\int^{\infty}_0\,\mathrm{d}s\expo^{-i\omega s-s/\Delta\tau}\left(\log{\left[2X+\Gamma s\,\right]} +O\left(\frac{\expo^{-\alpha s}}{P\left(2X+\Gamma s\right)}\right)\right)\\
&=\frac{\Realpart}{2\pi}\int^{\infty}_0\,\mathrm{d}s\,\expo^{-i\omega s-s/\Delta\tau}\left(\log{(2X)}+\log{\left[1+\frac{\Gamma s}{2X}\right]} +O\left(\frac{\expo^{-\alpha s}}{P\left(2X+\Gamma s\right)}\right)\right)\\
&=\frac{\Realpart}{2\pi}\int^{\infty}_0\,\mathrm{d}s\,\expo^{-i\omega s-s/\Delta\tau}\left(\log{\left[1+\frac{\Gamma s}{2X}\right]} +O\left(\frac{\expo^{-\alpha s}}{P\left(2X+\Gamma s\right)}\right)\right)\,.
\end{aligned}
\label{eq:Schw:2d:movMirr:P3_alt1}
\end{equation}
The final equality is obtained because $X$ is a positive (although infinite) constant in the limit of $\tau\to-\infty$ and, thus, $\log{(2X)}$ is a real constant, which vanishes against the Langlois cut-off. After changing variables to $y=\Gamma s/2X$, the dominant term from this contribution to the transition rate is
\begin{equation}
\mathcal{\dot{F}}^{3}(\omega)\approx\frac{X \Realpart}{\pi\Gamma}\int^{\infty}_0\,\mathrm{d}y\,\expo^{-y(a+ib)}\log{(1+y)}\,,
\label{eq:Schw:2d:movMirr:P3_alt2}
\end{equation}
with $a:=2X/\Gamma \Delta\tau$ and $b:=2\omega X/\Gamma$. 
\par Integrating~\eqref{eq:Schw:2d:movMirr:P3_alt2} by parts and then changing variables, first to $z=1+y$ and, finally, to $t=(a+ib)z$, we have
\begin{equation}
\begin{aligned}
\mathcal{\dot{F}}^{3}(\omega)&\approx \frac{X \Realpart}{\pi\Gamma} \left[\frac{(a-ib)\expo^{a+ib}}{a^2+b^2} \int^{\infty}_{a+ib}\,\mathrm{d}t\,t^{-1} \expo^{-t}\right]\\
&=\frac{X \Realpart}{\pi\Gamma} \left[\frac{(a-ib)\expo^{a+ib}}{a^2+b^2} \Gamma(0,a+ib)\right]\,,
\end{aligned}
\label{eq:Schw:2d:movMirr:P3_alt3}
\end{equation}
where $\Gamma(0,a+ib)$ is the incomplete Gamma function.
\par Taking the cut-off $\Delta\tau\to\infty$ (or $a\to 0$), we arrive at
\begin{equation}
\begin{aligned}
\mathcal{\dot{F}}^{3}(\omega)&\approx \frac{X}{\pi\Gamma} \Realpart\left[\frac{\expo^{ib}}{ib}\Gamma(0,ib)\right]\,.
\end{aligned}
\label{eq:Schw:2d:movMirr:P3_alt4}
\end{equation}
For $b\in\BbbR$ and $b\ne0$, let 
\begin{align}
f(b) 
:= \frac{\expo^{ib}}{ib} 
\Gamma(0,ib) 
\ . 
\end{align}
Equation (8.21.1) and the branch discussion in \S8.21(ii) of~\cite{dlmf} give 
\begin{align}
f(b) 
= \frac{\expo^{ib}}{ib} 
\left[ 
\ci(0,|b|) - i \sgn(b) \si(0,|b|)
\right] \, . 
\end{align}
Equation (8.21.10) of~\cite{dlmf} gives 
\begin{align}
f(b) 
&= \frac{\expo^{ib}}{b} 
\left[ 
\sgn(b) \si(|b|) + i \Ci(|b|)
\right] 
\nonumber 
\\
& = 
\frac{\cos b}{|b|} \si(|b|) - \frac{\sin b}{b} \Ci(|b|)
+ \text{imaginary}
\ \,, 
\end{align}
so that 
\begin{align}
\Realpart \left[ f(b) \right]  
& = 
\frac{\cos b}{|b|} \si(|b|) - \frac{\sin{|b|}}{|b|} \Ci(|b|)
\ . 
\end{align}

Therefore, the transition rate contribution is

\begin{equation}
\mathcal{\dot{F}}^{3}(\omega) \approx \frac{1}{2\pi|\omega|} \left[\cos{\left(\frac{2X|\omega|}{\Gamma}\right)}\si{\left(\frac{2X|\omega|}{\Gamma}\right)}-\sin{\left(\frac{2X|\omega|}{\Gamma}\right)}\Ci{\left(\frac{2X|\omega|}{\Gamma}\right)}\right]\,.
\label{eq:Schw:2d:movMirr:P3_res}
\end{equation}
\par Using the asymptotic expansions for the sine and cosine integrals for large argument, noting that as $\tau \to-\infty$ then $X \to \infty$, we see that 
\begin{equation}
\mathcal{\dot{F}}^{3}(\omega) \to 0\,,\qquad\qquad\ \tau\to-\infty\,.
\label{eq:Schw:2d:movMirr:P3fin}
\end{equation}
\par We would next like to consider the $\tau\to+\infty$ limit of~\eqref{eq:Schw:2d:movMirr:P3_init}, but we have a dilemma. One could take the $\tau\to\infty$ limit --- prior to the $\Delta\tau\to\infty$ limit --- and initially make expansions of the logarithms in the integrand for \emph{fixed} $s$, and this results in an integral of the form
\be
\mathcal{\dot{F}}^{3}(\omega)=\frac{\Realpart}{2\pi}\int^{\infty}_0\,\mathrm{d}s\,\expo^{-i\omega s-s/\Delta\tau}\log{\left(1+O\left(\frac{P\expo^{\alpha s}}{\kappa V} \right)\right)}\,,
\label{eq:Schw:2d:movMirr:P3_tauinf_alt0}
\ee
which does have a vanishing integrand for fixed $s$ when $\tau\to\infty$, but whose integral we have not been able to show to vanish, using the monotone convergence theorem or otherwise. Alternatively, one could consider taking the limit of the cut-off, $\Delta\tau\to\infty$, in advance of taking the limit $\tau\to\infty$, but this leaves the problem of directly evaluating~\eqref{eq:Schw:2d:movMirr:P3_init} without the aid of any expansions, a task we have been unsuccessful at completing.
\par We are forced to leave the part-III contribution to the $\tau\to\infty$ limit of the transition rate as an unknown, despite suspecting that it is vanishing.
\subsubsection{Part-IV}
Finally, we consider the $p(u)-v'$ factor in the Wightman function~\eqref{eq:Schw:2d:movMirr:inW}, which gives a contribution to the transition rate of
\bea
\mathcal{\dot{F}}^{4}(\omega)&=\frac{\Realpart}{2\pi}\int^{\infty}_0\,\mathrm{d}s\,\expo^{-i\omega s-s/\Delta\tau}\log{\left[R+\frac{s}{\Gamma}\right]}\,,
\label{eq:Schw:2d:movMirr:P4_init}
\eea
where
\be
R:=-\frac{1}{\kappa}\log{\left[\expo^{\kappa\left(\tau/\Gamma+x_0\right)}+\expo^{\kappa\left(2x_0-2\nu\tau/\sqrt{1-\nu^2}\right)}\right]}.
\label{eq:Schw:2d:movMirr:P4_Rdef}
\ee
\par Defining $A:=\expo^{\kappa x_0}$ and $g:=\expo^{-\kappa\tau}$, we can write the constant $R$ as
\be
R=-\frac{1}{\kappa}\log{\left[Ag^{-1/\Gamma}+A^2 g^{\Gamma-1/\Gamma}\right]} \leq 0\,.
\label{eq:Schw:2d:movMirr:P4_Rineq}
\ee
To show that the final inequality holds, consider the argument of the logarithm
\be
f(g):=Ag^{-1/\Gamma}+A^2 g^{\Gamma-1/\Gamma}\,,
\ee
and use the lower bound established for $x_0$, equation~\eqref{eq:Schw:2d:movMirr:x0bound}, to obtain
\be
A\geq \sqrt{\frac{1-\nu}{1+\nu}}\left(\frac{2\nu}{\sqrt{1-\nu^2}}\right)^{\nu}\,,
\ee
and, hence,
\bea 
f(g)&\geq \sqrt{\frac{1-\nu}{1+\nu}}\left(\frac{2\nu}{\sqrt{1-\nu^2}}\right)^{\nu} \exp{\left[\kappa\tau\sqrt{\frac{1-\nu}{1+\nu}}\right]}+\left(\frac{1-\nu}{1+\nu}\right)\left(\frac{2\nu}{\sqrt{1-\nu^2}}\right)^{2\nu}\exp\left[-\frac{2\kappa\tau\nu}{\sqrt{1-\nu^2}}\right]\\
&\geq 1\,,
\label{eq:Schw:2d:movMirr:P4_fgt}
\eea
where the final inequality can be verified by analysing the turning points of the right-hand side of the first line of~\eqref{eq:Schw:2d:movMirr:P4_fgt}. With~\eqref{eq:Schw:2d:movMirr:P4_fgt}, the validity of~\eqref{eq:Schw:2d:movMirr:P4_Rineq} is established.
\par Now that we have established that $R\leq 0$, we can rewrite the contribution from part-IV to the transition rate as
\bea
\mathcal{\dot{F}}^{4}(\omega)&=\frac{\Realpart}{2\pi}\int^{\infty}_0\,\mathrm{d}s\,\expo^{-i\omega s-s/\Delta\tau}\left(\log{|R|}-i\pi+\log{\left[1-\frac{s}{\Gamma|R|}\right]}\right)\\
&=-\frac{1}{2\omega}+\frac{\Realpart}{2\pi}\int^{\infty}_0\,\mathrm{d}s\,\expo^{-i\omega s-s/\Delta\tau}\log{\left[1-\frac{s}{\Gamma|R|}\right]}\,,
\eea
where the branch choice $\log{(-|R|)}=\log{|R|}-i\pi$ has been made in order to have consistency with the finite imaginary constant one obtains in the $m\to 0$ limit of the Minkowski half-space massive field Wightman function, which correspondingly, is also the finite imaginary constant chosen in Section~\ref{sec:2d:Schw:Unruh}.
\par Next, we change variables to $y=s/\left(|R|\Gamma\right)$ to obtain
\begin{equation}
\mathcal{\dot{F}}^{4}(\omega)=-\frac{1}{2\omega}+\frac{\Gamma|R|}{2\pi}\Realpart\int^{\infty}_0\,\mathrm{d}y\,\expo^{-y(c+id)}\log{\left(1-(y-i\epsilon)\right)}\,,
\label{eq:Schw:2d:movMirr:P4_preint}
\end{equation}
where $c:=\Gamma|R|/\Delta\tau$, $d:=\omega\Gamma|R|$ and we have restored the $i\epsilon$-regulator.
\par Our task now is the evaluation of 
\begin{equation}
I:=\Realpart\int^{\infty}_0\,\mathrm{d}z\expo^{-z(c+id)}\log{\left(1-z+i\epsilon\right)}\,,
\end{equation}
and we do so by contour methods.  First, we integrate by parts to obtain
\begin{equation}
I=\Realpart\left[\frac{id-c}{c^2+d^2}\int^{\infty}_0\,\mathrm{d}z\,\frac{\expo^{-z(c+id)}}{1-z+i\epsilon}\right]\,.
\end{equation}
For $\omega >0$ ($d>0$), we deform the contour by closing in the lower-half of the complex-plane to obtain
\begin{equation}
\begin{aligned}
I&=\Realpart\left[\frac{ic+d}{c^2+d^2}\int^{\infty}_0\,\mathrm{d}t\,\frac{1-it}{1+t^2}\expo^{-t(d-ic)}\right]\\
&=\Realpart\left[\frac{ic+d}{c^2+d^2}\int^{\infty}_0\,\mathrm{d}t\,\frac{1}{1+t^2}\expo^{-t(d-ic)}+\frac{c-id}{c^2+d^2}\int^{\infty}_0\,\mathrm{d}t\,\frac{t}{1+t^2}\expo^{-t(d-ic)}\right]\,.
\end{aligned}
\end{equation}
Taking the cut-off to infinity (which corresponds to $c\to 0$), we find
\begin{equation}
\begin{aligned}
I&=\Realpart\left[\frac{1}{d}\int^{\infty}_0\,\mathrm{d}t\,\frac{1}{1+t^2}\expo^{-dt}-\frac{i}{d}\int^{\infty}_0\,\mathrm{d}t\,\frac{t}{1+t^2}\expo^{-dt}\right]\\
&=\frac{1}{d}\Realpart \int^{\infty}_0\,\mathrm{d}t\,\frac{1}{1+t^2}\expo^{-dt}\\
&=\frac{1}{\omega\Gamma|R|}\left[\Ci{\left(\omega\Gamma|R|\right)}\sin{\left(\omega\Gamma|R|\right)}-\si{\left(\omega\Gamma|R|\right)}\cos{\left(\omega\Gamma|R|\right)}\right]\,,
\label{eq:Schw:2d:movMirr:P4_IomPos}
\end{aligned}
\end{equation}
where the final equality results from the standard integral (5.2.12) of~\cite{abramowitz}.
\par Next, we consider the integral in the case when $\omega <0$ (or rather $d<0$); in this case, we deform the contour to close in the upper half-space and pick up a contribution from the residue of the pole at $z=1$. We find
\begin{equation}
\begin{aligned}
I&=\Realpart\left[\frac{id-c}{c^2+d^2}\left(-2\pi i\expo^{-(c+id)}+i\int^{\infty}_0\,\mathrm{d}t\,\frac{1+it}{1+t^2}\expo^{-t(ic-d)}\right)\right]\,.
\end{aligned}
\end{equation}
After taking the cut-off to infinity (or equivalently $c\to 0$), we obtain
\begin{equation}
\begin{aligned}
I&=\Realpart\left[\frac{i}{d}\left(-2\pi i\left(\cos{|d|}+i\sin{|d|}\right)+i\int^{\infty}_0\,\mathrm{d}t\,\frac{1}{1+t^2}\expo^{-t|d|}-\int^{\infty}_0\,\mathrm{d}t\,\frac{t}{1+t^2}\expo^{-t|d|}\right)\right]\\
&=\frac{2\pi}{d}\cos{(d)}-\frac{1}{d}\int^{\infty}_0\,\mathrm{d}t\,\frac{\expo^{-t|d|}}{1+t^2}\\
&=\frac{2\pi}{\omega\Gamma|R|}\cos{(\omega\Gamma|R|)}-\frac{1}{\omega\Gamma|R|}\left[\Ci{\left(|\omega|\Gamma|R|\right)}\sin{\left(|\omega|\Gamma|R|\right)}-\si{\left(|\omega|\Gamma|R|\right)}\cos{\left(|\omega|\Gamma|R|\right)}\right]\,.
\label{eq:Schw:2d:movMirr:P4_IomNeg}
\end{aligned}
\end{equation}
\par Combining our results for $\omega>0$ and $\omega <0$, equations~\eqref{eq:Schw:2d:movMirr:P4_IomPos} and~\eqref{eq:Schw:2d:movMirr:P4_IomNeg}, we obtain the final result for the transition rate contribution from part-IV:
\begin{equation}
\begin{aligned}
\mathcal{\dot{F}}^{4}(\omega)=-\frac{1}{2\omega}&+\Theta(-\omega)\frac{\cos{(\omega\Gamma|R|)}}{\omega}\\
&+\frac{1}{2\pi|\omega|}\left[\Ci{\left(|\omega|\Gamma|R|\right)}\sin{\left(|\omega|\Gamma|R|\right)}-\si{\left(|\omega|\Gamma|R|\right)}\cos{\left(|\omega|\Gamma|R|\right)}\right]\,.
\label{eq:Schw:2d:movMirr:P4_res}
\end{aligned}
\end{equation}
\par We are interested in the limits $\tau\to \pm \infty$, and we first note that $|R|\to \infty$ in both of these limits. Thus, using the large-argument asymptotics of the sine and cosine integrals, we see that the term in the square brackets of~\eqref{eq:Schw:2d:movMirr:P4_res} goes to zero, and we have
\begin{equation}
\mathcal{\dot{F}}^{4}(\omega)=-\frac{1}{2\omega}+\Theta(-\omega)\frac{\cos{(\omega\Gamma|R|)}}{\omega}\,\, , \,\, \tau\to \pm\infty \,.
\label{eq:Schw:2d:movMirr:P4_res_asy}
\end{equation}
\subsection[Transition rate result]{Final result for the transition rate of the detector in two-dimensional Minkowski spacetime with receding mirror}
We now combine the results from parts-I, -II and -IV, first, for the $\tau\to\infty$ limit.
\par Recall that in the $\tau\to\infty$ limit, we were unable to prove that the part-III contribution vanishes. Nevertheless, we suspect this to be the case, and with this assumption our tentative result for the transition rate, combining  equations~\eqref{eq:Schw:2d:movMirr:P1_finRes},~\eqref{eq:Schw:2d:movMirr:P2_res}~and~\eqref{eq:Schw:2d:movMirr:P4_res_asy} is
\begin{equation}
\mathcal{\dot{F}}(\omega)\to \Theta(-\omega)\frac{\cos{(\omega\Gamma|R|)}}{\omega}+\frac{\tilde{T}}{2\omega^2}-\frac{\Theta(-\omega)}{2\omega}+\frac{1}{2\omega}\frac{1}{\expo^{\omega/\tilde{T}}-1}\,, \qquad\qquad \tau\to\infty\,,
\label{eq:Schw:2d:movMirr:finishedTauInf}
\end{equation}
where $\tilde{T}:=\kappa\Gamma/2\pi$.
 
In the $\tau\to-\infty$ limit, by combining the contributions from parts I-IV: equations~\eqref{eq:Schw:2d:movMirr:P1_result},~\eqref{eq:Schw:2d:movMirr:P2_res},~\eqref{eq:Schw:2d:movMirr:P3fin}
~and~\eqref{eq:Schw:2d:movMirr:P4_res_asy}, the result we arrive at is
\begin{equation}
\mathcal{\dot{F}}(\omega)\to -\frac{\Theta(-\omega)}{\omega}\left[1-\cos{\left(\omega\Gamma|R|\right)}\right]\,, \qquad\qquad \tau\to-\infty\,.
\label{eq:Schw:2d:movMirr:finishedTauMinusInf}
\end{equation}
\par The result~\eqref{eq:Schw:2d:movMirr:finishedTauMinusInf} is for the transition rate in the $\tau\to -\infty$ limit, and recalling the definition of $R$,~\eqref{eq:Schw:2d:movMirr:P4_Rdef}, we see that as $\tau\to-\infty$
\begin{equation}
|R|\to 2\left|x_0-\nu\tau/\sqrt{1-\nu^2}\right|\,.
\end{equation}
In particular, if we consider the static detector with $\nu=0$, then $|R|\to 2 x_0$ (noting that if $\nu=0$ then necessarily $x_0>0$), and we obtain 
\begin{equation}
\mathcal{\dot{F}}(\omega)\to -\frac{\Theta(-\omega)}{\omega}\left[1-\cos{\left(P\omega\right)}\right]\,, \qquad\qquad \tau\to-\infty\, ,
\label{eq:Schw:2d:movMirr:finishedTauMinusInf_nu0}
\end{equation}
where now $P:=2x_0$.
\par In this case,~\eqref{eq:Schw:2d:movMirr:finishedTauMinusInf} can be compared to the transition rate for a static detector in the Minkowski half-space (at $\tau\to-\infty$ the mirror is effectively a static boundary sat at the origin); indeed, the result is identical to~\eqref{eq:Schw:2d: masslessTRfinal}.
\par Consider the result~\eqref{eq:Schw:2d:movMirr:finishedTauInf} for the transition rate of an inertial detector in two-dimensional Minkowski spacetime with a receding mirror as $\tau\to\infty$. In this late-time limit, we expect the field modes~\eqref{eq:Schw:2d:movMirr:inModes} to have a close resemblance to field modes in a collapsing star spacetime, which the Unruh vacuum in two-dimensional Schwarzschild spacetime is designed to mimic. Indeed, comparing~\eqref{eq:Schw:2d:movMirr:finishedTauInf} with the transition rate we obtained for the Unruh vacuum,~\eqref{eq:Schw:2d:TransRateUnruh}, we see that they are closely related; the major difference is that in the flat-spacetime case we have an extra cosine term if $\omega<0$. 
\section{Summary}
\label{sec:2d:summary}
\par An interesting feature is that both~\eqref{eq:Schw:2d:TransRateUnruh} and~\eqref{eq:Schw:2d:movMirr:finishedTauInf} have the unexpected term of the form $T/2\omega^2$, and we deduce that this is likely to be an artefact of the Langlois cut-off being insufficient to control the divergences that occur in the two-dimensional massless Wightman function.
\par The unanticipated terms suggest a more robust approach may be needed in two dimensions; one such approach may be to study a detector that is coupled to the proper-time derivative of the massless scalar field, for which one would no longer have the troublesome infrared divergences to deal with.

\chapter[Four-dimensional Schwarzschild spacetime]{Four-dimensional Schwarzschild spacetime}
\chaptermark{$4d$ Schwarzschild}
\label{ch:4DSchw}
In four-dimensional Schwarzschild spacetime the Wightman function is not known analytically, and in this chapter we use numerical methods to study a detector coupled to a massless scalar field in the Hartle-Hawking, Boulware and Unruh vacuum states. 
\par After presenting the necessary analytic ground work and describing the numerical methods used, the static detector and detector on a circular geodesic exterior to the hole are studied in turn, and numerical results are presented. Our static results can be compared to the asymptotic results found in~\cite{Candelas:1980zt}.
\par We investigate the analogy between the right-hand Rindler wedge and the exterior of the Schwarzschild spacetime by comparing the transition rate of a static detector with the transition rate of a Rindler detector with an appropriately chosen proper acceleration. Similarly, the transition rate of a circular-geodesic detector is compared to the transition rate of a Rindler detector but with additional transverse drift.
\par Finally, we present the necessary analytic setup to allow investigation of the transition rate of a detector that falls radially on a geodesic into the black hole. This chapter presents work completed in collaboration with Jorma Louko and Adrian Ottewill. 
\section{Four-dimensional Schwarzschild spacetime}
\label{sec:4DSchw}
The metric of the Schwarzschild spacetime is given by
\begin{equation}
ds^2=-\left(1-\frac{2M}{r}\right)dt^2+\left(1-\frac{2M}{r}\right)^{-1}dr^2+r^2\left(d\theta^2+\sin^2{\theta}d\phi^2\right)\,,
\label{eq:Schw:4d:metric}
\end{equation}
where we assume the mass parameter $M$ to be positive, the black hole exterior is covered by $2M<r<\infty$, and the horizon is at $r\to 2M$.
\par Mode solutions of the Klein-Gordon equation in the Schwarzschild spacetime have the form~\cite{byd}
\begin{equation}
\frac{1}{\sqrt{4\pi\omega}}r^{-1}\rho_{\omega\ell}(r)Y_{\ell m}\left(\theta,\phi\right)\expo^{-i\omega t}\,,
\label{eq:Schw:4d:modeSolns}
\end{equation}
where $\omega>0$, $Y_{\ell m}$ is a spherical harmonic and the radial function $\rho_{\omega\ell}$ satisfies 
\begin{equation}
\frac{d^2 \rho_{\omega\ell}}{dr^{*2}}+\left\{\omega^2-\left(1-\frac{2M}{r}\right)\left[\frac{\ell(\ell+1)}{r^2}+\frac{2M}{r^3}\right]\right\}\rho_{\omega\ell}=0 \,,
\label{eq:Schw:4d:radModTort}
\end{equation}
with $r^{*}$ being the tortoise co-ordinate defined as
\begin{equation}
r^{*}=r+2M\log{\left(r/2M-1\right)} \,.
\label{eq:Schw:4d:tortoise}
\end{equation}
\par Alternatively, one can work with the Schwarzschild radial co-ordinate $r$ and define the function $\phi_{\omega\ell}(r):=\rho_{\omega\ell}(r)/r$, which satisfies
\begin{equation}
\phi^{\prime\prime}_{\omega\ell}(r)+\frac{2\left(r-M\right)}{r\left(r-2M\right)}\phi^{\prime}_{\omega\ell}(r)+\left(\frac{\omega^2 r^2}{(r-2M)^2}-\frac{\lambda}{r(r-2M)}\right)\phi_{\omega\ell}(r)=0\, ,
\label{eq:Schw:4d:radModPhi}
\end{equation}
with $\lambda:=\ell(\ell+1)$. Solutions to neither~\eqref{eq:Schw:4d:radModTort} nor~\eqref{eq:Schw:4d:radModPhi} can be found analytically, and as such we seek the solutions $\phi_{\omega\ell}(r)$ numerically using code written in Mathematica (TM)~\cite{Mathematica}. 

In the asymptotic limit of $r\to\infty$, equation~\eqref{eq:Schw:4d:radModPhi} has solutions $\phi(r)\approx \expo^{\pm i\omega r*}/r$. The mode solutions with the simple form $\expo^{+i\omega r*}/r$ as the leading order term at infinity are known as `up-modes', and despite being of this simple outgoing form at infinity they are a linear superposition of ingoing and outgoing modes at the horizon. Conversely, we have mode solutions known as `in-modes' that take on a simple ingoing form at the horizon, $\expo^{-i\omega r*}/r$, but because of scattering from the potential term in~\eqref{eq:Schw:4d:radModTort} they are a linear superposition of ingoing and outgoing modes at infinity.
Our first task is to find boundary conditions for both the in-modes and up-modes. With these boundary conditions for $\phin,\phup$ and $\phin^{\prime},\phup^{\prime}$, we can numerically solve the ODE~\eqref{eq:Schw:4d:radModPhi} to high precision using the Mathematica (TM) function `NDSolve' .
\subsection{Numerical methods for obtaining the boundary conditions}
\label{sec:4DSchw:bcs}
\subsection{Boundary conditions for the up-modes}
The up-modes take on the simple form $\phup \sim \expo^{i\omega r^{*}}/r$ as $r\to\infty$. To numerically obtain their value at a given suitably large radius, which we denote by $\rinf$, we substitute the ansatz
\begin{equation}
\phup\sim\frac{\expo^{i\omega r^{*}}}{r}\expo^{v(r)}\, ,
\label{eq:Schw:4d:upAnsatz}
\end{equation}
with
\begin{equation}
v(r):=\sum_{n=1}^{\infty}\frac{c_n}{r^n}\,,
\label{eq:Schw:4d:v}
\end{equation}
into~\eqref{eq:Schw:4d:radModPhi}. This leads to an equation for $v(r)$:
\begin{equation}
r^2(r-2M)v^{\prime\prime}(r)+r^2(r-2M)(v^{\prime}(r))^2+2r\left(M+i\omega r^2\right)v^{\prime}(r)-\left(\ell(\ell+1)r+2M\right)=0\, .
\label{eq:Schw:4d:veq}
\end{equation} 
We substitute~\eqref{eq:Schw:4d:v} into~\eqref{eq:Schw:4d:veq} and collect inverse powers of $r$. The coefficient of each power of $r$ must be set equal to zero. The lowest power leads to an equation only involving $c_1$, the next power only involves $c_1$ and $c_2$, the next only $c_1,~c_2,~c_3$, and so on. We iteratively solve for the $c_i$ by substituting the previous result into the next equation to be solved. In practice, the upper limit in the sum~\eqref{eq:Schw:4d:v} is replaced by some suitable cut-off, denoted as $\ninf$. This means that the highest power we can trust in the $r^{-1}$ expansion of~\eqref{eq:Schw:4d:veq} is $r^{-\left(\text{ninf}-2\right)}$, and the highest coefficient obtained is $c_{\ninf}$. The values of $\ninf$ and $\rinf$ are determined by the desired numerical accuracy.
\par The initial conditions for $\phup$ and $\phup^{\prime}$ are computed using these $c_i$ and by evaluating at $\rinf$:
\begin{equation}
\begin{aligned}
\phup(\rinf)&=\frac{\expo^{i\omega r^{*}(\rinf)}}{\rinf}\expo^{v(\rinf)}\,,\\
\phup^{\prime}(\rinf)&=\frac{d}{dr}\left[\frac{\expo^{i\omega r^{*}(r)}}{r}\expo^{v(r)}\right]_{r=\rinf}\, .
\label{eq:Schw:4d:inits}
\end{aligned}
\end{equation}
These initial conditions become more accurate as $\rinf$ and $\text{ninf}$ increase.
\par
We computed the initial conditions~\eqref{eq:Schw:4d:inits} in Mathematica (TM) for $\ninf=100$ and $\rinf=15000M$, where we set $M=1$ in the code and re-inserted the appropriate factors of $M$ in the computed physical answers by dimensional analysis. 
Having computed the boundary conditions, we then used Mathematica's `NDSolve' function to generate our up-modes $\phup$ for a given $(\omega,\ell)$. 
We sought a result for the transition rate that was accurate to around 3 or 4 decimal places. As we shall see later, the Wightman function is constructed using tens of thousands of points in $(\omega,\ell)$ parameter space, and a high precision in the individual $\phup,\phin$ modes is essential. This is particularly true if the detector is on a radially-infalling trajectory. In order to get the desired accuracy results for the transition rate, we used very high precision settings in `NDSolve'; we set `WorkingPrecision' to around 40, `AccuracyGoal' to around 32 and `PrecisionGoal' to around 20. With these settings, the results for $\phup,\phin$ did not change to around 10 decimal places upon further increases to the `NDSolve' precision settings. 
\subsection{Boundary conditions for the in-modes}
The in-modes are the modes that take on a simple ingoing form at the horizon, $\expo^{-i\omega r^{*}}/r$, but a complicated superposition of ingoing and outgoing plane waves at radial infinity because of the scattering from the gravitational potential. Thus, our strategy is to compute the initial conditions of the in-modes at the horizon, taking
\begin{equation}
\phin\sim \frac{\expo^{-i\omega r^{*}}}{r}w(r)\, 
\label{eq:Schw:4d:inAnsatz}
\end{equation} 
as our ansatz, with 
\begin{equation}
w(r):=\sum^{\infty}_{n=0} b_n \left(r-2M\right)^n\, ,
\label{eq:Schw:4d:w}
\end{equation}
and $b_0=1$. 

We substitute~\eqref{eq:Schw:4d:inAnsatz} into~\eqref{eq:Schw:4d:radModPhi} to obtain an equation in $w(r)$ that reads
\begin{equation}
r^2 (r-2M)w^{\prime\prime}(r)+2r\left(M-ir^2\omega\right)w^{\prime}(r)-(\ell(\ell+1)r+2M)w(r)=0\, .
\label{eq:Schw:4d:weq}
\end{equation} 
Using~\eqref{eq:Schw:4d:w} in~\eqref{eq:Schw:4d:weq}, a recursion relation can be obtained~\cite{leaver}:
\begin{equation}
\begin{aligned}
&b_0=1\,, b_{-1}=b_{-2}=0\,,\\
&b_n=-\frac{\left[-12i\omega M(n-1)+(2n-3)(n-1)-(\ell(\ell+1)+1)\right]}{2M\left(n^2-i4Mn\omega\right)}b_{n-1}\\
&-\frac{\left[(n-2)(n-3)-i12M\omega(n-2)-\ell(\ell+1)\right]}{4M^2\left(n^2-i4Mn\omega\right)}b_{n-2}\\
&+\frac{i\omega(n-3)}{2M^2\left(n^2-i4Mn\omega\right)}b_{n-3}\, .
\label{eq:Schw:4d:recursion}
\end{aligned}
\end{equation}
We are now in a position to compute the initial conditions for $\phin$ and $\phin^{\prime}$. We use these $b_i$ with the upper limit of the sum~\eqref{eq:Schw:4d:w} replaced by some finite integer $\nh$, determined by the accuracy requirements, and we evaluate at the near horizon radius $\rh$, obtaining
\begin{equation}
\begin{aligned}
\phin(\rh)&=\frac{\expo^{-i\omega r^{*}(\rh)}}{\rh}w(\rh)\,,\\
\phin^{\prime}(\rh)&=\frac{d}{dr}\left[\frac{\expo^{-i\omega r^{*}(r)}}{r}w(r)\right]_{r=\rh}\, .
\label{eq:Schw:4d:HorizonInits}
\end{aligned}
\end{equation}
In practice, the initial conditions~\eqref{eq:Schw:4d:HorizonInits} were computed in Mathematica (TM) for $\nh=200$ and $\rh=20,000,001/10,000,000M$, where $M=1$. Given these boundary conditions, we used Mathematica's `NDSolve' function to generate our in-modes $\phin$ for a given $(\omega,\ell)$, with the same precision settings as for the up-modes.
\section{Normalisation}
\label{sec:4DSchw:norm}
%
\begin{figure}[t] \centering
\tikzset{->-/.style={decoration={
  markings,
  mark=at position #1 with {\arrow[scale=1]{>}}},postaction={decorate}}}

\begin{center}
\begin{tikzpicture}[>=open triangle 90, scale=0.7]
\node (I)    at ( 4,0) {};
\node (II)   at (-5,0) {};

\path 
   (I) +(90:4)  coordinate[label=90:$i^+$]  (Itop)
       +(-90:4) coordinate[label=-90:$i^-$] (Ibot)
       +(180:4) coordinate (Ileft)
       +(0:4)   coordinate[label=0:$i^0$] (Iright)
       +(45: 2.825)   coordinate (Imidtr)
       +(225:2.825)   coordinate (Imidbl)
       +(135:2.825)   coordinate (Imidtl)
       +(-45:2.825)   coordinate (Imidbr)
       ;
       ;
\draw  (Ileft)  -- 
          node[midway, below, sloped] {$H^+$}
       (Itop)   -- 
          node[midway, above right]    {$\scri^+$}
          node[midway, below, sloped] {$\bar{v}=\infty$}
       (Iright) --
          node[midway, below right]    {$\scri^-$}
          node[midway, above, sloped] {$\bar{u}=-\infty$}
       (Ibot)   --
          node[midway, below, sloped]    {$H^-$}
       (Ileft)  -- cycle;
       
\draw[->-=.5] (Imidbr.center) -- (I.center);
\draw[->-=.5] (I.center) -- (Imidtl.center);   
\draw[->-=.5] (I.center) -- (Imidtr.center);    

\path 
   (II) +(90:4)  coordinate[label=90:$i^+$]  (IItop)
       +(-90:4) coordinate[label=-90:$i^-$] (IIbot)
       +(180:4) coordinate (IIleft)
       +(0:4)   coordinate[label=0:$i^0$] (IIright)
       +(45: 2.825)   coordinate (IImidtr)
       +(225:2.825)   coordinate (IImidbl)
       +(135:2.825)   coordinate (IImidtl)
       +(-45:2.825)   coordinate (IImidbr)
       ;
\draw  (IIleft)  -- 
          node[midway, below, sloped] {$H^+$}
       (IItop)   -- 
          node[midway, above right]    {$\scri^+$}
          node[midway, below, sloped] {$\bar{v}=\infty$}
       (IIright) --
          node[midway, below right]    {$\scri^-$}
          node[midway, above, sloped] {$\bar{u}=-\infty$}
       (IIbot)   --
          node[midway, below, sloped]    {$H^-$}
       (IIleft)  -- cycle;

\draw[->-=.5] (IImidbl) -- (II.center);
\draw[->-=.5] (II.center) -- (IImidtl);   
\draw[->-=.5] (II.center) -- (IImidtr);

\end{tikzpicture}
\caption{Illustrating the `up' and `in' modes on the right-hand wedge of the Penrose diagram representing the region exterior to the four-dimensional Schwarzschild black hole. The `up' modes are shown on the left-hand side and `in' modes on the right-hand side.}  
\end{center}
\end{figure}
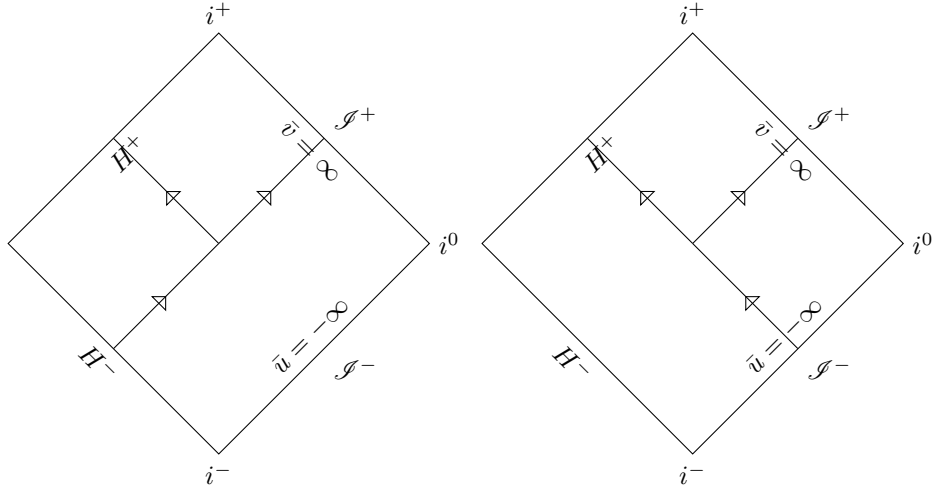
We choose a basis whose asymptotic behaviour as $r^{*}\to\pm\infty$ is
\begin{equation}
\Phin(r)\sim 
\begin{cases}
B^{\text{in}}_{\omega\ell}\expo^{-i\omega r^{*}}, & \,\, r\to 2M \, , \\
r^{-1}\expo^{-i\omega r^{*}}+A^{\text{in}}_{\omega\ell}r^{-1}\expo^{+i\omega r^{*}},  & \,\, r\to \infty \, ,
\label{eq:Schw:4d:asyIn}
\end{cases}
\end{equation}
and 
\begin{equation}
\Phup(r)\sim 
\begin{cases}
A^{\text{up}}_{\omega\ell}\expo^{-i\omega r^{*}}+\expo^{+i\omega r^{*}}, & \,\, r\to 2M \, , \\
B^{\text{up}}_{\omega\ell}r^{-1}\expo^{+i\omega r^{*}},  & \,\, r\to \infty \, .
\label{eq:Schw:4d:asyUp}
\end{cases}
\end{equation}
The reflection and transmission coefficients satisfy the following Wronskian relations:
\begin{equation}
\begin{aligned}
\Bup&=(2M)^2\Bin\,,\\   
|\Ain|^2&=1-4M^2|\Bin|^2\,,\\    
|\Ain|^2&=|\Aup|^2\,,\\
|\Aup|^2&=1-\frac{|\Bup|^2}{4M^2}\, ,  
\label{eq:Schw:4d:Wronsk}
\end{aligned}
\end{equation}
which we verify in equations~\eqref{ch:appendix:4DSchwTransRefl:AupVsAin},~\eqref{ch:appendix:4DSchwTransRefl:BupVsBin}
and~\eqref{ch:appendix:4DSchwTransRefl:AupVsB} of Appendix~\ref{ch:appendix:4DSchwTransRefl}.
In the Mathematica code, we compute the reflection and transmission coefficients using~\eqref{ch:appendix:4DSchwTransRefl:coeffsInCode}.
\par If we represent the normalised modes with a tilde, we have
\begin{equation}
\begin{aligned}
\tilde{\Phi}^{\text{in}}_{\omega\ell}&=\frac{\Phin}{N^{\text{in}}}\,,\\
\tilde{\Phi}^{\text{up}}_{\omega\ell}&=\frac{\Phup}{N^{\text{up}}}\, ,
\end{aligned}
\end{equation}
with the normalisation constants, $N^{\text{up}}$ and $N^{\text{in}}$, given by
\begin{equation}
\begin{aligned}
N^{\text{in}}&:=\sqrt{\frac{1}{2}+\frac{1}{2}|\Ain|^2+2M^2|\Bin|^2}\\
&=1\, 
\end{aligned}
\label{eq:Schw:4d:NinNormConst}
\end{equation}
and
\begin{equation}
\begin{aligned}
N^{\text{up}}&:=\sqrt{2M^2+\frac{1}{2}|\Bup|^2+2M^2|\Aup|^2}\\
&=2M\, ,
\end{aligned}
\label{eq:Schw:4d:NupNormConst}
\end{equation}
where we have used the Wronskian relations~\eqref{eq:Schw:4d:Wronsk} to perform the simplifications. This normalisation is such that $\tilde{R}_{\omega,\ell}:=r \tilde{\Phi}_{\omega,\ell}$ is normalised in the Schr\"{o}dinger way:
\begin{equation}
\int^{\infty}_{-\infty}\,\mathrm{d}r^{*}\,\tilde{R}_{\omega_1}(r)\tilde{R}_{\omega_2}^{*}(r)=2\pi\delta(\omega_1-\omega_2)\,.
\label{eq:Schw:4d:NormCond}
\end{equation}
Now and throughout the remainder of this thesis, the tilde will be suppressed.  
The normalised modes in this basis can be expressed in terms of the modes that we explicitly solve for in Mathematica (TM), $\phup$ and $\phin$, which were discussed in Section~\ref{sec:4DSchw:bcs}. The result is
\begin{equation}
\begin{aligned}
\Phin &=\frac{\Bup}{2M}\phin(r)\,,\\
\Phup &=\frac{\Bup}{2M}\phup(r)\, .
\label{eq:Schw:4d:relnAdrianModesNormModes}
\end{aligned}
\end{equation}
\par With this solution, we introduce the basis functions $u^{\text{in}}_{\omega\ell m}$ and $u^{\text{up}}_{\omega\ell m}$ by
\begin{equation}
\begin{aligned}
u^{\text{in}}_{\omega\ell m}(\x)&=\frac{1}{\sqrt{4\pi\omega}}\Phi^{\text{in}}_{\omega\ell}(r)Y_{\ell m}(\theta,\phi)\expo^{-i\omega t}\,,\\
u^{\text{up}}_{\omega\ell m}(\x)&=\frac{1}{\sqrt{4\pi\omega}}\Phi^{\text{up}}_{\omega\ell}(r)Y_{\ell m}(\theta,\phi)\expo^{-i\omega t}\, ,
\label{eq:Schw:4d:BoulwareModes}
\end{aligned}
\end{equation}
where $\omega>0$. These modes are positive frequency with respect to the Schwarzschild time translation Killing vector $\partial_t$.
\par Using the Wronskian relations~\eqref{eq:Schw:4d:Wronsk}, it can be verified that these modes satisfy the orthonormality relations
\bea 
\left(u^{\text{up}}_{\omega\ell m},u^{\text{up}}_{\omega^{\prime}\ell^{\prime} m^{\prime}}\right)&=\delta_{\ell \ell^{\prime}}\delta_{m m^{\prime}}\delta\left(\omega-\omega^{\prime}\right)\,,\\
\left(u^{\text{in}}_{\omega\ell m},u^{\text{in}}_{\omega^{\prime}\ell^{\prime} m^{\prime}}\right)&=\delta_{\ell \ell^{\prime}}\delta_{m m^{\prime}}
\delta\left(\omega-\omega^{\prime}\right)\,,\\
\left(u^{\text{in}}_{\omega\ell m},u^{\text{up}}_{\omega^{\prime}\ell^{\prime} m^{\prime}}\right)&=0\,,
\eea
where the Klein-Gordon (indefinite) inner product on a spacelike hyperplane of simultaneity at instant $t$ is defined by
\be 
( \phi, \chi)=-i\int_{2M}^{\infty}\,\mathrm{d}r\,\frac{\,r^2}{(1-2M/r)}\int^{\pi}_0\,\mathrm{d}\theta\,\sin{\theta}\,\int^{2\pi}_0\,\mathrm{d}\phi\, \left[\phi\partial_t \chi^{*}-\left(\partial_t \phi\right)\chi^{*}\right]\,.
\ee
The complex conjugate modes satisfy similar orthonormality relations with a minus sign, and the inner product relation between the modes~\eqref{eq:Schw:4d:BoulwareModes} and the complex conjugates vanish.
\section{Unruh modes and the Hartle-Hawking vacuum}
\par We shall wish to look at the transition rate of a detector when the field is in the Hartle-Hawking vacuum state, which is the vacuum state that is regular across the horizon. The modes that have the analytic properties of positive-frequency plane waves with respect to the horizon generators take the form~\cite{cyf,byd}
\begin{equation}
\begin{aligned}
w^{\text{in}}_{\omega\ell m}&=\frac{1}{\sqrt{2\sinh{\left(4\pi M\omega\right)}}}\left(\expo^{2\pi M\omega}u^{\text{in}}_{\omega\ell m}+\expo^{-2\pi M\omega}v^{\text{in}*}_{\omega\ell m}\right)\,,\\
\bar{w}^{\text{in}}_{\omega\ell m}&=\frac{1}{\sqrt{2\sinh{\left(4\pi M\omega\right)}}}\left(\expo^{-2\pi M\omega} u^{\text{in}*}_{\omega\ell m}+\expo^{2\pi M\omega}v^{\text{in}}_{\omega\ell m}\right)\,,\\
w^{\text{up}}_{\omega\ell m}&=\frac{1}{\sqrt{2\sinh{\left(4\pi M\omega\right)}}}\left(\expo^{2\pi M\omega}u^{\text{up}}_{\omega\ell m}+\expo^{-2\pi M\omega}v^{\text{up}*}_{\omega\ell m}\right)\,,\\
\bar{w}^{\text{up}}_{\omega\ell m}&=\frac{1}{\sqrt{2\sinh{\left(4\pi M\omega\right)}}}\left(\expo^{-2\pi M\omega} u^{\text{up}*}_{\omega\ell m}+\expo^{2\pi M\omega}v^{\text{up}}_{\omega\ell m}\right)\,,
\label{eq:Schw:4d:UnruhModes}
\end{aligned}
\end{equation}
where the $v$ are functions analogous to $u$ but on the second exterior region of the Kruskal manifold.
%


We can expand the quantum field $\psi(\x)$ in terms of these modes:
\begin{equation}
\begin{aligned}
\psi=\sum^{\infty}_{\ell=0}\sum^{+\ell}_{m=-\ell}\int^{\infty}_0\,\mathrm{d}\omega\,\Bigg[d^{\text{up}}_{\omega\ell m}w^{\text{up}}_{\omega\ell m}+\bar{d}^{\text{up}}_{\omega\ell m}\bar{w}^{\text{up}}_{\omega\ell m}+d^{\text{in}}_{\omega\ell m}w^{\text{in}}_{\omega\ell m}+\bar{d}^{\text{in}}_{\omega\ell m}\bar{w}^{\text{in}}_{\omega\ell m}\Bigg]+\text{h.c.}\,.
\label{eq:Schw:4d:FieldExp}
\end{aligned}
\end{equation}
The $d^{a}$ and $\bar{d}^{a}$ ($d^{a\,\dag}$ and $\bar{d}^{\text{a}\,\dag}$) operators, with $\text{a}:=\{\text{in},\text{up}\}$, are the annihilation (creation) operators with respect to the $w$ and $\bar{w}$ modes, and satisfy
\bea 
\left[d^{\text{a}}_{\omega\ell m},d^{\text{a}\,\dag}_{\omega^{\prime}\ell^{\prime} m^{\prime}}\right]&=\delta\left(\omega-\omega^{\prime}\right)\delta_{\ell\ell^{\prime}}
\delta_{m m^{\prime}}\,,\\
\left[\bar{d}^{\text{a}}_{\omega\ell m},\bar{d}^{\text{a}\,\dag}_{\omega^{\prime}\ell^{\prime} m^{\prime}}\right]&=\delta\left(\omega-\omega^{\prime}\right)\delta_{\ell \ell^{\prime}}
\delta_{m m^{\prime}}\,
\label{eq:Schw:4d:commd}
\eea
and
\be
d^{\text{a}}_{\omega\ell m}|0_K\rangle =\bar{d}^{\text{a}}_{\omega\ell m}|0_K\rangle =0\,.
\ee
The state $|0_K \rangle$ is the Hartle-Hawking vacuum state, and it is normalised such that
\be 
\langle 0_K|0_K \rangle=1\,.
\ee
In the exterior region of the hole, the modes~\eqref{eq:Schw:4d:UnruhModes} reduce to a simple form because the $v$ functions vanish, and if we compute the Wightman function for the Hartle-Hawking vacuum in the exterior region, we find
\begin{equation}
\begin{aligned}
&W(\x,\x'):=\langle 0_K|\psi(\x)\psi(\x')|0_K\rangle\\
&=\sum^{\infty}_{\ell=0}\sum^{+\ell}_{m=-\ell}\int^{\infty}_0\,\mathrm{d}\omega\, \frac{1}{8\pi\omega\sinh{\left(4\pi M\omega\right)}}\times\\
&\Bigg[\expo^{4\pi M\omega-i\omega\Delta t}Y_{\ell m}(\theta,\phi)Y^{*}_{\ell m}(\theta',\phi')\left(\Phup(r)\Phup^{*}(r')+\Phin(r)\Phin^{*}(r')\right)\\
&+\expo^{-4\pi M\omega+i\omega\Delta t}Y^{*}_{\ell m}(\theta,\phi)Y_{\ell m}(\theta',\phi')\left(\Phup^{*}(r)\Phup(r')+\Phin^{*}(r)\Phin(r')\right)\Bigg]\,,
\label{eq:Schw:4d:HHWightman}
\end{aligned}
\end{equation}
with $\Delta t:=t-t'$.
\section{Static detector}
\label{sec:4DSchw:staticDet}
\subsection[Hartle-Hawking vacuum]{Transition rate of the static detector in the Hartle-Hawking vacuum}
\par Consider a detector sat at fixed radius $r>2M$. Without loss of generality, we also choose the detector to be sat at the co-ordinates $\theta=\phi=0$. When the detector is static, the Wightman function of the Hartle-Hawking vacuum in the exterior region~\eqref{eq:Schw:4d:HHWightman} reduces to the form
\begin{equation}
\begin{aligned}
&W(\x,\x')\\
&=\sum_{\ell,m}\int^{\infty}_0\,\mathrm{d}\omega\, \frac{|Y_{\ell m}(0,0)|^2}{4\pi\omega\sinh{\left(4\pi M\omega\right)}}
\left(|\Phup(R)|^2+|\Phin(R)|^2\right)\cosh{\left[4\pi M\omega-\frac{i\omega \Delta\tau}{\sqrt{1-\frac{2M}{R}}}\right]}\,\\
&=\sum_{\ell}\int^{\infty}_0\,\mathrm{d}\omega\, \frac{|Y_{\ell 0}(0,0)|^2}{4\pi\omega\sinh{\left(4\pi M\omega\right)}}
\left(|\Phup(R)|^2+|\Phin(R)|^2\right)\cosh{\left[4\pi M\omega-\frac{i\omega \Delta\tau}{\sqrt{1-\frac{2M}{R}}}\right]}\,,
\label{eq:Schw:4d:HHWightmanStatic}
\end{aligned}
\end{equation}
where the second equality follows from (14.30.4) in~\cite{dlmf}.
\par We now substitute~\eqref{eq:Schw:4d:HHWightmanStatic} into the expression for the transition rate~\eqref{eq:techIntro:transRate-stat}, which is valid for static situations.
After interchanging the order of the $s$- and $\omega$-integrals and taking the regulator to zero, we arrive at
\begin{equation}
\begin{aligned}
\mathcal{\dot{F}}\left(E\right)=\int^{\infty}_0\,\mathrm{d}\omega\,\sum^{\infty}_{l=0}\frac{|Y_{\ell 0}(0,0)|^2}{4\pi\omega\sinh{\left(4\pi M\omega\right)}}&\left(|\Phup(R)|^2+|\Phin(R)|^2\right)\times\\
&\int^{\infty}_{-\infty}\,\mathrm{d}s\,\expo^{-iEs}\cosh{\left[4\pi M\omega-\frac{i\omega s}{\sqrt{1-\frac{2M}{R}}}\right]}\,.
\end{aligned}
\end{equation}
The $s$-integral can be done analytically resulting in
\begin{equation}
\begin{aligned}
\mathcal{\dot{F}}\left(E\right)=\int^{\infty}_0\,\mathrm{d}\omega\,\sum^{\infty}_{l=0}&\frac{|Y_{\ell 0}(0,0)|^2}{4\omega\sinh{\left(4\pi M\omega\right)}}\left(|\Phup(R)|^2+|\Phin(R)|^2\right)\times\\
&\left[\expo^{4\pi M\omega}\delta\left(E+\frac{\omega}{\sqrt{1-\frac{2M}{R}}}\right)+\expo^{-4\pi M\omega}\delta\left(E-\frac{\omega}{\sqrt{1-\frac{2M}{R}}}\right)\right]\,.
\end{aligned}
\end{equation}
The factors $|\Phup(R)|$ and $|\Phin(R)|$ can be extended to negative values of $\omega$ by symmetry. This allows one to write the transition rate as
\begin{equation}
\begin{aligned}
&\mathcal{\dot{F}}\left(E\right)=\sum^{\infty}_{l=0}|Y_{\ell 0}(0,0)|^2\sqrt{1-2M/R}\left(|\Phi^{\text{up}}_{\tilde{\omega}\ell}(R)|^2+|\Phi^{\text{in}}_{\tilde{\omega}\ell}(R)|^2\right)\times\\
&\Bigg[\frac{\expo^{-4\pi ME \sqrt{1-2M/R}}\Theta(-E)}{-4E\sqrt{1-2M/R}\sinh{\left(-4\pi ME\sqrt{1-2M/R}\right)}}\\
&\quad\quad\quad\quad\quad\quad\quad\quad+\frac{\expo^{-4\pi ME \sqrt{1-2M/R}}\Theta(E)}{4E\sqrt{1-2M/R}\sinh{\left(4\pi ME\sqrt{1-2M/R}\right)}}\Bigg]\,,
\end{aligned}
\end{equation}
where $\tilde{\omega}:=E\sqrt{1-2M/R}$. This can further be simplified to
\begin{equation}
\begin{aligned}
&\mathcal{\dot{F}}\left(E\right)=\frac{1}{2E}\frac{1}{\expo^{E/T_{\text{loc}}}-1}\sum^{\infty}_{l=0}|Y_{\ell 0}(0,0)|^2\left(|\Phi^{\text{up}}_{\tilde{\omega}\ell}(R)|^2+|\Phi^{\text{in}}_{\tilde{\omega}\ell}(R)|^2\right)\, ,
\label{eq:Schw:4d:staticHH_TR_result}
\end{aligned}
\end{equation}
with $T_{\text{loc}}:= T_0/\sqrt{1-2M/R}$, $T_0:=\kappa/2\pi$ and the surface gravity $\kappa=1/4M$.
\par The result~\eqref{eq:Schw:4d:staticHH_TR_result} manifestly obeys the KMS condition by virtue of the fact that the modes $\Phi^{\text{up}}_{\tilde{\omega}\ell}$ and $\Phi^{\text{in}}_{\tilde{\omega}\ell}$ only depend on the absolute value of $\tilde{\omega}:=E\sqrt{1-2M/R}$; hence, the modes only depend on the absolute value of excitation energy. Thus, the condition
\begin{equation}
\mathcal{\dot{F}}\left(E\right)=\expo^{-E/T_{\text{loc}}}\mathcal{\dot{F}}\left(-E\right)
\end{equation}
is obeyed, and the transition rate is thermal in the temperature $T_{\text{loc}}$.
\subsection[Boulware vacuum]{Transition rate of the static detector in the Boulware vacuum}
The Boulware vacuum is analogous to the Rindler vacuum in Rindler spacetime, and it is not regular across the black hole horizon. To construct the Wightman function for the Boulware vacuum, the quantum scalar field is expanded in terms of the modes~\eqref{eq:Schw:4d:BoulwareModes}, i.e.
\begin{equation}
\begin{aligned}
\psi=\sum^{\infty}_{\ell=0}\sum^{+\ell}_{m=-\ell}\int^{\infty}_0\,\mathrm{d}\omega\,\Bigg[b^{\text{up}}_{\omega\ell m}u^{\text{up}}_{\omega\ell m}+b^{\text{in}}_{\omega\ell m}u^{\text{in}}_{\omega\ell m}\Bigg]+\text{h.c.}\,,
\end{aligned}
\end{equation}
where the $b$ and $b^{\dag}$ operators are respectively the annihilation and creation  operators for the $u$ modes that satisfy the commutation relations
\be
\left[b^{\text{a}}_{\omega\ell m},b^{\text{a}\,\dag}_{\omega^{\prime}\ell^{\prime} m^{\prime}}\right]=\delta\left(\omega-\omega^{\prime}\right)\delta_{\ell\ell^{\prime}}
\delta_{m m^{\prime}}\,,
\label{eq:Schw:4d:commb}
\ee
and
\be
b^{\text{a}}_{\omega\ell m}|0_B\rangle =0\,,
\ee
with $\text{a}:=\{\text{in},\text{up}\}$. The state $|0_B \rangle$ is the Boulware vacuum, and it is normalised such that
\be 
\langle 0_B|0_B \rangle=1\,.
\ee
Hence, in the exterior region, the Wightman function of a scalar field in the Boulware vacuum state can be expressed as
\begin{equation}
\begin{aligned}
&W(\x,\x'):=\langle 0_B|\psi(\x)\psi(\x')|0_B\rangle\\
&=\sum^{\infty}_{\ell=0}\sum^{+\ell}_{m=-\ell}\int^{\infty}_0\,\mathrm{d}\omega\,\frac{Y_{\ell m}(\theta,\phi)Y^{*}_{\ell m}(\theta',\phi')}{4\pi\omega}\expo^{-i\omega\Delta t}\left(\Phup(r)\Phup^{*}(r')+\Phin(r)\Phin^{*}(r')\right)\,.
\label{eq:Schw:4d:preBoulwareStaticW}
\end{aligned}
\end{equation}
We specialise to a static detector: $r=r'=R$, $\Delta t=\Delta\tau/\sqrt{1-2M/R}$, and we can take $\theta=\theta'=\phi=\phi'=0$ without loss of generality. For this trajectory the Wightman function reduces to
\begin{equation}
W(\x,\x')=\sum^{\infty}_{\ell=0}\int^{\infty}_0\,\mathrm{d}\omega\,\frac{|Y_{\ell 0}(0,0)|^2}{4\pi\omega}\expo^{-i\omega\Delta\tau/\sqrt{1-2M/R}}\left(|\Phup(R)|^2+|\Phin(R)|^2\right)\,,
\label{eq:Schw:4d:BoulwareStaticW}
\end{equation}
where again (14.30.4) in~\cite{dlmf} has been used.

We substitute the Wightman function~\eqref{eq:Schw:4d:BoulwareStaticW} into transition rate~\eqref{eq:techIntro:transRate-stat}, which is valid for static situations. This allows us to switch the order the $s$- and $\omega$-integrals to obtain
\begin{equation}
\begin{aligned}
\mathcal{\dot{F}}\left(E\right)=\int^{\infty}_0\,\mathrm{d}\omega\,\sum^{\infty}_{l=0}\frac{|Y_{\ell 0}(0,0)|^2}{4\pi\omega}&\left(|\Phup(R)|^2+|\Phin(R)|^2\right)\times\\
&\int^{\infty}_{-\infty}\,\mathrm{d}s\,\expo^{-iEs}\expo^{-i\omega s/\sqrt{1-2M/R}}\,,
\end{aligned}
\end{equation}
and performing the $s$-integral gives
\begin{equation}
\begin{aligned}
\mathcal{\dot{F}}\left(E\right)=\int^{\infty}_0\,\mathrm{d}\omega\,\sum^{\infty}_{l=0}\frac{|Y_{\ell 0}(0,0)|^2}{2\omega}&\left(|\Phup(R)|^2+|\Phin(R)|^2\right)\times\\
&\delta\left(E+\frac{\omega}{\sqrt{1-\frac{2M}{R}}}\right)\, ,
\end{aligned}
\end{equation}
which can be simplified to
\begin{equation}
\begin{aligned}
\mathcal{\dot{F}}\left(E\right)=\sum^{\infty}_{l=0}\frac{|Y_{\ell 0}(0,0)|^2}{2|E|}\left(|\Phi^{\text{up}}_{\tilde{\omega}\ell}(R)|^2+|\Phi^{\text{in}}_{\tilde{\omega}\ell}(R)|^2\right)\Theta(-E)\,,
\label{eq:Schw:4d:staticBoulware_TR_result}
\end{aligned}
\end{equation}
where $\tilde{\omega}:=E\sqrt{1-2M/R}$. We note that when the field is in the Boulware vacuum, the transition rate for the static detector is only non-zero for negative energies of the detector, i.e. de-excitations. The result~\eqref{eq:Schw:4d:staticBoulware_TR_result} is very similar to the transition rate for the inertial detector in flat spacetime, $-E \Theta(-E)/2\pi$, only with modifications due to the curvature of spacetime. This is what one would expect for the Boulware vacuum.
\subsection[Unruh vacuum]{Transition rate of the static detector in the Unruh vacuum}
The Unruh vacuum mimics the geometric effects of a collapsing star, and it represents a time-asymmetric flux of radiation from the black hole. The Unruh mode construction,~\eqref{eq:Schw:4d:UnruhModes}, is applied only to the up-modes that originate on $\scrh^{-}$ and not to the in-modes originating on $\scri^{-}$. Hence, the Wightman function in the Unruh vacuum is defined by first expanding the quantum scalar field as
\begin{equation}
\begin{aligned}
\psi=\sum^{\infty}_{\ell=0}\sum^{+\ell}_{m=-\ell}\int^{\infty}_0\,\mathrm{d}\omega\,\Bigg[d^{\text{up}}_{\omega\ell m}w^{\text{up}}_{\omega\ell m}+\bar{d}^{\text{up}}_{\omega\ell m}\bar{w}^{\text{up}}_{\omega\ell m}+b^{\text{in}}_{\omega\ell m}u^{\text{in}}_{\omega\ell m}\Bigg]+\text{h.c.}\,,
\end{aligned}
\end{equation}
where now
\begin{equation}
b^{\text{in}}_{\omega\ell m}|0_U\rangle=d^{\text{up}}_{\omega\ell m}|0_U\rangle =\bar{d
}^{\text{up}}_{\omega\ell m}|0_U\rangle=0\,,
\end{equation}
with $|0_U \rangle$ the Unruh vacuum state. The annihilation and creation operators $b$,~$d$ and $b^{\dag}$,~$d^{\dag}$ satisfy the commutation relations given in~\eqref{eq:Schw:4d:commd} and~\eqref{eq:Schw:4d:commb}.
Hence, the Wightman function of a scalar field in this vacuum state can be expressed as
\begin{equation}
\begin{aligned}
&W(\x,\x'):=\langle 0_U|\psi(\x)\psi(\x')|0_U\rangle\\
&=\sum^{\infty}_{\ell=0}\sum^{+\ell}_{m=-\ell}\int^{\infty}_0\,\mathrm{d}\omega\,\left[w^{\text{up}}_{\omega\ell m}(\x)w^{\text{up}*}_{\omega\ell m}(\x')+\bar{w}^{\text{up}}_{\omega\ell m}(\x)\bar{w}^{\text{up}*}_{\omega\ell m}(\x')+u^{\text{in}}_{\omega\ell m}(\x)u^{\text{in}*}_{\omega\ell m}(\x')\right]\,.
\end{aligned}
\end{equation}
In the exterior region, this reduces to
\begin{equation}
\begin{aligned}
W(\x,\x')=\sum^{\infty}_{\ell=0}\sum^{+\ell}_{m=-\ell}\int^{\infty}_0\,\mathrm{d}\omega & \Bigg[\frac{\expo^{4\pi M\omega-i\omega\Delta t}Y_{\ell m}(\theta,\phi)Y^{*}_{\ell m}(\theta',\phi')\Phup(r)\Phup^{*}(r')}{8\pi\omega\sinh{\left(4\pi M\omega\right)}}\\
&+\frac{\expo^{-4\pi M\omega+i\omega\Delta t}Y^{*}_{\ell m}(\theta,\phi)Y_{\ell m}(\theta',\phi')\Phup^{*}(r)\Phup(r')}{8\pi\omega\sinh{\left(4\pi M\omega\right)}}\\
&+\frac{\expo^{-i\omega\Delta t}Y_{\ell m}(\theta,\phi)Y^{*}_{\ell m}(\theta',\phi')\Phin(r)\Phin^{*}(r')}{4\pi\omega}\Bigg]\,,\\
\label{eq:Schw:4d:UnruhWightman}
\end{aligned}
\end{equation}
with $\Delta t:=t-t'$.
\par We specialise to a static detector: $r=r'=R$, $\Delta t=\Delta\tau/\sqrt{1-2M/R}$, and we can take $\theta=\theta'=\phi=\phi'=0$ without loss of generality. On this trajectory, the Wightman function reduces to
\begin{equation}
\begin{aligned}
&W(\x,\x')=\sum^{\infty}_{\ell=0}\int^{\infty}_0\,\mathrm{d}\omega\,|Y_{\ell 0}(0,0)|^2\times\\
& \Bigg[\frac{|\Phup(R)|^2}{8\pi\omega\sinh{\left(4\pi M \omega\right)}}\left(\expo^{4\pi\omega-i\omega\Delta\tau/\sqrt{1-2M/R}}+\expo^{-4\pi\omega+i\omega\Delta\tau/\sqrt{1-2M/R}}\right)\\
&\,\,\,\,\,\,\,\quad\quad+\frac{|\Phin(R)|^2\expo^{-i\omega\Delta\tau/\sqrt{1-2M/R}}}{4\pi\omega}\Bigg]\,,
\label{eq:Schw:4d:UnruhStaticW}
\end{aligned}
\end{equation}
where again (14.30.4) in~\cite{dlmf} has been used.
We substitute the Wightman function~\eqref{eq:Schw:4d:UnruhStaticW} into transition rate~\eqref{eq:techIntro:transRate-stat}, and after commuting the $\omega$- and $s$-integrals, we obtain
\begin{equation}
\begin{aligned}
&\mathcal{\dot{F}}\left(E\right)=\int^{\infty}_0\,\mathrm{d}\omega\,\sum^{\infty}_{l=0}|Y_{\ell 0}(0,0)|^2\times\\
&\Bigg[\frac{|\Phup(R)|^2}{8\pi\omega\sinh{\left(4\pi M \omega\right)}}\left(\expo^{4\pi\omega}\int^{\infty}_{-\infty}\,\mathrm{d}s\,\expo^{-is\left(E+\omega/\sqrt{1-\frac{2M}{R}}\right)}+\expo^{-4\pi\omega}\int^{\infty}_{-\infty}\,\mathrm{d}s\,\expo^{-is\left(E-\omega/\sqrt{1-\frac{2M}{R}}\right)}\right)\\
&+\frac{|\Phin(R)|^2}{4\pi\omega}\int^{\infty}_{-\infty}\,\mathrm{d}s\,\expo^{-is\left(E+\omega/\sqrt{1-2M/R}\right)}\Bigg]\, .
\end{aligned}
\end{equation}
The $s$-integrals can be done analytically, as in the Hartle-Hawking and Boulware vacua static calculations, and the result for the transition rate is
\begin{equation}
\mathcal{\dot{F}}\left(E\right)=\sum^{\infty}_{l=0}|Y_{\ell 0}(0,0)|^2\Bigg[\frac{|\Phi^{\text{up}}_{\tilde{\omega}\ell}(R)|^2}{2E\left(\expo^{E/T_{\text{loc}}}-1\right)}-\frac{|\Phi^{\text{in}}_{\tilde{\omega}\ell}(R)|^2}{2E}\Theta(-E)
\Bigg]\,,
\label{eq:Schw:4d:staticUnruh_TR_result}
\end{equation}
where $\tilde{\omega}:=E\sqrt{1-2M/R}$ and $T_{\text{loc}}:=T_0/\sqrt{1-2M/R}$, with $T_0:=\kappa/2\pi$.
\section[Circular geodesic]{Transition rate for detector on a circular geodesic}
\label{sec:4DSchw:circ}
In this section, we investigate the transition rate of a detector orbiting the Schwarzschild black hole on a circular geodesic. Explicitly, the detector trajectory to be considered is
\begin{equation}
r=R\,,\quad\theta=\frac{\pi}{2}\,,\quad t=a\tau\,,\quad \phi = b\tau\,,
\label{eq:Schw:4d:circ:traj}
\end{equation}
where $R>3M$ and 
\begin{equation}
\begin{aligned}
&a:=\sqrt{\frac{R}{R-3M}}\,,\\
&b:=\sqrt{\frac{M}{R^2(R-3M)}}\,.
\end{aligned}
\label{eq:Schw:4d:circ:traj2}
\end{equation}
\subsection[Hartle-Hawking vacuum]{Transition rate of a detector in the Hartle-Hawking vacuum on a circular geodesic}
We first obtain the Wightman function for a detector on a circular geodesic in the Hartle-Hawking vacuum by substituting~\eqref{eq:Schw:4d:circ:traj} into \eqref{eq:Schw:4d:HHWightman} and expanding the spherical harmonics. We obtain
\begin{equation}
\begin{aligned}
&W(\x,\x')=\sum^{\infty}_{\ell=0}\sum^{+\ell}_{m=-\ell}\int^{\infty}_0\,\mathrm{d}\omega\, \frac{(\ell-m)!(2\ell+1)|P^m_{\ell}(0)|^2}{32\pi^2\omega(l+m)!\sinh{\left(4\pi M\omega\right)}}\times\\
&\left(|\Phup(R)|^2+|\Phin(R)|^2\right)\left[\expo^{4\pi M\omega-ia \omega s+imbs}+\expo^{-4\pi M\omega+i a\omega s-imbs}\right]\,,
\label{eq:Schw:4d:circ:W}
\end{aligned}
\end{equation}
where $s=\Delta\tau$.
Additionally, one can use (14.7.17) of~\cite{dlmf} to see that the contribution to the Wightman function will vanish unless $\ell+m$ is even. This means that for a given $\ell$ we can set $m \equiv \ell \,(\text{mod} 2)$. 
We use~\eqref{eq:Schw:4d:circ:W} in~\eqref{eq:techIntro:transRate-stat}, and as in the static section, we can evaluate the $s$-integral analytically. The resulting expression reads
\begin{equation}
\begin{aligned}
&\mathcal{\dot{F}}\left(E\right)=\sum^{\infty}_{\ell=0}\sum^{\ell}_{m=-\ell}\int^{\infty}_0\,\mathrm{d}\omega\, \frac{(\ell-m)!(2\ell+1)|P^m_{\ell}(0)|^2}{16\pi\omega(l+m)!\sinh{\left(4\pi M\omega\right)}}\times\\
&\left(|\Phup(R)|^2+|\Phin(R)|^2\right)\left[\expo^{4\pi M\omega}\delta\left(E+a\omega-mb\right)+\expo^{-4\pi M\omega}\delta\left(E-a\omega+mb\right)\right]\,.
\label{eq:Schw:4d:circ:HHrate1}
\end{aligned}
\end{equation}
Evaluating the integral over $\omega$, we finally obtain
\begin{equation}
\begin{aligned}
&\mathcal{\dot{F}}\left(E\right)=\sum^{\infty}_{\ell=0}\sum^{+\ell}_{m=-\ell} \frac{(\ell-m)!(2\ell+1)|P^m_{\ell}(0)|^2}{16\pi(l+m)!}\times\\
&\Bigg[\frac{\left(|\Phi^{\text{up}}_{\omega_{-}\ell}(R)|^2+|\Phi^{\text{in}}_{\omega_{-}\ell}(R)|^2\right)\expo^{4\pi M\omega_-}}{a\omega_{-}\sinh{\left(4\pi M\omega_{-}\right)}}\Theta(mb-E)\\
&\quad\quad\quad\quad+\frac{\left(|\Phi^{\text{up}}_{\omega_{+}\ell}(R)|^2+|\Phi^{\text{in}}_{\omega_{+}\ell}(R)|^2\right)\expo^{-4\pi M\omega_+}}{a\omega_{+}\sinh{\left(4\pi M\omega_{+}\right)}}\Theta(mb+E)\Bigg]\, ,
\label{eq:Schw:4d:circ:HHrateRes}
\end{aligned}
\end{equation}
with
\begin{equation}
\begin{aligned}
\omega_{\pm}:=\frac{mb\pm E}{a}\,.
\end{aligned}
\end{equation}
\subsection[Boulware vacuum]{Transition rate for detector on a circular geodesic in the Boulware vacuum}
We start by substituting~\eqref{eq:Schw:4d:circ:traj} into \eqref{eq:Schw:4d:preBoulwareStaticW}, and we expand the spherical harmonics. The Wightman function then reads
\begin{equation}
\begin{aligned}
W(\x,\x')=\sum^{\infty}_{\ell=0}\sum^{\ell}_{m=-\ell}\int^{\infty}_0\,\mathrm{d}\omega\,\frac{(\ell-m)!(2\ell+1)|P^{m}_{\ell}(0)|^2}{16\pi^2\omega(\ell+m)!}&\expo^{i mb\Delta\tau-ia\omega\Delta\tau}\times\\
&\left(|\Phup(R)|^2+|\Phin(R)|^2\right)\,.
\end{aligned}
\end{equation}
We substitute this Wightman function into~\eqref{eq:techIntro:transRate-stat}, and we evaluate the $s$-integral analytically. The resulting expression for the transition rate is
\begin{equation}
\begin{aligned}
\mathcal{\dot{F}}\left(E\right)=\sum^{\infty}_{\ell=0}\sum^{\ell}_{m=-\ell}\int^{\infty}_0\,\mathrm{d}\omega\,\frac{(\ell-m)!(2\ell+1)|P^{m}_{\ell}(0)|^2}{8\pi\omega(\ell+m)!}&\left(|\Phup(R)|^2+|\Phin(R)|^2\right)\times\\
&\delta\left(a\omega-(mb-E)\right)\,.
\end{aligned}
\end{equation}
Evaluating the $\omega$-integral yields
\begin{equation}
\begin{aligned}
\mathcal{\dot{F}}\left(E\right)=\frac{1}{a}\sum^{\infty}_{\ell=0}\sum^{\ell}_{m=-\ell}\frac{(\ell-m)!(2\ell+1)|P^{m}_{\ell}(0)|^2}{8\pi\omega_{-}(\ell+m)!}&\left(|\Phi^{\text{up}}_{\omega_{-}\ell}(R)|^2+|\Phi^{\text{in}}_{\omega_{-}\ell}(R)|^2\right)\times\\
&\Theta\left(mb-E\right)\,,
\label{eq:Schw:4d:circ:BoulwarerateRes}
\end{aligned}
\end{equation}
with
\begin{equation}
\omega_{-}:=\frac{mb- E}{a}\,.
\end{equation}
\subsection[Unruh vacuum]{Transition rate for detector on a circular geodesic in the Unruh vacuum}
This time we substitute~\eqref{eq:Schw:4d:circ:traj} into \eqref{eq:Schw:4d:UnruhWightman}, and we expand the spherical harmonics. The Wightman function then reads
\begin{equation}
\begin{aligned}
&W(\x,\x')=\sum^{\infty}_{\ell=0}\sum^{\ell}_{m=-\ell}\int^{\infty}_0\,\mathrm{d}\omega\,\frac{(\ell-m)!(2\ell+1)|P^{m}_{\ell}(0)|^2}{16\pi^2(\ell+m)!}\times\\
&\Bigg[\frac{|\Phup(R)|^2\left(\expo^{4\pi M\omega-ia\omega\Delta\tau+imb\Delta\tau}+\expo^{-4\pi M\omega+ia\omega\Delta\tau-imb\Delta\tau}\right)}{2\omega\sinh{\left(4\pi M\omega\right)}}+\frac{|\Phin(R)|^2\expo^{-ia\omega\Delta\tau+imb\Delta\tau}}{\omega}\Bigg]\,.
\end{aligned}
\end{equation}
Substituting this Wightman function into~\eqref{eq:techIntro:transRate-stat} and evaluating the $s$-integral analytically, the transition rate is
\begin{equation}
\begin{aligned}
&\mathcal{\dot{F}}\left(E\right)=\sum^{\infty}_{\ell=0}\sum^{\ell}_{m=-\ell}\int^{\infty}_0\,\mathrm{d}\omega\,\frac{(\ell-m)!(2\ell+1)|P^{m}_{\ell}(0)|^2}{8\pi(\ell+m)!}\times\\
&\Bigg[\frac{|\Phup(R)|^2}{2\omega\sinh{\left(4\pi M\omega\right)}}\left(\expo^{4\pi M\omega}\delta\left(E+a\omega-mb\right)+\expo^{-4\pi M\omega}\delta\left(E-a\omega+mb\right)\right)\\
&\quad\quad\quad+\frac{|\Phin(R)|^2}{\omega}\delta\left(E+a\omega-mb\right)\Bigg]\,.
\end{aligned}
\end{equation}
Evaluating the $\omega$-integral yields
\begin{equation}
\begin{aligned}
&\mathcal{\dot{F}}\left(E\right)=\frac{1}{a}\sum^{\infty}_{\ell=0}\sum^{\ell}_{m=-\ell}\frac{(\ell-m)!(2\ell+1)|P^{m}_{\ell}(0)|^2}{8\pi(\ell+m)!}\times\\
&\Bigg[\left(\frac{|\Phi^{\text{up}}_{\omega_{-}\ell}(R)|^2}{2\omega_{-}\sinh{\left(4\pi M\omega_{-}\right)}}\expo^{4\pi M\omega_{-}}+\frac{|\Phi^{\text{in}}_{\omega_{-}\ell}(R)|^2}{\omega_{-}}\right)\Theta(mb-E)\\
&\quad\quad\quad\quad\quad\quad\quad\quad+\frac{|\Phi^{\text{up}}_{\omega_{+}\ell}(R)|^2}{2\omega_{+}\sinh{\left(4\pi M\omega_{+}\right)}}\expo^{-4\pi M\omega_{+}}\Theta(mb+E)\Bigg]\,,
\label{eq:Schw:4d:circ:UnruhrateRes}
\end{aligned}
\end{equation}
with
\begin{equation}
\omega_{\pm}:=\frac{mb\pm E}{a}\,.
\end{equation}
\section{Comparison with a Rindler observer}
The analogy between the right-hand Rindler wedge and the exterior Schwarzschild spacetime is well known~\cite{byd}. It seems a natural question to ask if the experience of the static detector, which we have described in the previous sections, is related to the experience of a detector in Rindler spacetime on a Rindler trajectory. Similarly, we ask if the experience of a detector on a circular geodesic in Schwarzschild spacetime is related to that of a detector on a Rindler trajectory but given some boost in the transverse direction.
\par The Rindler observer's trajectory in $(3+1)$-dimensional Minkowski spacetime is specified  by
\begin{equation}
\x(\tau)=\frac{1}{a}\left(\sinh{\left(a\tau\right)},\cosh{\left(a\tau\right)},L,0\right)\,,
\label{eq:Schw:4d:circ:rindTraj}
\end{equation}
where $a$ is the proper acceleration and $L$ is a real constant. The transition rate for a detector on such a trajectory is given by~\cite{byd}
\begin{equation}
\begin{aligned}
\mathcal{\dot{F}}\left(E\right)=\frac{E}{2\pi}\frac{1}{\expo^{E/T}-1}\,,
\label{eq:Schw:4d:circ:rindRate}
\end{aligned}
\end{equation}
where $T:=a/2\pi$ is the temperature. Recalling that the local temperature we found for the static detector in Schwarzschild is $T_{\text{loc}}:=\kappa/(2\pi\sqrt{1-2M/R})$, this suggests that the transition rate for the static detector at radius $R$ should be compared with the transition rate of a Rindler detector with proper acceleration
\begin{equation}
a=\frac{\kappa}{\sqrt{1-\frac{2M}{R}}}=\frac{1}{4M\sqrt{1-\frac{2M}{R}}}\,.
\label{eq:Schw:4d:circ:propAccel}
\end{equation}
The results of the comparison with the static detector at radius $R$ in Schwarzschild and the Rindler detector with proper acceleration~\eqref{eq:Schw:4d:circ:propAccel} will be examined in Section~\ref{sec:4DSchw:results:stat}.
\par Next, consider the Rindler observer with proper acceleration $a$ but with constant drift-velocity in the transverse $y$-direction:
\begin{equation}
\x(\tau')_{\text{drift}}=\frac{1}{a}\left(\sinh{\left(q\tau'\right)},\cosh{\left(q\tau'\right)},p\tau',0\right)\,,
\label{eq:Schw:4d:circ:driftTraj}
\end{equation}
where $q$ and $p$ are real constants. Note that in order for the four-velocity to be correctly normalised, we require that
\begin{equation}
a^2=q^2-p^2\,.
\label{eq:Schw:4d:circ:paqRel}
\end{equation}
\par In Schwarzschild spacetime, the static detector has four-velocity given by
\be 
\U_{\text{static}}=\left(\sqrt{\frac{R}{R-2M}},0,0,0\right)\, ,
\ee
and the circular-geodesic trajectory, specified by~\eqref{eq:Schw:4d:circ:traj} and~\eqref{eq:Schw:4d:circ:traj2}, has four-velocity 
\be
\U_{\text{circ}}=\left(\sqrt{\frac{R}{R-3M}},0,0,\sqrt{\frac{M}{R^2(R-3M)}}\right)\, .
\ee
It follows that
\begin{equation}
\U_{\text{circ}}\cdot\U_{\text{static}}=-\sqrt{\frac{R-2M}{R-3M}}\,.
\label{eq:Schw:4d:circ:dotProd4velSchw}
\end{equation}
\par We want to compare the Rindler detector with transverse drift to the circular-geodesic detector in Schwarzschild. In order to make this comparison, we demand that the four-velocity inner product $\U_{\text{drift}}\cdot\U_{\text{Rindler}}$ matches that of~\eqref{eq:Schw:4d:circ:dotProd4velSchw}, where $\U_{\text{Rindler}}$ and $\U_{\text{drift}}$ are the four-velocity of the Rindler detector and Rindler detector with drift in the transverse direction respectively. Taking this four velocity dot product must be done when the Rindler and Rindler plus drift observers are at the same spacetime point. Comparison of~\eqref{eq:Schw:4d:circ:rindTraj} and~\eqref{eq:Schw:4d:circ:driftTraj} shows that in order to be at the same point we must demand $a\tau=q\tau'$ and $\tau'=L/p$.
This means that at this spacetime point
\begin{equation}
\begin{aligned}
&\U_{\text{Rindler}}=\left(\cosh{\left(qL/p\right)},\sinh{\left(qL/p\right)},0,0\right)\,,\\
&\U_{\text{drift}}=\left(\frac{q}{a}\cosh{\left(qL/p\right)},\frac{q}{a}\sinh{\left(qL/p\right)},p,0\right)\,,
\end{aligned}
\end{equation}
so that
\begin{equation}
\U_{\text{Rind}}\cdot\U_{\text{drift}}=-\frac{q}{a}\,.
\label{eq:Schw:4d:circ:dotProd4velFlat}
\end{equation}
This means that we want
\begin{equation}
q=\frac{1}{4M}\sqrt{\frac{R}{R-3M}}\,,
\label{eq:Schw:4d:circ:q}
\end{equation}
and by virtue of~\eqref{eq:Schw:4d:circ:propAccel} and~\eqref{eq:Schw:4d:circ:paqRel}, we have
\begin{equation}
p=\frac{1}{4M}\sqrt{\frac{MR}{(R-3M)(R-2M)}}\,.
\label{eq:Schw:4d:circ:p}
\end{equation}
The transition rate for the Rindler plus drift detector can now easily be computed. By~\eqref{eq:Schw:4d:circ:driftTraj} we first note that the Minkowski interval is 
\begin{equation}
\Delta\x^2=\frac{p^2}{a^2}\Delta\tau^2-\frac{4}{a^2}\sinh^2{\left(\frac{q\Delta\tau}{2}\right)}\,.
\label{eq:Schw:4d:circ:interval}
\end{equation}
This can be substituted into the transition rate found in~\cite{louko-satz:profile}, and the results of the comparison with the detector on a circular geodesic in Schwarzschild will be examined in Section~\ref{sec:4DSchw:results:circ}.
\par
We also would like to see if the comparison between the detector on a circular geodesic in Schwarzschild and the detector on a Rindler trajectory plus drift becomes better if we make the transverse direction, in which the Rindler detector is drifting, periodic. The proper-time period for the circular-geodesic detector in Schwarzschild to complete a loop is
\begin{equation}
P:=2\pi\sqrt{\frac{R^2(R-3M)}{M}}\,.
\label{eq:Schw:4d:circ:period}
\end{equation}
We wish to identify the transverse direction of Minkowski spacetime that our Rindler plus drift detector exists on by the same period in proper time. This means identifying the points
\begin{equation}
\begin{aligned}
y(\tau) &\sim y(\tau+nP)\\
&= y(\tau)+npP/a\,,
\end{aligned}
\end{equation}
where $n$ is an integer.
In order to get the transition rate of the Rindler plus drift detector on flat spacetime with periodic boundary conditions in the transverse drift direction, we employ the method of images. This results in the square interval
\begin{equation}
\Delta\x^2_n=-\frac{4}{a^2}\sinh^2{\left(\frac{q\Delta\tau}{2}\right)}+\left[\frac{p\Delta\tau}{a}+\frac{npP}{a}\right]^2\,,\quad n \in \BbbZ\,.
\label{eq:Schw:4d:circ:methImInterval}
\end{equation}
We substitute this interval into the transition rate~\eqref{eq:techIntro:transRate-stat} and perform the image sum over $n$. Because the periodicity could lead to singularities at $\Delta\tau\neq 0$, not dealt with by the Hadamard short distance form, we need the form of the transition rate with regulator intact. The exception, of course, is the $n=0$ term for which we can use the form of the transition rate found in~\cite{louko-satz:profile} with the regulator already taken to zero, see also Chapter~\ref{ch:btz}, where such singularities were also encountered and dealt with.
For the $n\neq 0$ terms, the transition rate can be written as
\begin{equation}
\begin{aligned}
&\mathcal{\dot{F}}\left(E\right)=-\frac{a^2}{2q}\sum^{\infty}_{n=-\infty}\int^{\infty}_{-\infty}\,\mathrm{d}r\, \expo^{-2iEr/q}\frac{1}{\left[\sinh^2{(r)}-(\frac{rp}{q}+\frac{npP}{2})^2\right]}\\
&=-\frac{a^2}{4q}\sum^{\infty}_{n=-\infty}\int^{\infty}_{-\infty}\, \frac{\mathrm{d}r\,\expo^{-2iEr/q}}{\left(\frac{rp}{q}+\frac{npP}{2}\right)}\left(\frac{1}{\left[\sinh{(r)}-(\frac{rp}{q}+\frac{npP}{2})\right]}-\frac{1}{\left[\sinh{(r)}+(\frac{rp}{q}+\frac{npP}{2})\right]}\right)\,,
\label{eq:Schw:4d:circ:n_not0_rate}
\end{aligned}
\end{equation}
where the $i\epsilon$ prescription amounts to giving $r$ a small, negative, imaginary part near the singularities on the real axis.
\par We evaluate~\eqref{eq:Schw:4d:circ:n_not0_rate} numerically. We first use Mathematica's `FindRoot' function to solve the transcendental equations that specify the singularities in the integrand. With the singularities known, we compute the integral in~\eqref{eq:Schw:4d:circ:n_not0_rate} by using Mathematica's `CauchyPrincipalValue' method of `NIntegrate' and adding the contribution from the small semi-circle contours that pass around the singularities in the lower half-plane. The sum is cut off at some suitable value of $|n|$ when convergence has occurred to the desired precision.
\section[Radially-infalling detector]{Radially-infalling detector in Schwarzschild}
%
%
\begin{figure}[t]\centering
\tikzset{->-/.style={decoration={
  markings,
  mark=at position #1 with {\arrow[scale=3]{>}}},postaction={decorate}}}

\begin{tikzpicture}[scale=0.8]
\node (I)    at ( 4,0)   {I};
\node (II)   at (-4,0)   {II};
\node (III)  at (0, 2.5) {III};
\node (IV)   at (0,-2.5) {IV};

\path  
  (II) +(90:4)  coordinate[label=90:$i^+$]  (IItop)
       +(-90:4) coordinate[label=-90:$i^-$] (IIbot)
       +(0:4)   coordinate                  (IIright)
       +(180:4) coordinate[label=180:$i^0$] (IIleft)
       ;
\draw (IIleft) -- 
          node[midway, above left]    {$\scri^+$}
          node[midway, below, sloped] {$\bar{u}=\infty$}
      (IItop) --
          node[midway, below, sloped] {$\bar{v}=0$}
      (IIright) -- 
          node[midway, below, sloped] {$\bar{u}=0$}
      (IIbot) --
          node[midway, above, sloped] {$\bar{v}=-\infty$}
          node[midway, below left]    {$\scri^-$}    
      (IIleft) -- cycle;

\path 
   (I) +(90:4)  coordinate[label=90:$i^+$]  (Itop)
       +(-90:4) coordinate[label=-90:$i^-$] (Ibot)
       +(180:4) coordinate (Ileft)
       +(0:4)   coordinate[label=0:$i^0$] (Iright)
       ;
\draw  (Ileft)  -- 
          node[midway, below, sloped] {$H^+$}
       (Itop)   -- 
          node[midway, above right]    {$\scri^+$}
          node[midway, below, sloped] {$\bar{v}=\infty$}
       (Iright) --
          node[midway, below right]    {$\scri^-$}
          node[midway, above, sloped] {$\bar{u}=-\infty$}
       (Ibot)   --
          node[midway, below, sloped]    {$H^-$}
       (Ileft)  -- cycle;

\draw[decorate,decoration=zigzag] (IItop) -- (Itop)
      node[midway, above, inner sep=2mm] {$r=0$};

\draw[decorate,decoration=zigzag] (IIbot) -- (Ibot)
      node[midway, below, inner sep=2mm] {$r=0$};
      
\draw[->-=.5] (Ibot) .. controls (7,0) and (4,3) .. (3.3,4);

\end{tikzpicture}
\caption{Example radial-infall trajectory shown on the conformal diagram of the Kruskal manifold.}  
\end{figure}
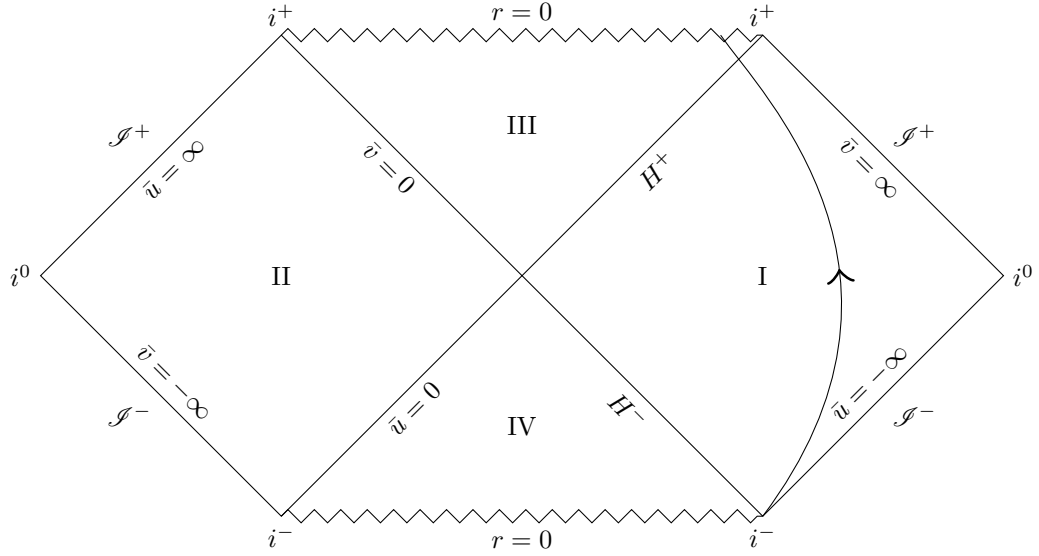
%
%
\par In this section, we shall examine a detector that falls radially on a geodesic into the Schwarzschild black hole. We are interested in the case where the detector starts at rest at $r\to\infty$ and then falls radially inward toward the hole. 
\par The equations that specify the motion are
\bea
&\frac{dt}{d\tau}=\frac{1}{1-2M/r}\,,\\
&\frac{dr}{d\tau}=-\sqrt{2M/r}\,,\\
&\theta=0\,,\\
&\phi=0\,.
\eea
Solving these, we find the trajectory is given by
\bea
&\left(\frac{r}{r_H}\right)=\left(\frac{\tau}{\tau_H}\right)^{2/3}\,,\\
&t=3\tau_H\left(\frac{\tau}{\tau_H}\right)^{1/3}+\tau-3\tau_H\arctanh{\left[\left(\frac{\tau}{\tau_H}\right)^{-1/3}\right]}\,,\\
&\theta=0\,,\\
&\phi=0\,,
\label{eq:Schw:4d:radial:trajSchw}
\eea
with
\bea
&r_H:=2M\,,\\
&\tau_H:=-\frac{4M}{3}\,.
\eea
The additive constant in $\tau$ has been chosen so that $-\infty<\tau<0$, with $\tau\to 0$ as $r\to 0$.
In terms of the Kruskal co-ordinates~\eqref{eq: Schw:2d:Kruskals}, the trajectory reads
\bea
&\bar{u}=-4M\left(\left(\frac{\tau}{\tau_H}\right)^{1/3}-1\right)\exp{\left[\left(\frac{\tau}{\tau_H}\right)^{1/3}+\frac{1}{2}\left(\frac{\tau}{\tau_H}\right)^{2/3}+\frac{1}{3}\left(\frac{\tau}{\tau_H}\right)\right]}\,,\\
&\bar{v}=4M\left(\left(\frac{\tau}{\tau_H}\right)^{1/3}+1\right)\exp{\left[-\left(\frac{\tau}{\tau_H}\right)^{1/3}+\frac{1}{2}\left(\frac{\tau}{\tau_H}\right)^{2/3}-\frac{1}{3}\left(\frac{\tau}{\tau_H}\right)\right]}\,.
\eea
\par If we start by restricting our attention to the exterior ($\bar{v}>0, \bar{u}<0$) of the black hole, we find that the modes~\eqref{eq:Schw:4d:UnruhModes} may be written in terms of the Kruskal co-ordinates as
\begin{equation}
\begin{aligned}
w^{\text{in}}_{\omega\ell m}&=\frac{\expo^{2\pi M\omega}\left[\expo^{i\omega r^*}\Phin(r)\right]Y_{\ell m}(\theta,\phi)}{\sqrt{8\pi\omega\sinh{\left(4\pi M\omega\right)}}}\left(\frac{\bar{v}}{4M}\right)^{-i4M\omega}\,,\\
\bar{w}^{\text{in}}_{\omega\ell m}&=\frac{\expo^{-2\pi M\omega}\left[\expo^{i\omega r^*}\Phin(r)\right]^{*}Y^{*}_{\ell m}(\theta,\phi)}{\sqrt{8\pi\omega\sinh{\left(4\pi M\omega\right)}}}\left(\frac{\bar{v}}{4M}\right)^{i4M\omega}\,,\\
w^{\text{up}}_{\omega\ell m}&=\frac{\expo^{2\pi M\omega}\left[\expo^{-i\omega r^*}\Phup(r)\right]Y_{\ell m}(\theta,\phi)}{\sqrt{8\pi\omega\sinh{\left(4\pi M\omega\right)}}}\left(\frac{-\bar{u}}{4M}\right)^{i4M\omega}\,,\\
\bar{w}^{\text{up}}_{\omega\ell m}&=\frac{\expo^{-2\pi M\omega}\left[\expo^{-i\omega r^*}\Phup(r)\right]^{*}Y^{*}_{\ell m}(\theta,\phi)}{\sqrt{8\pi\omega\sinh{\left(4\pi M\omega\right)}}}\left(\frac{-\bar{u}}{4M}\right)^{-i4M\omega}\, .
\label{eq:Schw:4d:radial:UnruhModesExtKrusk}
\end{aligned}
\end{equation}
Also note that by virtue of~\eqref{eq:Schw:4d:asyIn}, the combination $\expo^{i\omega r^*}\Phin(r)$ remains regular as we cross the future horizon; we shall use this fact when we come to examine trajectories that cross this horizon.
\par We work in the Hartle-Hawking vacuum state (which is regular across all the horizons) and use the modes~\eqref{eq:Schw:4d:radial:UnruhModesExtKrusk} in~\eqref{eq:Schw:4d:FieldExp}, along with the radial-infall trajectory equations, to form the Wightman function for the Hartle-Hawking vacuum state that is valid in the exterior region. This Wightman function reads
\bea
W(\x,\x')&:=\langle 0_K|\psi(\x)\phi(\x')|0_K\rangle\\
&=\sum^{\infty}_{\ell=0}\sum^{\ell}_{m=-\ell}\int^{\infty}_0\,\mathrm{d}\omega\,\Bigg[w^{\text{up}}_{\omega\ell m}(\x)w^{\text{up}*}_{\omega\ell m}(\x')+\bar{w}^{\text{up}}_{\omega\ell m}(\x)\bar{w}^{\text{up}*}_{\omega\ell m}(\x')\\
&\quad\quad\quad\quad\quad\quad+w^{\text{in}}_{\omega\ell m}(\x)w^{\text{in}*}_{\omega\ell m}(\x')+\bar{w}^{\text{in}}_{\omega\ell m}(\x)\bar{w}^{\text{in}*}_{\omega\ell m}(\x')\Bigg]\\
&=\sum^{\infty}_{\ell=0}\sum^{\ell}_{m=-\ell}\int^{\infty}_0\,\mathrm{d}\omega\,\frac{(2\ell+1)}{32\pi^2\omega\sinh{\left(4\pi M \omega\right)}}\times\\
&\Bigg[\expo^{4\pi M\omega}\left(\tPhup(r)\tPhup^{*}(r')\left(\frac{\bar{u}}{\bar{u}'}\right)^{i4M\omega}+\tPhin(r)\tPhin^{*}(r')\left(\frac{\bar{v}}{\bar{v}'}\right)^{-i4M\omega}  \right)\\
&+\expo^{-4\pi M\omega}\left(\tPhup^{*}(r)\tPhup(r')\left(\frac{\bar{u}}{\bar{u}'}\right)^{-i4M\omega}+\tPhin^{*}(r)\tPhin(r')\left(\frac{\bar{v}}{\bar{v}'}\right)^{i4M\omega}  \right)\Bigg]\,,
\label{eq:Schw:4d:radial:radialW}
\eea
where $\tPhup(r):=\expo^{-i\omega r^{*}}\Phup(r)$ and $\tPhin(r):=\expo^{i\omega r^{*}}\Phin(r)$.
\subsection{Continuing the modes across the horizon}
The in-modes take a simple form near the horizon, and we can easily continue them across the horizon. The up-modes, on the other hand, are a linear combination of ingoing and outgoing modes near the horizon. We can use~\eqref{eq:Schw:4d:asyIn} and~\eqref{eq:Schw:4d:asyUp} in order to write the up-modes in terms of the in-modes, which we know how to continue across the horizon; therefore, we can also continue the up-modes through to the interior of the black hole. Recalling the normalisation factors~\eqref{eq:Schw:4d:NinNormConst} and~\eqref{eq:Schw:4d:NupNormConst}, the relation between the normalised up-modes and in-modes is
\be 
\Phup(r)=\frac{1}{2M}\left[\frac{A^{\text{up}}_{\omega\ell}}{B^{\text{in}}_{\omega\ell}}\Phin(r)+\frac{1}{B^{\text{in}*}_{\omega\ell}}\Phin^{*}(r)\right]\,.
\ee
As mentioned earlier, the combination $\tPhin:=\expo^{i\omega r^{*}}\Phin$ is regular through the horizon; hence, for example, the up-mode in the exterior region can be written as
\bea
w^{\text{up}}_{\omega\ell m}=\frac{\expo^{2\pi M\omega}Y_{\ell m}(\theta,\phi)}{2M\sqrt{8\pi  \omega\sinh{(4\pi M\omega)}}}&\Bigg[\frac{\Aup}{\Bin}\tPhin \left(\frac{\bar{v}}{4M}\right)^{-i4M\omega}\\
&\quad\quad\quad\quad+\frac{\tPhin^{*}}{B^{\text{in}*}_{\omega\ell}}\left(\frac{-\bar{u}}{4M}\right)^{i4M\omega}\Bigg]\,, \quad \bar{u}<0\,,
\eea
and we analytically continue this into the interior via the lower half of the complex $\bar{u}$ plane. We obtain
\bea
w^{\text{up}}_{\omega\ell m}=\frac{Y_{\ell m}(\theta,\phi)}{2M\sqrt{8\pi  \omega\sinh{(4\pi M\omega)}}}&\Bigg[\expo^{2\pi M\omega}\frac{\Aup}{\Bin}\tPhin \left(\frac{\bar{v}}{4M}\right)^{-i4M\omega}\\
&\quad\quad\quad\quad+\expo^{-2\pi M\omega}\frac{\tPhin^{*}}{B^{\text{in}*}_{\omega\ell}}\left(\frac{\bar{u}}{4M}\right)^{i4M\omega}\Bigg]\,, \quad \bar{u}>0\,.
\eea
Similarly, we analytically continue the $\bar{w}^{\text{up}}$ mode:
\bea
\bar{w}^{\text{up}}_{\omega\ell m}=\frac{\expo^{-2\pi M\omega}Y^{*}_{\ell m}(\theta,\phi)}{2M\sqrt{8\pi  \omega\sinh{(4\pi M\omega)}}}&\Bigg[\frac{A^{\text{up}*}_{\omega\ell}}{B^{\text{in}*}_{\omega\ell}}\tPhin^{*} \left(\frac{\bar{v}}{4M}\right)^{i4M\omega}\\
&\quad\quad\quad\quad+\frac{\tPhin}{\Bin}\left(\frac{-\bar{u}}{4M}\right)^{-i4M\omega}\Bigg]\,, \quad \bar{u}<0\,,
\eea
into
\bea
\bar{w}^{\text{up}}_{\omega\ell m}=\frac{Y^{*}_{\ell m}(\theta,\phi)}{2M\sqrt{8\pi  \omega\sinh{(4\pi M\omega)}}}&\Bigg[\expo^{-2\pi M\omega}\frac{A^{\text{up}*}_{\omega\ell}}{B^{\text{in}*}_{\omega\ell}}\tPhin^{*} \left(\frac{\bar{v}}{4M}\right)^{i4M\omega}\\
&\quad\quad\quad\quad+\expo^{+2\pi M\omega}\frac{\tPhin}{\Bin}\left(\frac{\bar{u}}{4M}\right)^{-i4M\omega}\Bigg]\,, \quad \bar{u}>0\,.
\eea
The in-modes are simply 
\bea
w^{\text{in}}_{\omega\ell m}&=\frac{\expo^{2\pi M\omega}\tPhin(r)Y_{\ell m}(\theta,\phi)}{\sqrt{8\pi\omega\sinh{\left(4\pi M\omega\right)}}}\left(\frac{\bar{v}}{4M}\right)^{-i4M\omega}\,,\\
\bar{w}^{\text{in}}_{\omega\ell m}&=\frac{\expo^{-2\pi M\omega}\tPhin^{*}(r)Y^{*}_{\ell m}(\theta,\phi)}{\sqrt{8\pi\omega\sinh{\left(4\pi M\omega\right)}}}\left(\frac{\bar{v}}{4M}\right)^{i4M\omega}\,, 
\label{eq:Schw:4d:radial:inModes}
\eea
valid for $\bar{u}\in \BbbR$.
Numerically, these modes and their continuations can be used to form the Wightman function by using Mathematica's pattern constraints to test whether $\x$ and $\x'$ are in the interior or exterior of the hole and then to choose the appropriate form of the modes to substitute into the first equality of~\eqref{eq:Schw:4d:radial:radialW}, which is valid in both the interior and exterior regions of the hole.
\subsection{Evaluation of the transition rate}
In four-dimensional curved spacetime, the transition rate takes the form~\cite{satz-louko:curved}
\be
\mathcal{\dot{F}}_{\tau_f}\left(E\right) =-\frac{E}{4\pi}+2\int^{\Delta\tau}_0\,\mathrm{d}s\,\Realpart\left[\expo^{-iEs}W_0(\tau_f,\tau_f-s)+\frac{1}{4\pi^2 s^2}\right]+\frac{1}{2\pi^2 \Delta\tau}\,,
\label{eq:Schw:4d:radial:satzCurvedTR}
\ee
where $\tau_f$ and $\tau_i$ are respectively the switch-off and switch-on times of the detector, $\Delta\tau:=\tau_f-\tau_i$ and $W_0$ is the Wightman function where the regulator has been taken point-wise to zero. 
\par Computationally, it will be most efficient if we can commute the $s$-integral appearing in the transition rate~\eqref{eq:Schw:4d:radial:satzCurvedTR} with the $\omega$-integral and $\ell$-sum appearing in the Wightman function. This way we can use Mathematica's `NDSolve' function to solve the ordinary differential equation~\eqref{eq:Schw:4d:radModPhi} for a given $(\omega,\ell)$ only once, and then compute the $s$-integral over the entire range of flight for this $(\omega,\ell)$. However, if we  substitute~\eqref{eq:Schw:4d:radial:radialW} into~\eqref{eq:Schw:4d:radial:satzCurvedTR} and attempt, na\"{\i}vely, to commute the integral order, we face  potential issues at small $s$; the $1/4\pi^2 s^2$ term currently cancels the singularity arising at small $s$, but if we switch the $s$-integral and $\omega$-integral in the $W_0$ term, this cancellation will no longer occur.
\par We shall explain how to deal with these small-$s$ issues in Sections~\ref{sec:Schw:4d:rad:smallsExt} and~\ref{sec:Schw:4d:rad:smallsInt}. Once we have dealt with these issues, we fix the detector's excitation energy $E$, and we fix the initial radius of the detector --- effectively by fixing the switch-on proper time, $\tau_i$. We then evaluate the transition rate by using `NDSolve' to solve the ordinary differential equation~\eqref{eq:Schw:4d:radModPhi} for each given $(\omega,\ell)$ over the desired range of radial flight. We take $\omega$ and the proper time at which the detector is switched off, $\tau_f$, to be on a grid, for example, $\omega=1/10,~2/10,~\ldots,~\omega_\text{cutoff}$ and $\tau_f=-80,~-79,~\ldots,~-40$. As $\ell$ becomes larger, the contributions these modes make becomes increasingly negligible, and we only gather data values in $\ell$ up to an $\omega$-dependent upper-limit (as $\omega$ increases this cut-off must be at larger and larger values of $\ell$). For each $\omega,~\ell,~\tau_f$ point, we then numerically perform the $s$-integral using `NIntegrate'. With this data gathered, we next numerically sum over $\ell$ for each $\omega,~\tau_f$ point on the grid. With the $\ell$-sum complete, we use Mathematica to interpolate the resulting integrand to produce a function of $\omega$ for each $\tau_f$ point. Finally, we use `NIntegrate' to numerically evaluate the $\omega$-integral. The result is the transition rate as a function of the switch-off time, $\tau_f$.

\subsection[Small-$s$ divergence in exterior]{Method to deal with small-$s$ divergence in exterior}
\label{sec:Schw:4d:rad:smallsExt}
\par To deal with these small-$s$ issues, we note that it is possible to write the $1/s^2$ factor as the Wightman function of a massless scalar field in 3+1 Minkowski spacetime, which can be expressed as a mode sum in spherical co-ordinates, with an integral over $\omega$ and a sum over $\ell,~m$. The flat spacetime $s=0$ divergence and curved spacetime $s=0$ divergence match mode by mode. Written in terms of spherical co-ordinates, the Wightman function of the massless scalar field in 3+1 Minkowski spacetime  reads
\be
W_M(\x,\x')=\int^{\infty}_0\sum^{\infty}_{\ell=0}\,\mathrm{d}\omega\, \frac{(2\ell+1)\omega}{4\pi^2}j_{\ell}(R\omega)j_{\ell}(R'\omega) \expo^{-i\omega\Delta T}\,,
\label{eq:Schw:4d:radial:mig}
\ee
where the $j_{\ell}$ are spherical Bessel functions, $T$ and $R$ are the Minkowski co-ordinates and we have used the fact that we are on a radial-infall trajectory to eliminate the $m$ dependence. The Wightman function~\eqref{eq:Schw:4d:radial:mig} can be shown to be equal to (see Chapter~\ref{ch:by4d}) the alternative form:
\be
W_M(\x,\x')=\frac{1}{4\pi^2}\frac{1}{-\Delta T^2+\Delta R^2}\,.
\label{eq:Schw:4d:radial:migAlt}
\ee
We shall use this fact to replace the $1/s^2$ term. First, we note that $s>0$ and we are on a timelike trajectory; we would like to choose $T,~T',~R,~R'$ as functions of $s$ such that
\bea 
\frac{1}{4\pi^2 s^2}&=\frac{1}{4\pi^2\left[\Delta T^2-\Delta R^2\right]}\\
&=-\frac{1}{4\pi^2\left[-\Delta T^2+\Delta R^2\right]}\\
&=-W_M(T,R; T',R')\, ,
\label{eq:Schw:4d:radial:replacing1over4pi2s2}
\eea
where here the Minkowski co-ordinates, $T$ and $R$, are some yet to be determined functions of $s$. Thus, we demand that
\be 
\Delta T=\sqrt{s^2+\Delta R^2}\,.
\label{eq:Schw:4d:radial:constraint}
\ee
\par The relation~\eqref{eq:Schw:4d:radial:constraint} does not uniquely specify $T,~T',~R,~R'$ as functions of $s$; that is, we have some choice in their form. We find the appropriate functions of $s$ by comparing the large-$\omega$ asymptotics of the curved spacetime Wightman function with those of the Minkowski spacetime Wightman function. Consider first the large-$\omega$ asymptotics of the Minkowski spacetime Wightman function~\eqref{eq:Schw:4d:radial:mig}. Using the relation (10.47.3) and the large-argument asymptotic expansion (10.17.3) of~\cite{dlmf}, we find that at large $\omega$, to leading order and for a given $\ell$, the summand in the mode sum of the Wightman function~\eqref{eq:Schw:4d:radial:mig} has the asymptotic form
\be
\int^{\infty}_0\,\mathrm{d}\omega\,\frac{(2\ell+1)\expo^{-i\omega\Delta T}}{8\pi^2\omega R R'}
\begin{cases}
&2\cos{(R\omega)}\cos{(R'\omega)}\,,\qquad\ell~\text{odd}\\
&2\sin{(R\omega)}\sin{(R'\omega)}\,,\qquad \ell~\text{even}\,.
\end{cases}
\label{eq:Schw:4d:radial:migAsy}
\ee
\par To get the large-$\omega$ asymptotics of the curved spacetime Wightman function in the exterior~\eqref{eq:Schw:4d:radial:radialW}, we first note that the radial equation written in terms of the tortoise co-ordinate~\eqref{eq:Schw:4d:radModTort}, along with~\eqref{eq:Schw:4d:relnAdrianModesNormModes}, shows us that as $\omega\to\infty$ then
\bea
\Phup(r)&\to\frac{\Bup}{2M}\frac{\expo^{i\omega r*}}{r}\,,\\
\Phin(r)&\to\frac{\Bup}{2M}\frac{\expo^{-i\omega r*}}{r}\,.
\eea
As $\omega \to \infty$, these high energy waves can penetrate the gravitational potential, and we see from~\eqref{eq:Schw:4d:asyIn} and~\eqref{eq:Schw:4d:asyUp} that this means $\Bin\to 1/2M$. Thus, by the Wronskian relations~\eqref{eq:Schw:4d:Wronsk}, $\Bup/2M \to 1$ as $\omega \to \infty$. Hence, we can write the asymptotic form of the summand, for fixed $\ell$, in the Wightman function in the exterior region as
\bea
&\int^{\infty}_0\,\mathrm{d}\omega\,\frac{(2\ell+1)}{16rr'\pi^2\omega}
\Bigg[\left(\frac{\bar{u}}{\bar{u}'}\right)^{i4M\omega}+\left(\frac{\bar{v}}{\bar{v}'}\right)^{-i4M\omega}\Bigg]\\
&=\int^{\infty}_0\,\mathrm{d}\omega\,\frac{(2\ell+1)\expo^{-i\omega\Delta t}}{8\pi^2\omega rr'}
\cos{\left(\omega\Delta r*\right)}\,.
\label{eq:Schw:4d:radial:radialWasy}
\eea
\par Subject to the constraint~\eqref{eq:Schw:4d:radial:constraint}, we want to choose $T,T',R,R'$ as functions of $s$ such that the asymptotic forms of the Minkowski spacetime Wightman function~\eqref{eq:Schw:4d:radial:migAsy} and curved spacetime Wightman function~\eqref{eq:Schw:4d:radial:radialWasy} agree.
Considering~\eqref{eq:Schw:4d:radial:migAsy}, we see that we can write the trigonometric factor as
\be
\begin{cases}
&\cos{(\omega(R+R'))}+\cos{(\omega(R-R'))}\,,\qquad \ell~\text{odd}\\
&\cos{(\omega(R-R'))}-\cos{(\omega(R+R'))}\,,\qquad \ell~\text{even}\,.
\end{cases} 
\ee
We can neglect the $\cos{(\omega(R+R'))}$ terms because these lead to the piece of the integrand having the form $\cos{(\omega(R+R'))}/\omega$, whose integral gets suppressed as $\omega\to\infty$. On the other hand, we are interested in the small-$s$ behaviour because it is in the coincidence limit that we expect problems of divergence to arise; thus, we cannot neglect the $\cos{(\omega(R-R'))}$ terms. These go to unity in the coincidence limit, which leads to the relevant part of the integrand having the form $1/\omega$ and thus a logarithmically divergent integral as $\omega\to\infty$. Hence, for the purposes of comparison with the curved spacetime Wightman function, we can write~\eqref{eq:Schw:4d:radial:migAsy} as 
\bea
&\int^{\infty}_0\,\mathrm{d}\omega\,\frac{(2\ell+1)\expo^{-i\omega\Delta T}}{8\pi^2\omega R R'}
\cos{\left(\omega(R-R')\right)}\\
&=\int^{\infty}_0\,\mathrm{d}\omega\,\frac{(2\ell+1)\expo^{-iq\omega\Delta T}}{8\pi^2\omega R R'}
\cos{\left(q\omega(R-R')\right)}\, ,
\label{eq:Schw:4d:radial:migAsyAgain}
\eea
where to obtain the second equality, we changed variables as $\omega=q\Omega $, where $q$ is a real constant, before changing the dummy variable $\Omega$ back to $\omega$. The reason for doing this is to avoid problems with over-constraint as we shall shortly see. 
Comparing the integrand of~\eqref{eq:Schw:4d:radial:migAsyAgain} and~\eqref{eq:Schw:4d:radial:radialWasy}, we see that we must demand
\be
\frac{1}{RR'}\left[\expo^{-iq\omega(\Delta T-\Delta R)}+\expo^{-iq\omega(\Delta T+\Delta R)}\right]  =\frac{1}{rr'}\left[\expo^{-i\omega(\Delta t-\Delta r*)}+\expo^{-i\omega(\Delta t+\Delta r*)}\right] \,.
\ee
We perform this matching at small $s$. Recalling that $\Delta t$ and $\Delta r*$ are known functions of $s$ given by the trajectory equations~\eqref{eq:Schw:4d:radial:trajSchw}, we can obtain the small-$s$ expansions
\bea
\Delta t-\Delta r*&=\alpha s+O\left(s^2\right)\,,\\
\Delta t+\Delta r*&=\beta s+O\left(s^2\right)\, ,
\eea
where $\alpha$ and $\beta$ are defined by
\bea 
\alpha&:=\frac{\left(\frac{\tau_f}{\tau_H}\right)^{1/3}}{\left(\left(\frac{\tau_f}{\tau_H}\right)^{1/3}-1\right)}\,,\\
\beta&:=\frac{\left(\frac{\tau_f}{\tau_H}\right)^{1/3}}{\left(\left(\frac{\tau_f}{\tau_H}\right)^{1/3}+1\right)}\, ,
\eea
with $\tau_H:=-4M/3$ and $\tau_f$ being the switch-off time of the detector.

Thus, using the constraint~\eqref{eq:Schw:4d:radial:constraint}, we see that we must demand that to leading order
\bea
\alpha s&=q\sqrt{\Delta R^2+s^2}-q\Delta R\,,\\
\beta  s&=q\sqrt{\Delta R^2+s^2}+q\Delta R\, ,
\eea
and if we make the choice that $\Delta R=k s+O(s^2)$, where $k$ is a real constant, we then have 
\bea
\alpha &=q\left[\sqrt{1+k^2}-k\right]\,,\\
\beta  &=q\left[\sqrt{1+k^2}+k\right]\, .
\label{eq:Schw:4d:radial:alphbetakq}
\eea
Thus, we see that 
\be 
\alpha\beta=q^2\,.
\label{eq:Schw:4d:radial:alphbetaq}
\ee 
For radial-infall, the Schwarzschild metric can be written as 
\be 
\Delta\tau^2=F(\tau_f)\left(\Delta t+\Delta r*\right)\left(\Delta t-\Delta r*\right)\,,
\ee
where $\Delta\tau:=\tau_f-\tau=:s$ and $F:=1-2M/r$. By the definition of $\alpha$ and $\beta$, this leads us to conclude that
\be 
\alpha\beta=\frac{1}{F(\tau_f)}\,,
\ee 
and by comparison with~\eqref{eq:Schw:4d:radial:alphbetaq}, we find that
\be 
q=\frac{1}{\sqrt{F(\tau_f)}}=\frac{1}{\sqrt{1-2M/r(\tau_f)}}\,.
\label{eq:Schw:4d:radial:q} 
\ee 
Next, we define $k:=\sinh{(\lambda)}$ and substitute this, along with~\eqref{eq:Schw:4d:radial:q}, into~\eqref{eq:Schw:4d:radial:alphbetakq}. We find that
\be
\lambda=-\log{\left(\alpha\sqrt{F(\tau_f)}\right)}= \log{\left(\beta\sqrt{F(\tau_f)}\right)}\,.
\ee
Thus, we have the relation 
\bea
k&=\sinh{\left[\log{\left(\beta\sqrt{F(\tau_f)}\right)}\right]}\\
&=-\frac{1}{\sqrt{\left(\tau_f/\tau_H\right)^{2/3}-1}}\,
\eea
and, correspondingly,
\be
\Delta R=-\frac{s}{\sqrt{\left(\tau_f/\tau_H\right)^{2/3}-1}}+O\left(s^2\right)\,.
\label{eq:Schw:4d:radial:DR}
\ee
One choice of $R,~R'$ that would satisfy~\eqref{eq:Schw:4d:radial:DR} is
\bea
&R:=r(\tau_f)\,,\\
&R':=r(\tau_f)+\frac{s}{\sqrt{\left(\tau_f/\tau_H\right)^{2/3}-1}}\,.
\label{eq:Schw:4d:radial:RRpFin}
\eea
Using~\eqref{eq:Schw:4d:radial:RRpFin} in the constraint equation~\eqref{eq:Schw:4d:radial:constraint}, we find also that we need
\be
\Delta T=s\frac{\left(\tau_f/\tau_H\right)^{1/3}}{\sqrt{\left(\tau_f/\tau_H\right)^{2/3}-1}}\,.
\ee
The form of the Minkowski spacetime Wightman function that we must take is therefore
\bea
W_{\text{M}}(\tau_f,\tau_f-s)&=\int^{\infty}_0\,\mathrm{d}\omega\,\sum^{\infty}_{\ell=0} \frac{(2\ell+1)\omega}{4\pi^2}j_{\ell}\Big(\omega r(\tau_f)\Big)j_{\ell}\left(\omega\left(r(\tau_f)+\frac{s}{\sqrt{\left(\frac{\tau_f}{\tau_H}\right)^{2/3}-1}}\right)\right)\times\\
&\quad\quad\quad\quad\quad\quad\quad\quad\quad\quad\quad\quad\exp{\left[-i\omega s\frac{\left(\tau_f/\tau_H\right)^{1/3}}{\sqrt{\left(\tau_f/\tau_H\right)^{2/3}-1}}\right]}\,.
\label{eq:Schw:4d:radial:migWithSfuncs}
\eea
We are finally in a position to replace the $1/4\pi^2 s^2$ term of~\eqref{eq:Schw:4d:radial:satzCurvedTR} with the Minkowski Wightman function in spherical co-ordinates. Using~\eqref{eq:Schw:4d:radial:replacing1over4pi2s2}, we find
\be
\mathcal{\dot{F}}_{\tau_f}\left(E\right) =-\frac{E}{4\pi}+2\int^{\Delta\tau}_0\,\mathrm{d}s\,\Realpart\left[\expo^{-iEs}W_0(\tau_f,\tau_f-s)-W_{\text{M}}(\tau_f,\tau_f-s)\right]+\frac{1}{2\pi^2 \Delta\tau}\,,
\label{eq:Schw:4d:radial:TRwithMink}
\ee
and we replace $W_M$ with~\eqref{eq:Schw:4d:radial:migWithSfuncs}.
\subsection[Small-$s$ divergence in interior]{Method to deal with small-$s$ divergence in interior}
\label{sec:Schw:4d:rad:smallsInt}
\par The relation~\eqref{eq:Schw:4d:radial:migWithSfuncs} is valid in the exterior region of the black hole. In the interior region of the hole and for fixed $\ell$, the Wightman function has large-$\omega$ asymptotic form, which to leading order is given by
\bea
W(\x,\x')&\to\int^{\infty}_0\,\mathrm{d}\omega\,\frac{(2\ell+1)}{16rr'\pi^2\omega}
\Bigg[\left(\frac{\bar{u}}{\bar{u}'}\right)^{-i4M\omega}+\left(\frac{\bar{v}}{\bar{v}'}\right)^{-i4M\omega}\Bigg]\\
&=\int^{\infty}_0\,\mathrm{d}\omega\,\frac{(2\ell+1)}{16rr'\pi^2\omega}
\Bigg[\expo^{i\omega(\Delta t-\Delta r*)}+\expo^{-i\omega(\Delta t+\Delta r*)}\Bigg]\, .
\label{eq:Schw:4d:radial:radialWasyInt}
\eea
We mimic the procedure carried out in the exterior of the hole to determine the required functions of $s$ for $T,~T',~R,~R'$; the major difference now is that we find
\be
\alpha\beta=-q^2 \, 
\ee
and
\be 
q=\frac{1}{\sqrt{2M/r(\tau_f)-1}}\,.
\ee
Ultimately, this results in the choices
\bea
R&:=r(\tau_f)\,,\\
R'&:=r(\tau_f)+\frac{s}{\sqrt{\left(\tau_H/\tau_f\right)^{2/3}-1}}\,,\\
\Delta T&:=s\frac{\left(\tau_H/\tau_f\right)^{1/3}}{\sqrt{\left(\tau_H/\tau_f\right)^{2/3}-1}}\,,
\eea
and similarly, we use these to obtain the Minkowski spacetime Wightman function that can be used to replace the $1/4\pi^2 s^2$ term in~\eqref{eq:Schw:4d:radial:satzCurvedTR}. 
\section{Results}
\label{sec:4DSchw:results}
\subsection{Static detector}
\label{sec:4DSchw:results:stat}
First, we look at the numerical results for the transition rate of a static detector at fixed radius $R$. We use the results~\eqref{eq:Schw:4d:staticHH_TR_result},~\eqref{eq:Schw:4d:staticBoulware_TR_result} and~\eqref{eq:Schw:4d:staticUnruh_TR_result} to numerically obtain the transition rates in the Hartle-Hawking, Boulware and Unruh vacua respectively. 
\par In practice, it is the factors
\bea
|\Phup|^2&=\frac{|\Bup|^2}{4M^2}|\phup|^2\,,\\
|\Phin|^2&=\frac{|\Bup|^2}{4M^2}|\phin|^2\,
\eea
that pose a challenge to evaluate. As previously discussed, the transmission coefficient is evaluated using Wronskian methods (see Appendix~\ref{ch:appendix:4DSchwTransRefl} for more details), and the modes $\phin,\phup$ are obtained using Mathematica's `NDSolve'. 
\par We imposed a suitable cut-off in the $\ell$-sum that increased with excitation energy (through $\tilde{\omega}$) and also increased with increasing radius, $R$. Considering $R=3M$, for example, we evaluated the transition rate at the points $E=-20/100,~-19/100,~\ldots,~19/100,~20/100$, excluding the $E=0$ point. The point $E=0$ is problematic because it would involve solving for the modes at $\omega=0$, which proves difficult numerically. For $R=3M$ and $|E|=1/100$, we cut off the $\ell$-sum at $\ell=12$, whereas at $|E|=20/100$ we cut off the sum at $\ell=20$ (one could have used much lower cut-off values quite adequately here, but in the static case computation is fast and we could afford to use a larger value for the cut-off than strictly necessary). For $R=100M$, we found that at $|E|=20/100$ a cut-off of $\ell=40$ was more than adequate as these contributions had become vanishingly small.
\par A final point to note is that because the equation for the modes~\eqref{eq:Schw:4d:radModPhi} only depends on $\omega^2$ and in the static case we seek to evaluate modes at $\tilde{\omega}=E\sqrt{1-2M/R}$, the values of the modes $|\phi^{\text{in}}_{\tilde{\omega}\ell}|^2,|\phi^{\text{up}}_{\tilde{\omega}\ell}|^2$ only depends on the modulus of the detector's excitation energy, $|E|$; hence, we can just evaluate over the positive range: $E=1/100,~2/100,...,~20/100$, and then we immediately have the values of $|\phi^{\text{in}}_{\tilde{\omega}\ell}|^2,|\phi^{\text{up}}_{\tilde{\omega}\ell}|^2$ over the corresponding negative energies too.
\par
\begin{figure}[p]  
  \centering
  \includegraphics[width=0.8\textwidth]{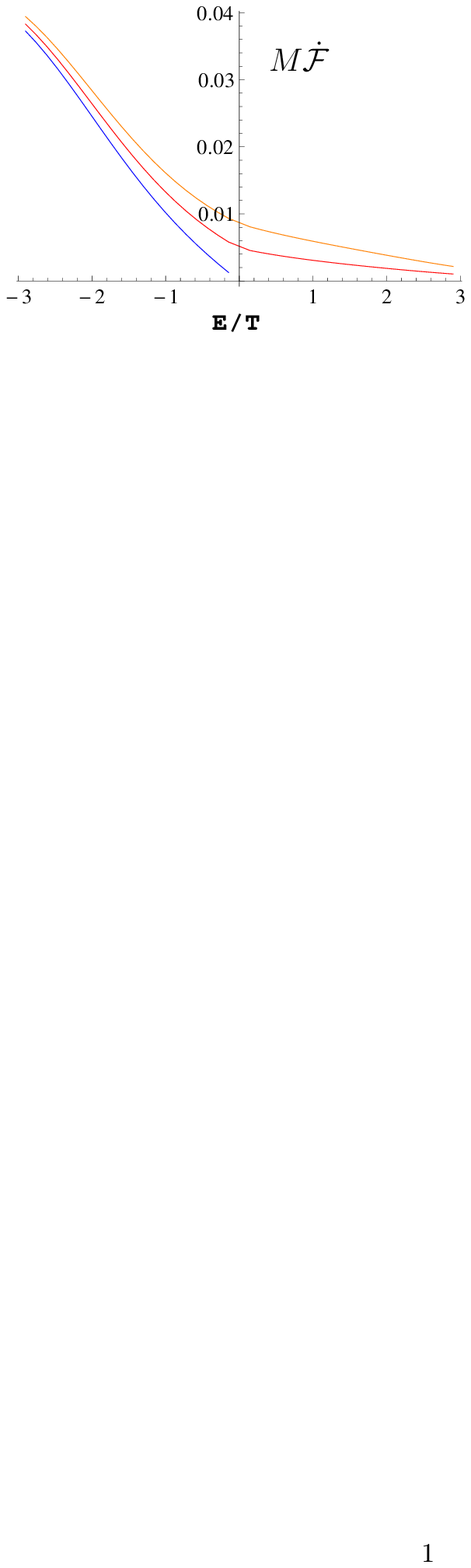}          
\caption{$M\dot{\mathcal{F}}$ as a function of $E/T_{\text{loc}}$ for the static detector at $R=3M$. Figure showing the results for the Hartle-Hawking vacuum (orange) computed from~\eqref{eq:Schw:4d:staticHH_TR_result}, Boulware vacuum (blue) computed from~\eqref{eq:Schw:4d:staticBoulware_TR_result} and the Unruh vacuum (red) computed from~\eqref{eq:Schw:4d:staticUnruh_TR_result}.}
\label{fig:static_R3M_3vacs}
\end{figure}

\begin{figure}[p]  
  \centering
  \includegraphics[width=0.8\textwidth]{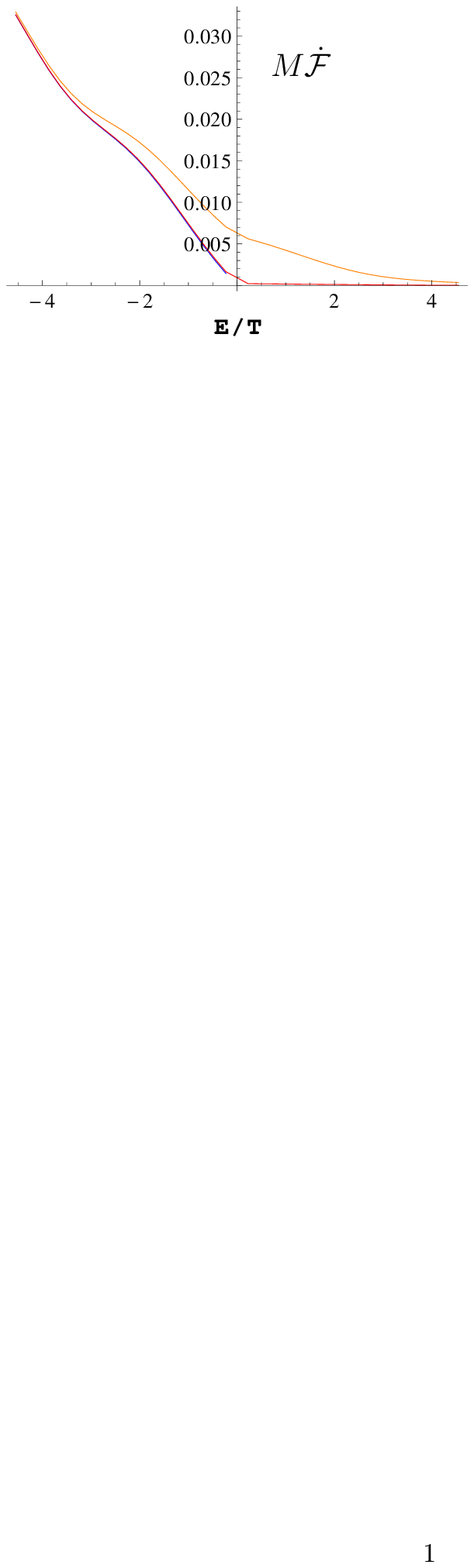}          
\caption{As in Figure~\ref{fig:static_R3M_3vacs} but with $R=11M$.}
\label{fig:static_R11M_3vacs}
\end{figure}

\begin{figure}[p]  
  \centering
  \includegraphics[width=0.8\textwidth]{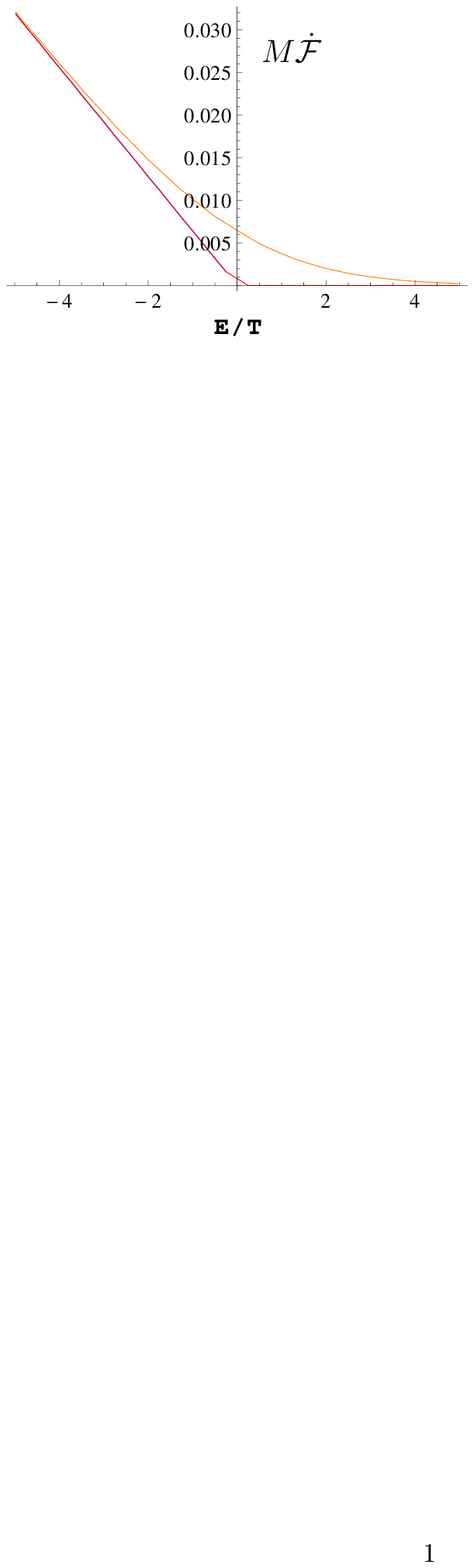}          
\caption{As in Figure~\ref{fig:static_R3M_3vacs} but with $R=100M$.}
\label{fig:static_R100M_3vacs}
\end{figure}

 Figures~\ref{fig:static_R3M_3vacs},~\ref{fig:static_R11M_3vacs} and~\ref{fig:static_R100M_3vacs} show the transition rate against the excitation energy of the detector divided by the local temperature $T_{\text{loc}}:= T_0/\sqrt{1-2M/R}$, with $T_0:=\kappa/2\pi$ and the surface gravity $\kappa=1/4M$. The horizon is at $R=2M$, and we see that as we move away from the horizon, far from the hole at $R=100M$, the transition rates for the Boulware and Unruh vacua align for negative energy gap.
\par
\begin{figure}[p]  
  \centering
  \includegraphics[width=0.8\textwidth]{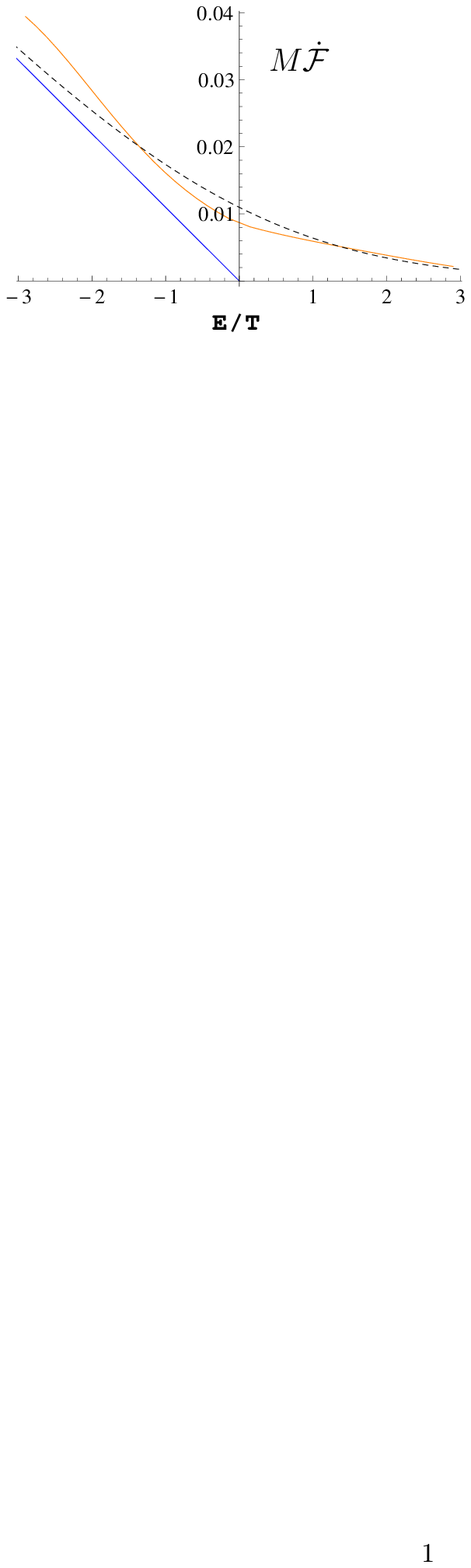}
 \caption{$M\dot{\mathcal{F}}$ as a function of $E/T_{\text{loc}}$ for the static detector at $R=3M$. Figure showing the results for the Hartle-Hawking vacuum (orange), computed from~\eqref{eq:Schw:4d:staticHH_TR_result}, alongside the response rate for an inertial detector in $3+1$ Minkowski spacetime (blue), $-\Theta\left(-E\right)E/2\pi$, and the response rate of a Rindler detector (black-dashed), computed from~\eqref{eq:Schw:4d:circ:rindRate} with a proper acceleration chosen to be~\eqref{eq:Schw:4d:circ:propAccel} with $R=3M$.}
\label{fig:static_R3M_HHMinkRind}
\end{figure}

\begin{figure}[p]  
  \centering
  \includegraphics[width=0.8\textwidth]{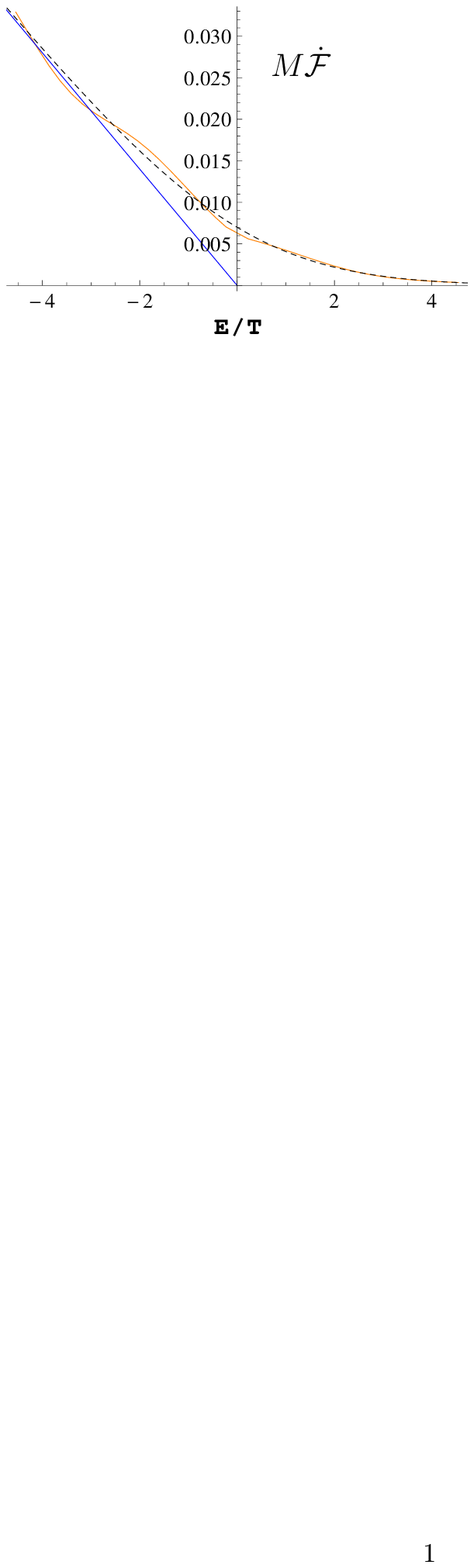}
 \caption{As in Figure~\ref{fig:static_R3M_HHMinkRind} but with $R=11M$.}
\label{fig:static_R11M_HHMinkRind}
\end{figure}

\begin{figure}[p]  
  \centering
  \includegraphics[width=0.8\textwidth]{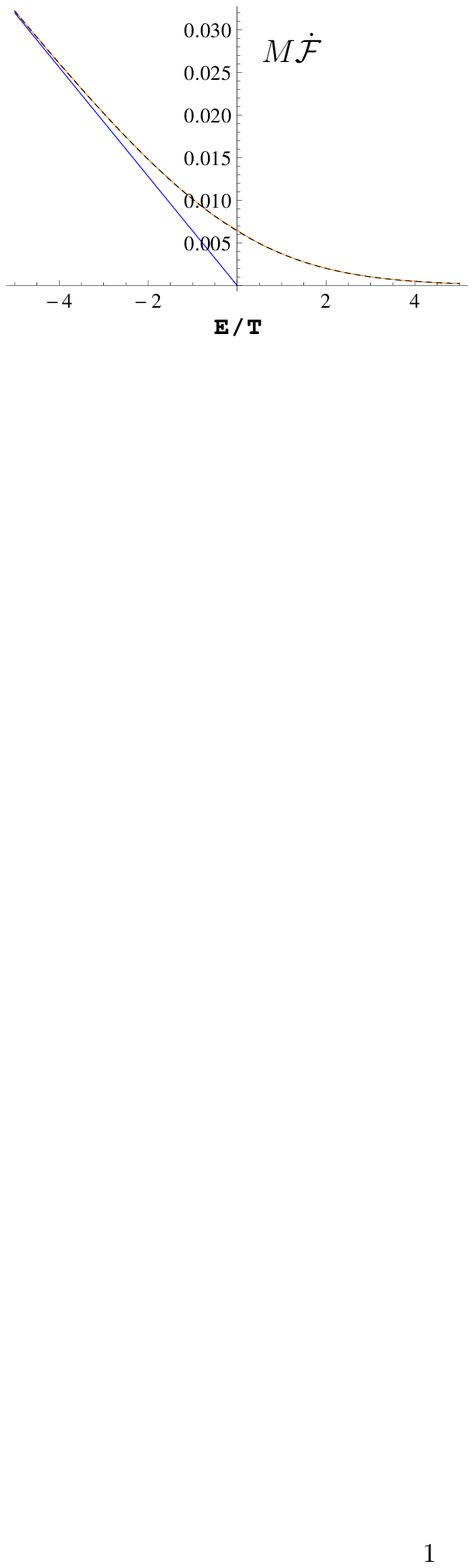}
 \caption{As in Figure~\ref{fig:static_R3M_HHMinkRind} but with $R=100M$.}
\label{fig:static_R100M_HHMinkRind}
\end{figure}

 Figures~\ref{fig:static_R3M_HHMinkRind},~\ref{fig:static_R11M_HHMinkRind} and~\ref{fig:static_R100M_HHMinkRind} show the transition rate of the static detector coupled to a scalar field in the Hartle-Hawking vacuum compared with the transition rate of the inertial detector in 3+1 Minkowski spacetime and a Rindler detector with proper acceleration given by~\eqref{eq:Schw:4d:circ:propAccel}. First, we see that at large, negative energies the transition rate of the detector coupled to the scalar field in the Hartle-Hawking vacuum, in the black hole spacetime, asymptotes to that of the inertial detector, in 3+1 Minkowski spacetime. Second, we observe that as $R$ increases, the Hartle-Hawking rate agrees to an increasing extent with the Rindler detector in flat spacetime. This is to be expected because as one moves further from the black hole the spacetime is asymptotically flat.
\par
\begin{figure}[p]  
  \centering
  \subfloat[$R=3M$]{\label{fig:static_R3M_ratioHHtoU}\includegraphics[width=0.5\textwidth]{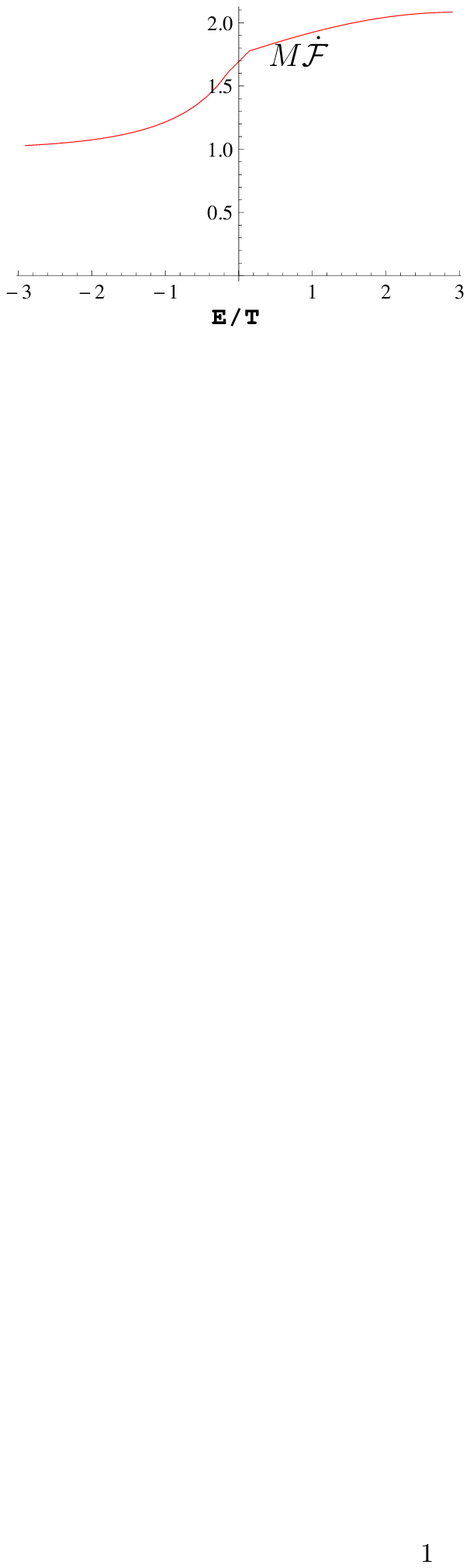}}   \\[2em]          
  \subfloat[$R=11M$]{\label{fig:static_R11M_ratioHHtoU}\includegraphics[width=0.5\textwidth]{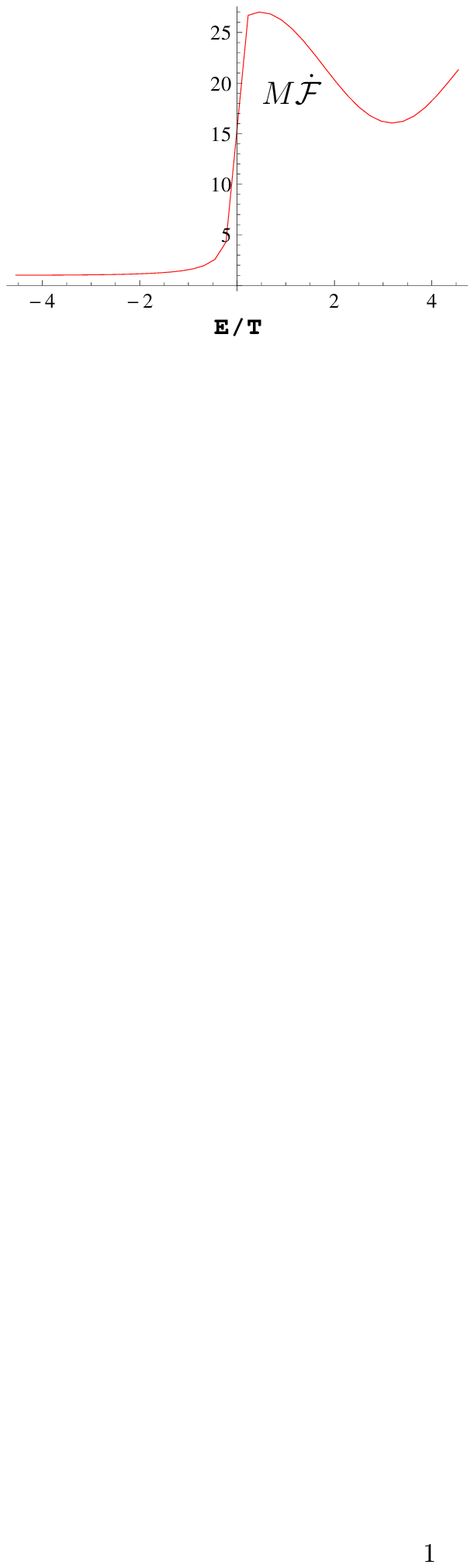}}  \\[2em]
  \subfloat[$R=100M$]{\label{fig:static_R100M_ratioHHtoU}\includegraphics[width=0.5\textwidth]{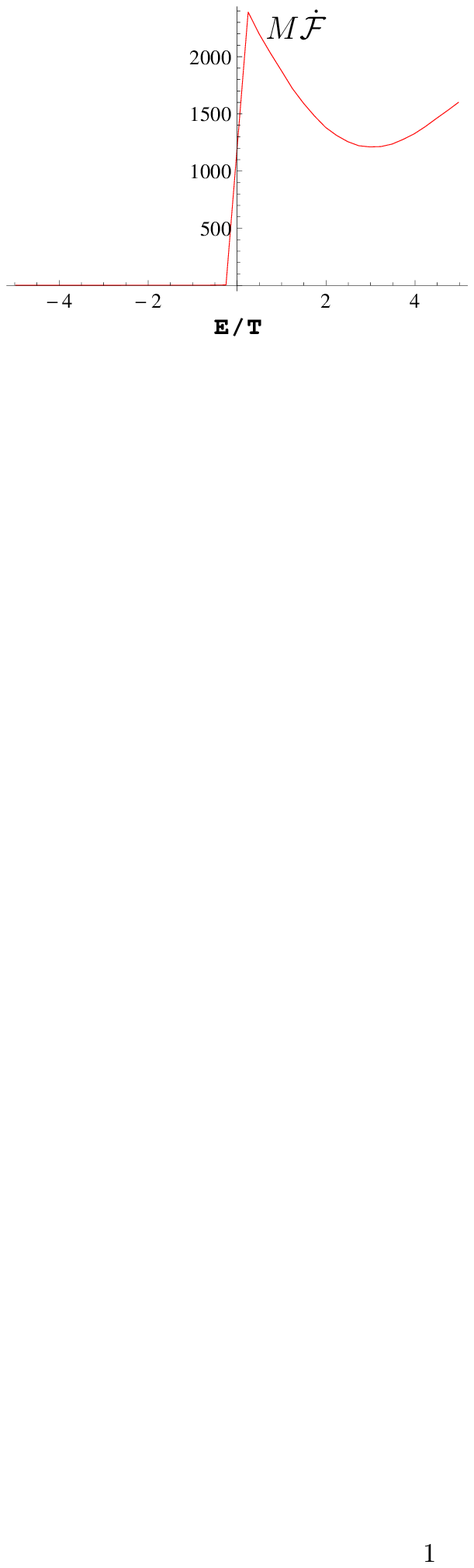}}  
\caption{Ratio of $M\dot{\mathcal{F}}$, as a function of $E/T_{\text{loc}}$, for the static detector in the Hartle-Hawking vacuum to that of the static detector in the Unruh vacuum. The discontinuity near the origin is caused by the numerical difficulty in computing the modes at small $\omega$.}
\label{fig:transrate_static_ratioHHtoU}
\end{figure}

 Finally, Figure~\ref{fig:transrate_static_ratioHHtoU} shows the ratio of the transition rate of the static detector coupled to a field in the Hartle-Hawking vacuum to the transition rate of the same detector coupled to a field in the Unruh vacuum. We see that this ratio becomes larger at positive excitation energies and when the radius increases. 
The Unruh vacuum represents a radiating black hole and this radiation will die off by an $r^{-2}$ power law, whereas the Hartle-Hawking vacuum state represents a constant heat bath at spatial infinity; therefore, it is to be expected that the ratio between the Hartle-Hawking and Unruh vacua becomes large as $R\to \infty$. The discontinuity that appears in the curves of Figure~\ref{fig:transrate_static_ratioHHtoU} is a numerical artefact caused by the fact that solving the ODE~\eqref{eq:Schw:4d:radModPhi} becomes difficult for small $\omega$. By the relation $\tilde{\omega}=E\sqrt{1-2M/R}$ that we found in Section~\ref{sec:4DSchw:staticDet}, this means computing the transition rate near $E=0$ is difficult and we did not attempt this.
\FloatBarrier
\subsection[Circular detector]{Circular detector results}
\label{sec:4DSchw:results:circ}
In this section, we present the results obtained for the detector on a circular-geodesic in Schwarzschild spacetime. These results are computed from the numerical evaluation of the transition rates~\eqref{eq:Schw:4d:circ:HHrateRes},~\eqref{eq:Schw:4d:circ:BoulwarerateRes} and~\eqref{eq:Schw:4d:circ:UnruhrateRes}.
\par For the circular-geodesic detector's transition rate, we had the double $\ell$-,~$m$-sum to compute, but as we noted in Section~\ref{sec:4DSchw:circ}, we can demand that $m\equiv \ell (\text{mod}2)$ to reduce the workload by half. 
\par Additionally, it proves only necessary to compute $\Phi^{\text{up}}_{\omega_{\pm},\ell},\Phi^{\text{in}}_{\omega_{\pm},\ell}$, where $\omega_{\pm}:=(mb \pm E)/a$,  over the positive range $E>0,~m\geq 0$ in order to have all the data we need to reconstruct the full transition rate over both negative and positive $E$ and $m$. The reason for this is the fact that the absolute square of the modes only depends on the absolute value of $\omega$, and $\omega_{\pm}(m,E)$ can always be related to $\pm\omega_{\pm}(|m|,|E|)$. For example, assuming we wished to compute the $|\Phi^{\text{up}}_{\omega_{+},\ell}|^2,|\Phi^{\text{in}}_{\omega_{+},\ell}|^2$ for a term in the sum where both $E, m <0$, we can observe that
\bea
\omega_{+}(-|m|, -|E|)&=\frac{-|m|b-|E|}{a}\\
&=-\frac{|m|b+|E|}{a}\\
&=-\omega_{+}(|m|, |E|)\, .
\eea
Thus, if we have already computed the modes at $\omega_{+}(|m|, |E|)$, then by the fact that $|\omega_{+}(-|m|, -|E|)|=|\omega_{+}(|m|, |E|)|$ and the independence of $|\Phi^{\text{up}}_{\omega_{+},\ell}|^2,|\Phi^{\text{in}}_{\omega_{+},\ell}|^2$ on the overall sign of $\omega$, we see that we also have the value of the absolute value squared of the modes over the range where both $E,~m <0$. Further relations are 
\bea
\omega_{+}(-|m|,|E|)&=-\omega_{-}(|m|,|E|)\,,\\
\omega_{+}(|m|,-|E|)&=\omega_{-}(|m|,|E|)\,,\\
\omega_{-}(-|m|,-|E|)&=-\omega_{-}(|m|,|E|)\,,\\
\omega_{-}(-|m|,|E|)&=-\omega_{+}(|m|,|E|)\,,\\
\omega_{-}(|m|,-|E|)&=\omega_{+}(|m|,|E|)\,.
\eea
We cut off the $\ell$-sum in the transition rate when the contributions at large $\ell$ become negligible. As with the static case, this cut-off is increased as $\omega$ or $R$ increases. 
\par
\begin{figure}[p]  
  \centering
  \includegraphics[width=0.8\textwidth]{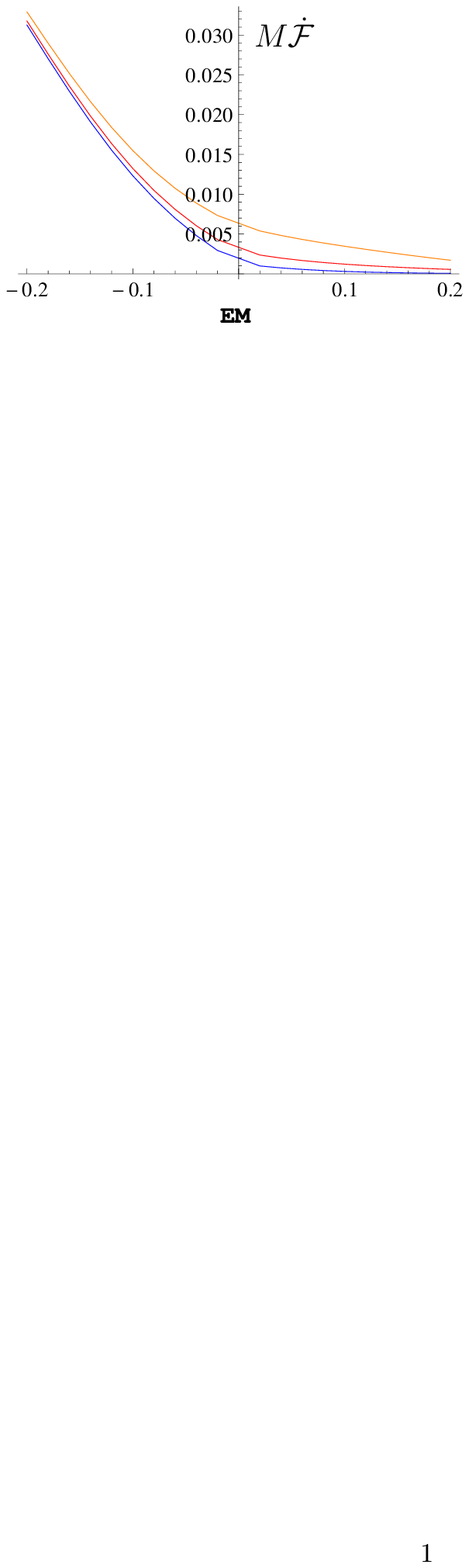}            
  \caption{$M\dot{\mathcal{F}}$ as a function of $EM$ for the circular detector at $R=4M$. Figure showing the results for the Hartle-Hawking vacuum (orange) computed from~\eqref{eq:Schw:4d:circ:HHrateRes}, Boulware vacuum (blue) computed from~\eqref{eq:Schw:4d:circ:BoulwarerateRes} and Unruh vacuum (red) computed from~\eqref{eq:Schw:4d:circ:UnruhrateRes}.}
\label{fig:circ_R4M_3vacs}
\end{figure}

\begin{figure}[p]  
  \centering
  \includegraphics[width=0.8\textwidth]{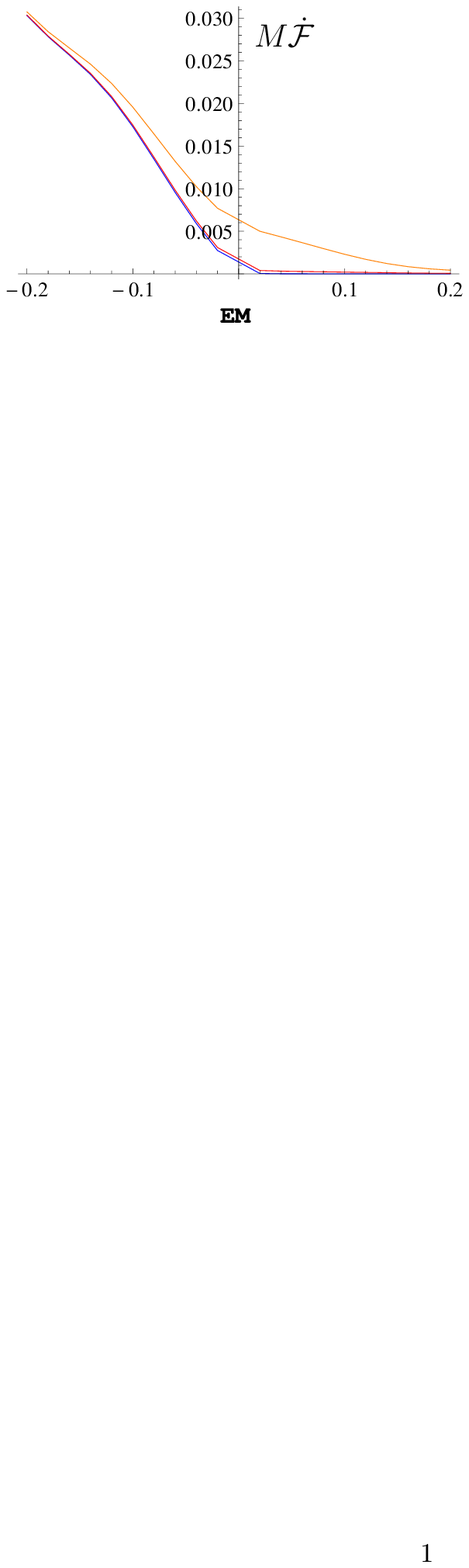}            
  \caption{As in Figure~\ref{fig:circ_R4M_3vacs} but with $R=8M$.}
\label{fig:circ_R8M_3vacs}
\end{figure}

\begin{figure}[p]  
  \centering
  \includegraphics[width=0.8\textwidth]{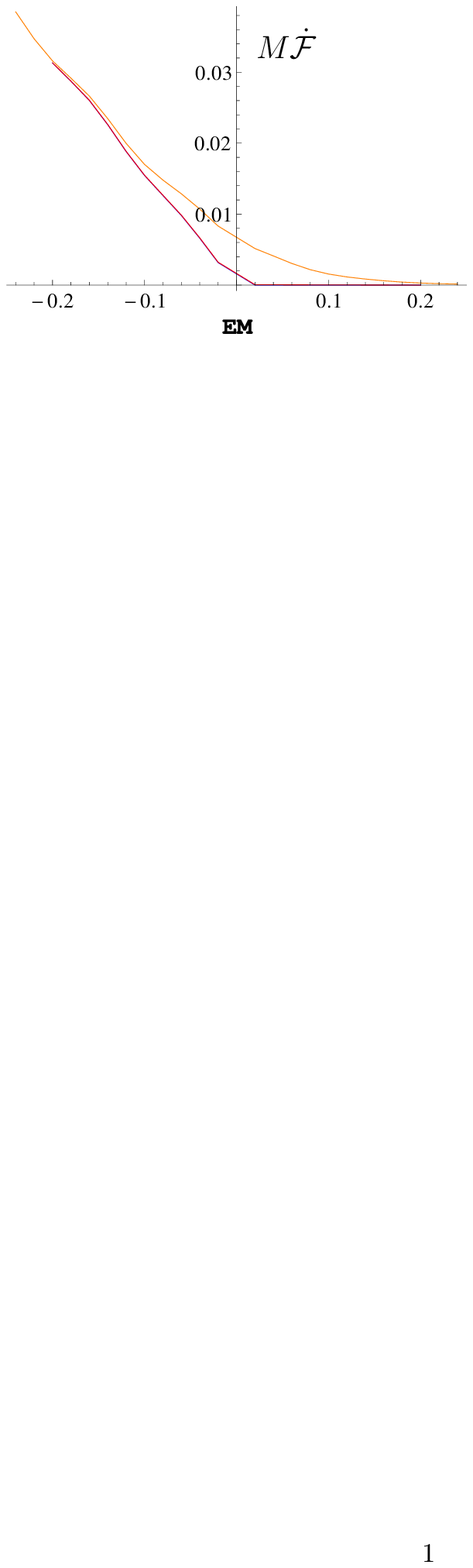}            
  \caption{As in Figure~\ref{fig:circ_R4M_3vacs} but with $R=20M$.}
\label{fig:circ_R20M_3vacs}
\end{figure}

 
Figures~\ref{fig:circ_R4M_3vacs},~\ref{fig:circ_R8M_3vacs} and~\ref{fig:circ_R20M_3vacs} show the transition rate against the excitation energy of the detector, made dimensionless by the multiplication by the mass of the black hole, $M$. The horizon is at $R=2M$, and we see that as we move away from the horizon, far from the hole at $R=20M$, the transition rates for the Boulware and Unruh vacua align for negative excitation energies. Below $R=6M$, the circular orbits are unstable but this seems to have no qualitative effect on the transition rate of the detector.
\par
\begin{figure}[p]  
  \centering
  \includegraphics[width=0.8\textwidth]{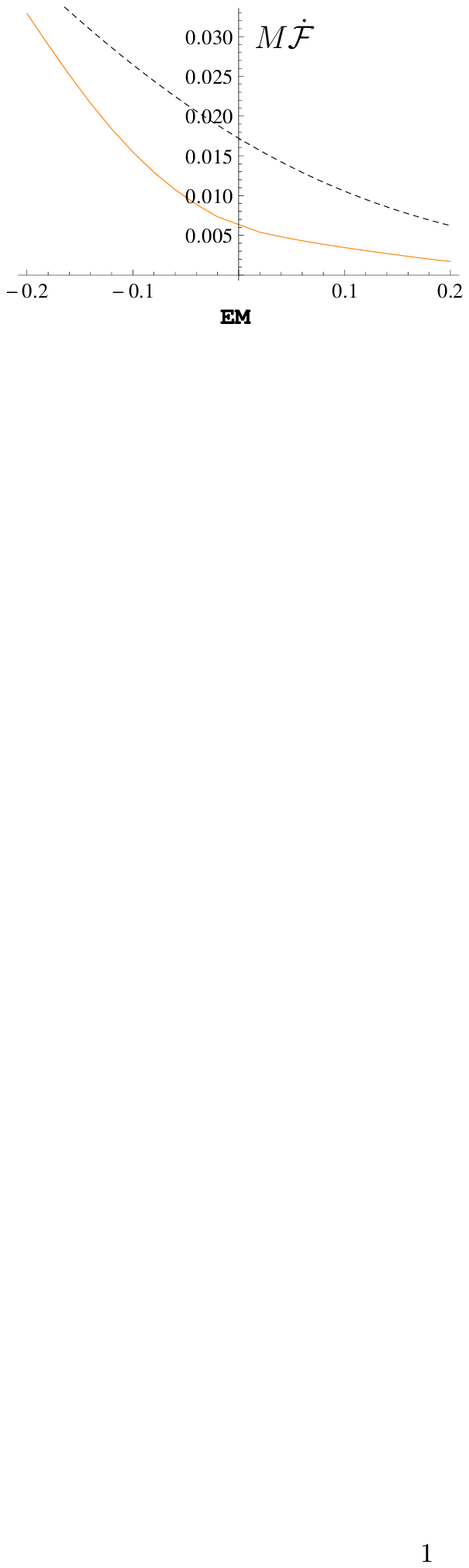}
\caption{$M\dot{\mathcal{F}}$ as a function of $EM$ for the circular detector at $R=4M$. Figure showing the results for the Hartle-Hawking vacuum (orange), computed from~\eqref{eq:Schw:4d:circ:HHrateRes}, alongside the response rate for a Rindler detector with transverse drift (black-dashed); the Rindler rate is computed by substituting the interval~\eqref{eq:Schw:4d:circ:interval} into the regulator-free transition rate found in~\cite{louko-satz:profile} and then numerically evaluating.}
\label{fig:circ_R4M_HHRind}
\end{figure}
\begin{figure}[p]  
  \centering
  \includegraphics[width=0.8\textwidth]{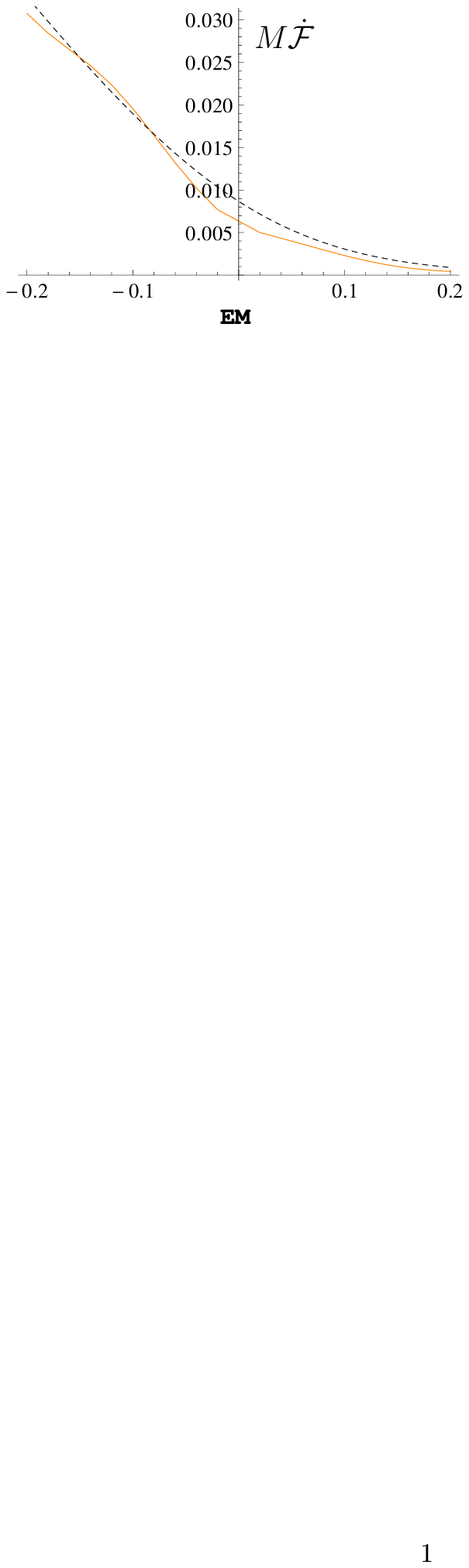} 
\caption{As in Figure~\ref{fig:circ_R4M_HHRind} but with $R=8M$.}
\label{fig:circ_R8M_HHRind}
\end{figure}
\begin{figure}[p]  
  \centering
  \includegraphics[width=0.8\textwidth]{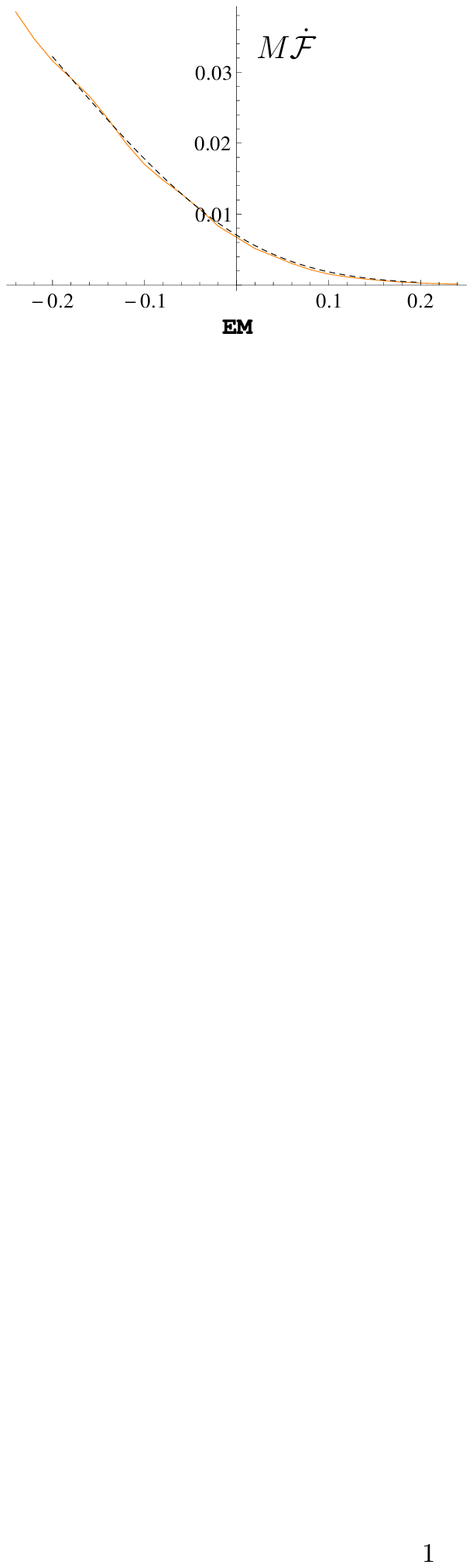}
\caption{As in Figure~\ref{fig:circ_R4M_HHRind} but with $R=20M$.}
\label{fig:circ_R20M_HHRind}
\end{figure}

Figures~\ref{fig:circ_R4M_HHRind},~\ref{fig:circ_R8M_HHRind} and ~\ref{fig:circ_R20M_HHRind} show the transition rate of the detector on the Schwarzschild black hole coupled to a scalar field in the Hartle-Hawking vacuum compared with a detector in Rindler spacetime, moving on a Rindler trajectory but drifting with constant velocity in the transverse $Y$-dimension; that is to say, the trajectory is given by~\eqref{eq:Schw:4d:circ:driftTraj}, with~\eqref{eq:Schw:4d:circ:paqRel},~\eqref{eq:Schw:4d:circ:q} and~\eqref{eq:Schw:4d:circ:p}.  We see that as the radius $R$ increases the agreement becomes better. As $R\to\infty$, the circular detector is becoming asymptotically a static detector, so the agreement should not be surprising considering our results in Section~\ref{sec:4DSchw:results:stat}.
\par
\begin{figure}[p]  
  \centering
  \subfloat[$R=4M$]{\label{fig:circ_R4M_PBC}\includegraphics[width=0.5\textwidth]{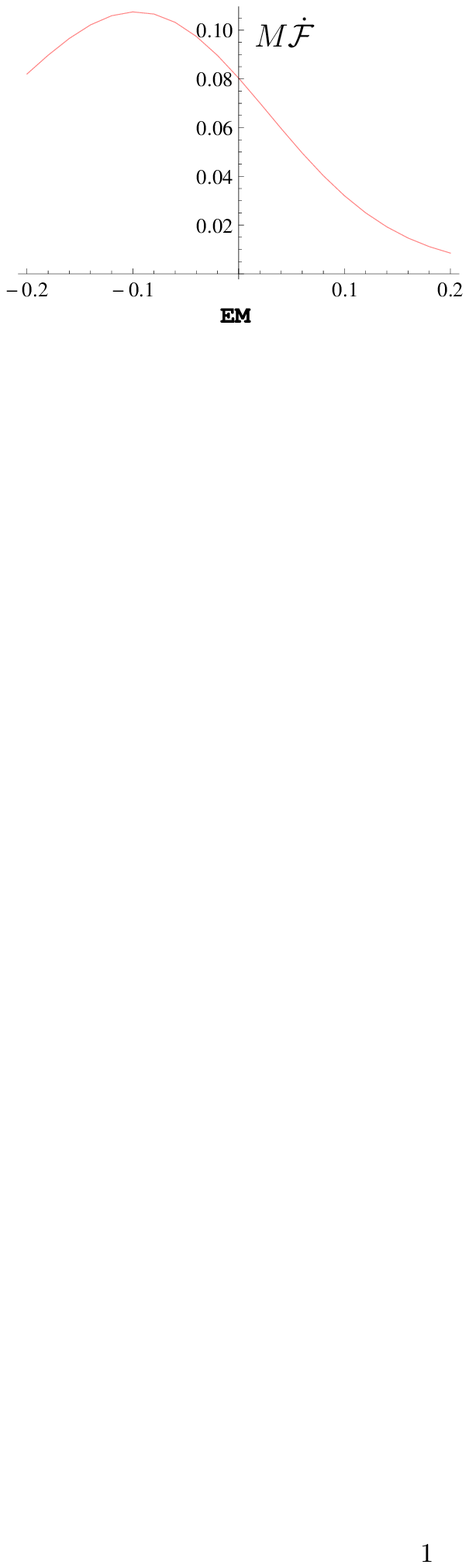}}      
  \\[2em]      
  \subfloat[$R=8M$]{\label{fig:circ_R8M_PBC}\includegraphics[width=0.5\textwidth]{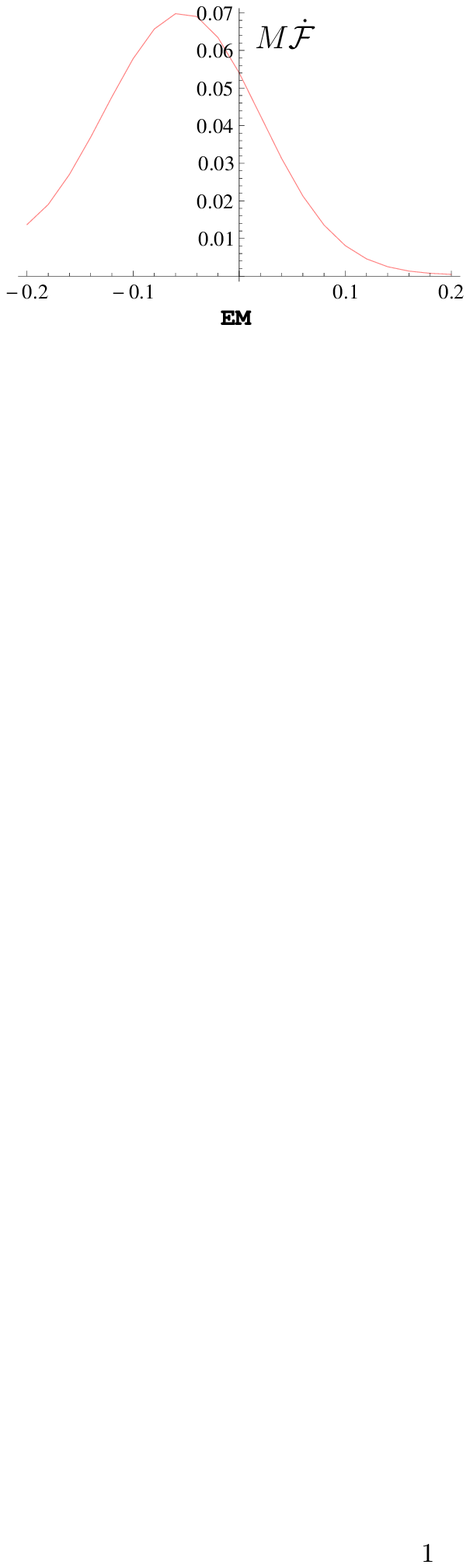}}  
  \\ [2em]
  \subfloat[$R=20M$]{\label{fig:circ_R20M_PBC}\includegraphics[width=0.5\textwidth]{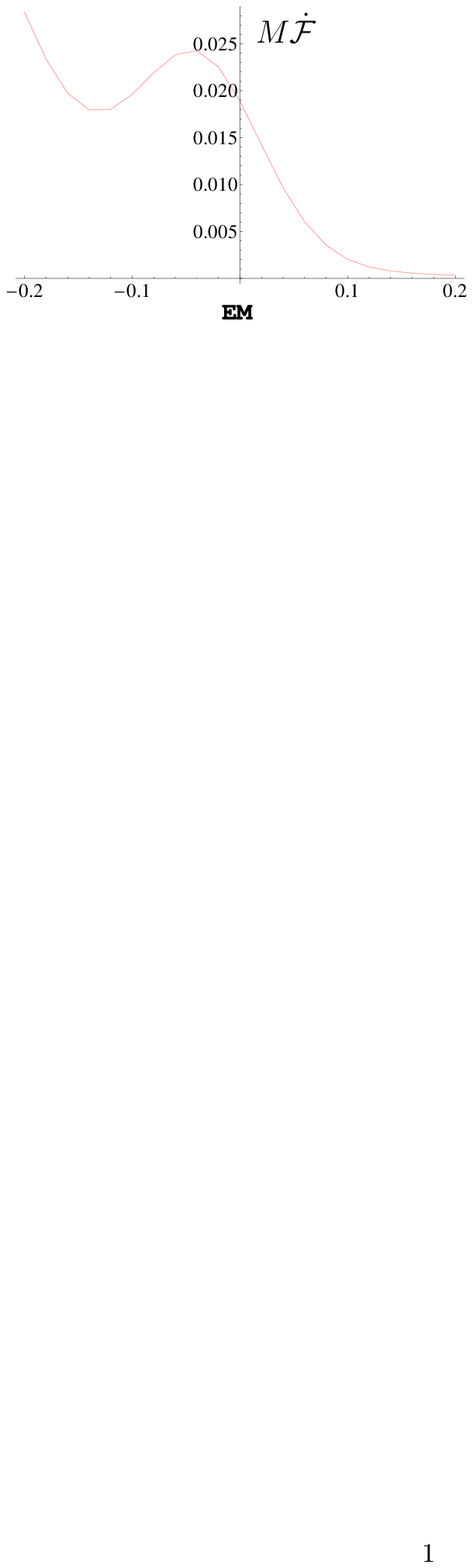}}  
\caption{Transition rate of a Rindler detector with drift in the transverse direction where the transverse direction has been periodically identified. Computed from~\eqref{eq:Schw:4d:circ:n_not0_rate} with $|n|$ cut off at 500.}
\label{fig:transrate_circ_PBC}
\end{figure}

 
Figure~\ref{fig:transrate_circ_PBC} shows the results that we obtained by making the transverse direction that the drifting Rindler detector's drift occurs in periodic, such that the period matches the period in proper time needed for the circular-geodesic detector, in Schwarzschild spacetime, to complete an orbit. The method of images sum~\eqref{eq:Schw:4d:circ:n_not0_rate} was cut off at $|n|=500$, by which point the sum had converged. We see by comparing Figures~\ref{fig:circ_R4M_HHRind},~\ref{fig:circ_R8M_HHRind} and ~\ref{fig:circ_R20M_HHRind} with Figure~\ref{fig:transrate_circ_PBC} that the agreement with the Schwarzschild detector is actually made worse by enforcing periodicity. We note that the oscillation at large, negative energies seen in Figure~\ref{fig:transrate_circ_PBC} is reminiscent of that seen for the co-rotating detector in Chapter~\ref{ch:btz}.
\par
\begin{figure}[p]  
  \centering
  \subfloat[$R=4M$]{\label{fig:circ_R4M_ratioHHtoU}\includegraphics[width=0.5\textwidth]{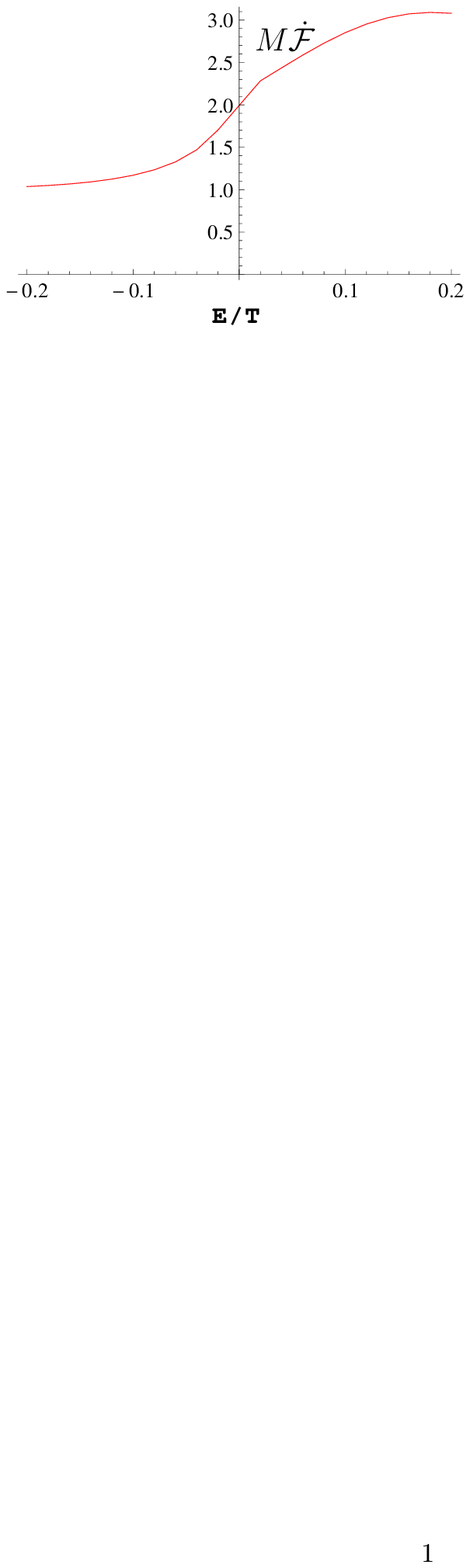}}     
  \\[1.8em]       
  \subfloat[$R=8M$]{\label{fig:circ_R8M_ratioHHtoU}\includegraphics[width=0.5\textwidth]{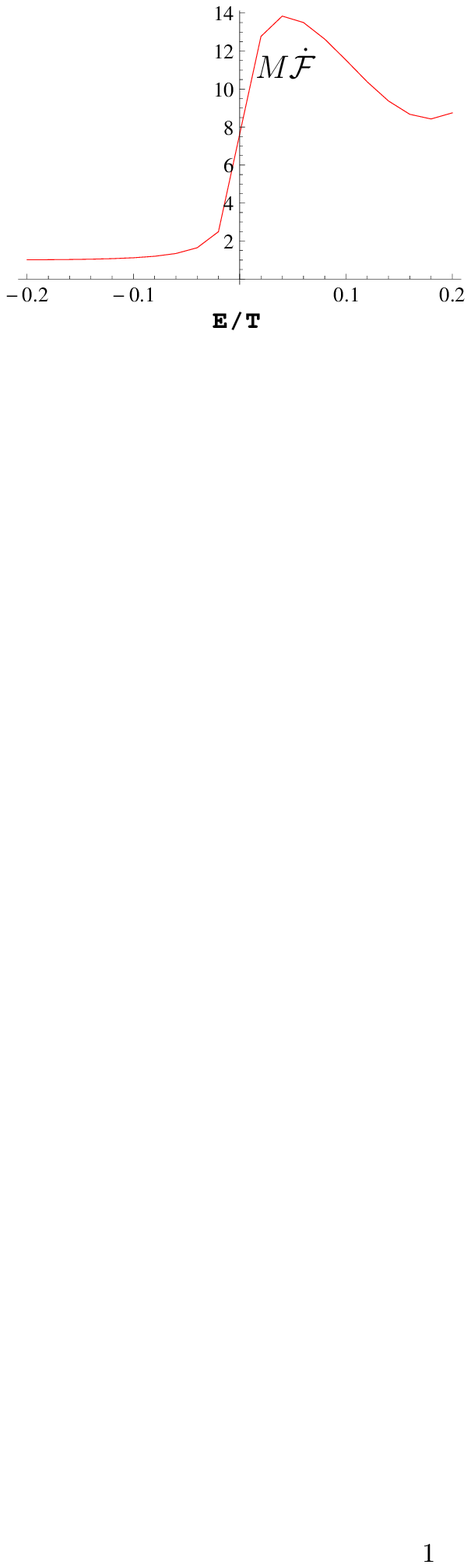}} 
  \\[1.8em] 
  \subfloat[$R=20M$]{\label{fig:circ_R20M_ratioHHtoU}\includegraphics[width=0.5\textwidth]{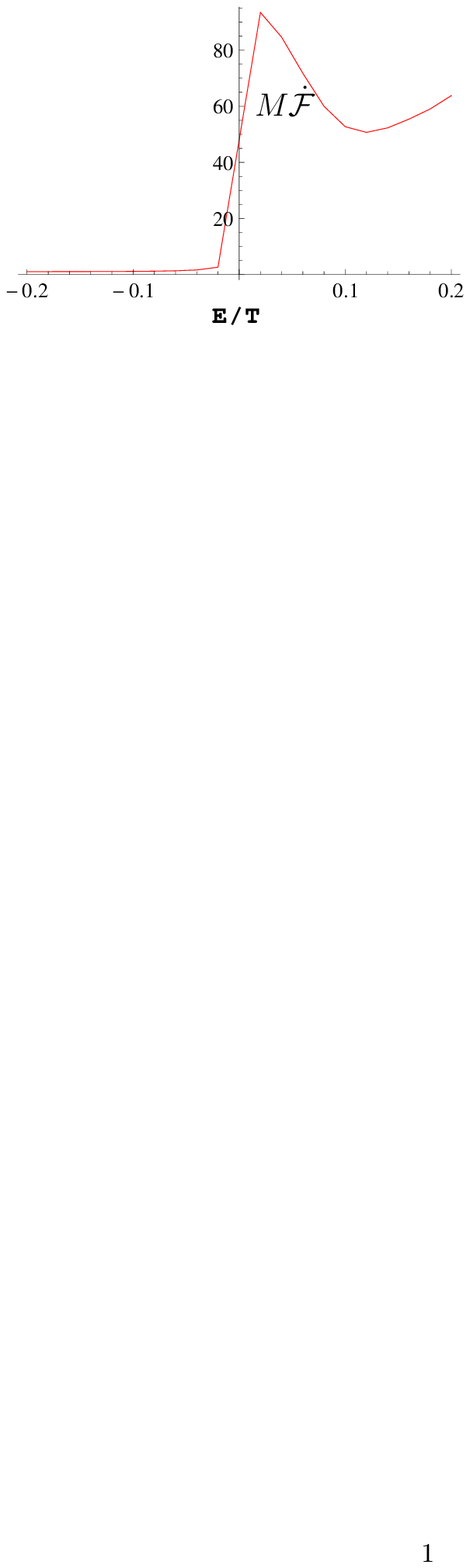}}  
\caption{Ratio of $M\dot{\mathcal{F}}$, as a function of $EM$, for the circular-geodesic detector in the Hartle-Hawking vacuum, to the transition rate of the circular-geodesic detector in the Unruh vacuum. The discontinuity that appears near the origin is a numerical artefact owing to the fact that solving the ODE~\eqref{eq:Schw:4d:radModPhi} becomes difficult at small $\omega$.}
\label{fig:transrate_circ_ratioHHtoU}
\end{figure}

 
Finally, Figure~\ref{fig:transrate_circ_ratioHHtoU} shows the ratio of the transition rate of the detector on a circular geodesic coupled to a field in the Hartle-Hawking vacuum, to the transition rate of the circular-geodesic detector coupled to a field in the Unruh vacuum. We see that just like in the static case, this ratio becomes larger at positive excitation energies and when the radius increases. 
\section{Summary}
\label{sec:4DSchw:summary}
\par In this chapter, we have analysed the response of an Unruh-DeWitt detector coupled to a massless scalar field on the four-dimensional Schwarzschild black hole using numerical methods.
\par For the static detector sat external to the hole's event horizon, we analysed the response when the field was in the Hartle-Hawking, Boulware and Unruh vacuum states. At a variety of radii, the results  were presented in the form of plots of the detector's transition rate, plotted against the detector's energy gap scaled by the local Hawking temperature.  For the field in the Hartle-Hawking vacuum state, we found that the response of the detector was thermal, in the KMS sense, with local Temperature given by $T=1/\left(8\pi M \sqrt{1-2M/R}\right)$, as expected. For a static detector and with the field in the Boulware vacuum state, the plots showed that the response of the detector consists only of de-excitation and that the excitation rate is vanishing; this is consistent with the fact that the static detector is on an orbit of the $\partial_t$ Killing vector, where $t$ is the Schwarzschild time co-ordinate. We also observed from the plots that as the radius increased, the Boulware and Unruh rates tended to become equal. This is consistent with the fact that the Unruh rate represents an outgoing flux of radiation from the hole that diminishes by $r^{-2}$ as the radius, $r$, tends to infinity, combined with the fact that the Boulware vacuum tends to the Minkowski vacuum as the radius tends to infinity. The Hartle-Hawking vacuum represents a thermal heat bath as the radius tends to infinity, and we plotted the ratio of the transition rate in the Hartle-Hawking vacuum to the transition rate in the Unruh vacuum, for the static detector, finding that the ratio of the excitation rates increases rapidly with radius.
\par We also presented results for a detector on a variety of circular geodesics, some stable and some unstable, outside the event horizon of the hole. The results were once again in the form of plots of the transition rate against the detector's energy gap, this time scaled to be dimensionless by multiplying by the mass of the black hole, $M$. Results were presented for the massless scalar field in the Hartle-Hawking, Boulware and Unruh vacuum states. The stability of the orbit seemed to have no qualitative effect on the transition rate of the detector. The Boulware vacuum in this case has a non-vanishing excitation component, and this component increases as the radius decreases. This is consistent with the fact that at large radius the circular-geodesic detector asymptotes to a static detector, so the detector becomes approximately on a $\partial_t$ orbit, but at small radius the detector is no longer on such an orbit, and there is room for positive energy excitations to occur. Similarly to the static case, the circular-geodesic plots also show that as the radius increases, the Boulware and Unruh vacuum states tend to become equal and that the ratio of the Hartle-Hawking rate to Unruh rate becomes large.
\par For the static detector coupled to a field in the Hartle-Hawking state, a comparison was made to the plot of the transition rate of the Rindler detector in the Minkowski vacuum state, with the proper acceleration chosen appropriately. Similarly, for the circular-geodesic detector a comparison was made to a Rindler detector with appropriately chosen proper acceleration, but this time also given a constant velocity drift in the transverse direction; the idea was that this would serve as an analogue to the angular motion of the circular geodesic. The results in both cases showed that as the radius increased, the Hartle-Hawking and Rindler rates aligned.
\par Finally in this chapter, we presented the necessary analytic setup and numerical methods needed to compute the transition rate of a detector on a radially-infalling geodesic to the Schwarzschild black hole. At the time of writing, data was still in the process of being gathered, so no numerical results were presented.

\chapter{Conclusions}
\label{ch:discussion}

This thesis has been concerned with the response of an Unruh-DeWitt particle detector in a variety of time-independent and time-dependent situations. Throughout, we have been careful with the regularisation procedure: ensuring that we switched on (off) our detector smoothly to obtain a regulator-free detector response function, \emph{before} taking the sharp-switching limit and only then, finally, differentiating with respect to the proper time to obtain the instantaneous transition rate.
\par 
We first motivated and provided the necessary background for the Unruh-DeWitt particle detector model using first-order perturbation theory. We then introduced the key concepts of the \emph{detector response function} and the \emph{instantaneous transition rate}, with emphasis on the problems that arise if one maintains the point-like detector regularisation and is simultaneously cavalier about the switching on (off) of the detector. We gave a brief overview of the work of Schlicht,  wherein the point-like coupling is replaced by a ``spatially-smeared'' coupling, effectively an alternative regularisation scheme, which in a sense, models a more realistic detector with finite extent.  We discussed the work of Satz that showed that the issues Schlicht observed could be traced to the distributional nature of the Wightman function; specifically, unless one integrates the Wightman function against smooth, compactly-supported test functions there is no guarantee of obtaining a unique result. Next, we discussed the Satz procedure that we adopted throughout the thesis; the reason for using this approach is that we consider it easier to adapt to general curved spacetime than the approach of Schlicht, which relies on finding a Lorentzian profile-function defined in the detector's hyperplane of simultaneity.
\par 
Satz had previously shown that in four-dimensional Minkowski spacetime for a detector on an arbitrary trajectory coupled to a scalar field in the Minkowski vacuum, the detector response function, as well as the total transition probability, diverges as $\log{\delta}$, where $\delta$ is the switching duration, but the transition rate, as derived via the smooth-switching method, remains finite. In Chapter~\ref{ch:by4d}, we set out to extend this result to Minkowski spacetimes of dimension other than four. The two-dimensional case is trivial because the Wightman function has only a logarithmic singularity, and the $\epsilon\to 0$ limit can be taken point-wise under the integral in the response function. This leads to the response function and the transition rate being finite. For other dimensions up to and including six, we followed closely the procedure set out in~\cite{satz:smooth} for taking the $\epsilon\to 0$ limit, finding first the regulator-free forms of the response functions, which were well defined, non-singular and contained no Lorentz-noncovariant terms. For each dimension, we then took the sharp-switching limit and differentiated with respect to the proper time to obtain the instantaneous transition rate. In three dimensions, we found that both the response function and transition rate remained finite in the sharp-switching limit. In five dimensions, the response function had a $1/\delta$ divergence in the sharp-switching limit; nevertheless, the transition rate remained finite.  In six dimensions, both the response function and transition rate diverged. The response function diverged as $1/\delta^2$, and the transition rate contained a term proportional to $\ddot{\x}\cdot\x^{(3)} \log{\delta}$. Moreover, the coefficient of $\ddot{\x}\cdot\x^{(3)}$ in the divergent term depended on the switch-on (off) profile, which goes against one of the original aspirations of the smooth-switching program. The presence of the $\ddot{\x}\cdot\x^{(3)} \log{\delta}$ in six dimensions means that the transition rate diverges for generic trajectories, but it remains finite for trajectories on which the scalar proper acceleration is constant, including all stationary trajectories. 
\par We believe the divergence can be explained by the fact that as the dimension increases the singularity of the Wightman function is becoming stronger, and hence, it is very likely that if we pushed the computations to higher-dimensional Minkowski spacetimes the transition rate would continue to diverge on all but perhaps a special class of trajectories.
\par 
Chapter~\ref{ch:by4d} closed with an application of our six-dimensional results to the GEMS approach. We considered a particle detector in four-dimensional Schwarzschild spacetime by embedding this spacetime into six-dimensional Minkowski spacetime and specifying the six-dimensional quantum field to be initially in the Minkowski vacuum. We expected, from~\cite{satz-louko:curved}, a well-defined transition rate for all stationary trajectories in Schwarzschild; however, we found that the only Schwarzschild geodesics that lifted to trajectories of constant scalar proper acceleration in the six-dimensional Minkowski embedding were the circular geodesics. Thus, given the fact that our six-dimensional Minkowski transition rate diverged on all but constant scalar proper acceleration trajectories we saw a contradiction that may suggest GEMS methods have limited validity for non-stationary trajectories whenever the embedding spacetime has dimension higher than five.
\par 
In Chapter~\ref{ch:btz}, we generalised the three-dimensional results of Chapter~\ref{ch:by4d} from the Minkowski vacuum to an arbitrary Hadamard state in an arbitrary three-dimensional spacetime, using similar techniques to~\cite{satz-louko:curved}. The transition probability and the transition rate were shown to remain well defined when the switching limit became sharp. In the special case of the detector in three-dimensional Minkowski spacetime coupled to a field in the Minkowski vacuum, this result reduced to that found in Chapter~\ref{ch:by4d}.
\par 
We next specialised the three-dimensional spacetime to that of the BTZ black hole, and we analysed the case of the detector coupled to a massless conformally-coupled scalar field in the Hartle-Hawking like vacuum state. This spacetime is asymptotically $\text{AdS}$ with a timelike infinity, as such it was necessary to impose boundary conditions in order to build a sensible quantum field theory. We considered the cases of transparent, Dirichlet or Neumann boundary conditions at the infinity. With the spacetime and quantum state specified, we next specified the trajectory of the detector.  A stationary detector external to the outer event horizon, co-rotating with the angular-velocity of the horizon and switched on in the asymptotic past was considered first. As a special case, the static detector external to a non-rotating hole was also considered. For the co-rotating detector, thermality, in the sense of the KMS property in the local Hawking temperature, was verified. We note that we did not consider a stationary non-corotating detector in detail; the primarily reason for this was that, as we showed in Appendix~\ref{app:non-corotating}, the parameter space has at least some regimes for which the response of such a detector does not have the KMS property. Analytic results for the transition rate in a number of asymptotic regimes of the parameter space were obtained, including those of large and small black hole mass, and we complemented these with numerical results in the interpolating regimes
\par
We also considered a detector that falls into a non-rotating BTZ hole along a radial geodesic. The trajectory is now non-stationary and the switch-on of the detector cannot be pushed to the asymptotic past without colliding with the white hole singularity. Unlike the co-rotating detector, for the radially-infalling detector, thermality, in the KMS sense, was not found, not even near the moment of maximum radius on a trajectory, and we traced the reasons for this to the properties of $\text{AdS}_3$ geodesics that have been previously analysed
from GEMS considerations~\cite{Deser:1997ri,Deser:1998bb,Deser:1998xb,Russo:2008gb}. Namely, that detectors with \emph{sub-critical} accelerations, $a<1/\ell$, have no well-defined temperature. We obtained analytic results for the transition rate when the black hole mass is large, 
and we evaluated the transition rate numerically for small values of the black hole mass provided the switch-on and switch-off take place in the exterior. 
\par 
In Chapters~\ref{ch:2DSchw} and~\ref{ch:4DSchw}, we investigated a detector on the Schwarzschild spacetime. For four-dimensional Schwarzschild the Wightman function, even for a detector coupled to a massless scalar field, is not known analytically. In an attempt to gain insight into a detector in the full four-dimensional spacetime, we first dropped the angular co-ordinates in Chapter~\ref{ch:2DSchw} and investigated detectors in $(1+1)$-dimensional Schwarzschild spacetime, where the conformal triviality was exploited to make analytic progress. We attempted to regularise the infrared divergence that arises when considering the Wightman function of a massless scalar field in a $(1+1)$-dimensional spacetime by invoking a temporal cut-off, of the kind used by Langlois~\cite{Langlois}, and pushing the detector switch-on to the infinite past. 
\par First, in an attempt to gain confidence in this infrared regularisation scheme we considered a static detector on the $(1+1)$-dimensional Minkowski half-space. We explicitly compared the $m\to 0$ limit of the transition rate of a static detector coupled to the massive scalar field, to the transition rate obtained for the detector coupled to the massless scalar field ---from the outset--- with the infrared sickness treated by a temporal-window cut-off. The results, ~\eqref{eq:Schw:2d: masslessTRfinal} and~\eqref{eq: Schw:2d:staticMassiveTm0lim}, agree exactly, which gave us confidence in this cut-off procedure.
\par 
Reassuringly, using the Langlois cut-off and analysing the static detector external to the $(1+1)$-Schwarzschild black hole coupled to a massless scalar field in the Hartle-Hawking vacuum, we found a transition rate that was Planckian and thermal in the local Hawking temperature. However, when we looked at the static detector coupled to a massless scalar field in the Unruh vacuum, in addition to the expected terms (the average of the Hartle-Hawking and Boulware rates) we found an unexpected term of the form $T/2\omega^2$, with $T$ being the local Hawking temperature. 
\par 
This unexpected term was also found when we next looked at the transition rate of a detector in $(1+1)$-Minkowski spacetime with a receding mirror, whose in-vacuum in the late-time limit is a close analogue of the Unruh vacuum. It would have been interesting calculate the transition rate of the detector coupled to a massive scalar field in the receding-mirror spacetime to see if the unexpected term arises also in that case when the $m\to 0$ limit is taken, but owing to the fact that the left- and right-movers no longer decouple in this case, the calculation proves prohibitive. As such, we must speculate that the Langlois infrared regularisation scheme is not sufficient in these examples.
\par 
In Chapter~\ref{ch:4DSchw}, we investigated detectors on the full four-dimensional Schwarzschild black hole numerically. We coupled the detector to a massless scalar field and considered the Hartle-Hawking, Boulware and Unruh vacuum states. For a static detector external to the black hole, thermality in the sense of the KMS property was recovered when the field was in the Hartle-Hawking vacuum. Numerical results were also presented for a detector on a circular geodesic at a variety of radii, both stable and unstable. We compared the static and circular-geodesic detectors to a Rindler detector and Rindler detector with transverse drift respectively, where the proper scalar acceleration was chosen appropriately. We found good agreement as the radius of the detector around the Schwarzschild black hole increased. Finally, we presented the necessary analytic setup to compute the transition rate of the detector radially-infalling on a geodesic. All numerical work was carried out using the software package Mathematica (TM).
\par 
We regard the main achievements of this thesis to be a contribution to the growing evidence that
Unruh-DeWitt detectors are a conceptually well-motivated and
computationally efficient tool for probing the physical content of
states in quantum field theory, in flat and curved spacetimes, in both
stationary and non-stationary settings. The sharp-switching results
establish new bounds on situations where an instantaneous transition
rate can still be meaningfully defined even though the detector's
response is not stationary. The applications to black hole spacetimes
provide new information on how black hole radiation is experienced by
observers in various states of motion in the spacetime, particularly on
the interplay between the observer's motion and the thermal character of
the radiation.
\par 
There are several future directions that are worth mentioning here. It may be interesting to extend our Chapter~\ref{ch:by4d} results in Minkowski spacetime to dimension $d>6$. Does the transition rate continue to diverge as we expect? What is the exact nature of this divergence? Are there any trajectories for which this divergence vanishes? Second, using similar techniques to Chapter~\ref{ch:by4d} combined with those that led to~\eqref{eq:resp3d:sharp:rate} in Chapter~\ref{ch:btz} and to the four-dimensional equivalent in~\cite{satz-louko:curved}, it should be possible to compute the instantaneous transition rate for a detector coupled to a field in an arbitrary Hadamard state in dimension $d \geq 5$.
\par
Regarding $(1+1)$-dimensional Schwarzschild spacetime: it seems to us that the Langlois temporal-window style cut-off is not robust enough to deal with the infrared sickness in more complicated examples than the inertial or uniformly accelerated detectors in flat spacetime. An alternative way to deal with this infrared divergence would be to look at a detector with \emph{derivative coupling}; that is to say, one could couple the detector to the proper-time derivative of the scalar field, rather than the field itself. This would lead to an analogue of the Wightman function, $\langle 0| \dot{\phi}(\x)\dot{\phi}(\x^{\prime})|0\rangle$, in the transition rate, which has a $1/\sigma^2$ singularity structure (with $\sigma$ being the spacetime interval), similar to the four-dimensional Wightman function. This would eliminate the infrared divergence in the transition rate and the need to use Langlois style cut-offs altogether.
\par 
Another obvious future direction is to complete the numerical analysis of the detector radially-infalling on a geodesic to the four-dimensional Schwarzschild hole. At the time of writing, all analytic and coding work is complete but data gathering at the University of Nottingham High Performance Cluster (HPC) is still ongoing. Upon successfully obtaining results for this radially-infalling detector, which we assumed to start at radial infinity with zero initial velocity, we may also be interested in considering alternative infall trajectories, such as those with some initial velocity. 
\par 
Further in the future, it would be an interesting application of the Satz transition rate formula to investigate the recent proposals of Hartle-Hawking and Boulware like vacua on the Kerr Black Hole~\cite{Casals:2012es}. We foresee investigation with detectors to be useful in probing if these vacua really do possess the expected properties, such as regularity as one crosses the horizon and thermality in the Hartle-Hawking case. We believe the main challenges in pursuing such investigations would be extending the code to the more complicated Kerr geometry and the fact that the field of interest is now Fermionic. Nevertheless, the four-dimensional Schwarzschild work presented here should provide an excellent starting point.
\par 
Finally, a more ambitious project would be the use of Unruh-DeWitt detector models to investigate the recent proposal of firewalls~\cite{Almheiri:2012rt}. One would need first to construct the appropriate Wightman function for a black hole that had been decaying for a significant time and had shrunk in size, perhaps in $(1+1)$ dimensions to simplify matters, but with this obstacle overcome, detector models could be hoped to provide an insight into the nature of these firewalls, if indeed they prove to be a true feature of black holes in nature.

\appendix
\chapter[Six-dimensional Minkowski sharp-switching limit]{Six-dimensional Minkowski sharp-switching limit }
\chaptermark{$6d$ Minkowski sharp-switching limit}
\label{ch:appendix:6dSSL}
In this appendix, we analyse the third and fourth terms of the six-dimensional response function~\eqref{eq:6d:detresp} in the sharp-switching limit. We shall find that the third term diverges as $\delta^{-2}$ in the $\delta\to 0$ limit, and its derivative with respect to proper time in this limit goes, after including the $1/2\pi^3$ pre-factor, as
\be 
-\frac{1}{6\pi^3\Delta\tau^3}+O\left(\delta\right)\,.
\label{eq:app6dSSL:term3res}
\ee 
We shall find that the proper-time derivative of the fourth term of~\eqref{eq:6d:detresp} diverges logarithmically in the $\delta\to 0$ limit, and in this limit, after restoring the $-1/12\pi^3$ pre-factor, reads
\be
\frac{\ddot{\x}(\tau)\cdot\x^{(3)}(\tau)}{12\pi^3}\left(\ln{\left(\frac{\Delta\tau}{\delta}\right)+C^{'}_{+}}\right)+\frac{\ddot{\x}^2(\tau)}{12\pi^3 \Delta\tau}+O\left(\delta\ln{\left(\frac{\Delta\tau}{\delta}\right)}\right)\,,
\label{eq:app6dSSL:term4res}
\ee
where the constant $C^{'}_{+}$ is defined by
\begin{align}
C^{'}_{+}&=-2\int^1_0\,\mathrm{d}r\,\frac{1}{r^2}
\left(\int^1_0\,\mathrm{d}v\,h_2(1-v)\left[h_2(1-v+r)-h_2(1-v)\right]-\tfrac{1}{2}r\right)
\notag 
\\
&\hspace{8ex}-2\int^1_0\,\mathrm{d}v\,h_2(v)\left[1-h_2(v)\right]
\ . 
\label{eq:app6dSSL:Cprimeplus}
\end{align}
We note here, it is also possible to show using alternative methods that the fourth term of~\eqref{eq:6d:detresp} itself diverges logarithmically in the sharp-switching limit.
\section{Third term of~\eqref{eq:6d:detresp}}
\par We first consider the third term of~\eqref{eq:6d:detresp}, ignoring the $1/2\pi^3$ pre-factor:
\be
I=\int^{\infty}_{0}\,\frac{\mathrm{d}s}{s^4}\int^{\infty}_{-\infty}\,\mathrm{d}u\, \chi(u)
\bigl[
\chi(u-s)-\chi(u)
-\tfrac12 s^2\ddot{\chi}(u)\,
\bigr]\,.
\ee
If we substitute in the switching function~\eqref{eq:ss:switchingfunc}, we obtain
\bea 
&I=\int^{\infty}_{0}\,\frac{\mathrm{d}s}{s^2}\int^{\infty}_{-\infty}\,\mathrm{d}u\, \huOne\huTwo\times \\
&\Bigg[\frac{1}{s^2}\left(\hsOne\hsTwo-\huOne\huTwo\right)\\
&-\frac{1}{2\delta^2}\ddhuOne\huTwo+\frac{1}{\delta^2}\dhuOne\dhuTwo\\
&\quad\quad\quad\quad\quad\quad\quad\quad\quad\quad\quad\quad-\frac{1}{2\delta^2}\huOne\ddhuTwo\Bigg]\,,
\eea
which can be expressed as
\bea 
&I=\int^{\infty}_{0}\,\frac{\mathrm{d}r}{r^2}\int^{\infty}_{-\infty}\,\mathrm{d}v\, \hvOne\hbvTwo\times\\
&\Bigg[\frac{\hvrOne\hbvrTwo-\hvOne\hbvTwo}{\delta^2 r^2}-\frac{\ddhvOne\hbvTwo}{2\delta^2}\\
&\quad\quad\quad\quad\quad\quad\quad\quad\quad\quad\quad\quad+\frac{\dhvOne\dhbvTwo}{\delta^2}-\frac{\hvOne\ddhbvTwo}{2\delta^2}\Bigg]\,
\label{eq:app6dSSL:vrSSresp}
\eea 
after the change of variables
\bea
&v=\frac{u-\tau_0+\delta}{\delta}\,,\\
&r=\frac{s}{\delta}\,
\label{eq:app6dSSL:ssVarChange}
\eea
and the definition $b:=1+\Delta\tau/\delta$. In the response function~\eqref{eq:app6dSSL:vrSSresp}, only the range $(0,b+1)$ of the $v$-integral can make a contribution. To evaluate this expression, we mimic the techniques used for three- and five-dimensional spacetime in Chapter~\ref{ch:by4d}, and in four dimensions in~\cite{satz:smooth,satz-louko:curved}; we split the $r$-integral into five sub-integrals over the intervals $(0,~1),~(1,~b-1),~(b-1,~b),~(b,~b+1),~(b+1,~\infty)$, which we shall label as $I_{1,2,3,4,5}$ respectively. Moreover, in each of these sub-integrals we shall further split the $v$-integral range.
\par Recalling that $h_1$ and $h_2$ are smooth, non-negative functions satisfying 
$h_i(x)=0$ for $x\le0$ and $h_i(x)=1$ for $x\ge1$, and using the fact that $b=1+\Delta\tau/\delta>>1$ in the $\delta\to 0$ limit, we find that the $I_1$-integral collapses to
\bea
I_1&=\frac{1}{\delta^2}\int^{1}_0\,\frac{\mathrm{d}r}{r^2}\Bigg[\int^{1}_0\,\mathrm{d}v\,\hvOne\left(\frac{\hvrOne-\hvOne}{r^2}-\frac{\ddhvOne}{2}\right)\\
&+\int^{1+r}_1\mathrm{d}v\,\frac{\hvrOne-1}{r^2}\\
&+\int^{b+1}_b\,\mathrm{d}v\,\hbvTwo\left(\frac{\hbvrTwo-\hbvTwo}{r^2}-\frac{\ddhbvTwo}{2}\right)\Bigg]\,.
\label{eq:app6dSSL:I1fin}
\eea
By changing variables as $v\to b+1-v$ in the last $v$-integral, we see that $I_1$ is a constant, independent of $b$; thus, $I_1$ is independent of the switch-off time and vanishes upon taking the derivative with respect to $\tau$. As a consequence, $I_1$, although itself $\delta^{-2}$ divergent in the $\delta\to 0$ limit, makes no contribution to the transition rate.
\par Similarly, for the $I_2$-integral we have 
\bea 
I_2&=\frac{1}{\delta^2}\left(\frac{1}{2(b-1)^2}-\frac{1}{3(b-1)^3}-\frac{1}{6}\right)\\
&+\frac{1}{\delta^2}\int^{b-1}_1\,\frac{\mathrm{d}r}{r^2}\Bigg[-\int^{1}_0\,\mathrm{d}v\,\left(\frac{h_1^2(v)}{r^2}+\frac{\hvOne\ddhvOne}{2}\right)\\
&+\int^{r+1}_r\,\mathrm{d}v\,\left(\frac{\hvrOne-1}{r^2}\right)\\
&+\int^{b+1}_b\,\mathrm{d}v\,\hbvTwo\left(\frac{1-\hbvTwo}{r^2}-\frac{\ddhbvTwo}{2}\right)\Bigg]\,.
\label{eq:app6dSSL:I2alt1}
\eea
Making the change of variables to $v\to v-r$ in the second $v$-integral in the square brackets, and to $v\to b+1-v$ in the last $v$-integral, we can write 
\bea 
I_2&=\frac{1}{\delta^2}\left(\frac{1}{2(b-1)^2}-\frac{1}{3(b-1)^3}-\frac{1}{6}\right)\\
&-\frac{1}{2\delta^2}\int^{b-1}_1\,\frac{\mathrm{d}r}{r^2}\int^{1}_0\,\mathrm{d}v\,\Big(\hvOne\ddhvOne+\hvTwo\ddhvTwo\Big)\\
&+\frac{1}{\delta^2}\int^{b-1}_1\,\frac{\mathrm{d}r}{r^4}\int^{1}_0\,\mathrm{d}v\,\Big(\hvOne+\hvTwo-h_1^2(v)-h_2^2(v)-1\Big)\,.
\label{eq:app6dSSL:I2alt2}
\eea
Performing the $r$-integrals in the final two terms of~\eqref{eq:app6dSSL:I2alt2}, we find
\bea 
I_2&=\frac{1}{\delta^2}\left(\frac{1}{2(b-1)^2}-\frac{1}{3(b-1)^3}-\frac{1}{6}\right)\\
&-\frac{1}{2\delta^2}\left(1-\frac{1}{(b-1)}\right)\int^{1}_0\,\mathrm{d}v\,\Big(\hvOne\ddhvOne+\hvTwo\ddhvTwo\Big)\\
&+\frac{1}{3\delta^2}\left(1-\frac{1}{(b-1)^3}\right)\Bigg[-1+\int^{1}_0\,\mathrm{d}v\,\Big(\hvOne+\hvTwo-h_1^2(v)-h_2^2(v)\Big)\Bigg]\,,
\eea
and this means that we can write $I_2$ as
\bea 
I_2&=\frac{1}{\delta^2}\left(-\frac{1}{2}-A+B-\frac{C}{3}\right)\\
&+\frac{1}{2\delta^2 \left(b-1\right)^2}+\frac{A}{\delta^2 \left(b-1\right)}-\frac{B}{\delta^2 (b-1)^3}+\frac{C}{3\delta^2 (b-1)^3}\,,
\eea
where $A$, $B$ and $C$ are constants, independent of $b$, and defined by
\bea
A&:=\frac{1}{2}\int^{1}_0\,\mathrm{d}v\,\Big(\hvOne\ddhvOne+\hvTwo\ddhvTwo\Big)\,,\\
B&:=\frac{1}{3}\int^{1}_0\,\mathrm{d}v\,\Big(\hvOne+\hvTwo\Big)\,,\\
C&:=\int^1_0\,\mathrm{d}v\,\Big(h_1^2(v)+h_2^2(v)\Big)\,.
\label{eq:app6dSSL:AB}
\eea
Alternatively, using $b=1+\Delta\tau/\delta$, we can express $I_2$ in the form
\be
I_2=\frac{1}{\delta^2}\left(-\frac{1}{2}-A+B-\frac{C}{3}\right)+\frac{1}{2\Delta\tau^2}+\frac{A}{\delta \Delta\tau}-\frac{\delta B}{\Delta\tau^3}+\frac{\delta C}{3\Delta\tau^3}\,.
\label{eq:app6dSSL:I2fin}
\ee
\par Next we analyse $I_3$:
\bea
I_3 &=\frac{1}{2\delta^2b^2}-\frac{1}{3\delta^2b^3}-\frac{1}{2\delta^2(b-1)^2}+\frac{1}{3\delta^2(b-1)^3}\\
&+\frac{1}{\delta^2}\int^{b}_{b-1}\,\frac{\mathrm{d}r}{r^2}\Bigg[-\int^1_0\,\mathrm{d}v\,\left(\frac{h_1^2(v)}{r^2}+\frac{\hvOne\ddhvOne}{2}\right)\\
&+\int^b_r\,\mathrm{d}v\,\frac{\hvrOne-1}{r^2}\\
&+\int^{b+1}_b\,\mathrm{d}v\,\hbvTwo\left(\frac{\hvrOne-\hbvTwo}{r^2}-\frac{\ddhbvTwo}{2}\right)\Bigg]\,,
\eea
where the terms on the first line come from explicitly evaluating the integral that results from the portion of the $v$-integral range $1<v<r$. After making the change of variables $v\to v-r$ in the second $v$-integral in the square brackets, and $v\to b+1-v$ in the third $v$-integral, we get
\bea
I_3 &=\frac{1}{2\delta^2b^2}-\frac{1}{3\delta^2b^3}-\frac{1}{2\delta^2(b-1)^2}+\frac{1}{3\delta^2(b-1)^3}\\
&-\frac{1}{2\delta^2}\int^{b}_{b-1}\,\frac{\mathrm{d}r}{r^2}\int^1_0\,\mathrm{d}v\,\Big(\hvOne\ddhvOne+\hvTwo\ddhvTwo\Big)\\
&+\frac{1}{\delta^2}\int^{b}_{b-1}\,\frac{\mathrm{d}r}{r^4}\Bigg[-\int^1_0\,\mathrm{d}v\,h_1^2(v)
+\int^{b-r}_0\,\mathrm{d}v\,\Big(\hvOne-1\Big)\\
&\quad\quad\quad\quad\quad\quad\quad\quad+\int^1_0\,\mathrm{d}v\,\hvTwo\Big(h_1(b+1-v-r)-\hvTwo\Big)\Bigg]\,,
\eea
which can be written as
\bea
I_3 &=\frac{(b-1)}{3\delta^2 b^3}-\frac{1}{3\delta^2(b-1)^2}-\frac{A}{\delta^2 b(b-1)}\\
&+\frac{C}{\delta^2}\left(\frac{1}{3b^3}-\frac{1}{3(b-1)^3}\right)\\
&+\frac{1}{\delta^2}\int^b_{b-1}\,\frac{\mathrm{d}r}{r^4}\,\int^{b-r}_0\,\mathrm{d}v\,\hvOne\\
&+\frac{1}{\delta^2}\int^b_{b-1}\,\frac{\mathrm{d}r}{r^4}\,\int^1_0\,\mathrm{d}v\,\hvTwo h_1(b+1-v-r)\,,
\label{eq:app6dSSL:I3fin}
\eea
where $A$ and $C$ are the constants defined in~\eqref{eq:app6dSSL:AB}.
\par Now we turn to the integral $I_4$. Proceeding in a similar manner, we have
\bea 
I_4&=\frac{(b-1)}{3\delta^2(b+1)^3}-\frac{(b-1)}{3\delta^2 b^3}\\
&+\frac{1}{\delta^2}\int^{b+1}_{b}\,\frac{\mathrm{d}r}{r^2}\Bigg[-\int^1_0\,\mathrm{d}v\left(\frac{h_1^2(v)}{r^2}+\frac{\hvOne\ddhvOne}{2}\right)\\
&+\int^{b+1}_b\,\mathrm{d}v\,\hbvTwo\left(\frac{\hvrOne-\hbvTwo}{r^2}-\frac{\ddhbvTwo}{2}\right)\Bigg]\,,
\eea
where the terms on the first line come from explicitly evaluating the integral that results from the portion of the $v$-integral range $1<v<b$. After changing variables as $v \to b+1-v$ in the last $v$-integral, we can write
\bea 
I_4&=\frac{(b-1)}{3\delta^2(b+1)^3}-\frac{(b-1)}{3\delta^2 b^3}\\
&+\frac{1}{\delta^2}\int^{b+1}_{b}\,\frac{\mathrm{d}r}{r^2}\Bigg[-\int^1_0\,\mathrm{d}v\left(\frac{h_1^2(v)}{r^2}+\frac{\hvOne\ddhvOne}{2}\right)\\
&+\int^{1}_0\,\mathrm{d}v\,\hvTwo\left(\frac{\hbvrOne-\hvTwo}{r^2}-\frac{\ddhvTwo}{2}\right)\Bigg]\,,
\eea
which can be written as
\bea 
I_4&=\frac{(b-1)}{3\delta^2(b+1)^3}-\frac{(b-1)}{3\delta^2 b^3}\\
&-\frac{1}{2\delta^2}\int^{b+1}_{b}\,\frac{\mathrm{d}r}{r^2}\,\int^1_0\,\mathrm{d}v\,\Big(\hvOne\ddhvOne+\hvTwo\ddhvTwo\Big)\\
&+\frac{1}{\delta^2}\int^{b+1}_{b}\,\frac{\mathrm{d}r}{r^4}\,\int^1_0\,\mathrm{d}v\,\Big(-h_1^2(v)-h_2^2(v)+\hvTwo\hbvrOne \Big)\,.
\eea
Evaluating the $r$-integral on the second line, we obtain the result, in terms of the constants $A$ and $C$,
\bea
I_4&=\frac{(b-1)}{3\delta^2(b+1)^3}-\frac{(b-1)}{3\delta^2 b^3}-\frac{A}{\delta^2}\frac{1}{b(b+1)}\\
&-\frac{C}{\delta^2}\left(\frac{1}{3b^3}-\frac{1}{3(b+1)^3}\right)\\
&+\frac{1}{\delta^2}\int^{b+1}_{b}\,\frac{\mathrm{d}r}{r^4}\,\int^1_0\,\mathrm{d}v\,\hvTwo\hbvrOne\,.
\label{eq:app6dSSL:I4fin}
\eea
\par Finally, we look at the integral $I_5$:
\bea 
I_5&=-\frac{(b-1)}{3\delta^2(b+1)^3}\\
&-\frac{1}{\delta^2}\int^{\infty}_{b+1}\,\frac{\mathrm{d}r}{r^2}\,\int^1_0\,\mathrm{d}v\,\left(\frac{h_1^2(v)}{r^2}+\frac{\hvOne\ddhvOne}{2}\right)\\
&-\frac{1}{\delta^2}\int^{\infty}_{b+1}\,\frac{\mathrm{d}r}{r^2}\,\int^{b+1}_b\,\,\mathrm{d}v\,\left(\frac{h_2^2(b+1-v)}{r^2}+\frac{\hbvTwo\ddhbvTwo}{2}\right)\,,
\eea
where the first line comes from explicitly evaluating the integrals that arise from the $1<v<b$ portion of the $v$-integral. After changing variables to $v\to b+1-v$ in the second $v$-integral, we can write
\be 
I_5=-\frac{(b-1)}{3\delta^2(b+1)^3}-\frac{A}{\delta^2}\int^{\infty}_{b+1}\frac{\mathrm{d}r}{r^2}-\frac{C}{\delta^2}\int^{\infty}_{b+1}\frac{\mathrm{d}r}{r^4}\,,
\ee
where $A$ and $C$ are the constants defined in~\eqref{eq:app6dSSL:AB}.
Evaluation of the $r$-integrals yields
\be 
I_5=-\frac{(b-1)}{3\delta^2(b+1)^3}-\frac{A}{\delta^2(b+1)}-\frac{C}{3\delta^2(b+1)^3}\,.
\label{eq:app6dSSL:I5fin}
\ee
\par Combining the integrals $I_{1,2,3,4,5}$ in~\eqref{eq:app6dSSL:I1fin},~\eqref{eq:app6dSSL:I2fin},~\eqref{eq:app6dSSL:I3fin},~\eqref{eq:app6dSSL:I4fin} and~\eqref{eq:app6dSSL:I5fin}, and then performing the cancellations that occur, we find that aside from the terms leading to the $b$-independent constant $\Lambda$, defined below, terms with coefficient $A$ or $C$ completely vanish and we are left with
\bea
I&=\frac{\Lambda}{\delta^2}+\frac{1}{6\delta^2(b-1)^2}-\frac{1}{3\delta^2(b-1)^3}\int^1_0\,\mathrm{d}v\,\Big(\hvOne+\hvTwo\Big)\\
&+\frac{1}{\delta^2}\int^{b}_{b-1}\,\frac{\mathrm{d}r}{r^4}\,\int^{b-r}_0\,\mathrm{d}v\,\hvOne+\frac{1}{\delta^2}\int^{b+1}_{b-1}\,\frac{\mathrm{d}r}{r^4}\,\int^1_0\,\mathrm{d}v\,\hvTwo h_1(b+1-v-r)\,,
\eea
where $\Lambda$ is a constant, independent of $b$, defined by
\bea
\Lambda&=-\frac{1}{2}-A+B-C/3\\
&+\int^{1}_0\,\frac{\mathrm{d}r}{r^2}\Bigg[\int^{1}_0\,\mathrm{d}v\,\hvOne\left(\frac{\hvrOne-\hvOne}{r^2}-\frac{\ddhvOne}{2}\right)\\
&+\int^{1+r}_1\mathrm{d}v\,\frac{\hvrOne-1}{r^2}\\
&+\int^{1}_0\,\mathrm{d}v\,h_2(v)\left(\frac{h_2(v-r)-h_2(v)}{r^2}-\frac{h^{\prime\prime}_2(v)}{2}\right)\Bigg]\,.
\eea
Changing variables to $r\to r-b+1$ in the remaining $r$-integrals gives
\bea
I&=\frac{\Lambda}{\delta^2}+\frac{1}{6\delta^2(b-1)^2}-\frac{1}{3\delta^2(b-1)^3}\int^1_0\,\mathrm{d}v\,\Big(\hvOne+\hvTwo\Big)\\
&+\frac{1}{\delta^2}\int^{1}_{0}\,\frac{\mathrm{d}r}{(r+(b-1))^4}\,\int^{1-r}_0\,\mathrm{d}v\,\hvOne\\
&+\frac{1}{\delta^2}\int^{2}_{0}\,\frac{\mathrm{d}r}{(r+(b-1))^4}\,\int^1_0\,\mathrm{d}v\,\hvTwo h_1(2-v-r)\,.
\eea
Finally, using $1/(b-1)=\delta/\Delta\tau$ we can express $I$ as
\bea
I&=\frac{\Lambda}{\delta^2}+\frac{1}{6\Delta\tau^2}-\frac{\delta}{3\Delta\tau^3}\int^1_0\,\mathrm{d}v\,\Big(\hvOne+\hvTwo\Big)\\
&+\frac{\delta^2}{\Delta\tau^4}\int^1_0\,\frac{\mathrm{d}r}{\left(1+\frac{r\delta}{\Delta\tau}\right)^4}\int^{1-r}_0\,\mathrm{d}v\,\hvOne\\
&+\frac{\delta^2}{\Delta\tau^4}\int^2_0\,\frac{\mathrm{d}r}{\left(1+\frac{r\delta}{\Delta\tau}\right)^4}\,\int^{1}_0\,\mathrm{d}v\,\hvTwo h_1(2-v-r)\\
&=\frac{\Lambda}{\delta^2}+\frac{1}{6\Delta\tau^2}+O(\delta)\,.
\label{eq:app6dSSL:Ifin}
\eea
Hence, in the sharp-switching limit, the derivative with respect to the proper time $\tau$ is 
\be 
\frac{dI}{d\tau}=-\frac{1}{3\Delta\tau^3}+O(\delta)\,.
\label{eq:app6dSSL:dIfin}
\ee
Restoring the pre-factor that we dropped, $1/2\pi^3$, completes our derivation of the result,~\eqref{eq:app6dSSL:term3res}.
\section{Fourth term of~\eqref{eq:6d:detresp}}
Consider the fourth term of~\eqref{eq:6d:detresp}, ignoring the $-1/12\pi^3$ and using the same change of variables that we used for the third term,~\eqref{eq:app6dSSL:ssVarChange}, this term becomes
\bea
&J:=\int^{\infty}_0\,\frac{\mathrm{d}r}{r}\,\int^{\infty}_{-\infty}\,\mathrm{d}v\,\hvOne\hbvTwo\times\\
&\Bigg[\left(\frac{\hvrOne\hbvrTwo-\hvOne\hbvTwo}{r}\right)\ddot{\x}^2\left[(v-1)\delta+\tau_0\right]\\
&-\delta\hvrOne\hbvrTwo\ddot{\x}\left[(v-1)\delta+\tau_0\right]\x^{(3)}\left[(v-1)\delta+\tau_0\right]
\Bigg]\,.
\eea
With this form, we split the $r$-integral as $(0,1),~(1,b-1),~(b-1,b),~(b,b+1),~(b+1,\infty)$, labelling these integrals as $J_{1,2,3,4,5}$ respectively, and we note that the $v$-integral only contributes in the range $(0,b+1)$.
Looking first at $J_1$, we have 
\bea
J_1&=\int^1_0\,\frac{\mathrm{d}r}{r}\,\Bigg[\int^{r+1}_0\,\mathrm{d}v\,\hvOne\Bigg(\frac{\hvrOne-\hvOne}{r}\,\ddot{\x}^2\left[(v-1)\delta+\tau_0\right]\\
&-\delta\hvrOne\ddot{\x}\left[(v-1)\delta+\tau_0\right]\x^{(3)}\left[(v-1)\delta+\tau_0\right]\Bigg)\\
&-\delta\int^b_{r+1}\, \mathrm{d}v\,\xTwoThreeZero\\
&+\int^{b+1}_{b}\,\mathrm{d}v\,\hbvTwo\Bigg(\frac{\hbvrTwo-\hbvTwo}{r}\,\ddot{\x}^2\left[(v-1)\delta+\tau_0\right]\\
&-\delta\hbvrTwo\ddot{\x}\left[(v-1)\delta+\tau_0\right]\x^{(3)}\left[(v-1)\delta+\tau_0\right]\Bigg)\Bigg]\,.
\eea
After a change of variables $v\to b+1-v$ in the last two $v$-integrals and using $b=1+\Delta\tau/\delta$, we can express $J_1$ as
\bea
J_1&=\int^1_0\,\frac{\mathrm{d}r}{r}\,\Bigg[\int^{r+1}_0\,\mathrm{d}v\,\hvOne\Bigg(\frac{\hvrOne-\hvOne}{r}\,\ddot{\x}^2\left[(v-1)\delta+\tau_0\right]\\
&-\delta\hvrOne\ddot{\x}\left[(v-1)\delta+\tau_0\right]\x^{(3)}\left[(v-1)\delta+\tau_0\right]\Bigg)\\
&-\delta\int^{b-r}_{1}\, \mathrm{d}v\,\xTwoThree\\
&+\int^{1}_{0}\,\mathrm{d}v\,\hvTwo\Bigg(\frac{h_2(v+r)-\hvTwo}{r}\,\ddxx\\
&-\delta h_2(v+r)\xTwoThree\Bigg)\Bigg]\,.
\eea
The first $v$-integral will lead to a constant, with respect to $\tau$, so we can safely neglect it because it will not contribute to the transition rate. If we explicitly evaluate the second $v$-integral, we obtain
\bea
J_1&=C_1+\int^1_0\,\frac{\mathrm{d}r}{r}\,\Bigg[\left(-\frac{\ddot{\x}^2(\tau)}{2}\right)\\
&+\int^{1}_{0}\,\mathrm{d} v\,\hvTwo\Bigg(\frac{h_2(v+r)-\hvTwo}{r}\,\ddxx\\
&-\delta h_2(v+r)\xTwoThree\Bigg)\Bigg]\,,
\eea
where $C_1$ is a constant, independent of switch-off time $\tau$. Note that any divergences that appear in $J$ as $r\to 0$ are a result of the range splitting and will cancel when we come to recombine these pieces.
\par We now take the derivative of $J_1$ with respect to the switch-off time $\tau$:
\bea 
\frac{dJ_1}{d\tau}&=-\ddot{\x}(\tau)\x^{(3)}(\tau)\int^1_0\,\frac{\mathrm{d}r}{r}\\
&+\int^1_0\,\frac{\mathrm{d}r}{r}\,\int^1_0\,\mathrm{d}v\,\hvTwo\Bigg[2\frac{h_2(v+r)-\hvTwo}{r}\,\xTwoThree\\
&\qquad\qquad-\delta h_2(v+r)\left(\left(\x^{(3)}[(1-v)\delta+\tau]\right)^2+\ddot{\x}[(1-v)\delta+\tau]\x^{(4)}[(1-v)\delta+\tau]\right)\Bigg]\,,
\eea
before making a small-$\delta$ Taylor expansion, obtaining
\bea 
\frac{dJ_1}{d\tau}&=-\ddot{\x}(\tau)\x^{(3)}(\tau)\int^1_0\,\frac{\mathrm{d}r}{r}\\
&+2\ddot{\x}(\tau)\x^{(3)}(\tau)\int^1_0\,\frac{\mathrm{d}r}{r}\,\int^1_0\,\mathrm{d}v\,\hvTwo \left(\frac{h_2(v+r)-\hvTwo}{r}\right)+O(\delta)\\
&=2\ddot{\x}(\tau)\x^{(3)}(\tau)\int^1_0\,\frac{\mathrm{d}r}{r^2}\,\left[\int^1_0\,\mathrm{d}v\,\hvTwo \Big(h_2(v+r)-\hvTwo\Big)-\frac{r}{2}\right]+O(\delta)\,.
\label{eq:app6dSSL:dJ1}
\eea
The final $r$-integrand of $dJ_1/d\tau$ is finite as $r\to 0$.
\par Proceeding similarly, we now look at the integral $J_2$, which is given by 
\bea 
J_2&=\int^{b-1}_1\,\frac{\mathrm{d} r}{r}\,\Bigg[-\int^1_0\,\mathrm{d}v\,\frac{h^2_1(v)\,\ddxxZero}{r}-\int^r_1\,\mathrm{d} v\,\frac{\ddxxZero}{r}\\
&+\int^{r+1}_r\,\mathrm{d}v\,\Bigg(\frac{\hvrOne-1}{r}\,\ddxxZero\\
&\qquad\qquad\qquad\qquad-\delta\hvrOne\xTwoThreeZero\Bigg)\\
&-\delta\int^b_{r+1}\,\mathrm{d} v\,\xTwoThreeZero\\
&+\int^{b+1}_b\,\mathrm{d} v\,\hbvTwo\Bigg(\frac{1-\hbvTwo}{r}\,\ddxxZero\\
&\quad\quad\quad\quad\quad\quad\quad\quad\quad\quad\quad\quad-\delta\xTwoThreeZero\Bigg)\Bigg]\,.
\eea
First, we perform the $r$-integral that is the coefficient of the first $v$-integral. Next, we change variables as $v\to v-r$ in the third $v$-integral. Then we explicitly perform the fourth $v$-integral. Finally, in the last $v$-integral, we first change variables as $v\to b+1-v$, before performing the $r$-integral that is the coefficient of the last $v$-integral. This procedure results in an expression for $J_2$ of the form
\bea 
&J_2=-\frac{(b-2)}{(b-1)}\int^1_0\,\mathrm{d} v\,h^2_1(v)\ddxxZero-\int^{b-1}_1\,\frac{\mathrm{d} r}{r^2}\,\int^r_1\,\mathrm{d} v\,\ddxxZero\\
&+\int^{b-1}_1\,\frac{\mathrm{d} r}{r}\,\int^{1}_0\,\mathrm{d} v\,\Bigg(\frac{\hvOne-1}{r}\,\ddot{\x}^2\left[(v+r-1)\delta+\tau_0\right]\\
&\quad\quad\quad\quad-\delta\hvOne\ddot{\x}\left[(v+r-1)\delta+\tau_0\right]\x^{(3)}\left[(v+r-1)\delta+\tau_0\right]\Bigg)\\
&-\int^{b-1}_1\,\frac{\mathrm{d} r}{r}\,\left(\frac{\ddot{\x}^2(\tau)}{2}-\frac{\ddot{\x}^2(r\delta+\tau_0)}{2}\right)\\
&+\frac{(b-2)}{(b-1)}\int^1_0\,\mathrm{d} v\, \hvTwo\Big(1-\hvTwo\Big)\ddxx\\
&-\delta\log{(b-1)}\int^1_0\,\mathrm{d} v\,\hvTwo\xTwoThree\,.
\eea
Taking the derivative of $J_2$ with respect to $\tau$ and using $b=1+\Delta\tau/\delta$ gives
\bea
&\frac{dJ_2}{d\tau}=-\frac{\delta}{\Delta\tau^2}\int^1_0\,\mathrm{d} v\,h^2_1(v)\,\ddxxZero-\frac{\delta}{\Delta\tau^2}\int^{\Delta\tau/\delta}_1\,\mathrm{d} v\,\ddxxZero\\
&+\frac{\delta}{\Delta\tau}\int^1_0\,\mathrm{d} v\,\Bigg(\frac{\hvOne-1}{\Delta\tau}\,\ddot{\x}^2\left[(v-1)\delta+\tau\right]-\hvOne\ddot{\x}\left[(v-1)\delta+\tau\right]\x^{(3)}\left[(v-1)\delta+\tau\right]\Bigg)\\
&-\ddot{\x}(\tau)\x^{(3)}(\tau)\log{\left(\frac{\Delta\tau}{\delta}\right)}+\left(\frac{\delta}{\Delta\tau^2}\right)\int^{1}_{0}\,\mathrm{d}v\,\hvTwo\Big(1-\hvTwo\Big)\ddot{\x}^2[(1-v)\delta+\tau]\\
&+2\left(1-\frac{\delta}{\Delta\tau}\right)\int^{1}_{0}\,\mathrm{d}v\, \hvTwo\Big(1-\hvTwo\Big)\,\ddot{\x}[(1-v)\delta+\tau]\x^{(3)}[(1-v)\delta+\tau]\\
&-\frac{\delta}{\Delta\tau}\int^1_0\,\mathrm{d} v\, \hvTwo\ddot{\x}[(1-v)\delta+\tau]\x^{(3)}[(1-v)\delta+\tau]\\
&-\delta\log{\left(\frac{\Delta\tau}{\delta}\right)}\int^1_0\,\mathrm{d} v\,\hvTwo\,\left(\left(\x^{(3)}[(1-v)\delta+\tau]\right)^2+\ddot{\x}[(1-v)\delta+\tau]\x^{(4)}[(1-v)\delta+\tau]\right)\,.
\eea
We now perform a small-$\delta$ Taylor expansion of the derivative, $dJ_2/d\tau$, to obtain 
\bea 
\frac{dJ_2}{d\tau}&=-\ddot{\x}(\tau)\x^{(3)}(\tau)\log{\left(\frac{\Delta\tau}{\delta}\right)}-\frac{1}{\Delta\tau^2}\int^{\tau}_{\tau_0}\,\mathrm{d} u\,\ddot{\x}^2(u)\\
&+2\ddot{\x}(\tau)\x^{(3)}(\tau)\int^1_0\,\mathrm{d} v\,\hvTwo\Big(1-\hvTwo\Big)+O\left(\delta\log{\left(\frac{1}{\delta}\right)}\right)\,.
\label{eq:app6dSSL:dJ2}
\eea
\par Next, we turn to the integral $J_3$:
\bea 
&J_3=\int^b_{b-1}\,\frac{\mathrm{d} r}{r}\,\Bigg[-\int^1_0\,\mathrm{d} v\,\frac{h_1^2(v)\,\ddxxZero}{r}-\int^r_1\,\mathrm{d} v\,\frac{\ddxxZero}{r} \\
&+\int^b_r\,\mathrm{d} v \,\Bigg( \frac{\hvrOne-1}{r}\ddxxZero\\
&\qquad\qquad\qquad\qquad-\delta\hvrOne\xTwoThreeZero \Bigg)\\
&+\int^{r+1}_b\,\mathrm{d} v\, \hbvTwo\Bigg(\frac{\hvrOne-\hbvTwo}{r}\ddxxZero\\
&\qquad\qquad\qquad\qquad-\delta\hvrOne\xTwoThreeZero\Bigg)\\
&+\int^{b+1}_{r+1}\,\mathrm{d} v\,\hbvTwo\Bigg(\frac{1-\hbvTwo}{r}\ddxxZero\\
&\quad\quad\quad\quad\quad\quad\quad\qquad\qquad\quad\quad\quad\quad\quad-\delta\,\xTwoThreeZero\Bigg)\Bigg]\,.
\eea
We change variables in the $r$-integral to $r\to r-b+1$, and in the last three $v$-integrals we change variables to $v \to b+1-v$. This leads to
\bea 
J_3&=\int^1_{0}\,\frac{\mathrm{d} r}{r+b-1}\,\Bigg[-\int^1_0\,\mathrm{d} v\,\frac{h_1^2(v)\ddxxZero}{r+b-1}-\int^{(r+b-1)}_1\,\mathrm{d} v\,\frac{\ddxxZero}{(r+b-1)} \\
&+\int^{2-r}_1\,\mathrm{d} v \,\Bigg( \frac{h_1(2-v-r)-1}{(r+b-1)}\ddot{\x}^2[(1-v)\delta+\tau]\\
&\qquad\qquad\qquad-\delta h_1(2-v-r)\ddot{\x}[(1-v)\delta+\tau]\x^{(3)}[(1-v)\delta+\tau] \Bigg)\\
&+\int^{1}_{1-r}\,\mathrm{d} v\, \hvTwo\Bigg(\frac{h_1(2-r-v)-\hvTwo}{(r+b-1)}\ddxx\\
&\quad\quad\quad\quad\quad\qquad\qquad\quad\quad\quad\quad\quad-\delta h_1(2-r-v)\xTwoThree\Bigg)\\
&+\int^{1-r}_{0}\,\mathrm{d} v\,\hvTwo\Bigg(\frac{1-\hvTwo}{(r+b-1)}\ddxx\\
&\quad\quad\quad\quad\quad\quad\quad\qquad\qquad\quad\quad\quad\quad\quad-\delta\,\xTwoThree\Bigg)\Bigg]\,.
\eea
If we differentiate $J_3$ with respect to the switch-off time $\tau$ and then perform a small-$\delta$ Taylor expansion, this leads to 
\be 
\frac{dJ_3}{d\tau}=O\left(\delta\right)\,.
\label{eq:app6dSSL:dJ3}
\ee
%
%
\par Next, we consider the integral labelled $J_4$, which reads
\bea
J_4&=\int^{b+1}_b\,\frac{\mathrm{d} r}{r}\,\Bigg[-\int^b_0\,\mathrm{d} v\,\frac{h_1^2(v)\,\ddxxZero}{r}\\
&+\int^{b+1}_b\,\mathrm{d} v\,\hbvTwo\Bigg(\frac{\hvrOne-\hbvTwo}{r}\ddxxZero\\
&\qquad\qquad\qquad\qquad-\delta\hvrOne\xTwoThreeZero\Bigg)\Bigg]\,.
\eea
First, we evaluate the $r$-integral associated with the first $v$-integral and change variables to $v\to b+1-v$ in the final $v$-integral, along with $r\to r-b+1$ in its associated $r$-integral, to obtain
\bea
J_4&=-\frac{1}{b(b+1)}\int^b_0\,\mathrm{d} v\,h_1^2(v)\,\ddxxZero\\
&+\int^2_1\,\frac{\mathrm{d} r}{(r+b-1)}\int^{1}_0\,\mathrm{d} v\,\hvTwo\Bigg(\frac{h_1(2-v-r)-\hvTwo}{(r+b-1)}\ddxx\\
&\qquad\qquad\qquad\qquad-\delta h_1(2-v-r)\xTwoThree\Bigg)\,.
\eea
Differentiation with respect to the switch-off time, followed by a small-$\delta$ Taylor expansion, leads to
\be
\frac{d J_4}{d\tau}=O\left(\delta\right)\,.
\label{eq:app6dSSL:dJ4}
\ee
\par Finally, we analyse the integral $J_5$:
\bea 
J_5&=\int^{\infty}_{b+1}\,\frac{\mathrm{d} r}{r}\,\Bigg[-\int^1_0\,\frac{h_1^2(v)\,\ddxxZero}{r}-\int^b_1\,\mathrm{d} v\,\frac{\ddxxZero}{r}\\
&-\int^{b+1}_b\,\mathrm{d} v\, \frac{h_2^2(b+1-v)\,\ddxxZero}{r}\Bigg]\,,
\eea
for which we evaluate the $r$-integral and change variables to $v \to b+1-v$ in the final $v$-integral. This results in 
\bea 
J_5&=-\frac{1}{(b+1)} \Bigg[\int^1_0\,\mathrm{d} v\,h_1^2(v)\,\ddxxZero+\int^b_1\,\mathrm{d} v\,\ddxxZero\\
&+\int^{1}_0\,\mathrm{d} v\, h_2^2(v)\,\ddxx\Bigg]\,.
\eea
Differentiating $J_5$ and making use of the change of variables $ v \to v\delta-\delta+\tau_0$, we find
\bea 
\frac{d J_5}{d\tau}&=\frac{\delta}{\Delta\tau^2}\frac{1}{\left(1+\frac{2\delta}{\Delta\tau}\right)^2}\int^1_0\,\mathrm{d} v\,h_1^2(v)\,\ddxxZero\\
&-\frac{\ddot{\x}^2(\tau)}{\Delta\tau}+\frac{1}{\Delta\tau^2}\frac{1}{\left(1+\frac{2\delta}{\Delta\tau}\right)^2}\int^{\tau-\delta}_{\tau_0}\,\mathrm{d} v\,\ddxxZero\\
&+\frac{\delta}{\Delta\tau^2}\frac{1}{\left(1+\frac{2\delta}{\Delta\tau}\right)^2}\int^1_0\,\mathrm{d} v\, h_2^2(v)\,\ddxx\\
&-\frac{2\delta}{\Delta\tau}\frac{1}{\left(1+\frac{2\delta}{\Delta\tau}\right)}\int^1_0\,\mathrm{d} v\,h_2^2(v)\,\xTwoThree\,,
\label{eq:app6dSSL:dJ5preexp}
\eea
and Taylor expanding~\eqref{eq:app6dSSL:dJ5preexp} in the parameter $\delta$, which we take to zero in the sharp-switching limit, gives
\bea
\frac{dJ_5}{\delta\tau}=-\frac{\ddot{\x}^2(\tau)}{\Delta\tau}+\frac{1}{\Delta\tau^2}\int^{\tau}_{\tau_0}\,\mathrm{d} u\,\ddot{\x}^2(u)+O(\delta)\,.
\label{eq:app6dSSL:dJ5}
\eea
We are now in a position to combine equations~\eqref{eq:app6dSSL:dJ1}, ~\eqref{eq:app6dSSL:dJ2}, ~\eqref{eq:app6dSSL:dJ3}, ~\eqref{eq:app6dSSL:dJ4} and~\eqref{eq:app6dSSL:dJ5}. The result is
\bea
\frac{dJ}{d\tau}&=2\ddot{\x}(\tau)\x^{(3)}(\tau)\int^1_0\,\frac{\mathrm{d} r}{r^2}\,\left[\int^1_0\,\mathrm{d} v\,\hvTwo \left(h_2(v+r)-\hvTwo\right)-\frac{r}{2}\right]\\
&-\ddot{\x}(\tau)\x^{(3)}(\tau)\log{\left(\frac{\Delta\tau}{\delta}\right)}-\frac{\ddot{\x}^2(\tau)}{\Delta\tau}\\
&+2\ddot{\x}(\tau)\x^{(3)}(\tau)\int^1_0\,\mathrm{d}v\,\hvTwo\Big(1-\hvTwo\Big)+O\left(\delta\log{\left(\frac{\Delta\tau}{\delta}\right)}\right)\,.
\eea
Restoring the pre-factor $-1/12\pi^3$ that we dropped, we obtain the desired result,~\eqref{eq:app6dSSL:term4res}.
\chapter{Small-$s$ convergence of groupings in~\eqref{eq:6d:detresp}}
\label{ch:appendix:6dgroups}
In this appendix, we show an example of how the groupings in~\eqref{eq:6d:detresp} lead to well defined expressions in the small-$s$ limit. The regularity of the other terms in this limit follows similarly.
\par Consider the integral with the $-E^2/4\pi^3$ coefficient in~\eqref{eq:6d:detresp}:
\bea
&\int^{\infty}_{0}\,\mathrm{d}s\,\frac{1}{s^2}\int^{\infty}_{-\infty}\,\mathrm{d}u\, \chi(u)\left[\chi(u-s)-\chi(u)\right]\\
&=\int^{\infty}_{0}\,\mathrm{d}s\,\frac{F(s)}{s^2}\,,
\label{eq:6dapp:E2term}
\eea
where this coefficient has been suppressed, and where we have defined
\be
F(s):=\int^{\infty}_{-\infty}\,\mathrm{d}u\, \chi(u)\left[\chi(u-s)-\chi(u)\right]\,.
\ee
The first and second derivatives of $F(s)$ with respect to $s$ are
\bea
F^{\prime}(s)&=-\int^{\infty}_{-\infty}\,\mathrm{d}u\, \chi(u)\frac{d\chi(u-s)}{du}\,,\\
F^{\prime\prime}(s)&=\int^{\infty}_{-\infty}\,\mathrm{d}u\, \chi(u)\frac{d^2\chi(u-s)}{du^2}\,.
\eea
Hence,
\bea 
F(0)&=0\,,\\
F^{\prime}(0)&=-\int^{\infty}_{-\infty}\,\mathrm{d}u\, \chi(u)\frac{d\chi(u)}{du}=-\left[\frac{\chi^2(u)}{2}\right]^{\infty}_{-\infty}=0\,,\\
F^{\prime\prime}(0)&=\int^{\infty}_{-\infty}\,\mathrm{d}u\, \chi(u)\frac{d^2\chi(u)}{du^2}=-\int^{\infty}_{-\infty}\,\mathrm{d}u\,\dot{\chi}^2(u)\,,
\eea
where the final equalities are obtained after integrating by parts and using the compact support of $\chi$.
Therefore, the small-$s$ Taylor expansion of $F(s)$ has the form
\be 
F(s)=-\frac{s^2}{2}\int^{\infty}_{-\infty}\,\mathrm{d}u\,\dot{\chi}^2+O\left(s^4\right)\,,
\ee
which shows that the integral over $s$ in~\eqref{eq:6dapp:E2term} converges at small $s$. Convergence at large $s$ follows because $F(s)$ vanishes for sufficiently large $s$, by the compact support of $\chi$.
\par  The other groupings in~\eqref{eq:6d:detresp} can be shown to be non-divergent in a similar manner. 
\chapter{Evaluation of the integral~\eqref{eq:3d:lowerevenfinal}}
\label{ch:appendix:3dint}
In this appendix, we use a series of variable changes to evaluate the integral~\eqref{eq:3d:lowerevenfinal}, encountered in Chapter~\ref{ch:by4d}. Ignoring the $\chi(u)$ coefficient, this integral reads
\be
I:=\int^{\infty}_0\,\mathrm{d}r\,\sqrt{\frac{N}{P}}\,, 
\label{eq:App3dMinkInt:I}
\ee
where
\be
P:=1+2Ar^2+r^4\,,
\ee
with $A\geq 1$, and
%
\be 
N:=\sqrt{P}+1-r^2\,.
\ee
The result we shall establish is
\be 
I=\frac{\pi}{\sqrt{2}}\,.
\label{eq:App3dMinkInt:I-fin}
\ee
\par We begin by writing integral~\eqref{eq:App3dMinkInt:I} as
\be 
I=\sqrt{2(A+1)}\int^{\infty}_0\,\frac{r}{\sqrt{\sqrt{P}-(1-r^2)}}\frac{\,\mathrm{d}r}{\sqrt{P}}\,,
\ee
and we then change variables to $s=r^2$ to give
\be 
I=\sqrt{\frac{(A+1)}{2}}\int^{\infty}_0\,\frac{1}{\sqrt{\sqrt{P}+s-1}}\frac{\mathrm{d}s}{\sqrt{P}}\,,
\ee
where now $P=1+2As+s^2$. Using the fact that $A$ is a positive, real constant, we can express it in terms of a parameter $\alpha>0$ as $A=\cosh{(\alpha)}$, leading to
\be 
P=(s+\expo^{\alpha})(s+\expo^{-\alpha})\,.
\ee
Next we write
\bea 
s=u\sinh{(\alpha)}-\cosh{(\alpha)}\,
\eea
in order to express $P$ as
\bea
P&=\Big(u\sinh{(\alpha)}+\sinh{(\alpha)}\Big)\Big(u\sinh{(\alpha)}-\sinh{(\alpha)}\Big)\\
&=\sinh^2{(\alpha)}\left(u^2-1\right)\,.
\eea
Another change of variables to $u=\cosh{(v)}$ allows us express the integral as
\be 
I=\sqrt{\frac{(A+1)}{2}}\int^{\infty}_{v_0}\,\frac{\mathrm{d}v}{\sqrt{\sinh{(\alpha)}\expo^v-(1+\cosh{(\alpha)})}}\,,
\ee
where $v_0=\arccosh{\left(\coth{(\alpha)}\right)}$.
\par If we now use the change of variables $v=v_0+w$ and the fact that
\be 
\expo^{v_0}=\frac{\cosh{(\alpha)}+1}{\sinh{(\alpha)}}\,
\ee
then the integral collapses to the simple form
\be 
I=\frac{1}{\sqrt{2}}\int^{\infty}_0\,\frac{\mathrm{d}w}{\sqrt{\expo^w-1}}\,.
\ee
This form is easily evaluated by making a final change of variables $t=\sqrt{\expo^{w}-1}$, which leads to the standard integral
\be
I=\sqrt{2}\int^{\infty}_0\,\frac{\mathrm{d}t}{t^2+1}\,,
\ee
the evaluation of which establishes~\eqref{eq:App3dMinkInt:I-fin}.
\chapter{Expressions used in BTZ}
\label{ch:appendixBTZ}

In this appendix, we calculate of some of the expressions
appearing in Chapter~\ref{ch:btz}.

\section{Derivation of \eqref{eq:corot:rate.evald.alt} and \eqref{eq:corot:rate.evald}}
\label{app:A} 

In this appendix, we verify the passage from 
\eqref{eq:corot:rate} to \eqref{eq:corot:rate.evald.alt} 
and~\eqref{eq:corot:rate.evald}. 

\subsection{$n=0$ term}

Let 
\begin{align}
I(a,P) := 
\Realpart
\int_0^\infty \frac{\mathrm{e}^{-iax} \, \mathrm{d}x}{\sqrt{P - \sinh^2 \!x}} \ , 
\label{eq:app:int:Idef}
\end{align}
where $a\in\BbbR$, $P\ge0$, and the square root is 
positive for positive argument and positive imaginary for negative argument. 
We shall show that 
\begin{subequations}
\label{eq:both-singfree}
\begin{align}
I(a,0) &= 
- \frac{\pi \tanh(\pi a/2)}{2}
\ , 
\label{eq:00singfree}
\\[1ex]
I(a,P) &= 
\mathrm{e}^{-\pi a/2}
\int_0^\infty \frac{\cos(ay) \, \mathrm{d}y}{\sqrt{P + \cosh^2 \!y}}
\hspace{5ex} \text{for $P>0$}\ . 
\label{eq:singfree}
\end{align}
\end{subequations}
Applying 
\eqref{eq:both-singfree}
and 
\eqref{eq:00singfree.alt}
to the $n=0$ term in 
\eqref{eq:corot:rate} 
yields the corresponding terms 
in \eqref{eq:corot:rate.evald.alt} 
and~\eqref{eq:corot:rate.evald}. 

\begin{figure}\centering
\tikzset{-->-/.style={decoration={
  markings,
  mark=at position #1 with {\arrow[scale=2]{>}}},postaction={decorate}}}

\begin{tikzpicture}[scale=3]
\node (x)    at (1.5,0)  [label={[label distance=0.001cm]-135:$C_1$}]   {};

\node (orig)   at (0,0)  [label={[label distance=0.001cm]-135:O}]    {};  
\node (tr)  at (3.0,0)   {};
\node (br)  at  (3,-1.0)   {};
\node (bl)  at  (0,-1.0)   {};
\node (C2)  at  (0,-.5)   []  {};
\node (C3)  at  (1.5,-1)   []  {};

\path  
  (x) +(180:1mm)  coordinate   (xl)
       +(0:1mm) coordinate  (xr)
  (xr) +(0:1)  coordinate  (TopZigl)
  (TopZigl) +(0:2mm) coordinate  (TopZigr)
  (br) +(180:0.2)  coordinate  (BottZigr)
  (BottZigr) +(180:2mm) coordinate (BottZigl)
        ;      
     
\draw[-triangle 90] (-0.1,0) -- (3.5,0);
\draw[-triangle 90] (0,-1.5) -- (0,0.5);

\node[below=0.2cm] at (3.5,0) {Re(z)};
\node[left=0.2cm] at (0,0.5) {Im(z)};

\foreach \y/\ytext in {-1/{-\frac{i\pi}{2}}} \draw (1pt,\y cm) -- (-1pt,\y cm) node[anchor=east] {$\ytext$};

\draw [-->-=.75,thick] (orig.center)--(xl);
\draw[-->-=.5,thick] (xl) arc (-180:0: 1mm);
\draw [-->-=.75,thick] (xr)--(TopZigl);
\draw[decorate,decoration=zigzag, thick] (TopZigl)--(TopZigr) ; 
\draw [thick] (TopZigr)--(tr.center);
\draw [-->-=.5, dashed] (tr.center)--(br.center);
\draw [thick] (br.center)--(BottZigr);
\draw[decorate,decoration=zigzag, thick] (BottZigr)--(BottZigl);
\draw [-->-=.5, thick] (BottZigl)--(bl.center);
\draw [-->-=.5, thick] (bl.center)--(orig.center);

\filldraw [gray] (x) circle (0.5pt)
                  ;
\end{tikzpicture}
\caption{Contour deformation made in the evaluation of~\eqref{eq:app:int:Idef} when $P>0$.}  
\label{fig:AppBTZA:Contour}
\end{figure}
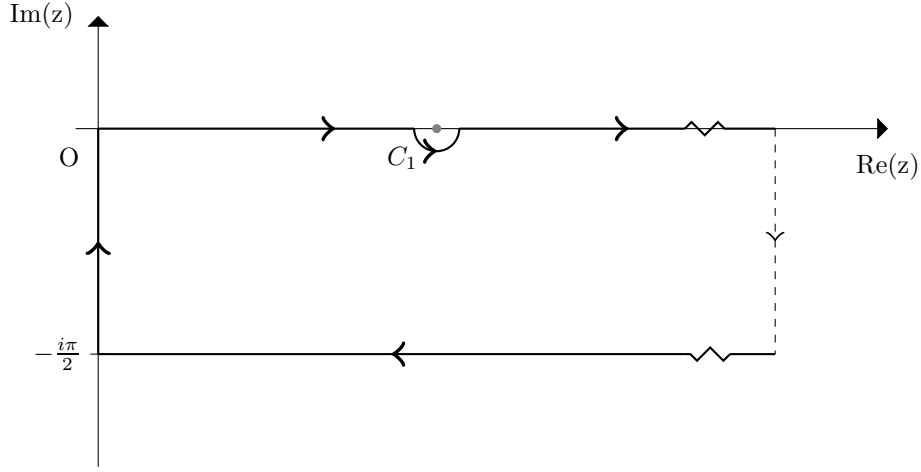

Suppose first $P=0$. For $P=0$,  
\eqref{eq:app:int:Idef} reduces to 
$
I(a,0) = 
- \int_0^\infty  
\sin(ax)/\sinh x 
$, 
which evaluates to \eqref{eq:00singfree} (3.981.1)~\cite{gradshteyn}. 

We note in passing the relation 
\begin{align}
I(a,0) = 
-\frac{\pi}{2}
+ 
\mathrm{e}^{-\pi a/2}
\int_0^\infty \frac{\cos(ay) \, \mathrm{d}y}{\cosh y}
\ , 
\label{eq:00singfree.alt}
\end{align}
which follows by 
evaluating the 
integral in \eqref{eq:00singfree.alt} (3.981.2)~\cite{gradshteyn} and using~\eqref{eq:00singfree}. 
Comparison of 
\eqref{eq:singfree} and \eqref{eq:00singfree.alt} shows that 
$I(a,P)$ is not continuous at $P=0$. 


Suppose then $P>0$. We rewrite \eqref{eq:app:int:Idef} as the contour integral 
\begin{align}
I(a,P) := 
\Realpart \int_{C_1} \frac{\mathrm{e}^{-iaz} \, \mathrm{d}z}{\sqrt{P - \sinh^2 \!z}}
\ , 
\end{align}
where the contour $C_1$ goes from $z=0$ to $z=\infty$ along the positive real axis, 
with a dip in the lower half-plane near the branch point $z=\arcsinh\sqrt{P}$. 
The square root denotes the branch that is positive for small, positive~$z$.

We deform $C_1$ into the union of $C_2$ and~$C_3$, where 
$C_2$ goes from $z=0$ to $z= -i\pi/2$
along the negative imaginary axis 
and 
$C_3$ consists of the half-line 
$z=y-i\pi/2$ with 
$0\le y < \infty$, as shown in Figure~\ref{fig:AppBTZA:Contour}.
Owing to the integrand having no singularities within the strip $- \pi/2 \le \Imagpart z <0$ 
and to the fact that it falls off exponentially within this strip as $\Realpart z \to +\infty$, 
the deformation does not change the value of the integral. 
The contribution from $C_2$ is purely imaginary and 
vanishes on taking the real part. 
The contribution from $C_3$ yields~\eqref{eq:singfree}. 



\subsection{$n\ne0$ terms}

Let 
\begin{align}
I_b(a,P) := \Realpart 
\int_0^\infty 
\frac{\mathrm{e}^{-iax} \, \mathrm{d}x}{\sqrt{P - \sinh^2 (x+b)}} \ , 
\label{eq:bprotoint}
\end{align}
where $a\in\BbbR$, $P>0$, $b \in \BbbR$ and the square root is 
positive for positive argument and analytically continued to 
negative values of the argument by giving $x$ a small, negative imaginary part. 

We shall show that 
\begin{align}
I_b(a,P) + I_{-b}(a,P)
=2\cos{(ab)} \,  I(a,P) \ , 
\label{eq:app:int:pairfinal}
\end{align}
where $I(a,P)$ is given in~\eqref{eq:singfree}. 
Applying 
\eqref{eq:app:int:pairfinal}
with 
\eqref{eq:singfree}
to the $n\ne0$ terms in 
\eqref{eq:corot:rate} 
yields the corresponding terms 
in \eqref{eq:corot:rate.evald.alt}
and~\eqref{eq:corot:rate.evald}. 

For $b=0$, \eqref{eq:app:int:pairfinal} follows from~\eqref{eq:singfree}. 
Both sides of \eqref{eq:app:int:pairfinal} are even in~$b$, and it hence suffices to 
consider \eqref{eq:app:int:pairfinal} for $b>0$. 

Let $b>0$. Changing the integration variable in \eqref{eq:bprotoint} to $y = x+b$ yields 
\begin{align}
I_b(a,P) + I_{-b}(a,P)
& =2\cos{(ab)}\Realpart\int^{\infty}_{0} \frac{\mathrm{e}^{-iay} 
\,\mathrm{d}y}{\sqrt{P-\sinh^2 \! y}}
\nonumber
\\[1ex]
& \hspace{1ex}
-\Realpart\left(
\mathrm{e}^{iab}\int^{b}_{0} \frac{\mathrm{e}^{-iay} \,
\mathrm{d}y}{\sqrt{P-\sinh^2 \! y}}
+\mathrm{e}^{-iab}\int^{-b}_{0} \frac{\mathrm{e}^{-iay} \,
\mathrm{d}y}{\sqrt{P-\sinh^2 \! y}}
\right) , 
\label{eq:app:int:pair}
\end{align}
where the branches of the square roots
are as inherited from~\eqref{eq:bprotoint}: 
positive when the argument is positive and continued to negative argument 
by giving $y$ a small, negative imaginary part. 
As we shall now show, examination of the branches shows that the last two terms in \eqref{eq:app:int:pair}
cancel on taking the real part.
Defining 
\begin{align}
S_b(a,P) = \Realpart\left(
\mathrm{e}^{iab}\int^{b}_{0} \frac{\mathrm{e}^{-iay} \,
\mathrm{d}y}{\sqrt{P-\sinh^2 \! (y-i\epsilon)}}
+\mathrm{e}^{-iab}\int^{-b}_{0} \frac{\mathrm{e}^{-iay} \,
\mathrm{d}y}{\sqrt{P-\sinh^2 \! (y-i\epsilon)}}
\right) , 
\label{eq:app:int:Salt1}
\end{align}
where we have been explicit about the placement of the regulator, and then changing
variables to $y=-x$ in the second integral gives
\begin{align}
S_b(a,P) = \Realpart\left(
\mathrm{e}^{iab}\int^{b}_{0} \frac{\mathrm{e}^{-iay} \,
\mathrm{d}y}{\sqrt{P-\sinh^2 \! (y-i\epsilon)}}
-\mathrm{e}^{-iab}\int^{b}_{0} \frac{\mathrm{e}^{iay} \,
\mathrm{d}y}{\sqrt{P-\sinh^2 \! (y+i\epsilon)}}
\right) . 
\label{eq:app:int:Salt2}
\end{align}
If we now denote the singularity in the denominator by $y_0:=\arcsinh{\left(\sqrt{P}\right)}$, then for the case that $b<y_0$ the argument of the square root is uniformly positive across the range of integration and~\eqref{eq:app:int:Salt2} vanishes upon taking the real part. If $b>y_0$ then the portion of the integral for which $y<y_0$ once more vanishes on taking the real part and the remaining part of $S_b(a,P)$ is
\begin{align}
S_b(a,P) &=\Realpart\left(
\mathrm{e}^{iab}\int^{b}_{y_0} \frac{\mathrm{e}^{-iay} \,
\mathrm{d}y}{i\sqrt{\sinh^2 \! y-P}}
-\mathrm{e}^{-iab}\int^{b}_{y_0} \frac{\mathrm{e}^{iay} \,
\mathrm{d}y}{-i\sqrt{\sinh^2 \! y-P}}
\right)\nonumber \\
&=\Realpart\left(-2i\int^{b}_{y_0} \frac{\mathrm{d}y}{\sqrt{\sinh^2 \! y-P}} \cos{\left[a\left(b-y\right)\right]}
\right) \\
&=0\,.
\label{eq:app:int:Salt3}
\end{align}
Using \eqref{eq:app:int:Idef} in the first term of~\eqref{eq:app:int:pair} leads to~\eqref{eq:app:int:pairfinal}.

\section{Derivation of \eqref{eq:corot:larger+}}
\label{app:A2}

In this appendix, we verify the asymptotic formula~\eqref{eq:corot:larger+}. 

Let 
\begin{align}
\label{eq:J-def}
J(a,P) := \int_{0}^{\infty} \frac{\cos(ay) \, \mathrm{d}y}{\sqrt{P + \cosh^2 \! y}}
\ , 
\end{align}
where $P>0$ and $a\in\BbbR$. Note from \eqref{eq:singfree} that 
$I(a,P) = \mathrm{e}^{-\pi a/2}J(a,P)$ for $P>0$. 
We shall show that as $P\to\infty$ with fixed~$a$, 
$J(a,P)$ has the asymptotic form 
\begin{subequations}
\begin{align}
J(a,P) &= 
\frac{1}{a \sqrt{\pi P}} 
\Imagpart \bigl[
{(4P)}^{ia/2}
\Gamma(1 + ia/2) \Gamma(\tfrac12 - ia/2) 
\bigr]
+ O\bigl(P^{-3/2}\bigr)
\hspace{3ex}
\text{for $a\ne0$}\,, 
\label{eq:Jall3}
\\[1ex]
J(0,P) &= 
\frac{1}{2\sqrt{P}} 
\left[
\ln(4P)
+ \psi(1) - \psi(\tfrac12)
\right]
+ O\bigl(P^{-3/2} \ln P\bigr)
\ , 
\label{eq:Jall30}
\end{align}
\end{subequations}
where $\psi$ is the 
digamma function~\cite{dlmf}. 

Starting from~\eqref{eq:J-def}, writing 
$\cos(ay) = \Realpart (\mathrm{e}^{iay})$ and making the substitution 
$y = \ln t$, we find 
\begin{align}
J(a,P) &= 2 \Realpart \int_{1}^{\infty}  
\frac{t^{ia} \, \mathrm{d}t}
{\sqrt{t^4 + B^2 t^2} \, {\displaystyle\sqrt{1 + \frac{1}{t^4 + B^2 t^2}}}}
\nonumber 
\\[1ex]
&= 2 \sum_{p=0}^\infty
b_p \Realpart \int_{1}^{\infty}  
\frac{t^{ia} \, \mathrm{d}t}
{t^{2p+1}{\bigl(t^2 + B^2\bigr)}^{p + (1/2)}}
\ , 
\label{eq:t-powers}
\end{align}
where $B = \sqrt{4 P + 2}$ and 
$b_p$ are the coefficients in the binomial expansion 
${(1+x)}^{-1/2} = \sum_{p=0}^\infty b_p x^p$. 
Because the $p>0$ terms in \eqref{eq:t-powers} are 
$O\bigl(B^{-2p - 1}\bigr) = O\bigl(P^{-p - (1/2)}\bigr)$ 
by dominated convergence, we have $J(a,P) = J_0(a,P) + O\bigl(P^{-3/2}\bigr)$, 
where the substitution $t = B v$ in the $p=0$ terms gives 
\begin{align}
J_0(a) &= \frac{2}{B} 
\Realpart \left(
B^{ia}
\int_{1/B}^{\infty}  
\frac{v^{ia-1} \, \mathrm{d}v}
{\sqrt{1+v^2}}
\right) \ . 
\label{eq:Jnought1}
\end{align}

When $a\ne0$, integrating \eqref{eq:Jnought1} by parts and 
extending the lower limit of the integral to zero gives 
\begin{align}
J_0(a,P) &= \frac{2}{Ba} 
\Imagpart \left[
B^{ia}
\int_{0}^{\infty} 
\frac{v^{1+ia} \, dv}
{{\bigl(1+v^2\bigr)}^{3/2}}
+ O\bigl(B^{-2}\bigr)\right]
\ . 
\label{eq:Jnought2}
\end{align}
The integral in \eqref{eq:Jnought2}
may be evaluated by writing
\be 
{(1+v^2)}^{-3/2} = {\bigl(\Gamma(3/2)\bigr)}^{-1}\int_0^\infty
\mathrm{d} y \, y^{1/2} \, \expo^{-(1+v^2)y}
\ee
and interchanging the order of the integrals, 
with the result~\eqref{eq:Jall3}. 
When $a=0$, similar manipulations lead to~\eqref{eq:Jall30}.

\section{Derivation of \eqref{eq:corot:smallr+}}
\label{app:C}

Let $p>0$, $q>0$, 
$a\in\BbbR$ and 
$\gamma \in\BbbR$. For $n\in\BbbZ$, let 
$K_n := p^2 \sinh^2 (nq)$, and define 
\begin{align}
F_n := \int_{0}^{\infty} 
\frac{\cos{(n\gamma q)}\cos(ay) \, \mathrm{d}y}{\sqrt{K_n + \cosh^2 \! y}} 
\ , 
\label{eq:Fn}
\end{align}
where we suppress the dependence of $F_n$ on $p$, $q$, 
$a$ and~$\gamma$. We shall show that the sum 
$S := \sum_{n=-\infty}^\infty F_n$
has the asymptotic form 
\begin{align}
S = 
\frac{2}{q}
\int_{0}^{\infty} \mathrm{d}r 
\int_{0}^{\infty} 
\frac{\cos{(r\gamma)}\cos(ay) \, \mathrm{d}y}{\sqrt{p^2 \sinh^2\!r + \cosh^2 \! y}} 
\ + \frac{o(1)}{q}
\label{eq:Fint}
\end{align}
as $q\to0$ with the other parameters fixed. 
Note that the leading term in \eqref{eq:Fint} 
diverges as $q\to0$. 

Let 
\begin{align}
G(r) := \cos(\gamma r) 
\int_{0}^{\infty} 
\frac{\cos(ay) \, \mathrm{d}y}{\sqrt{p^2 \sinh^2\!r + \cosh^2 \! y}} 
\ , 
\label{eq:Gr}
\end{align}
where we suppress the dependence of $G$ on $a$ and~$\gamma$. 
$S$~then equals $q^{-1}$ times the Riemann sum of 
$G$ with the sampling points $r = nq$, $n\in\BbbZ$. 
$G$~is continuous, and from Appendix \ref{app:A2} we see that 
$|G(r)|$ is exponentially small as $r\to\pm\infty$. 
The Riemann sum of $G$ therefore converges to the integral of $G$ as $q\to0$.  
Noting finally that $G$ is even, we recover~\eqref{eq:Fint}.

\section{Co-rotating response at $E\ell \to\pm\infty$}
\label{app:D}

In this appendix, we analyse the individual terms in 
the co-rotating detector response 
\eqref{eq:corot:rate.evald}
in the limit $E\ell\to\pm\infty$. 
These terms are of the form 
\begin{align}
\label{eq:tildeIdef}
\tilde{I}(\chi,a,P):=\cos(\chi a) \, \mathrm{e}^{-\pi a/2} J(a,P)
\ , 
\end{align}
where $\chi\in\BbbR$, $a\in\BbbR$, $P>0$
and $J(a,P)$ is given by \eqref{eq:J-def}. 
We shall show that when $a\to\pm\infty$ with fixed $\chi$ and~$P$, 
$\tilde{I}(\chi,a,P)$ has the asymptotic form
\begin{align}
\tilde{I}(\chi,a,P) = 
\begin{cases}
\displaystyle{\frac{2\sqrt\pi \, 
\mathrm{e}^{-\pi a} \cos{(\chi a)}\cos(\alpha a - \pi/4)}{\sqrt{a \sinh(2\alpha)}}
+ o \bigl({a}^{-1/2}\,\mathrm{e}^{-a\pi} \bigr)
\ , }
& a\to+\infty, 
\\[4ex]
\displaystyle{\frac{2\sqrt\pi \cos{(\chi a)}\cos(-\alpha a - \pi/4)}{\sqrt{-a \sinh(2\alpha)}}
+ o \bigl({(-a)}^{-1/2}\bigr)
\ , }
& a\to-\infty, 
\end{cases} 
\label{eq:Itilde-asymptotics}
\end{align}
where $\alpha = \arcsinh\sqrt{P}$. 

%

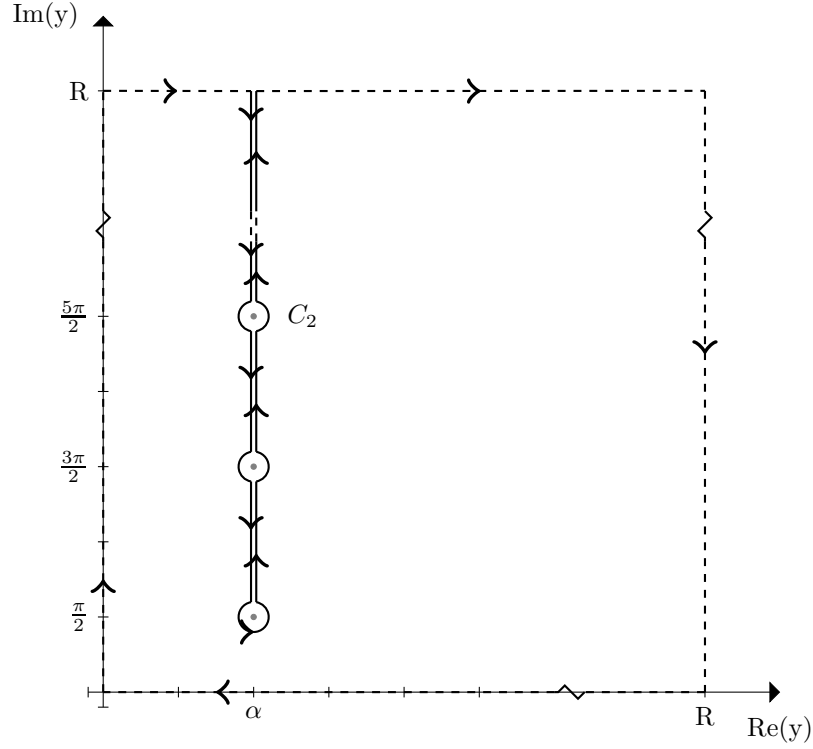
\begin{figure}[t] \centering
\tikzset{-->-/.style={decoration={
  markings,
  mark=at position #1 with {\arrow[scale=2]{>}}},postaction={decorate}}}
\begin{center}

\begin{tikzpicture}[scale=2]

\node (zI)    at (1,0.5)   {};
\node (zII)   at (1,1.5)   {};
\node (zIII)  at (1,2.5) [label={[label distance=0.2cm]0:$C_2$}]  {};

\path  
  (zI) +(80:1mm)  coordinate  (zItr)
       +(-260:1mm) coordinate  (zItl)
       +(-80:1mm)  coordinate  (zIbr)
       +(-100:1mm) coordinate  (zIbl)
  (zII) +(80:1mm)  coordinate  (zIItr)
       +(-260:1mm) coordinate  (zIItl)
       +(-80:1mm)  coordinate  (zIIbr)
       +(-100:1mm) coordinate  (zIIbl)
  (zIII) +(80:1mm)  coordinate  (zIIItr)
       +(-260:1mm) coordinate  (zIIItl)
       +(-80:1mm)  coordinate  (zIIIbr)
       +(-100:1mm) coordinate  (zIIIbl)
       ;
       
\path (zIIItl) +(90:1.40)  coordinate  (topLeft);
\path (zIIItr) +(90:1.40)  coordinate  (topRight); 
\path (zIIItl) +(90:0.4)  coordinate  (zigLeftLow);
\path (zigLeftLow) +(90:0.2)  coordinate  (zigLeftHigh);
\path (zIIItr) +(90:0.4)  coordinate  (zigRightLow);
\path (zigRightLow) +(90:0.2)  coordinate  (zigRightHigh);

\draw[-triangle 90] (-0.1,0) -- (4.5,0);
\draw[-triangle 90] (0,-0.1) -- (0,4.5);

\node[below=0.2cm] at (4.5,0) {Re(y)};
\node[left=0.2cm] at (0,4.5) {Im(y)};

\foreach \x/\xtext in {-0.1/{}, 0.5/{}, 1/\alpha, 1.5/{}, 2/{}, 2.5/{}, 4.0/{$R$}} \draw (\x cm,1pt) -- (\x cm,-1pt) node[anchor=north] {$\xtext$};
\foreach \y/\ytext in {-0.1/{}, 0.5/\frac{\pi}{2}, 1/{}, 1.5/{\frac{3\pi}{2}},2.0/{}, 2.5/{\frac{5\pi}{2}},4.0/{$R$}}
\draw (1pt,\y cm) -- (-1pt,\y cm) node[anchor=east] {$\ytext$};

\draw [-->-=.75,thick,dashed] (3,0)--(0,0);
\draw[decorate,decoration=zigzag, thick] (3.2,0)--(3,0) ; 
\draw [thick,dashed] (4,0)--(3.2,0) ;
\draw [-->-=.25, thick,dashed]  (4,3) --(4,0);
\draw[decorate,decoration=zigzag,  thick]  (4,3.2)--(4,3); 
\draw [thick,dashed] (4,4)--(4,3.2);
\draw [-->-=.5, thick,dashed] (topRight) -- (4,4);
\draw [-->-=.5, thick,dashed] (0,4)--(topLeft);
\draw [ thick,dashed] (0,3.2)--(0,4);
\draw[decorate,decoration=zigzag, thick] (0,3.2) -- (0,3); 
\draw [-->-=.25, thick,dashed] (0,0)--(0,3);

\filldraw [gray] (zI) circle (0.5pt)
                 (zII) circle (0.5pt)
                 (zIII) circle (0.5pt)
                 ;
\draw[-->-=.5,thick] (zItl) arc (-260:80: 1mm);
\draw[thick] (zIItr) arc (80: -80: 1mm);
\draw[thick] (zIItl) arc (-260: -100: 1mm);
\draw[thick] (zIIItr) arc (80: -80: 1mm);
\draw[thick] (zIIItl) arc (-260: -100: 1mm);

\draw[-->-=.4,thick] (zItr)--(zIIbr) ;
\draw[-->-=.4,thick] (zIIbl)--(zItl);
\draw[-->-=.4,thick] (zIItr)--(zIIIbr);
\draw[-->-=.4,thick] (zIIIbl)--(zIItl);

\draw[-->-=.25,thick]  (zigLeftLow)--(zIIItl);
\draw[dashed, thick] (zigLeftHigh)-- (zigLeftLow); 
\draw[-->-=.25,thick]  (topLeft)--(zigLeftHigh) ;

\draw[-->-=.5,thick]  (zigRightHigh)--(topRight);
\draw[dashed,thick]  (zigRightLow)--(zigRightHigh); 
\draw[-->-=.5, thick] (zIIItr)--(zigRightLow);

\end{tikzpicture}
\end{center}
\caption{The contour deformation that gives $C_2$. After we take the limit $R\to\infty$, the contributions to the contour integral from the top and right sides of the square are vanishing. }  
\label{fig:App2dSchw:ContourC2}
\end{figure}


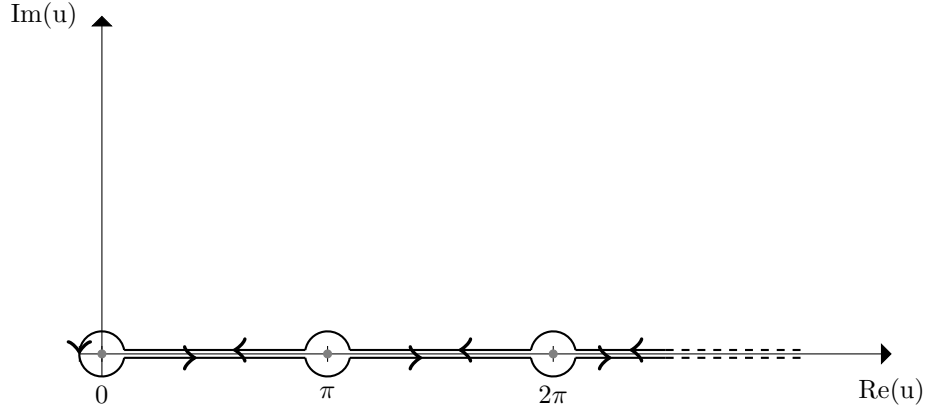
\begin{figure}[t] \centering
\tikzset{-->-/.style={decoration={
  markings,
  mark=at position #1 with {\arrow[scale=2]{>}}},postaction={decorate}}}
\begin{center}
\begin{tikzpicture}[scale=3]

\node (zI)    at (0,0)   {};
\node (zII)   at (1,0)   {};
\node (zIII)  at (2,0)   {};

\path  
  (zI) +(10:1mm)  coordinate  (zItr)
       +(170:1mm) coordinate  (zItl)
       +(-10:1mm)  coordinate  (zIbr)
       +(-170:1mm) coordinate  (zIbl)
  (zII) +(10:1mm)  coordinate  (zIItr)
       +(170:1mm) coordinate  (zIItl)
       +(-10:1mm)  coordinate  (zIIbr)
       +(-170:1mm) coordinate  (zIIbl)
  (zIII) +(10:1mm)  coordinate  (zIIItr)
       +(170:1mm) coordinate  (zIIItl)
       +(-10:1mm)  coordinate  (zIIIbr)
       +(-170:1mm) coordinate  (zIIIbl)
       ;
       
\path (zIIItr) +(0:.40)  coordinate  (zigTopLeft); 
\path (zIIIbr) +(0:.40)  coordinate  (zigBottLeft);
\path (zigTopLeft) +(0:0.6)  coordinate  (zigTopRight);
\path (zigBottLeft) +(0:0.6)  coordinate  (zigBottRight);

\draw[-triangle 90] (-0.1,0) -- (3.5,0);
\draw[-triangle 90] (0,-0.1) -- (0,1.5);

\node[below=0.2cm] at (3.5,0) {Re(u)};
\node[left=0.2cm] at (0,1.5) {Im(u)};

\foreach \x/\xtext in {-0.1/{}, 0/{0},  1/\pi, 2/{2\pi}} \draw (\x cm,1pt) -- (\x cm,-1pt) node[below=0.2cm] {$\xtext$};

\filldraw [gray] (zI) circle (0.5pt)
                 (zII) circle (0.5pt)
                 (zIII) circle (0.5pt)
                 ;
\draw[-->-=.5,thick] (zItr) arc (10:350: 1mm);
\draw[thick] (zIItr) arc (10: 170: 1mm);
\draw[thick] (zIIbr) arc (-10: -170: 1mm);
\draw[thick] (zIIItr) arc (10: 170: 1mm);
\draw[thick] (zIIIbr) arc (-10: -170: 1mm);

\draw[-->-=.4,thick] (zIItl)--(zItr) ;
\draw[-->-=.4,thick] (zIbr)--(zIIbl);
\draw[-->-=.4,thick] (zIIItl)--(zIItr);
\draw[-->-=.4,thick] (zIIbr)--(zIIIbl);

\draw[-->-=.4,thick] (zigTopLeft)--(zIIItr);
\draw[-->-=.4,thick] (zIIIbr)--(zigBottLeft);
\draw[dashed,thick] (zigTopLeft)--(zigTopRight);
\draw[dashed,thick] (zigBottLeft)--(zigBottRight);

\end{tikzpicture}
\end{center}
\caption{The contour $C_3$ as used in evaluation of~\eqref{eq:JC3}. }  
\label{fig:App2dSchw:ContourC3}
\end{figure}

Assuming $a\ne0$ and writing $\cos(ay) = \Realpart \bigl(\mathrm{e}^{i|a|y}\bigr)$, we 
start by rewriting 
$J(a,P)$ from \eqref{eq:J-def} as
\begin{align}
\label{eq:JC1}
J(a,P) = \Realpart \int_{C_1} \frac{\mathrm{e}^{i|a|y} \, 
\mathrm{d}y}{\sqrt{P + \cosh^2 \! y}}
\ , 
\end{align}
where the contour $C_1$ consists of the positive imaginary 
axis travelled downwards and the positive real axis travelled rightwards. 
The contribution from the imaginary axis vanishes on taking the real part. 

Writing $P = \sinh^2\!\alpha$ where $\alpha>0$ and 
factorising the quantity under the square root in~\eqref{eq:JC1}, we obtain 
\begin{align}
\label{eq:JC1prime}
J(a,P) = \Realpart \int_{C_1} \frac{\mathrm{e}^{i|a|y} \, \mathrm{d}y}
{\sqrt{\sinh(\alpha+y-i\pi/2) \sinh(\alpha-y+i\pi/2)}}
\ . 
\end{align}

The branch points of the integrand in \eqref{eq:JC1prime} are at 
$y = \pm\alpha + i\pi(n + \frac12)$, $n\in\BbbZ$. 
We may deform $C_1$ into the contour $C_2$ 
that comes down from $\alpha + i\infty$ 
at $\Realpart y =\alpha$, passing the branch points from the left,
encircles the branch point at $y=\alpha+i\pi/2$ counterclockwise, 
and finally goes back up to $\alpha + i\infty$ at $\Realpart y =\alpha$ 
but now passing the branch points from the right. 
\par Changing the integration variable by 
$y = \alpha + i\pi/2 + iu$, we then have  
\begin{align}
\label{eq:JC3}
J(a,P) = \mathrm{e}^{-|a|\pi/2}
\Realpart 
\left(i \mathrm{e}^{i\alpha |a|}
\int_{C_3} \frac{\mathrm{e}^{-|a|u} \, \mathrm{d}u}
{\sqrt{-i \sin(u) \sinh(2\alpha+iu)}}
\right)
\ , 
\end{align}
where contour $C_3$ comes from 
$u = +\infty$ to $u=0$ on the upper lip of the positive $u$ axis, 
encircles $u=0$ counterclockwise and goes back to $u = +\infty$ on the lower 
lip of the positive $u$ axis. The square root is positive at 
$u = \pi/2$ on the upper lip and 
it is analytically continued to the rest of~$C_3$. The contours are shown in Figures~\ref{fig:App2dSchw:ContourC2} and~\ref{fig:App2dSchw:ContourC3}.

We now note that $\sinh(2\alpha+iu) = \sinh(2\alpha)\cos(u) + i\cosh(2\alpha)\sin(u)$, 
and that the modulus of this expression is bounded below by~$\sinh(2\alpha)$. 
In~\eqref{eq:JC3}, 
the contribution from the two intervals in which $\pi/2 \le u \le \pi$ is 
therefore bounded above by 
$\mathrm{e}^{-|a|\pi}/\sqrt{\sinh(2\alpha)}$ times a numerical constant, 
and the contribution from the 
two intervals in which $n \pi  \le u \le (n+1)\pi$, $n=1,2,\ldots\,$, 
is bounded above by 
$\mathrm{e}^{-|a|\pi\left[n+(1/2)\right]}/\sqrt{\sinh(2\alpha)}$
times a numerical constant. 
The sum of all of these contributions is hence $O\bigl(\mathrm{e}^{-|a|\pi}\bigr)$. 
In the remaining contribution, coming from the two intervals in which $0 \le u \le \pi/2$, 
we combine the upper and lower lips and change the integration variable to $w = |a|u$. 
This gives 
\begin{align}
\label{eq:C4}
& 
J(a,P)=\frac{2\,\mathrm{e}^{-|a|\pi/2}}{\sqrt{|a|}} \times
\nonumber\\
&\times 
\Realpart\left(\mathrm{e}^{i(\alpha |a|- \pi/4)}
\int_0^{|a|\pi/2} \frac{\mathrm{e}^{-w} \, \mathrm{d}w}
{\sqrt{|a|\sin(w/|a|) 
\bigl[\sinh(2\alpha)\cos(w/|a|) + i\cosh(2\alpha)\sin(w/|a|)\bigr]}}
\right)
\nonumber\\
&
\hspace{2ex}
+ O\bigl(\mathrm{e}^{-|a|\pi}\bigr)
\ , 
\end{align}
where the square root denotes the branch that is positive in the limit $w\to0_+$. 

By Jordan's lemma, the modulus of the integrand in 
\eqref{eq:C4} 
is bounded from above in the range of integration 
by the function 
$g(w) := \sqrt{\frac{\pi}{2\sinh(2\alpha)}}
\, w^{-1/2} \, e^{-w}$. 
As $g(w)$
is integrable over $0< w < \infty$ and independent of~$a$, 
dominated convergence guarantees that when  
$|a|\to\infty$, the limit in the integrand in \eqref{eq:C4} 
can be taken under the integral. The integral that ensues in the limit is 
elementary, and we obtain 
\begin{align}
J(a,P) = 
\frac{2\sqrt\pi\,\mathrm{e}^{-|a|\pi/2} \cos(\alpha |a|- \pi/4)}
{\sqrt{|a| \sinh(2\alpha)}}
+ o \bigl({|a|}^{-1/2}\,\mathrm{e}^{-|a|\pi/2} \bigr)
\ . 
\label{eq:Ja-as}
\end{align}
\eqref{eq:Itilde-asymptotics} then follows by substituting 
\eqref{eq:Ja-as} in~\eqref{eq:tildeIdef}.


\section{Derivation of \eqref{eq:inertial:negE}}
\label{app:zeroterm-largenegenergy}

In this appendix, we verify the asymptotic expansions 
\begin{subequations}
\begin{align}
\int^{m}_{0}\,\mathrm{d}x\,\frac{\cos{(\beta x})}{\cos x} &=
\frac{\sin(m\beta)}{\beta\cos m}
+ O\bigl(\beta^{-2}\bigr)
\ , 
\label{eq:app:cosasymp}
\\[1ex]
\int^{m}_{0}\,\mathrm{d}x\,\frac{\sin{(\beta x})}{\sin x}&=
\frac{\pi\sgn\beta}{2}-\frac{\cos{(m\beta)}}{\beta\sin m }
+O\bigl(\beta^{-2}\bigr)
\ , 
\label{eq:app:f-asymp}
\end{align}
\end{subequations}
valid as $\beta\to\pm\infty$ with fixed~$m \in (0,\pi)$. 

\eqref{eq:app:cosasymp} follows by repeated integrations by parts 
that bring down inverse powers of $\beta$~\cite{wong}. 

In \eqref{eq:app:f-asymp}, we split the integral as 
\begin{align}
\label{eq:app:fasymp:C}
\int^m_0\,\mathrm{d}x\,\left(\frac{1}{\sin{x}} -\frac{1}{x} \right)\sin(\beta x)
- 
\int_m^{\infty}\,\mathrm{d}x\,\frac{\sin{\left(\beta x\right)}}{x} 
+ 
\int_0^\infty\,\mathrm{d}x\,\frac{\sin{\left(\beta x\right)}}{x} 
\ . 
\end{align}
Repeated integrations by parts now apply to the first two terms in~\eqref{eq:app:fasymp:C}, 
and the third term equals $\frac\pi2\sgn\beta$~\cite{gradshteyn}. 
Combining, we obtain~\eqref{eq:app:f-asymp}.

\section{Stationary but non-co-rotating detector}
\label{app:non-corotating}

In this appendix, we discuss briefly a detector that is stationary in the 
exterior region of the BTZ black hole but not co-rotating with the horizon. 
For the transparent boundary condition at the infinity, 
we show that the $n=0$ term in the transition rate 
\eqref{eq:BTZ:rate} breaks the KMS property 
already in second order in the difference between the horizon and detector angular velocities. 
As the $n=0$ term is expected to give the dominant contribution when the black hole mass is large, 
we take this as evidence that the transition rate does not satisfy the KMS property, 
in agreement with the GEMS prediction 
\cite{Deser:1997ri,Deser:1998bb,Deser:1998xb,Russo:2008gb}. 

Consider a detector that is stationary in the exterior region of the 
BTZ spacetime at exterior BTZ co-ordinate~$r$, 
but not necessarily co-rotating with the horizon. 
The tangent vector of the trajectory is a linear 
combination of $\partial_t$ and~$\partial_\phi$. 
By \eqref{eq:BoyerLind} and~\eqref{eq:alpha},  
the lift of the trajectory to $\text{AdS}_3$ reads 
\begin{align}
X_1&=\ell\cosh\chi\sinh(2ky) 
\ , 
\nonumber\\
T_1&=\ell\cosh\chi\cosh(2ky)
\ , 
\nonumber\\
X_2&=\ell\sinh\chi\cosh(2y)
\ , 
\nonumber\\
T_2&=\ell\sinh\chi\sinh(2y)
\ , 
\label{appF:XTXT}
\end{align}
where we have written $\sqrt\alpha = \cosh\chi$ with $\chi>0$, 
the constant $k$ is proportional to the difference 
of the detector and horizon angular velocities, and $y$ is a parameter along the trajectory. 
We assume $|k| < \tanh\chi$, which is the condition for the trajectory to be timelike. 
The proper time $\tau$ is related by $y$ by $\tau = 2 \ell \sinh\chi \sqrt{1 - k^2 \coth^2\!\chi} \, y$. 

Let $\dot{\mathcal{F}}^{n=0}$ denote the $n=0$ term in the transition rate~\eqref{eq:BTZ:rate}. 
Substituting \eqref{appF:XTXT} in~\eqref{eq:DXN2}, and specialising to the 
transparent boundary condition, $\zeta=0$, 
we find 
\begin{align}
\label{eq:transraten=0}
\dot{\mathcal{F}}^{n=0}(E) 
=\frac{1}{4}-\frac{1}{2\pi}\sqrt{1-k^2 \coth^2\!\chi}
\int^{\infty}_{0} \,\mathrm{d}y \, 
\frac{\sin \bigl(2 E\ell  \sinh\chi \, \sqrt{1-k^2 \coth^2\!\chi} \, y\bigr)}
{\sqrt{\sinh^2 \! y-\coth^2\!\chi\sinh^2(k y)}}
\ . 
\end{align}
It can be verified that the quantity under the square root 
in the denominator is positive for $0<y<\infty$. 

Expanding \eqref{eq:transraten=0} as a power series in $E$ and then expanding 
the coefficients as power series in~$k$, we find 
\begin{align}
\dot{\mathcal{F}}^{n=0}(E)
&=\frac{1}{4}
+ 
\left[- \frac{\pi}{4} \sinh\chi + \frac{\pi}{8}\left(\frac{\pi^2}{4}-1\right)
\frac{\cosh^2\!\chi}{\sinh\chi} k^2 + O\bigl(k^4\bigr)
\right]
E\ell
\nonumber
\\[1ex]
& 
\hspace{2ex}
+ \left[
\frac{\pi^3}{12}
\sinh^3\!\chi
+ \frac{\pi^3}{4}
\left(1 - \frac{\pi^2}{6}\right)
\sinh\chi \cosh^2\!\chi 
\, k^2 
+ O\bigl(k^4\bigr)
\right]
{(E\ell)}^3
+ O\bigl({(E\ell)}^5\bigr)
\ . 
\label{eq:transraten=0-expansion}
\end{align}
From \eqref{eq:transraten=0-expansion} it is seen that the
power series expansion of 
$\dot{\mathcal{F}}^{n=0}(-E) / \dot{\mathcal{F}}_{n=0}(E)$ 
in $E$ is incompatible with a pure exponential in~$E$, 
and the discrepancy arises in the coefficient of the ${(E\ell)}^3$ term in order~$k^2$.
$\dot{\mathcal{F}}^{n=0}$ \eqref{eq:transraten=0} hence 
does not satisfy the KMS property at small but non-zero~$k$.

\chapter{Two-dimensional Schwarzschild integral}
\label{ch:appendixB}

In this appendix, we shall show how to obtain the transition rate~\eqref{eq:Schw:2d:TRstaticFin} for the static detector in the Hartle-Hawking vacuum external to the two-dimensional Schwarzschild black hole, encountered in Chapter~\ref{ch:2DSchw}.

Our task is to evaluate the real part of the integral
\begin{equation}
I:=\int^{\infty}_0\,\mathrm{d}z\,\expo^{-z(a+ib)} \log{\left[\sinh{\left(z\right)}\right]}\,,
\label{eq:App2dSchwHH:I}
\end{equation}
where $a,b \in \BbbR$ and $a>0$ and where we shall take $a\to 0$ at the end of the calculation.
\par Note, the logarithmic divergence at $z=0$ in~\eqref{eq:Schw:2d:trHadamardAlt3} is integrable, and we have dropped the $i\epsilon$-regulator because we never cross $s=0$. We shall perform this computation by deforming the contour as shown in Figure~\ref{fig:App2dSchw:Contour}. The result we shall ultimately obtain in the $a\to 0$ limit is
\be 
\Realpart[I]=-\frac{\pi}{2b}\left[\frac{1+\expo^{-b\pi}}{1-\expo^{-b\pi}}\right]\,.
\label{eq:App2dSchw:ReIRes}
\ee
%

\tikzset{-->-/.style={decoration={
  markings,
  mark=at position #1 with {\arrow[scale=2]{>}}},postaction={decorate}}}
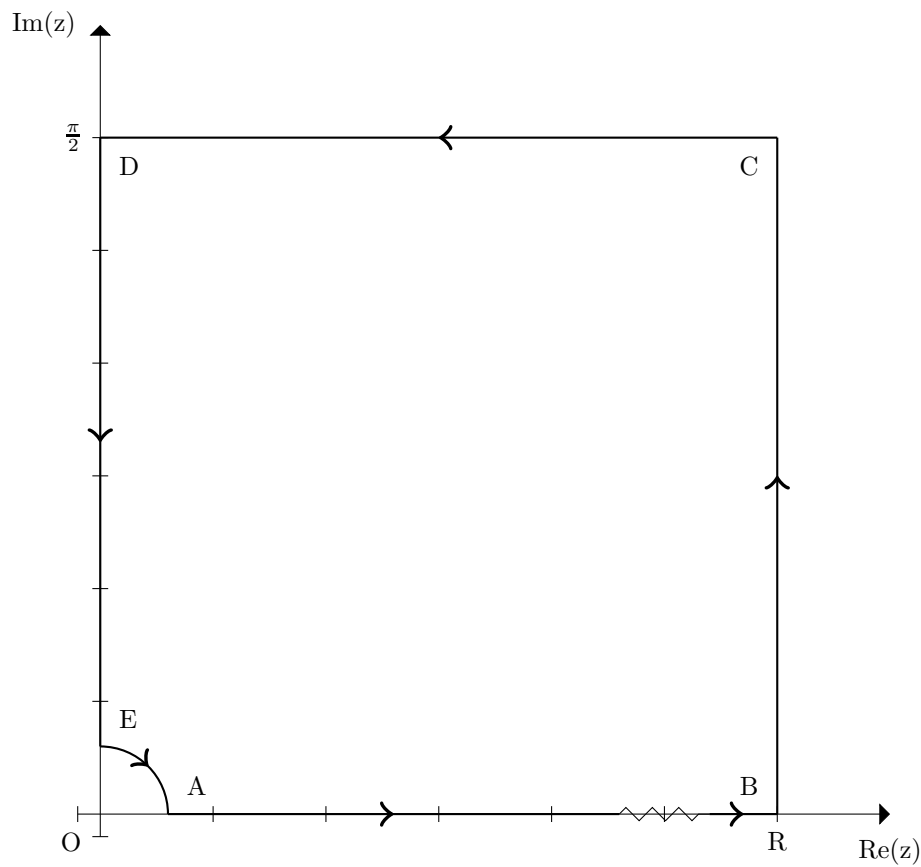
\begin{figure}[t]\centering
 \centering
\begin{tikzpicture}[scale=3]

\node (orig)    at (0,0)  [label={[label distance=0.001cm]-135:O}]   {};
\node (A)    at (3mm,0)  [label={[label distance=0.001cm]45:A}]   {};
\node (B)    at (3,0)     [label={[label distance=0.001cm] 135:B}] {};
\node (C)    at (3,3)     [label={[label distance=0.001cm]-135:C}] {};
\node (D)    at (0,3)     [label={[label distance=0.001cm]-45:D}] {};
\node (E)    at (0,3mm)   [label={[label distance=0.001cm]45:E}]{};

\path (A) +(0:2)  coordinate  (zigLeft); 
\path (zigLeft) +(0:.40)  coordinate  (zigRight);

\draw[-triangle 90] (-0.1,0) -- (3.5,0);
\draw[-triangle 90] (0,-0.1) -- (0,3.5);

\node[below=0.2cm] at (3.5,0) {Re(z)};
\node[left=0.2cm] at (0,3.5) {Im(z)};

\foreach \x/\xtext in {-0.1/{}, 0.5/{}, 1/{}, 1.5/{}, 2/{}, 2.5/{}, 3.0/{$R$}} \draw (\x cm,1pt) -- (\x cm,-1pt) node[anchor=north] {$\xtext$};
\foreach \y/\ytext in {-0.1/{}, 0.5/{}, 1/{}, 1.5/{},2.0/{}, 2.5/{},3/{\frac{\pi}{2}}}
\draw (1pt,\y cm) -- (-1pt,\y cm) node[anchor=east] {$\ytext$};

\draw [-->-=.5,thick] (A.center)--(zigLeft);
\draw[decorate,decoration=zigzag] (zigLeft) -- (zigRight);
\draw [-->-=.5,thick] (zigRight)-- (B.center);

\draw [-->-=.5,thick] (B.center)--(C.center);
\draw [-->-=.5,thick] (C.center)--(D.center);
\draw [-->-=.5,thick] (D.center)--(E.center);
\draw [-->-=0.5,thick] (E.center) arc (90:0:3mm);

\end{tikzpicture}
\caption{Contour deformation used in the evaluation of~\eqref{eq:App2dSchwHH:I}.}  
\label{fig:App2dSchw:Contour}
\end{figure}
\par It is straightforward to verify that the contribution from the part of the contour along `BC' is vanishing in the limit that $R\to \infty$, and we shall not need to analyse that piece of the contour explicitly in this appendix. We shall now examine the contributions to~\eqref{eq:App2dSchwHH:I} from the pieces of the contour path shown in Figure~\ref{fig:App2dSchw:Contour} labelled `CD' and `DE' in turn.
\section[Integral along CD part of contour]{Integral along CD part of contour%
              \sectionmark{CD part of contour}}
\sectionmark{CD part of contour}
Along this portion of the path $z=t+i\pi/2$, where $0 \leq t < \infty$. We then have
\begin{equation}
I_{CD}:=-\int^{\infty}_{0}\,\mathrm{d}t\,\expo^{-(t+i\pi/2)(a+ib)}\log{\left(\sinh{\left(t+i\pi/2\right)}\right)}\,.
\label{eq:Schw:2d:preICD}
\end{equation}
We use the identity $\sinh{(t+i\pi/2)}=i\cosh{(t)}$, followed by $\log{\left(i\cosh{(t)}\right)}=\log{\left(\cosh{(t)}\right)}+i\pi/2$, where the branch is chosen to give agreement with the small imaginary constant one obtains in $(1+1)$-dimensional Minkowski spacetime. After this, ~\eqref{eq:Schw:2d:preICD} reads
\begin{equation}
I_{CD}=-\int^{\infty}_{0}\,\mathrm{d}t\,\expo^{(b\pi/2-at)-i(bt+a\pi/2)}\left[\log{\left(\cosh{(t)}\right)}+i\pi/2\right]\,.
\label{eq:Schw:2d:ICD}
\end{equation}
\par To make progress evaluating~\eqref{eq:Schw:2d:ICD}, let us first focus on the simpler term, arising from the finite imaginary part of~\eqref{eq:Schw:2d:ICD}. We denote this term by $I_{CD2}$:
\begin{equation}
\begin{aligned}
I_{CD2}&:=-i\frac{\pi}{2}\int^{\infty}_{0}\,\mathrm{d}t\,\expo^{(b\pi/2-at)-i(bt+a\pi/2)}\\
&=-i\frac{\pi}{2}\expo^{\frac{\pi}{2}\left(b-ia\right)}\int^{\infty}_{0}\,\mathrm{d}t\,\expo^{-t(a+ib)}\\
&=-i\frac{\pi}{2}\expo^{\frac{\pi}{2}\left(b-ia\right)}\frac{1}{(a+ib)}\\
&=-i\frac{\pi}{2}\expo^{\frac{\pi}{2}\left(b-ia\right)}\frac{a-ib}{(a^2+b^2)}\,.
\end{aligned}
\end{equation}
Thus,
\begin{equation}
\Realpart\left[{I_{CD2}}\right]=-\frac{\pi}{2(a^2+b^2)}\expo^{\frac{b\pi}{2}}\left[b\cos{(a\pi/2)}+a\sin{(a\pi/2)}\right]\,,
\end{equation}
and if we take the limit $a\to 0$, we obtain
\begin{equation}
\Realpart\left[{I_{CD2}}\right]=-\frac{\pi}{2b}\expo^{\frac{b\pi}{2}}\,.
\label{eq:App2dSchw: ICD2Res}
\end{equation}
\par Next, we focus on the logarithmic term in the integrand of~\eqref{eq:Schw:2d:ICD}, whose integral we denote as $I_{CD1}$:
\begin{equation}
\begin{aligned}
I_{CD1}&=-\int^{\infty}_{0}\,\mathrm{d}t\,\expo^{(b\pi/2-at)-i(bt+a\pi/2)}\log{\left(\cosh{(t)}\right)}\\
&=-\expo^{\frac{\pi}{2}\left(b-ia\right)}\int^{\infty}_{0}\,\mathrm{d}t\,\expo^{-t(a+ib)}\log{\left(\cosh{(t)}\right)}\,.
\end{aligned}
\end{equation}
If we now integrate by parts, the boundary term vanishes and we are left with
\begin{equation}
\begin{aligned}
I_{CD1}&=-\expo^{\frac{\pi}{2}\left(b-ia\right)}\frac{1}{a+ib}\int^{\infty}_{0}\,\mathrm{d}t\,\expo^{-t(a+ib)}\tanh{(t)}\,.
\end{aligned}
\end{equation}
Given that $a>0$, this integral can be evaluated by using the standard integral (3.541.7) in~\cite{gradshteyn}, and after using (8.370) of~\cite{gradshteyn}, we obtain
\begin{equation}
\begin{aligned}
I_{CD1}=-\expo^{\frac{\pi}{2}\left(b-ia\right)}\frac{1}{a+ib}\Bigg[\frac{1}{2}\psi\left(\frac{1}{2}+\frac{(a+ib)}{4}\right)&-\frac{1}{2}\psi\left(\frac{(a+ib)}{4}\right)\\
&-\frac{1}{(a+ib)}\Bigg]\,.
\label{eq:Schw:2d:ICD1}
\end{aligned}
\end{equation}
Taking the $a\to 0$ limit, equation~\eqref{eq:Schw:2d:ICD1} reduces to
\begin{equation}
I_{CD1}=\frac{i}{b}\expo^{\frac{b\pi}{2}}\Bigg[\frac{1}{2}\psi\left(\frac{1}{2}+\frac{ib}{4}\right)-\frac{1}{2}\psi\left(\frac{ib}{4}\right)+\frac{i}{b}\Bigg]\,.
\label{eq:Schw:2d:ICD1alt}
\end{equation}
We take the real part of~\eqref{eq:Schw:2d:ICD1alt} using (6.3.11) and (6.3.12) of~\cite{abramowitz}:
\begin{equation}
\begin{aligned}
\Realpart\left[I_{CD1}\right]&=-\expo^{\frac{b\pi}{2}}\left[\frac{1}{b^2}+\frac{1}{2b}\Imagpart\Bigg(\psi\left(\frac{1}{2}+\frac{ib}{4}\right)-\psi\left(\frac{ib}{4}\right)\Bigg)\right]\\
&=-\expo^{\frac{b\pi}{2}}\Bigg[\frac{1}{b^2}+\frac{1}{2b}\Bigg(\frac{\pi}{2}\tanh{\left(\frac{\pi b}{4}\right)}-\frac{2}{b}-\frac{\pi}{2}\coth{\left(\frac{\pi b}{4}\right)}\Bigg)\Bigg]\\
&=-\expo^{\frac{b\pi}{2}}\Bigg[\frac{\pi}{4b}\Bigg(\tanh{\left(\frac{\pi b}{4}\right)}-\coth{\left(\frac{\pi b}{4}\right)}\Bigg)\Bigg]\\
&=-\expo^{\frac{b\pi}{2}}\Bigg[-\frac{\pi}{2b}\frac{1}{\sinh{\left(b\pi/2\right)}}\Bigg]\\
&=\frac{\pi}{b}\frac{1}{1-\expo^{-b\pi}}\,.
\label{eq:App2dSchw: ICD1Res}
\end{aligned}
\end{equation}
\par Combining~\eqref{eq:App2dSchw: ICD2Res} with~\eqref{eq:App2dSchw: ICD1Res}, we obtain the complete contribution along the `CD' section of the contour to be
\begin{equation}
\Realpart\left[I_{CD}\right]=-\frac{\pi}{2b}\expo^{\frac{b\pi}{2}}+\frac{\pi}{b}\frac{1}{1-\expo^{-b\pi}}\,.
\label{eq:App2dSchw:ICDfin}
\end{equation}
\section[Integral along DE part of contour]{Integral along DE part of contour%
              \sectionmark{DE part of contour}}
\sectionmark{DE part of contour}
Along this part of the path $z=it$, with $\epsilon \leq t \leq \pi/2$, where $\epsilon$ is a small, positive constant that we employ to avoid the logarithmic singularity at the origin. We shall take $\epsilon$ to zero at the end of the calculation. We have
\begin{equation}
\begin{aligned}
I_{DE}&:=-i\int^{\pi/2}_{\epsilon}\,\mathrm{d}t\,\expo^{-it(a+ib)}\log{\left(\sinh{\left(it\right)}\right)}\\
&=-i\int^{\pi/2}_{\epsilon}\,\mathrm{d}t\,\expo^{bt}\expo^{-iat}\log{\left(i\sin{\left(t\right)}\right)}\\
&=-i\int^{\pi/2}_{\epsilon}\,\mathrm{d}t\,\expo^{bt}\expo^{-iat}\left[\log{\left(\sin{(t)}\right)}+i\pi/2\right]\,.
\label{eq:App2dSchw:IDE}
\end{aligned}
\end{equation}
\par We start by evaluating the non-logarithmic term of~\eqref{eq:App2dSchw:IDE}, which we label $I_{DE2}$:
\begin{equation}
I_{DE2}:=\frac{\pi}{2}\int^{\pi/2}_{0}\,\mathrm{d}t\,\expo^{bt}\expo^{-iat}\,.
\end{equation}
After taking the cut-off to infinity (or $a\to 0$) and taking the real part, we find
\begin{equation}
\Realpart\left[I_{DE2}\right]=\frac{\pi}{2b}\left(\expo^{\frac{b\pi}{2}}-1\right)\,.
\end{equation}
Next, consider the logarithmic part of~\eqref{eq:App2dSchw:IDE}, which we denote by $I_{DE1}$:
\begin{equation}
I_{DE1}=-i\int^{\pi/2}_{\epsilon}\,\mathrm{d}t\,\expo^{bt}\expo^{-iat}\log{\left(\sin{(t)}\right)}\,.
\label{eq:App2dSchw:IDE1}
\end{equation}
In the limit $a\to 0$,~\eqref{eq:App2dSchw:IDE1} is purely imaginary and so upon taking the real part it vanishes. 
\par The total result for this section of the contour is thus
\begin{equation}
\Realpart[I_{DE}]=\frac{\pi}{2b}\left(\expo^{\frac{b\pi}{2}}-1\right)\,.
\label{eq:App2dSchw:IDEfin}
\end{equation}
\section{Result}
We have
\begin{equation}
I:=I_{AB}=-I_{BC}-I_{CD}-I_{DE}\,,
\end{equation}
and therefore by combining~\eqref{eq:App2dSchw:ICDfin} and~\eqref{eq:App2dSchw:IDEfin}, we find~\eqref{eq:App2dSchw:ReIRes}.
\chapter{Bound on $x_0$ from Chapter~\ref{ch:2DSchw}}
\label{ch:Appendix:x0bound}
In this appendix, we verify the no-collision bound~\eqref{eq:Schw:2d:movMirr:x0bound}.
We may assume $0<\nu<1$. The mirror trajectory~\eqref{eq:Schw:2d:movMirr:traj} can be expressed in Cartesian co-ordinates as
\begin{equation}
t=\frac{1}{\kappa}\log{\Big(2\sinh{(-\kappa x)}\Big)}\,.
\label{eq:Schw:2d:movMirr:Carttraj}
\end{equation}
Consider the function
\begin{equation}
f(x):= \frac{1}{\kappa}\log{\Big(2\sinh{(-\kappa x)}\Big)}-\left(\frac{x_0-x}{\nu}\right)\,,
\end{equation}
where $-\infty<x<0$. Geometrically, $f(x)$ is the time co-ordinate of the mirror's trajectory subtracted from the time co-ordinate of the inertial detector's trajectory, as functions of $x$. This function will tend to $-\infty$ when $x\to0$ because the detector will intersect the time axis at some finite value, whilst the mirror asymptotes to the time axis as $t\to-\infty$. Similarly, independent of the value of $x_0$, $f(x)$ will tend to $-\infty$ as $x\to-\infty$ because the mirror asymptotes to $t=-x$, whilst for the detector $0< \nu <1$. 
\par A direct calculation shows that $f$ has exactly one stationary point, at
\begin{equation}
x_t=\frac{1}{2\kappa}\log{\left(\frac{1-\nu}{1+\nu}\right)}\,,
\end{equation}
and
\begin{equation}
f(x_t)=\frac{1}{2\kappa\nu}\log{\left(\frac{1-\nu}{1+\nu}\right)}+\frac{1}{\kappa}\log{\left(\frac{2\nu}{\sqrt{1-\nu^2}}\right)}-\frac{x_0}{\nu}\,.
\label{eq:Schw:2d:movMirr:fTurn}
\end{equation}
The asymptotic considerations above imply that $x_t$ is the global maximum of $f$. The no-collision condition is where $f(x_t)<0$, which can be rewritten as~\eqref{eq:Schw:2d:movMirr:x0bound}.
\chapter{Four-dimensional Schwarzschild transmission and reflection coefficients}
\label{ch:appendix:4DSchwTransRefl}
In this appendix, we use the constancy of the Wronskian to compute the transmission and reflection coefficients and derive interrelations between them.
\par The Wronskian is defined as
\be
W[f,g]:=f\frac{dg}{dr*}-g\frac{df}{dr*}\,. 
\ee
Considering the \emph{unnormalised} modes, which in this appendix we denote by $\Phup$ and $\Phin$, specified by the asymptotic behaviour~\eqref{eq:Schw:4d:asyIn} and~\eqref{eq:Schw:4d:asyUp}. If we define $\Rin:=\Phin/r$ and $\Rup:=\Phup/r$, these have asymptotics of the form
\begin{equation}
\Rin(r)\sim 
\begin{cases}
(2M)B^{\text{in}}_{\omega\ell}\expo^{-i\omega r^{*}}, & \,\, r*\to -\infty \,,  \\
\expo^{-i\omega r^{*}}+A^{\text{in}}_{\omega\ell}\expo^{+i\omega r^{*}},  & \,\, r*\to \infty \, ,
\label{eq:Schw:4d:RasyIn}
\end{cases}
\end{equation}
and 
\begin{equation}
\Rup(r)\sim 
\begin{cases}
(2M)A^{\text{up}}_{\omega\ell}\expo^{-i\omega r^{*}}+(2M)\expo^{+i\omega r^{*}}, & \,\, r*\to -\infty \,, \\
B^{\text{up}}_{\omega\ell}\expo^{+i\omega r^{*}},  & \,\, r*\to \infty \, .
\label{eq:Schw:4d:RasyUp}
\end{cases}
\end{equation}
If we evaluate the Wronskian first at $r*\to-\infty$ and then at $r*\to\infty$, then by the constancy of the Wronskian the results must be equal. It is easy to verify that
\be
W[R^{\text{in}}_{\omega\ell},R^{\text{up}*}_{\omega\ell}]\to 
\begin{cases}
&-2i\omega\Ain B^{\text{up}*}_{\omega\ell}\,,\quad\quad r*\to\infty\,,\\
&  2i\omega A^{\text{up}*}_{\omega\ell} \Bup\,, \quad\quad r*\to-\infty\,,
\label{ch:appendix:4DSchwTransRefl:Winupstar}
\end{cases}
\ee
\be
W[\Rin,\Rup]\to 
\begin{cases}
&2i\omega\Bup\,,\quad\quad r*\to\infty\,,\\
& (2M)^2 2i\omega\Bin\,, \quad\quad r*\to-\infty\,,
\label{ch:appendix:4DSchwTransRefl:Winup}
\end{cases}
\ee
and
\be
W[R^{\text{in}}_{\omega\ell},R^{\text{up}*}_{\omega\ell}]\to 
\begin{cases}
& - 2i\omega |\Bup|^2\,, \quad\quad r*\to\infty\,,\\
& 2i\omega(2M)^2\left(|\Aup|^2-1\right)\,, \quad\quad r*\to-\infty\,.
\label{ch:appendix:4DSchwTransRefl:Wupupstar}
\end{cases}
\ee
From the $r*\to\infty$ and $r*\to-\infty$ behaviour of~\eqref{ch:appendix:4DSchwTransRefl:Winupstar} and  the constancy of the Wronskian, we see that it must hold that
\be
|\Ain|^2=|\Aup|^2\,.
\label{ch:appendix:4DSchwTransRefl:AupVsAin}
\ee
Similarly, from~\eqref{ch:appendix:4DSchwTransRefl:Winup} we find the relation
\be 
\Bup=(2M)^2\Bin\,,
\label{ch:appendix:4DSchwTransRefl:BupVsBin}
\ee
and from~\eqref{ch:appendix:4DSchwTransRefl:Wupupstar} we have 
\bea
|\Aup|^2&=1-\frac{|\Bup|^2}{(2M)^2}\\
&=1-(2M)^2|\Bin|^2\,,
\label{ch:appendix:4DSchwTransRefl:AupVsB}
\eea
where the second equality follows from~\eqref{ch:appendix:4DSchwTransRefl:BupVsBin}.
\par From~\eqref{ch:appendix:4DSchwTransRefl:Winup} we can also write
\be
\Bup=\frac{ W[\Rin,\Rup]}{2i\omega}\,,
\ee
and thus combining this with~\eqref{ch:appendix:4DSchwTransRefl:Winupstar}, we obtain
\be 
\Aup=\frac{W[R^{\text{in}}_{\omega\ell},R^{\text{up}*}_{\omega\ell}]^{*}}{W[\Rin,\Rup]^{*}}\,
\ee
and
\be 
\Ain=\frac{W[R^{\text{in}}_{\omega\ell},R^{\text{up}*}_{\omega\ell}]}{W[\Rin,\Rup]^{*}}\,.
\ee
\par Alternatively, we can express the transmission and reflection coefficients in terms of the modes associated with the solutions of~\eqref{eq:Schw:4d:radModPhi} $\phin,\phup$, which we denote by $\rin,\rup$ and are related to $\Rup,\Rin$ by
\bea 
\Rup&=\Bup \rup\,,\\
\Rin&=\frac{\Bup}{2M}\rin\,.
\eea
To verify this relation, recall that $\Rup$ and $\Rin$ are associated with $\Phin$ and $\Phup$, which in this appendix denote the unnormalised modes. 
In terms of these modes (which in practice are the modes we work with in the Mathematica code) the transmission and reflection coefficients take the form
\bea 
\Bup&=\frac{(2M)2i\omega}{W[\rin,\rup]}\,,\\
\Aup&=-\frac{W[\rin,\rup^{*}]^{*}}{W[\rin,\rup]}\,,\\
\Ain&=-\frac{W[\rin,\rup^{*}]}{W[\rin,\rup]}\,.
\label{ch:appendix:4DSchwTransRefl:coeffsInCode}
\eea
Note that by virtue of~\eqref{ch:appendix:4DSchwTransRefl:BupVsBin}, once we know $\Bup$ we can get $\Bin$. These Wronskians could be computed at any radius on the trajectory, but in practice we compute them at the most inward point, closest to the black hole, on the trajectory. The reason for this is that `NDSolve' computes the mode at this end point explicitly rather than interpolating it, giving us increased accuracy.
\chapter[Four-dimensional smoothly-switched transition rate for stationary detectors]{Four-dimensional smoothly-switched transition rate for stationary detectors}
\chaptermark{$4d$ transition rate for stationary detectors}
\label{ch:appendix:4dStatTR}
In this appendix we show that in four dimensions, the transition rate valid for stationary situations~\eqref{eq:techIntro:transRate-stat}, obtained by simply dropping the infinite, external $\tau^{\prime}$-integral in the detector response function, is equivalent to the instantaneous transition rate found in~\cite{satz-louko:curved}, which was obtained by smoothly switching the detector on (off) and only at the very end of the calculation taking the sharp-switching limit. 
\par 
The transition rate~\eqref{eq:techIntro:transRate-stat} reads
\bea
\mathcal{\dot{F}}\left(E\right)&= \int^{\infty}_{-\infty}\,\mathrm{d}s\,\expo^{-iEs} W_{\epsilon}(s)\\
&=\int^{\infty}_{-\infty}\,\mathrm{d}s\,\expo^{-iEs}\left( W_{\epsilon}(s)+\frac{1}{4\pi^2(s-i\epsilon)^2}\right)-\frac{1}{4\pi^2}\int^{\infty}_{-\infty}\,\mathrm{d}s\,\frac{\expo^{-iEs}}{(s-i\epsilon)^2}\,,
\label{ch:appendix:4dStatTR:TR}
\eea
where the limit $\epsilon\to0_+$ outside the integrals is understood. 
The first term in~\eqref{ch:appendix:4dStatTR:TR} equals
\bea
&\int^{\infty}_{-\infty}\,\mathrm{d}s\,\expo^{-iEs}\left( W_0(s)+\frac{1}{4\pi^2 s^2}\right)\\
&=2\Realpart\int^{\infty}_{0}\,\mathrm{d}s\,\left[\expo^{-iEs}W_0(s)+\frac{\cos{(Es)}}{4\pi^2 s^2}\right]\\
&=2\Realpart \int^{\infty}_{0}\,\mathrm{d}s\,\left[\expo^{-iEs}W_0(s)+\frac{1}{4\pi^2 s^2}\right]+\frac{1}{2\pi^2}\int^{\infty}_{0}\mathrm{d}s\,\left(\frac{\cos{(Es)}-1}{s^2}\right)\,,
\label{ch:appendix:4dStatTR:term1}
\eea
where we have first taken $\epsilon\to 0$ by the Hadamard property of $W_{\epsilon}$ and then used $W_0(s)=\overline{W}_0(-s)$. The last term on the last line of~\eqref{ch:appendix:4dStatTR:TR} and the 
last term on the last line of~\eqref{ch:appendix:4dStatTR:term1} can be evaluated by contour integration, with the result that their sum equals $-E/4\pi$. Combining we have
\be
\mathcal{\dot{F}}\left(E\right)= -\frac{E}{4\pi}+ 2\Realpart\int^{\infty}_{0}\,\mathrm{d}s\,\left[\expo^{-iEs} W_{0}(s)+\frac{1}{4\pi^2 s^2}\right]\,,
\label{ch:appendix:4dStatTR:fin}
\ee
which, for the special case of a detector on a stationary trajectory and switched on in the asymptotic past, $\Delta\tau\to -\infty$, is exactly the smoothly-switched instantaneous transition rate after the sharp-switching limit has been taken found in~\cite{satz-louko:curved}.

\newpage
\addcontentsline{toc}{chapter}
         {\protect\numberline{Bibliography\hspace{-96pt}}}

\end{document}